\documentstyle[eqsecnum,aps,epsfig,psfig,epic,eepic,amsmath]{revtex}
\newcommand{\nc}{\newcommand} 
\nc{\ba}{\begin{array}}
\nc{\ea}{\end{array}} 
\nc{\bml}{\begin{mathletters}}
\nc{\eml}{\end{mathletters}} 
\nc{\nn}{\nonumber}
\nc{\pp}{{\prime\prime}} 
\nc{\lav}{\left<\!\left<}
\nc{\rav}{\right>\!\right>} 
\nc{\hf}{\case{1}{2}}
\nc{\ptl}{\partial} 
\nc{\eg}{, {\em e.~g.},~} 
\nc{\ie}{, {\em i.~e.},~} 
\nc{\rhs}{r.\ h.\ s.\ } 
\nc{\lhs}{l.\ h.\ s.\ }
\nc{\wrt}{w.\ r.\ t.\ } 
\nc{\rd}{{\mathrm d}}
\nc{\rD}{{\mathrm D}}
\nc{\re}{{\mathrm e}}
\nc{\rs}{{\mathrm s}}
\nc{\rc}{{\mathrm c}}
\nc{\sv}{\mbox{\boldmath$S$}}
\nc{\vv}{\mbox{\boldmath$v$}}
\nc{\wv}{\mbox{\boldmath$w$}}
\nc{\xv}{\mbox{\boldmath$x$}}
\nc{\xvs}{\mbox{\scriptsize\boldmath$x$}}
\nc{\yv}{\mbox{\boldmath$y$}}
\nc{\yvs}{\mbox{\scriptsize\boldmath$y$}}
\nc{\qv}{\mbox{\boldmath$q$}}
\nc{\qvs}{\mbox{\scriptsize\boldmath$q$}}
\nc{\mv}{\mbox{\boldmath$m$}}
\nc{\mvs}{\mbox{\scriptsize\boldmath$m$}}
\nc{\zv}{\mbox{\boldmath$z$}}
\nc{\nv}{\mbox{\boldmath$n$}}
\nc{\svi}{\mbox{\scriptsize \boldmath$S$}}
\nc{\jv}{\mbox{\boldmath$J$}} 
\nc{\jvi}{\mbox{\scriptsize \boldmath$J$}}
\nc{\lamb}{\mbox{\boldmath $\lambda$}}
\nc{\bsigma}{\mbox{\boldmath $\sigma$}} 
\nc{\bsigmas}{\mbox{\scriptsize\boldmath $\sigma$}} 
\begin{document}

\draft 

\title{~ \\ ~ \\ ~\\ ~ \\ Techniques of replica symmetry breaking and
  the storage problem of the McCulloch-Pitts neuron \\~ \\~}
\author{G. Gy\"orgyi} 
\address{Institute of Theoretical Physics, E\"otv\"os University \\
1518 Budapest, Pf.\ 32, Hungary, e-mail: gyorgyi@glu.elte.hu}
\date{\today} \maketitle \widetext \vspace{15pt}

\begin{abstract}
  In this article we review the framework for spontaneous replica
  symmetry breaking.  Subsequently that is applied to the example of
  the statistical mechanical description of the storage properties of
  a McCulloch-Pitts neuron\ie simple perceptron.  It is shown that in
  the neuron problem the general formula appears that is at the core
  of all problems admitting Parisi's replica symmetry breaking ansatz
  with a one-component order parameter.  The details of Parisi's
  method are reviewed extensively, with regard to the wide range of
  systems where the method may be applied.  Parisi's partial
  differential equation and related differential equations are
  discussed, and a Green function technique introduced for the
  calculation of replica averages, the key to determining the averages
  of physical quantities.  The Green functions of the Fokker-Planck
  equation due to Sompolinsky turns out to play the role of the
  statistical mechanical Green function in the graph rules for replica
  correlators.  The subsequently obtained graph rules involve only
  tree graphs, as appropriate for a mean-field-like model.  The lowest
  order Ward-Takahashi identity is recovered analytically and shown to
  lead to the Goldstone modes in continuous replica symmetry breaking
  phases.  The need for a replica symmetry breaking theory in the
  storage problem of the neuron has arisen due to the thermodynamical
  instability of formerly given solutions.  Variational forms for the
  neuron's free energy are derived in terms of the order parameter
  function $x(q)$, for different prior distribution of synapses.
  Analytically in the high temperature limit and numerically in
  generic cases various phases are identified, among them one similar
  to the Parisi phase in long range interaction spin glasses.
  Extensive quantities like the error per pattern change slightly with
  respect to the known unstable solutions, but there is a significant
  difference in the distribution of non-extensive quantities like the
  synaptic overlaps and the pattern storage stability parameter.  A
  simulation result is also reviewed and compared to the prediction of
  the theory.

\end{abstract} 
\vspace{15pt}

\pacs{07.05.Mh, 61.43.-j, 75.10Nr, 84.35.+i}
\pacs{Keywords: neural networks, pattern storage, spin glasses,
replica symmetry breaking} 

\eject
\tableofcontents

\eject
\section{Introduction and Overview$^\ast$}
\label{intro}

\subsection{Introduction}

In the past one-and-a-half decade, statistical physical methods
yielded a rich harvest in theoretical and practical results in the
exploration of artificial neural network models.  In contrast to more
traditional mathematical approaches, such as combinatorics,
statistical data analysis, graph theory, or mathematical learning
theory, the main emphasis in statistical physics lies on
interconnected model neurons, considered as a physical many-body
problem, in the limit of large number of variables.  The latter
property renders the problems similar to statistical mechanical
systems in the thermodynamical limit, that is, when the number of
particles is very large.  This does not necessarily mean large number
of units in a neural network, the thermodynamic limit applies also in
the case of a single neuron if the number of adjustable variables, the
analog of synaptic strengths of biological neurons, is sufficiently
large.  A much studied type of network is constructed from the
McCulloch-Pitts model neuron \cite{mcc43}, called also single-layer,
or simple, perceptron if it is operating alone as a single unit
\cite{ros62}.  In this paper we will examine the single model neuron's
ability to store, i.e., to memorize, patterns, crucial for the
understanding of networked systems.

The paper is strictly about the artificial model neuron, and does not
imply biological relevance.  However, the notions neuron and synapses,
the latter designating coupling strength parameters, are biologically
inspired, and will use them throughout.

We shall apply the statistical mechanical framework introduced by
Gardner and Derrida \cite{eg87,eg88,gd88} in 1988-89, which gave birth
to a subfield of the theory of neural networks.  Since then, the
McCulloch-Pitts neuron has become well understood below the storage
capacity, where patters, or, examples, can be perfectly stored.  The
region beyond it, however, remained the subject of continuous research
\cite{gg91,mez93,et93,bou94,der94,ws96,wes97}.  If the number of
patterns exceeds the capacity then there is no way of storing all of
them.  One possible approach beyond capacity is to choose a quantity
to be optimized.  Examples for such a quantity are the stability of
the patters -- in other words, their resistance to errors during
retrieval --, or, the number of correctly stored patterns irrespective
of their stability.  Such problems can be formulated by means of a
cost, or, energy function, giving rise to a statistical mechanical
system.  In the case of minimization of the number of incorrectly
stored patterns, difficulties have arisen on every front where the
problem was attacked.  On the one hand, the analytical method
inherited from spin glass research is no longer applicable in its
simplest form, that is, the so-called replica symmetric (RS) ansatz
breaks down.  On the other hand, near and beyond capacity numerical
algorithms begin to require excessive computational power.  The
physical picture behind that is the roughening of the landscape of the
cost function the algorithms try to minimize.

Phases of similar complexity, wherein the optimum-finding algorithm,
the analog of the dynamics in the statistical mechanical system, slows
down to the extent that can be considered as breakdown of ergodicity,
were observed in combinatorial optimization problems and still keep
eluding analysis \cite{gar79,sgrev87,mon98}.  Several empirically hard
optimization problems \cite{gar79}, including minimization of error
beyond storage capacity for the McCulloch-Pitts neuron \cite{pap94},
are known to belong to the so-called non-deterministic polynomial (NP)
complete class.  It is of significance, if by means of statistical
physical methods some properties of the energy, or, free energy,
landscape of NP-complete problems can be clarified.  The statistical
physical equivalent of a few NP-complete systems were shown, in
averaged thermal equilibrium, to exhibit spin-glass-like behavior
\cite{sgrev87}.  That gives rise to the belief that there may be a a
general connection between NP-completeness and spin glass behavior.
Thus the identification and description of such thermodynamic phases
may be instructive from the algorithmic viewpoint as well.  It should
be emphasized that NP-complete optimization problems are of diverse
origin and many of their quantitative properties show little
resemblance.  Accordingly, those reformulated as statistical
mechanical systems exhibit different thermodynamic behavior\eg in
averaged equilibrium have different phase diagrams.  Nevertheless, by
the notion of glassy phases statistical physics may provide us with a
common concept for understanding at least some ingredients of
NP{\-}-completeness.

It is the region beyond capacity of a single McCulloch-Pitts neuron
that we claim to uncover in the present paper, within the averaged
statistical mechanical description of thermal equilibrium.  While the
theoretical framework is in some respects different from, rather a
generalization of, the techniques applied to the Ising spin glass, we
can now reinforce the so far vague expectation about the appearance of
a spin glass phase and deliver quantitative results.  Networks beyond
saturation are long known to have complex features, here we
demonstrate that even a single neuron can exhibit extreme complexity.

The present article grew out of the work with P. Reimann, presented in
a letter \cite{our-paper}.  A more extended article, still in many
respects a summary of the main results, has been accepted for
publication \cite{our-paper-00}.  The emphasis in the present paper is
twofold.  On the one hand, we give a comprehensive review of the
technical details of the replica symmetry breaking theory, including
the so-called continuous replica symmetry breaking.  In the core is
Parisi's original theory, which is here technically generalized to
incorporate also the neuron problem.  Furthermore, several extensions
of the theory are introduced here that are applicable also to spin
glasses.  Along the mathematical parts an educative and self-contained
line of reasoning is favored over a terse style.  By that we would
like to fill a hiatus in the literature on the theory of disordered
systems in that we present the technical details to those wishing to
understand Parisi's method and possibly to use it to other problems.
On the other hand, we apply the theory to the storage problem of a
single neuron.  Since the first statistical mechanical approach to
this question, several other neural functions have been treated by
statistical mechanical methods, and some of those may be more
important for applications than pattern storage.  However, even
storage represents a strong theoretical challenge.  Beside the
Little-Hopfield model of auto-associative memory \cite{hkp91}, it can
be considered as a point of entry of the statistical mechanical
approach into hard problems in the field of artificial neural
networks, and may open the way for further applications.
 
On the technical side, the paper is centered about Parisi's method,
successful in solving the mean equilibrium properties of the infinite
range interaction Ising spin glass, the Sherrington-Kirkpatrick
model \cite{sgrev87}.  It turns out that after some generalization
\cite{our-paper,our-paper-00} of the original method
\cite{par80a,par80b,par80c,par80d}, this becomes adaptable to the
statistical mechanical formulation by Gardner and Derrida
\cite{eg88,gd88} of the neuron problem.  The single neuron with a
general cost function\ie error measure, was introduced by Griniasty
and Gutfreund \cite{gg91} and called by them potential.  We show that
this model will give rise to the most general term that admits
Parisi's solution with one order parameter function.  Under Parisi
solution we understand for now the hierarchical structure of the order
parameter matrix that gives rise to the nonlinear partial differential
equation introduced by Parisi  in an auxiliary role,
allowing continuous replica symmetry breaking.

We would like to point out that all systems studied by means of the
Parisi ansatz with one order parameter matrix, like the multi-spin
interaction Ising \cite{gar85} and the Potts glass near criticality
\cite{gks85}, contained as special cases the aforementioned general
term.  Therefore, our results about the Parisi solution of the neuron,
go well beyond the scope of neural computation.  Here we call the
reader's attention to the fact that Parisi's method has been applied
to the study of metastable states in the Sherrington-Kirkpatrick model
\cite{bm81}, where in fact three order parameter matrices emerged.
That work indicates how the continuous replica symmetry breaking
solution is to be obtained there and implicitly suggests the
generalization to vector order parameters as we outline in this paper.
 
Beyond giving a comprehensive account of Parisi's framework, we shall
perform a concrete field theoretical study, including the calculation
of averages, graph rules involving Green functions for the evaluation
of correlation functions, analytic derivation of a Ward-Takahashi
identity, and integral expressions for the generalized
susceptibilities necessary to determine thermodynamic stability of the
solution.  The insightful works about the second and higher order
correlations of the magnetization in the continuous replica symmetry
breaking phase of the Sherrington-Kirkpatrick model present concrete
examples for field expectation values \cite{dl83,mv85}.  These are
generalized by our formulation in this article, new even in the
context of spin glass problems.  With the notable exception of the
Sherrington-Kirkpatrick model and the formally analogous
Little-Hopfield system, where the low temperature phase was also
extensively described \cite{dl83,mv85,sd84,mpstv84,tok94}, most
studies of long range interaction disordered systems concerned the
region near criticality.  The framework we present here is naturally
designed for application deeply within the glassy phase.
 
The differences between the Sherrington-Kirkpatrick and neuron models
are obvious at first sight.  The former is an Ising-type system, with
a multiplicative two-spin interaction.  In contrast, our main focus
here is a spherical model-neuron\ie whose microstates are
characterized by the synaptic couplings, continuous and arbitrary up
to an overall normalization factor.  The interaction between synapses
is mediated by the error measure potential of Griniasty-Gutfreund
\cite{gg91}, a function arbitrary to a large extent.  In this light
one may find the close analogy between disordered spin systems and the
neuron model somewhat surprising.  The similarity becomes, however,
apparent when the statistical mechanical system is reduced to a
variational problem in terms of a single order parameter function.
Such have been available for the Sherrington-Kirkpatrick model,
whereas we have constructed one for the single neuron.  The
variational framework is brief, it allows a quick derivation of the
stationarity relations, gives account of thermodynamic stability in a
subspace called longitudinal, and is of help in numerical
computations.  The differences between the Sherrington-Kirkpatrick and
the neuron problems may be small in the variational free energy
formula, but are still the cause of technical complications for the
neuron problem.  The physical reasons are that, firstly, the neuron
does not possess the spin flip symmetry of the spin glass without
external field, secondly, the neuron's error measure potential is more
complicated than the multiplicative spin exchange energy term.  Thus a
few special properties of the Sherrington-Kirkpatrick model that
allowed for some analytic results and simplified numerics
\cite{dl83,sd84} are absent in the neuron.  Generically similar
complications may arise in other spin glass variants, so the much
studied Sherrington-Kirkpatrick model is to be considered as a rather
special, simple case.

It is worth mentioning briefly two important areas among the many we
do not treat in this paper.  First and foremost, we do not discuss
here the dynamical evolution of disordered systems.  Since the
ground-breaking early works on the dynamics of the
Sherrington-Kirkpatrick model by Sompolinsky and Zippelius
\cite{som81a,som81c,som82,somp84}, and the path-integral formulation
for Ising spins by Sommers \cite{som87}, many aspects of the dynamics
of disordered systems have been clarified.  They proved essential for
the understanding also of numerical algorithms.  However, one has to
reckon that even averaged, stable equilibrium properties of complex
phases of disordered systems are still far from clarified.  The
existence of many metastable states, the signature of glassy systems,
and the ensuing complex nature of dynamical evolution, often termed as
breakdown of ergodicity, puts in doubt even the existence of thermal
equilibrium.  On several model systems, however, extensive numerical
simulations have demonstrated that equilibrium properties, averaged
over the quenched disorder, can carry physical meaning.  These
properties are the subject of the present article.  Secondly, from the
viewpoint of mathematical rigor, the replica method raises many a
question that we leave unanswered.  In fact, quite a few scientists
view this method with suspicion, partly because the limit of ``zero
number of replicated systems'' may seem to violate physical intuition.
However, the large number of simulations confirming replica symmetric
solutions, and the fewer ones supporting replica symmetry breaking, as
well as the absence of numerical results outright disproving the
theory to this date, should provide ground for confidence.
Theoretical physics often employs methods of seemingly shaky
mathematical foundations, whose confirmation may come from comparison
with real or numerical experiments.  Such a confirmation may then
trigger rigorous clarification.  
 
\subsection{Overview}

Here we give a review of what subsequent sections are about.  Section
\ref{history} introduces some fundamental concepts and gives a brief
historical review on neural modeling and, to a very basic extent, on
the Sherrington-Kirkpatrick model of spin glasses.  In Section
\ref{sec-thermodyn} the single McCulloch-Pitts neuron is described as
a statistical mechanical system following Gardner, Derrida
\cite{eg88,gd88} and Griniasty, Gutfreund \cite{gg91}.  Pattern
storage is interpreted as an optimization problem in the space of
synaptic coupling strengths, and the ensuing thermodynamic picture is
outlined.  The replica free energy for various prior distributions of
synapses is derived, such as the spherical constraint as well as
arbitrary distribution of independent synapses.  Highlighted is the
central role of the neural local stability parameter, whose
distribution gives through a simple formula the average error.  Most
of this section recites known concepts, with a few new details.
Sections \ref{sec-PA-Gen}, \ref{sec-corr-stab}, and
\ref{sec-special-properties} are devoted to the Parisi solution.  We
start out from the ``hard'' term in the replica free energy of the
neuron, that can be considered as a generalization of free energy
terms emerging from the classic long range interaction, disordered,
spin problems.  In \ref{sec-PA-Gen} a comprehensive presentation of
the Parisi solution is given, including the derivation of Parisi's
partial differential equation.  It is demonstrated that this equation
incorporates all finite replica symmetry breaking ans\"atze, besides
continuous replica symmetry breaking. Parisi's partial differential
equation gives rise to a collection of related partial differential
equations, they are reviewed here, and several useful Green functions
are presented, among them prominently the Green function for Parisi's
partial differential equation.  Section \ref{sec-corr-stab} contains
new results such as analytic expressions for expectation values and
correlation functions of replica variables.  The eigenvalues of the
Hessian of the replica free energy are discussed, determining
thermodynamic stability.  The Green function of Parisi's partial
differential equation turns out to be the field theoretical Green
function that correlators are composed of, and allows the introduction
of a graph technique.  Section \ref{sec-special-properties} discusses
a few aspects of the Parisi solution and two particular cases where
Parisi's partial differential equation can be explicitly solved.  We
return to the special problem of the model-neuron in Sections
\ref{ssec-spher-neur} and \ref{ssec-indep-synaps}, and apply the
rather abstract results of the preceding sections to it.  The case of
continuous synapses with the spherical constraint, including the
conditions of stationarity and thermodynamic stability, is analyzed in
detail in Section \ref{ssec-spher-neur}.  In the limit of high
temperature and large number of patterns the formalism becomes easily
manageable, while exhibiting a nontrivial phase diagram with three
different glassy states.  This section contains our variational
approach, the main result being a variational free energy whence
thermodynamic properties can be straightforwardly derived and
numerically explored.  By means of the various partial differential
equations several relations about the stationary state are uncovered.
The scaling required, when the temperature goes to zero, is also
described.  The variational free energy is numerically evaluated for
several characteristic parameter settings, together with the order
parameter function and the probability density of local stabilities.
Previous simulation data \cite{wen95b} were improved upon in Ref.\ 
\cite{our-paper-00}, whence we redisplay the comparison of simulation
results with the theoretical prediction.  The case of arbitrarily
distributed independent synapses is considered and the corresponding
variational framework presented in Section \ref{ssec-indep-synaps}.
Often used abbreviations are listed in Appendix \ref{app-sec-abbrev}.
Further appendixes contain more technical details.  Appendix
\ref{app-repl-fe} gives the derivation of the replica free energy for
synapses with spherical as well as with independent but otherwise
arbitrary normalization.  Appendix \ref{app-sec-deriv-r-rsb} bridges a
gap in the calculation of Section \ref{sec-PA-Gen}.  In Appendix
\ref{app-sec-duplantier} the short way of deriving Parisi's partial
differential equation is given, which requires the continuity of the
order parameter function.  Note that the this equation is valid even
in the case of discontinuities, but then the derivation, as shown in
Section \ref{sssec-ppde}, is more involved.  We do not pursue in the
paper the case of vector order parameters, but give a brief account of
how Parisi's partial differential equation for a vector field emerges
in Appendix \ref{app-high-dim}.  A technically useful identity between
Green functions is derived in \ref{app-sec-GF-ident}, and the high
temperature limit of some relevant partial differential equations are
presented in \ref{app-sec-highT-pde}.  The only case where we can show
longitudinal stability far from criticality is analyzed in Appendix
\ref{app-stab-high-T}.

As also stated in the Acknowledgment, sections with special
contributions by P. Reimann are marked by a $^\ast$.

\section{Artificial Neural Networks and Spin Glasses$^\ast$}
\label{history}

The purpose of this section is to put the often technical analysis of
later parts of the paper in the wider context of neural networks and
spin glasses.  The central issues of this work are the intricate
details of Parisi's continuous replica symmetry breaking (CRSB)
scheme, furthermore, the adaptation of the method to the equilibrium
storage properties of a McCulloch-Pitts neuron, or, simple perceptron.
We have made an attempt to cover the most relevant literature on these
two narrower themes.  On the other hand, we also mention other
subjects like learning algorithms, generalization and unsupervised
learning, layered perceptrons, and spin glass models, where our
selection of references is far from complete, and not necessarily even
representative. 
 
\subsection{The McCulloch-Pitts neuron and perceptrons}
\label{intro-mcp}
 
The model of a neuron as put forth by McCulloch and Pitts in a
ground-breaking paper \cite{mcc43} in 1943 has attracted since
continuous interest \cite{hkp91}.  While inspired by real neurons in
the brain, it is oversimplified from the biological viewpoint.  The
model neuron can assume two states, one ``firing'', the other
``quiescent''.  The state depends on input signals, obtained possibly
from other such units, and on the coupling parameters that weight the
inputs.  The couplings are often termed ``synaptic'' in reference to
the synapses, the connection points of biological neurons.
Mathematically speaking, the model neuron computes the projection of
an $N$-dimensional input ${\sv}$ along a vector ${\jv}$ of synaptic
couplings and outputs $\xi =1$ (say it fires) or $\xi= -1$ (it is
quiescent) according to the sign of this product ${\jv} \cdot {\sv}$
as
\begin{equation}
\xi = \mbox{sign}\left(\sum_{k=1}^NJ_k\,S_k\right) \ .
\label{1}
\end{equation}
The argument of the sign can be extended by a constant threshold,
which alternatively may be represented by $J=1$ if one only allows
$S_1=1$ as input.  Remarkably, as already McCulloch and Pitts noticed,
a sufficiently large collection of such model neurons, when properly
connected and the couplings properly set, can represent an arbitrary
Boolean function.  The model can be naturally extended to continuous
outputs, when the sign function is replaced by a continuous transfer
function, generally of sigmoid shape \cite{hkp91}.

The next major step forward was achieved with the introduction of the
perceptron concept by Rosenblatt \cite{ros62}.  The idea is to place a
number of McCulloch-Pitts neurons into different layers, with the
output of neurons in one layer serving as input for those in the next
layer, hence its name multi-layer feedforward network.  As it was
intended to model vision, such a network is also called multi-layer
perceptron.  The input to the network as a whole goes into the first
layer, while the final output is that of the last layer.  A widely
applied learning concept is to try to determine appropriate synaptic
couplings ${\jv}$ for all the neurons so as to satisfy a prescribed
set of input-output data, called training examples.  In other words,
the aim is to store the training examples.  One of the motivations for
doing so is that a possibly existing systematics behind the training
examples may be approximately reproduced also on previously unseen
inputs, that is, the network will be able to generalize.  The special
case of a single McCulloch-Pitts unit is a single-layer perceptron,
called also simple perceptron by Rosenblatt, and lately sometimes just
perceptron.  For the simple perceptron with binary outputs, as defined
in Eq.\ (\ref{1}), he proposed an explicit learning algorithm that
provably converges towards a vector of synaptic couplings ${\jv}$,
which correctly stores the training examples, provided such a ${\bf
  J}$ exists.  Simultaneously with Widrow and Hoff \cite{wid60,wid62},
he also studied two-layer perceptrons with an adaptive second layer,
while using the first layer as preprocessor with fixed (non-adaptive)
synaptic couplings, however, without being able to generalize his
learning algorithm to this case.
 
The field was driven into a crisis by the observation of Minsky and
Papert \cite{min69} that the simple perceptron (\ref{1}) is unable to
realize certain elementary logical tasks.  Confidence returned when
the so-called error back-propagation learning algorithm began to gain
wide acceptance (see \cite{rum86a} and further references in
\cite{hkp91}).  This algorithm performs training by examples of fully
adaptive multi-layer feedforward networks with generically
differentiable transfer function.  Such networks, if chosen
sufficiently large, are known to be capable to realize arbitrary
smooth input-output relations see\eg \cite{hor89a,har90}.  Though this
algorithm and its various descendants converge often quite slowly and
in principle one cannot exclude that they get stuck before reaching a
desired state, they have been successful in a great variety of
practical applications \cite{hkp91}.

\subsection{Associative memory}
\label{intro-assoc-memory}
 
Besides the layered feedforward perceptron architectures, a second
eminent problem in neural computation is the so-called associative
memory network or attractor network.  We limit our discussion to the
auto-associative case\ie the memory network is addressable by its own
content.  The concept can be traced back to Refs.\ \cite{tay56,ste61}
and rediscovered later (see \cite{hkp91} for further references).  The
recurrent (in contrast to feedforward) network of McCulloch-Pitts
model neurons, originally suggested by \cite{tem54,tem55,car61} was
especially suited for the task.  A recurrent network contains
interconnected units where signals pass through directed links that
can form loops.  Here the desired patterns to be stored correspond to
collective states of the units in the network, and the idea is to
define a discrete-time dynamics of the states so that the prescribed
patters (examples) are fixed point attractors.  For a collection of
$N$ model neurons (\ref{1}), the outputs $\xi_k$, $k=1,...,N$, at a
given time step $t$ are taken as new inputs $S_k(t)$ for the next time
step.  Denoting by $J_{ik}$ the synaptic coupling by which the $i$-th
neuron weights the signal $S_k(t)$ stemming from the $k$-th neuron, we
can write the discrete-time dynamics of neurons with binary output as
\begin{equation} 
S_i(t+1) = \mbox{sign}\left(\sum_{k=1}^N J_{ik}\, S_k(t)\right)\ , 
\label{2}  
\end{equation}
where self-interactions are usually excluded by setting $J_{ii}=0$.
Taking an input pattern ${\sv}={\sv}(0)$ as initial condition, we
understand in the definition (\ref{2}) that the update is done
sequentially, either by scanning through the $S_i$-s one after the
other, $i=1,2,\dots,N,1,2,\dots$, or by randomly selecting the sites
$i$ one after the other.  Such a dynamics is supposed to evolve
towards the closest attractor (hence the name attractor network).  If
this attractor is a fixed point, that it is if the couplings are
symmetric as $J_{ik}=J_{ki}$, for all $i, k$, a previously unseen
pattern ${\sv}$ can be associated with one of the stored examples,
assumed to be the ``most similar'' of all stored patterns, hence the
name associative memory network.  Note that for such an associative
memory network patterns ${\sv}$ have binary components $S_k=\pm 1$.
In case of units with continuous states one only requires that the
length $|{\sv}|$ goes like $N^{1/2}$ for $N\to\infty$.  We mention
that if the synaptic couplings are non-symmetric, convergence to a
fixed point is no longer certain and chaos can arise
\cite{scs88,kan96}.
  
Given the patterns to be stored, the aim is to construct a dynamics
with prescribed attractors.  This is the reverse of and possibly more
difficult than the more conventional problem of finding the attractors
for a given dynamical system.  If we accept the neural dynamics like
in (\ref{2}), the task is then to set the $J_{ik}$ couplings to such
values that lead to the desired attractors.  In his pioneering works
\cite{lit74,lit75} Little suggested an approach to this problem by
giving an explicit form for the synaptic couplings $J_{ik}$ of the
McCulloch-Pitts neurons as inspired by the ideas of Hebb \cite{heb49}
about the working of brain cells.  Little defined a parallel update
rule for (\ref{2}) and included a stochastic element characterized by
temperature.  Hopfield's milestone contribution \cite{hop82,hop84}
consisted in reformulating the dynamics (\ref{2}) as a sequential
update algorithm, which led to an optimization problem with an energy
function.  We will call the network with the dynamics (\ref{2})
associative memory, while in the special case, when the synaptic
couplings $J_{ik}$ are chosen according to the Hebb rule, the name
Little-Hopfield model will be used.  For a neuro-physiological
argument for a non-Hebbian learning rule we refer to \cite{hee99}.
 
The associative memory network (\ref{2}) may be appealing because it
models, albeit very crudely in details, a biological concept, its use
for practical purposes is, however, in doubt \cite{eg88}.  Indeed, the
required storage space for the synaptic couplings is comparable to
that for directly storing the patters, and the computational effort of
the retrieval dynamics (\ref{2}) is similar to a direct comparison of
a given input pattern with all the stored patterns.  Only with
appropriate modifications of the original setup\eg non-uniformly
distributed patters, may a digital implementation of the network 
become advantageous \cite{eg88}.  For various such modifications and
their possible practical use we refer to \cite{hkp91}.
 
\subsection{Sherrington-Kirkpatrick model}  
\label{intro-sk-model} 
 
Spin glasses are normal metals ({\em e. g.} Cu or Au) with dilute
magnetic impurities ({\em e. g.} Mn or Fe), or, lattices of random
mixtures of magnetic ions ({\em e.~g.} Eu$_x$Sr$_{1-x}$O) exhibiting a
freezing transition of the spin disorder at low temperatures
\cite{bin86}.  Due to spatial disorder, the spin interactions can be
considered as random.  The random sign of the interactions can be the
cause of one of the basic features of spin glasses, the effect of
frustration \cite{tou77}, when the interaction energies of all spin
pairs cannot be minimized simultaneously.  In a pioneering paper,
Edwards and Anderson \cite{ea75} introduced a simplified model of a
spin glass, essentially an Ising system with randomly selected, but
fixed, exchange couplings.  The infinite-range interaction version of
that is called Sherrington-Kirkpatrick (SK) model \cite{she75,kir78}
and is considered a realization of the mean field approximation.  The
theoretical analysis of the SK-model triggered the invention of novel
statistical mechanical concepts and methods which subsequently found
applications in modified spin glass models such as the random energy
\cite{der81} and $p$-spin interaction models \cite{gm84,gar85,cri92},
the Heisenberg \cite{gt81} and the Potts glass
\cite{erz83,eld83,gks85}, multi-$p$-spin and quantum spin glass models
\cite{nie95,nie95b,opp98}.  Methods inherited from spin glass theory
also provided insight into many other problems, several of them
originating from outside of physics.  Prominent examples are various
models of interfaces in random environment
\cite{vil83,kad87,mez91,eng93,dou95}, granular media \cite{are99},
combinatorial optimization (see \cite{sgrev87} for an early review and
\cite{mon98} for a new development), game theory \cite{be98}, protein
and nucleic acid folding \cite{gar88,sha89,sas92,tak97,pag00}, and
noise reduction in signal processing \cite{mou99}.  Last but not
least, as we will expound it in the present paper, methods first
introduced for describing the equilibrium properties of the SK model
are of paramount importance in the statistical mechanical approach to
neural networks.  We give, therefore, a brief account of the SK model,
concentrating on basic properties in thermal equilibrium.  The general
mathematical framework described in the main part of this paper covers
the SK model as a special case.  For pedagogical introductions into
the calculation techniques we also refer to \cite{eng93}, to Section
10 in \cite{hkp91} and Section 3 in \cite{sgrev91}.  A detailed
discussion of the physical content of the solution can be found\eg in
\cite{kir78,bin86,sgrev87,sgrev91}.  We only mention here that the
question, whether the solution of the SK model provides a
qualitatively appropriate description of short range interaction spin
glasses, is still debated.  See Refs.\ \cite{mar98b,bok99a,mar99a} and
\cite{moo98,mar99b,bok99b} for two exchanges on the subject, and
\cite{mar98} for a review and simulation results.
  
The state variables of the SK model are the Ising spins $S_i=\pm 1$,
interacting via random coupling strengths $J_{ik}$ ($i,k=1,2,...,N$).
In the absence of external magnetic field, the spin Hamiltonian
 is of the form 
\begin{equation}
{\cal H}_{{\sf J}}({\sv})=-\frac{1}{2}\sum_{i\not=
  k}J_{ik} S_i S_k 
\label{sk-ham} 
\end{equation} 
and the couplings $J_{ik}$ are independently sampled from an unbiased
Gaussian distribution with variance $1/N$.  The scaling by $N$
guarantees the extensivity of the energy in the thermodynamic limit
$N\to\infty$.  The feature that the interactions $J_{ik}$ are randomly
chosen but then frozen while the spins obey Boltzmannian
thermodynamics is summarized by our calling the $J$-s quenched
variables.  An important goal is then to calculate, in the large $N$
limit, the free energy per spin
\begin{equation}
 f_{SK} = -\lim_{N\to\infty} (N\beta)^{-1}\ln Z_{{\sf J}}. 
\label{sk1} 
\end{equation} 
Here $\beta=1/(k_B T)$ is the inverse thermal energy unit and $Z_{{\sf
    J}}$ is the partition sum $\sum_{{\svi}} \, e^{-\beta {\cal
    H}_{{\sf J}}({\svi})}$ over all spin configurations ${\sv}$. The
sum over the discrete spin states $\sum_{{\svi}}$ is often denoted by
a trace as $\text{Tr}_{{\svi}}$.  The interactions $J_{ik}$ being
quenched random variables, the expression (\ref{sk1}) as it stands is
analytically intractable.  Physically, one expects that two different
realizations of the random interactions $J_{ik}$ will exhibit the same
behavior at thermal equilibrium for $N\to\infty$.  Mathematically,
this means self-averaging of the free energy density $f_{SK}$, i.e.,
for any randomly sampled set of the $J_{ik}$, Eq.(\ref{sk1}) yields
the same result with probability $1$, allowing us to rewrite $\ln
Z_{{\sf J}}$ as an average $\langle \ln Z_{{\sf J}} \rangle_{{\rm
    qu}}$ over the quenched disorder ${\sf J}$.  Rigorous mathematical
discussions of this property for the SK and the related
Little-Hopfield model can be found in Refs.  \cite{pas91,fen95,bov95}.
 
The direct evaluation of $\langle \ln Z_{{\sf J}} \rangle_{{\rm qu}}$
is difficult, but it can be considerably simplified by means of the
replica method.  This was independently discovered several times (see
discussions in Refs.\ \cite{kir78,hem79}) but well known only since
its application to the spin glass problem by Edwards and Anderson
\cite{ea75}.  The first step of this method consists in what has
become known as the ``replica trick'',
\begin{equation}
\lim_{n\to 0}\frac{x^n -1}{n}=\ln x.
\label{ri} 
\end{equation} 
Thus Eq.(\ref{sk1}) can be rewritten as
\begin{equation} 
f_{SK} = \lim_{N\to\infty}\lim_{n\to 0}
\frac{1- \langle  Z^n_{{\sf J}} \rangle_{{\rm qu}}}{\beta\, N\, n}.
\label{sk2} 
\end{equation} 
The name replica refers to the fact that the $n$-th power of $Z_{{\sf
J}}$ is the partition function of $n$ non-interacting, identical
replicas of the original system.  The average $ \langle \dots
\rangle_{\rm qu}$ will create interactions between the replicated
systems.  As second step we interchange the two limits in (\ref{sk2}),
which has been proved valid for the SK model by van Hemmen and Palmer
\cite{hem79}.  A further step consists in the assumption that it is
sufficient to evaluate $\langle Z^n_{{\sf J}} \rangle_{{\rm qu}}$ for
integer $n$ and then interpret $n$ as real variable (``continuation'')
in order to evaluate the limit $n\to 0$.  In doing so, the point is
that the averaged partition sum $\langle Z^n_{{\sf J}} \rangle_{{\rm
qu}}$ with integer $n$-s can be technically tackled, while with
non-integer $n$ it is as intractable as the $\langle \ln Z_{{\sf J}}
\rangle_{{\rm qu}}$ of Eq.\ (\ref{sk1}).  The fourth step is the
evaluation of $\langle Z^n_{{\sf J}} \rangle_{{\rm qu}}$ by means of
a saddle point approximation, becoming exact as $N\to\infty$.  The
detailed calculations along this program are given in
\cite{she75,kir78} with the result
\begin{eqnarray}
& & f_{SK} = \lim_{n\to 0}\frac{1}{n} \min_{{\sf Q}} f_{SK}({\sf
Q})\label{sk3}\\ & & f_{SK}({\sf Q}) = -\frac{n\beta}{4} +
\frac{\beta}{4}\sum_{a\not = b} q^2_{ab} - \beta^{-1}\, \ln Z_{\beta
{\sf Q}} \label{sk4}.
\end{eqnarray}
Here the minimization -- stemming from the saddle point approximation
-- runs over all symmetric, $n\times n$ matrices ${\sf Q}$ with
elements $q_{aa}=1$ and $-1\leq q_{ab}\leq 1$ ($a,b=1,2,...,n$ being
the replica indices).  Furthermore, $Z_{\beta {\sf Q}}$ is formally
identical to $Z_{{\sf J}}$ if one sets $N=n$ and ${\sf J}=\beta {\sf
  Q}$, a specialty of the SK model.  The function $f_{SK}({\sf Q})$ is
often referred to as replica free energy.
 
The practical meaning of (\ref{sk3}) can be understood as follows.  A
direct analytical evaluation of the minimum in (\ref{sk3}) for
arbitrary integer $n$ is typically not feasible.  Therefore, one
introduces an ansatz for ${\sf Q}$ with a set of variational
parameters $\lamb$ that lead to formulas explicitly containing $n$,
and so continuation of formulas containing the elements of ${\sf Q}$
to real $n$-values becomes feasible.  Then (\ref{sk3}) is to be
understood as first a minimum condition for general $n$ by
differentiating the replica free energy $f_{SK}({\sf Q})$ with respect
to the matrix elements $q_{ab}$, the so-called stationarity condition,
and the requirement of at least the absence of negative eigenvalues of
the second derivative matrix, the Hessian, of $f_{SK}({\sf Q})$\ie the
condition of local thermodynamic stability.  (Here we disregarded the
border case when the minimum does not satisfy stationarity, and the
situation when there may be several locally stable states.
Interestingly, in the SK model these cases do not occur, but they do
in other systems.)  These relations cannot be continued to $n=0$
without further parametrization.  But insertion of the ansatz with the
variational parameters $\lamb$ allows for the limit $n\to 0$.  In this
light (\ref{sk3}) does not prescribe a customary minimization, rather
defines the minimum condition consisting of the aforementioned
stationarity and stability relations, which await parametrization. 
 
On the other hand, we can reverse the order of parametrizing and
minimum search.  The parametrization should allow us to construct
$f_{SK}({\sf Q}(\lamb ))$ for any $n$.  The minimization condition
for integer $n$ with respect to the variational parameters $\lamb$
implies, in the generic case, the vanishing of the derivatives, and is
supposed to admit continuation to real $n$-values.  Closer inspection
shows \cite{sgrev87,kir78} that after such a continuation, in the
limit $n\to 0$, the condition of local stability described above will
no longer correspond to a local minimum of $f_{SK}({\sf Q}(\lamb ))$
but rather to a local maximum.  
 
This can be crudely understood when one realizes that the second term
on the \rhs of (\ref{sk4}) contains $\binom{n}{2}$ independent terms,
equal the number of order parameters.  The $\binom{n}{2}$ changes sign
when $n$ passes from $n>1$ to $n<1$, so for $n<1$ one has formally a
negative number of order parameters $q_{ab}$.  This does not cause,
however, confusion, because due to the parametrization of the matrix
$\sf Q$ we do not need to work with the elements $q_{ab}$ for $n<1$.
A similar sign change of terms obtained by expanding the third term in
(\ref{sk4}) changes the nature of the extremum of the free energy from
minimum to maximum.  
 
The above reasoning thus leads, within a given parametrization, to
\begin{equation}
f_{SK} = \max_{\lamb } \lim_{n\to 0} 
\frac{1}{n} f_{SK}({\sf Q}(\lamb )).
\label{sk5}
\end{equation}
This formula prescribes global maximization in $\lamb$.  So if several
local maxima are found, the $f_{SK}$ values there should be compared
and the global maximum within the given parametrization is thus well
defined.  However, we are in principle still not allowed to bypass the
aforementioned local stability analysis, because a global maximum as
in (\ref{sk5}), within a given parametrization, may still be unstable
with respect to changes in the $q_{ab}$ matrix elements. Thus one
should evaluate the spectrum of the Hessian matrix of $f_{SK}({\sf
  Q})$ and require that no negative eigenvalues exist in the limit
$n\to 0$.  This leads to the at first sight contradictory
prescriptions, namely, the minimization in (\ref{sk3}), formulated as
the absence of negative eigenvalues of the Hessian of $f_{SK}({\sf
  Q})$, and the maximization of the parametrized free energy in
(\ref{sk5}).  Closer inspection shows, however, that there is no
logical contradiction.  Indeed, maximization in the restricted space
of the variational parameters requires generically the negative
semidefiniteness of another Hessian, the one for $f_{SK}({\sf Q}(\lamb
))|_{n=0}$.  In special cases one can show that some eigenvalues of
the Hessian of $f_{SK}({\sf Q})$ correspond to the eigenvalues that of
$f_{SK}({\sf Q}(\lamb ))|_{n=0}$, such that non-negativity for the
former ones implies non-positivity for the latter ones
\cite{kir78,at78,tam80,dom83b}.  Following the reasoning in Section 3.3
of Ref.\ \cite{sgrev91} this can be intuitively understood in the way
that the infinitezimal increment around an extremum of $f_{SK}({\sf
  Q})$ is the sum of contributions negative in number for $n<1$,
responsible for the reversal of the type of extremum.  For a more
recent discussion of the problem of maximization in a descendant of
the SK model see Ref.\ \cite{nie95c}.
 
Aiming at an exact solution of the original minimization problem
(\ref{sk3}), one should choose a variational ansatz so that it
includes the global solution.  In principle, a parametrization should
be adopted so that it gives a maximal $f_{SK}$ value over all possible
parametrizations.  Verification of the global nature of a maximum
found within a given parametrization is a hard problem, physical
intuition for the right parametrization and comparison with reliable
simulation data, if such exist, may be of guidance. 
  
Considering that the replicated partition sum in (\ref{sk2}) is
symmetric under permutation of the replicas, a first guess is that
also the minimizing ${\sf Q}$-matrix in (\ref{sk3}) -- characterizing
the state of the system at equilibrium -- exhibits this symmetry.
This leads us to the replica symmetric (RS) ansatz with a single
variational parameter $\lambda = q =q_{ab}\in[-1,1]$ for all $a\not=
b$, named Edwards-Anderson order parameter.  The explicit evaluation
of (\ref{sk5}) with such an ansatz and clarification of the physical
content of the resulting RS solution has been performed in Ref.\ 
\cite{she75,kir78}.  (For the sake of brevity we do not discuss the
inclusion of external magnetic field and that of a nonzero average of
the couplings $J_{ij}$, some main concepts can be presented without
them.)  The local stability conditions for the RS solution have been
worked out by de Almeida and Thouless (AT) in \cite{at78}.  It turns
out that the AT stability condition is fulfilled only for temperatures
beyond a critical $k_B\,T_c=1$, below that the RS solution is
AT-unstable, implying that the replica symmetry of the system in
(\ref{sk2}) must be spontaneously broken by the equilibrium state of
the system.  Intriguingly, this instability does not announce itself
at any integer $n$, it only appears as $n$ decreases from $1$ towards
$0$ \cite{at78,hem79,kon83}.  Further evidences about the the fact
that the RS solution is incorrect are the negative ground state
entropy \cite{she75} and magnetic susceptibility \cite{pyt79}, and its
predictions for the ground state energy and the probability density of
the local magnetic field that contradict simulations, see
\cite{sgrev87}.
 
In order to find a consistent description of the SK model at low
temperatures, several replica symmetry breaking (RSB) parametrization
for the ${\sf Q}$ matrix in (\ref{sk3}) have been proposed
\cite{bla78,bra78,bra79a,bla80,dom81,dom83a}.  In what can be viewed
as the generalization of Blandin's one-step RSB ($1$-RSB)
\cite{bla78}, Parisi formulated on physical grounds a hierarchical
structure for the ${\sf Q}$ matrix (see also Sect. III.3 in
\cite{sgrev87}) and introduced the so far only RSB ansatz compatible
with these conditions in an ingenious series of works
\cite{par79a,par79b,par80a,par80b,par80c,par80d,p83}.  Depending on
the number $2R+1$ of variational parameters $\lamb$ in this ansatz,
one speaks of an $R$-step, $R=0,1,2,\dots$, RSB ansatz ($R$-RSB), and
of continuous RSB (CRSB) in the limit $R\to\infty$.  The RS ansatz
corresponds to $R=0$, and each higher step contains the previous ones
as special cases.  Later in the paper the explicit form of Parisi's
ansatz for ${\sf Q}$ will be given and its consequences thoroughly
discussed.  Following Parisi's study, mostly focusing on the region
near criticality, the deep spin glass phase was also extensively
analyzed within the CRSB ansatz, see Refs.\ in \cite{sgrev87,sgrev91}.
We highlight among the non-perturbative approaches the work of Sommers
and Dupond \cite{sd84}, where a variational free energy especially
suited for numerical evaluation was constructed and used to resolve
ground state properties.  One of their successes was a theoretical
prediction for the probability density of the local field, that
favorably compared to the simulation of Palmer and Pond (see Fig.\ 
III.6 of \cite{sgrev87}).  The generalization of the AT stability
conditions for the case of an $R$-RSB solution has been developed in a
series of works by De Dominicis, Kondor, and Temesv\'ari, initiated
with Ref.\ \cite{dom83b} and presented in the most general form in
Ref.\ \cite{tdk94}.  Due to the complicated form of these stability
conditions, they could so far be verified for Parisi's solution only
slightly below the AT instability.  Yet it is widely believed that
Parisi's solution captures the correct behavior of the SK model in the
entire low temperature regime. 
 
The global stability of Parisi's RSB ansatz has not been verified by
rigorous mathematics.  It is physically supported in part by the
suggestive picture of hierarchical organization of states in the
glassy phase.  Furthermore, it shows none of the aforesaid
inconsistencies the RS solution was plagued by, and it compared
satisfactorily with simulations.  In fact, we do not know of any
instance, where the replica method with Parisi's ansatz has been
applied and at the same time well founded analytical or numerical
approaches are available and would yield incompatible results.
Neither is a case known to us which admits application of the replica
method but cannot be handled in a self-consistent way by Parisi's
ansatz with sufficiently many, possibly infinitely many, steps of RSB.

As an alternative to the replica method, Thouless, Anderson, and
Palmer \cite{tho77} have established a modified form of the
Bethe-Peierls method reproducing the RS results at high temperatures,
while differing from both the RS and Parisi's solution in the
AT-unstable region.  This approach has been further developed by
Sommers \cite{som78,som79} in a way that was later realized
\cite{dom81,bra80} to be equivalent, in a certain limit, to a
generalized version of the RSB ansatz by Blandin and coworkers
\cite{bla78,bla80}.  A second alternative method is the dynamical
approach of Sompolinsky and Zippelius \cite{som81a,som82},
capturing Parisi's solution in the static case \cite{som81c,dom82}.
The latter may in turn be reproduced by an iterative extension of the
Blandin-Sommers scheme \cite{dom81}, the first step towards the
correct Parisi solution.  A further modified form of the Bethe-Peierls
approach -- the so called cavity approach -- by M\'ezard, Parisi, and
Virasoro \cite{sgrev87,mez86} contains the Thouless, Anderson, and
Palmer equations as a special case but can also be extended to become
equivalent to a Parisi-ansatz with an arbitrary number of RSB steps.
Again, this is not a mathematically rigorous method but rather an
ansatz in combination with an intuitive physical line of reasoning,
verified by self-consistency in the end.  While the physical picture
is less elusive than that behind the formal $n\to 0$ limit, the
equivalent replica method in conjunction with Parisi's ansatz seems to
be in a higher developed status as far as applicability for practical
calculations is concerned.  For instance, the self-consistency
condition of the cavity approach, expected to be equivalent to the
thermodynamic stability conditions of the replica method
\cite{sgrev87}, has so far been explicitly worked out only in the
simplest case, corresponding to the AT stability condition for the RS
state.  Another formulation of the dynamics was given by Sommers
\cite{som87}, who devised a path-integral approach specially suited
for discrete variables like Ising spins.  His results are in
accordance with those of Sompolinsky and Zippelius, who used a
continuous spin model that, in a singular limit, also covered the case
of Ising spins.  A recently suggested alternative method \cite{db98}
studies the $n$-dependence of $\left< Z^n \right>$, reiterates that
different continuations to $n\to 0$ give the RS and RSB solutions,
without the need of explicitly inserting Parisi's ansatz. However, the
heuristics involved may cause that the exact solution is obtained only
in special cases.

There is a large family of spin glass models, consisting of various
generalizations of the SK model, that have also been successfully
treated by Parisi's ansatz, albeit mostly near criticality in a
perturbative manner \cite{sgrev91}.  A prominent exception is
Nieuwenhuizen's multi-$p$-spin interaction model with continuous,
spherical, spins \cite{nie95}.  The fixed $p=2$ case is the long known
spherical SK model, which can solved within RS \cite{sgrev91}, with
multi-$p$-spin interactions, however, it can exhibit RSB.  Remarkably,
in CRSB phases the continuously increasing part of Parisi's order
parameter function can be analytically calculated for any
temperatures.  Even with a fixed $p>2$, one can also have phases where
the $1$-RSB solution is exact, a situation discussed for the neuron
with Ising couplings in Section \ref{intro-single-neuron-ising}.  The
multi-$p$-spin model has also become a test bed for equilibrium
thermodynamic calculations meant to capture asymptotic states of
dynamics not maximizing the free energy \cite{nie95c}.

It is well known that for a ferromagnet, the symmetry of the system as
a whole, i.e., of the Hamiltonian, is spontaneously broken by the
state of the system at thermal equilibrium, accompanied by a
spontaneous breaking of ergodicity \cite{domb}.  Such a state can be
reached by the decreasing of the temperature, when the system
undergoes a transition from a paramagnetic phase, exhibiting
macroscopically spherical symmetry and ergodicity, to a ferromagnet,
with only axial symmetry and restricted ergodicity.  In the SK model
described by the replica free energy (\ref{sk3}), as temperature
decreases, an analogous phase transition from a paramagnetic into a
spin glass phase takes place \cite{kir78,mac82,hou83,bin86}, with a
concomitant spontaneous breaking of ergodicity and of RS.  The
transition can be monitored by Parisi's variational parameters $\lamb
$ at stationarity, thus playing the role of order parameters
\cite{par80a,p83,you83}.  The emerging intuitive picture of RSB is
that of a very complicated, rugged, free energy landscape in some
coarse grained state space, with a large number of local minima, many
of them nearly degenerate, as well as a number of global minima,
separated by free-energy barriers, whose height diverges in the
thermodynamic limit.  What in ordered systems thermal equilibrium
state is, corresponds here to a global minimum, also termed as ergodic
component, or pure thermodynamical state, or metastate.  Within the
Parisi solution pure states are organized according to a hierarchical,
so-called ultrametric topology \cite{mpstv84,mv85,mv85b}.  The
ultrametric decomposition of the state space into pure states, from
the practical viewpoint, helps in the calculation of
non-self-averaging quantities \cite{dl83,mv85}, and is also a basic
ingredient of the cavity approach in \cite{sgrev87,mez86}.  However,
so far it withstood rigorous mathematical treatment, and as to real
spin glasses, it is the subject of ongoing controversy
\cite{mar98,new98}.  We would like to add here that, in the context of
neural networks, examples are known \cite{mok94,eng94,bex95} where
there are multiple ground states, and they are grouped into
disconnected regions\ie ergodicity is broken, while the replica method
implies that RS is preserved.  The aforesaid physical picture about
RSB can be maintained by distinguishing between pure states and
ergodic components \cite{mok94}, furthermore, it is unclear whether it
is a spontaneous symmetry breaking that takes place in those networks.
In the present manuscript we do not deal with such subtleties, and
concentrate mainly on the replica method as a tool for calculation.
 
The replica approach in conjunction with Parisi's ansatz provides so
far the most complete description of the SK model in averaged thermal
equilibrium.  However, this scheme, as well as the equivalent cavity
approach and the static limit of the path-integral formulation,
involve certain procedures which, up to now, could not be put on a
rigorous mathematical basis.  On the one hand, there exists a number
of remarkable rigorous results concerning the SK model: in
Ref.\ \cite{aiz87} it was shown that the quenched average $N^{-1}
\langle \ln Z_{{\sf J}}\rangle_{{\rm qu}}$ approaches the so-called
annealed average $N^{-1} \ln \langle Z_{{\sf J}}\rangle_{{\rm qu}}$ in
the thermodynamical limit (termed strong self-averaging property) above
the AT-line and in the absence of an external magnetic field.  The
evaluation of $N^{-1} \ln \langle Z_{{\sf J}}\rangle_{{\rm qu}}$ is
straightforward and reproduces the RS solution.  The basic reason
behind these conclusions is the vanishing of the Edwards-Anderson
order parameter so that the usual effective coupling of the replicas
after averaging out the quenched disorder does not arise, i.e.,
$\langle Z^n_{{\sf J}}\rangle_{{\rm qu}} = \langle Z_{{\sf
    J}}\rangle^n_{{\rm qu}}$.  Furthermore, some explicit bounds
pertaining to the low temperature region have been obtained in
\cite{aiz87} which imply \cite{pas91} the existence of a phase
transition at the same temperature as predicted by the AT-stability
criterion.  In \cite{pas91} it was shown by means of a rigorous
version of the cavity procedure, called martingale method in the
mathematical physics literature, that if the Edwards-Anderson order
parameter is self-averaging then the RS solution is exact.  In
\cite{shc97} it was rigorously verified that this order parameter is
self-averaging and thus the RS solution is exact if the AT stability
condition is fulfilled without and external magnetic field, and also
under a slightly stronger than the AT condition in the presence of a
field.  In view of this theorem, it is suggestive that an AT-stable RS
solution will provide the correct result also in other systems.  It
furthermore confirms Parisi's RSB ansatz to the extent that this
ansatz reduces to the RS result if the AT condition is satisfied.
Finally, the previously discussed evidences as well as the rigorous
mathematical proof from \cite{aiz87} that the RS solution is incorrect
at low temperatures, it follows that the Edwards-Anderson order
parameter is not self-averaging.  This feature is indeed reproduced by
the Parisi solution.  Another interesting rigorous result has been
obtained in Refs.\cite{gue95a,gue95b} via the martingale method,
namely that there exists a set of ``order parameter functions'' $0\leq
x(q)\leq 1$ such that the SK free energy can be expressed in terms of
antiparabolic martingale equations, each of them involving one such
function $x(q)$ and being exactly of the same form as the non-linear
partial differential equation in Parisi's CRSB scheme.  The remaining
non-trivial step in order to complete a rigorous derivation of
Parisi's CRSB solution is to show that this set of functions is
effectively equivalent to a single function $x(q)$.  Finally, in
\cite{ghi98,gue98} certain rather strong conditions are derived that
should be satisfied by the order parameter of a class of spin glass
models -- including the SK but also short ranged models.  These
constraints are indeed fulfilled by Parisi's solution but still leave
room for other possibilities.

We remark that the replica method in combination with the Parisi
ansatz is not restricted to the SK model and its variants, this is
also one of the main reasons why this paper was written.
Nevertheless, most of the above rigorous results pertain to the SK
model, only some of them have so far been generalized to the
Little-Hopfield network, and none but the last one to even further
systems.

\subsection{Little-Hopfield network}
\label{intro-lh-network}

One of the main breakthroughs of the statistical physical approach to
other fields was achieved on the Little-Hopfield model by the replica
calculation of Amit, Gutfreund, and Sompolinsky
\cite{ami85,ami87,ami89}.  They considered $M$ randomly sampled
patterns ${\sv}^\mu$, $\mu=1,2,...,M$, each of dimension $N$, where
$N$ is the number of participating neurons, for a fixed value of the
so-called load parameter 
\begin{equation}
\alpha = M/N
\label{3}
\end{equation}
in the thermodynamical limit $N\to\infty$.  The starting point of the
statistical mechanical treatment is a canonical Boltzmannian
formulation of the problem.  A microstate is a configuration of the
neuron states $S_i,\, i=1,\dots,N$ and a pattern is considered as
stored, if it is a stable fixed point attractor of the dynamics
(\ref{2}).  The energy function for the random, sequential dynamics
(\ref{2}) is analogous to the Hamiltonian of the SK model in
(\ref{sk-ham}) \cite{hop82}.  The main difference is in the exchange
couplings, taken now as $J_{ij}= N^{-1}\, \sum_{\mu=1}^M S_i^\mu
S_j^\mu$, called Hebb rule, thus the patterns ${\sv}^\mu$ play the
role of the quenched disorder.  At positive temperatures the dynamics
(\ref{2}) the update rule for the selected neuron is
non-deterministic, usually Glauber's prescription is applied, see\eg
Ref.\ \cite{sgrev91}.  The original storage problem corresponds to the
zero temperature limit.

Within the RS ansatz Amit, Gutfreund, and Sompolinsky obtained as
central result that the maximal number $M_c$ of patters which can be
stored with an error of a few percent, scales as $M_c=\alpha_c\, N$ in
the thermodynamical limit $N\to\infty$ with a critical capacity
$\alpha_c\simeq 0.138$.  Criticality manifests itself by the drop of
the overlap of a generic stationary state with the desired pattern
from a value below, but close to, one to nearly zero.  It has been
immediately noticed \cite{ami85} that the AT stability condition is
violated at zero temperature for all $\alpha > 0$, thus for exact
results RSB is required, but already a quite small temperature
restores the AT stability and thus the validity of the RS solution.

Applying the $1$-RSB ansatz, Crisanti, Amit, and Gutfreund
\cite{cri86} obtained a modified critical capacity of $\alpha_c\simeq
0.144$.  The problem was reconsidered in the $R$-RSB, $R=0,1,2$,
analysis of Steffan and K\"uhn \cite{ste94}, who put forth a ground
state capacity $\alpha_c\simeq 0.1382$ based on several cross-checking
of their computation. The authors raise the possibility that the
Parisi-Toulouse hypothesis \cite{pt80}, implying that in a CRSB
solution the magnetization in the SK model does not depend on the
temperature, believed to be exact for vanishing magnetization, holds
also in the Little-Hopfield model, at least as a good approximation.
In that case, they conclude, the capacity is given by the intersection
of the AT line and the RS phase boundary, that is, the capacity is
essentially the one calculated form the RS solution.

A CRSB calculation, an extension of Parisi's solution of the SK model
within the formalism of Ref.\ \cite{dom82}, was performed by Tokita
\cite{tok94}.  The sophisticated numerical method applied to evaluate
the CRSB equations showed an instability near $\alpha_c\simeq 0.155\pm
0.002$, which he identified as the capacity.  Numerical simulations
\cite{hop82,ami85,cri86,hor89} gave estimates mostly between the
aforesaid finite $R$-RSB and Tokita's CRSB results.  However, a more
recent simulation \cite{st96}, including a finite-size scaling
specially adapted for a discontinuous transition in the presence of
quenched disorder, yielded $\alpha_c= 0.141\pm 0.0015$, in better
agreement with the former result.  Given the fact that the numerical
evaluation of the CRSB state to the required precision is a much more
formidable task than that of $R$-RSB, $R=0,1,2$, and that even $1$-RSB
computations were the subject of debate \cite{cri86,ste94}, the the
question of theoretical prediction may still be considered as open.
The main issue here is less the precise number, the interesting
questions are rather the salient features of the phase diagram like
reentrance, the validity of the Parisi-Toulouse hypothesis, or what
kind of RSB describes the various phases \cite{ste94,tok94}.
 
Tokita's framework involving the freedom of a gauge function is
closely related to the variational approach for the SK model
\cite{dom82,sd84}, inspired in turn by dynamical studies \cite{som81c}
where the static gauge function is related to the time-dependent
susceptibility.  The variational framework we present in Section
\ref{ssec-spher-neur} on a purely static ground, turns out to be very
similar to those, albeit without our resorting to the gauge function.
On the technical side, we are unaware of any non-perturbative CRSB
analyses, that aims at the ground state or at least regions with
frustration far from criticality, beyond those performed for the SK
model and descendants, as well as the related Little-Hopfield model.
Filling this hiatus was an important motivation for the present paper.
 
The RS results of Amit, Gutfreund, and Sompolinsky have been
re-derived in several different ways
\cite{ges90,bou94b,rie88,col93,shi93}, based on certain assumptions
which are possibly equivalent to that of RS.  Alternative methods
comparable to RSB, however, do not seem to be available yet.  The
authors of Ref.\ \cite{rie88} speculate that their framework may admit
such an extension, being based on Sommers' dynamical path-integral
approach \cite{som87} which successfully reproduced some RSB features
in the SK model.
 
The following mathematically rigorous results for the Little-Hopfield
model are so far available.  The self-averaging property of the free
energy density has been proven in \cite{fen95,bov95}.  In \cite{pas94}
the RS solution is rigorously derived under the assumption that the
Edwards-Anderson order parameter is self-averaging, and in
\cite{shc97} the latter assumption is shown to hold under a condition
similar to, but somewhat stronger than, the AT stability condition.
Finally, a constraint similarly to ultrametricity on the order
parameter has been derived in \cite{ghi98,gue98} which is indeed
satisfied by the RS solution at high temperatures and Tokita's CRSB
solution at low temperatures.

\subsection{Pattern storage by a single neuron}
\label{intro-single-neuron}
 
As we have seen, the McCulloch-Pitts model neuron is the elementary
building block of two prominent types of neural networks, the layered,
feedforward, perceptron and the associative memory.  Therefore the
detailed exploration of such a single neuron is an indispensable
pre-requisite for a satisfactory understanding of the collective
behavior of networked units.

\subsubsection{Continuous synaptic coupling}
\label{intro-single-neuron-cont}

Firstly we describe the case of continuous synaptic couplings, i.e.,
arbitrary vectors ${\jv}$ in (\ref{1}).  If their norm is fixed then
the term spherical couplings is often used.  Note that in Eq.\ 
(\ref{1}) the norm does not influence the output.  An early remarkable
results is due to Winder \cite{win61} and Cover \cite{cov65} regarding
the maximal number $M_c$ of input-patters for which a single
McCulloch-Pitts neuron can correctly reproduce the prescribed outputs
according to (\ref{1}). This is understood as a theoretical maximum,
i.e., without reference to any specific training algorithm that may be
necessary to find the right couplings.  For randomly sampled patters
${\sv}^\mu$, $\mu=1,2,...M$ their critical capacity $\alpha_c
=M_c/N$ in the limit $N\to\infty$ approaches, with probability $1$,
the value $\alpha_c =2$, a widely referenced result in artificial
neural networks.  An easy to follow account of Covers geometrical
proof, for arbitrary $N$, can be found in Sect 5.7 of \cite{hkp91},
and notable extensions have been worked out in
\cite{ven86,bal87,mit89}.

A central notion for adaptive networks is the version space.  This is
the set of coupling vectors ${\jv}$ compatible with the patterns,
or, examples.  Intuitively it is clear that the version space shrinks
as the number of patterns increases, and beyond the capacity the
version space is empty, at least with probability one in the
thermodynamical limit.  
 
A breakthrough was achieved when the space of synaptic couplings of a
single McCulloch-Pitts neuron was explored, following the proposition
of Gardner \cite{eg87}, by Gardner and Derrida within both the
microcanonical \cite{eg88} and canonical \cite{gd88} approaches.  A
main novelty of the concept was in reversing the traditional analogy
between spin systems and neural networks.  In the Little-Hopfield
model the states of the neurons form the ``spin space'', and the
synaptic couplings are the quenched parameters.  The new proposition
was to consider the couplings as configuration space for statistical
mechanics, with constraints represented by randomly generated patterns
to be stored\ie which should be reproduced by appropriate setting of
the couplings, that is, to consider the version space.  By the
introduction of an appropriate cost, or, energy function in coupling
space (further synonyms are Hamiltonian function, or, error measure),
the stage was set for the statistical mechanical treatment.  This does
not restricts the study to the version space, but also allows for
finite temperatures, so beyond capacity provides a framework to
describe states with a given error, including the minimal positive
error of the ground state.  The common ingredient in both the
Little-Hopfield and the Gardner-Derrida concepts is that patterns\ie
examples, represent the quenched disorder, else they are quite
different.  For example, while the energy function of the
Little-Hopfield network closely resembles that of the SK model, not
much formal analogy exist between spin systems and synaptic coupling
space.  In what was a novel application of the replica method, within
the RS ansatz, Gardner and Derrida reproduced, and generalized to
biased pattern distributions, the Winder-Cover result.  They
calculated many a characteristics for the region below the critical
capacity $\alpha_c$, and also proved convergence of training
algorithms.  We note here that the traditional problem of error-free
storage corresponds to the condition of zero energy in the ground
state.  If not all patterns can be accommodated by the couplings, that
is, the neuron is beyond capacity, then, depending on the choice of
the Hamiltonian, various positive ground state energies arise.
 
The thermodynamical stability of the RS solution via the AT condition
\cite{at78} was formulated here by Gardner and Derrida \cite{gd88} and
revised later by Bouten \cite{bou94,der94}.  It turned out that the RS
ansatz beyond the critical capacity $\alpha_c = 2 $ is unstable for
the much studied energy function that measures the number of patterns
that are not stored\ie of unstable patterns.  This is sometimes called
the Gardner-Derrida error measure and will be in our focus in the
present paper.  An improved $1$-RSB ansatz by Majer, Engel, and
Zippelius \cite{mez93} and by Erichsen and Theumann \cite{et93}, as
well as the subsequent 2-RSB calculation by Whyte and Sherrington
\cite{ws96}, turned out to be still plagued by similar instability
beyond capacity.  The latter authors could prove that no finite
$R$-RSB ansatz in the ground state, beyond capacity, may possibly be
locally stable.  In the present article we propose Parisi's CRSB
ansatz as an appropriate description of a single neuron beyond
capacity, within the limits of an equilibrium, averaged statistical
mechanical treatment. 
 
As shown in \cite{ama91}, the effect of frustration, manifesting
itself in the spontaneous breaking of RS beyond capacity, brings along
from the viewpoint of numerical simulations, a very hard,
NP{\-}-complete problem \cite{gar79,pap94}.  That means that whatever
algorithm is used to find an $N$-dimensional vector of synaptic
couplings ${\jv}$ with the smallest possible number of misclassified
examples, the time necessary for it is expected (a rigorous proof is
not known) to increase faster than any power law with $N$.  Simple
algorithms that minimize the number of misclassifications locally,
i.e., within a certain neighborhood of the initial choice for ${\bf
  J}$, are due to Wendemuth \cite{wen95a,wen95b}.  While his result on
the error measuring the number of unstable patterns significantly
overestimated the error, as demonstrated in Ref.\ \cite{our-paper-00}
and cited in the present paper his algorithm may still yield 
acceptable approximations for global minimization as predicted by the
CRSB theory.  We refer also to Section VII.3.\ in \cite{sgrev87} for
the analogous observations in the context of the SK model.  Returning
to generic NP-complete problems, by admitting some random element in
the algorithm, the numerical effort can be reduced to some power of
$N$, hence the name non-deterministic polynomial that NP stands for.
The price to be payed then is that the absolute minimum will be found
only with a certain probability \cite{ruj88,gal90,fre92,dit96}.  A
most widely used such method is simulated annealing \cite{kir83} and
its descendants \cite{wen99}.  As pointed out in \cite{mon98}, the
average time required for the numerical solution may undergo a
dramatic change if certain parameters are varied, without changing its
NP-completeness.  Therefore, the so-called worst-case scenario, on
which the classification as NP-complete is based, may in fact not
capture very well the typical behavior, occurring with probability $1$
as $N\to\infty$, of such algorithms in specific applications.
Conversely, a proof that a problem can be solved deterministically
within polynomial times may still allow very long times for an
algorithm to converge.  Nevertheless, NP-completeness is generally
considered as the signature of algorithmically hard tasks.

It is natural to expect that some, possibly most, of the rigorous
results and alternatives to the replica method for the SK and
Little-Hopfield model can be carried over to the simple perceptron.
However, so far available is only the cavity method in its simplest
form, equivalent to a RS solution, together with a self-consistency
condition equivalent to the AT stability condition of the RS solution
\cite{mez89,gri93,bou95,won95}.

Beyond the critical capacity $\alpha_c = 2$ RS spontaneously breaks,
entailing -- like in the SK model -- an ultrametric organization
\cite{mpstv84,mv85,mv85b} of the synaptic couplings ${\jv}$ that
minimize, in the ground state, the number of incorrect input-output
relations ${\sv}^\mu,\,\xi^\mu$, $\mu=1,2,...,M$ in (\ref{1}).
Below $\alpha_c$, a complementary picture arises by introducing
``cells'' on the $N$-dimensional sphere of synaptic couplings
\begin{equation}
C_{\bsigmas} =\{ {\jv}\, | \, {\jv}^2 = N , \ 
\mbox{sign}( {\jv} \cdot {\sv} ^\mu )=
\sigma^\mu , \ \mu=1,2,...,M \} \ ,
\label{V}
\end{equation}
labeled by the $2^M$ possible output sequences $\bsigma = \{
\sigma^\mu \}$.  The idea to study the simple perceptron in terms of
these cells $C_{\bsigmas}$ is to some extent already contained in
Cover's geometrical derivation of the storage capacity \cite{cov65}
and has been employed again in \cite{der91}.  An appropriate
quantitative framework has been elaborated by Monasson and co-workers
\cite{mok94,mon95,mon96,coc96} in the context of multi-layer networks
and has later been adapted to the simple perceptron in
\cite{wei97,rie97,bro98}.  Based on a replica calculation, this method
enables one to characterize the distribution of cell-sizes
$|C_{\bsigmas} |$ to exponentially leading order in $N$ in terms of a
so-called multifractal spectrum, similarly as in the thermodynamical
formalism for fractals \cite{hal86,tel88}.  This multifractal analysis
opens an interesting view on the storage as well as the generalization
properties of the simple perceptron.

\subsubsection{Ising couplings}
\label{intro-single-neuron-ising}
 
Storage properties change considerably, if one restricts the analysis
to so-called Ising couplings, where each component of ${\jv}$ can
take only the two possible values $\pm 1$.  This extra constraint is
partly motivated by the fact that in a digital computer the $J_i$-s
have a discrete representation.  It has been observed already by
Gardner and Derrida \cite{gd88} that a self-contained treatment by an
RS ansatz of the critical storage capacity with Ising couplings is not
possible within a canonical statistical mechanical approach.
 
Krauth and M\'ezard performed a $1$-RSB analysis with the prominent
result $\alpha_c\simeq 0.833$ for the critical storage capacity of the
Ising perceptron \cite{km89}.  Their $2$-RSB explorations furthermore
indicate that no new solution arises \wrt $1$-RSB.  The RS state
turns out to be globally stable up to the capacity limit, the latter
being signaled by a vanishing of the entropy.  This is an intriguing
coincidence that could not have been foreseen by the RS analysis,
because therein the point whence the entropy becomes negative is
obviously only an upper limit for the capacity.

The need for RSB to calculate the capacity should be contrasted with
the spherical case, see Section \ref{intro-single-neuron-cont}, where
the capacity could be determined within the RS solution.  The reason
for the difficulty here is in that the transition form perfect to
imperfect storage is discontinuous for Ising couplings.  Here the
order parameter exhibits a jump in the sense that one of the overlaps
in $1$-RSB is not the continuation of the RS value, when $\alpha$
passes $\alpha_c$.  From the viewpoint of the order parameter such a
transition can be termed first order.  On the other hand, since the
probability weight of the discontinuously appearing order parameter
value vanishes at $\alpha_c$, the first derivative of the free energy
remains continuous and only the second one jumps.  The Ising neuron
also demonstrates the importance of global stability of a state.  The
RS solution formally exists beyond the transition and stays locally\ie
AT-stable up to $\alpha=4/\pi$.  However, its free energy is smaller
than that from RS, so global stability appears to be taken over by the
$1$-RSB solution, like in first order transitions. It should be added
that here the locally stable but globally unstable RS solution should
be ruled out as a metastable state in the traditional sense because of
its negative entropy.  Furthermore, the $1$-RSB solution is not a
locally stable equilibrium state before the transition, so two
spinodal points collapse onto the transition point.

While a major part of the existing statistical mechanical
investigations -- including the SK model in (\ref{sk1}) and our
present study of the simple perceptron -- are based on a canonical
Boltzmannian formulation of the problem, Gardner's seminal
calculations in Refs.\ \cite{eg87,eg88} uses microcanonical ensemble.
For the Ising perceptron, this approach was adopted by Fontanari and
Meir \cite{fon93}, reproducing Krauth and M\'ezards results without
going beyond RS and verifying in particular the AT stability condition
\cite{at78} as well as the physical requirement of a non-negative
entropy. 
 
Computing the optimal vector ${\jv}$ of synaptic couplings for the
Ising perceptron is an NP-complete problem \cite{gar79,pap94} for any
positive load parameter $\alpha$, as demonstrated in Refs.\
\cite{pit88,bex96}.  The challenge of numerically estimating the
critical capacity $\alpha_c$ has been attacked by several groups, most
of them verifying $\alpha_c\simeq 0.833$, 
with the exception of Ref.\ \cite{nad97}, criticized by the comment in
\cite{sch98}.  Subsequent, more extensive computations in
\cite{mil97,bou98} appear to confirm the original critical value.
 
Below critical capacity, a multifractal analysis of the space of Ising
couplings ${\jv}$, inspired by the work on the spherical case
\cite{mok94} as discussed in the previous section, has been worked out
in \cite{wei97,eng96a}.  Beyond criticality, a thermodynamical stability
analysis \cite{gyoXX} suggests that $1$-RSB is locally stable at and
beyond $\alpha_c$.  On the other hand, also the microcanonical RS
approach of Fontanari and Meir \cite{fon93} continues to coincide with
Krauth and M\'ezards results and satisfies the local thermal stability
criterion of de Almeida and Thouless \cite{at78}. 
 
The above numerical and analytical findings have given rise to the
conjecture that the Ising perceptron beyond capacity behaves quite
similarly to Derrida's random energy model \cite{der81}.  This system
is the $p\to\infty$ limit of the $p$-spin interaction version of the
SK model.  In particular, the $1$-RSB ansatz yields \cite{gm84} indeed
the what is accepted as the exact solution of the problem within the
canonical Boltzmannian approach and the zero entropy condition marks
the transition from RS to $1$-RSB.  Interestingly, as it has been done
originally by Derrida, even the spin glass phase of the random energy
model can be described by the replica method, but without the need to
introduce the $1$-RSB ansatz.  There by a direct calculation the mean
free energy could be maximized, without dealing with spin overlaps, so
this can be considered as an independent confirmation of RSB as
applied later by \cite{gm84}.

In the case of the neuron with Ising couplings, like in the random
energy model, an overlap $q_1=1$ arises, with probability exactly
$1-x=1-T/T_c$.  The fact that the microcanonical formulation within RS
gave as minimal error the ground state error beyond capacity
\cite{fon93} as the canonical $1$-RSB result \cite{km89}, is a further
peculiarity of Ising synapses.  There is no technical contradiction,
however, because if $q_1=1$ is set then the $1$-RSB free energy
becomes equivalent to the RS microcanonical entropy.  This can be
understood, if one realizes that in the latter the temperature is
essentially an extra variational parameter, taking the role of $1/x$,
related to the aforesaid probability in $1$-RSB.  The special nature
of the microcanonical approach was interpreted, and exploited for
calculating the storage capacity of certain multi-layer perceptrons,
in Ref.\ \cite{saa94}.

Further systems where stable $1$-RSB phases arise, albeit generally
without the zero entropy condition, are the p-spin interaction SK
model \cite{gar85}, its spherical variant \cite{cri92}, the spherical,
multi-$p$-spin interaction model \cite{nie95}, the Potts glass
\cite{erz83,eld83,gks85}, and protein folding models
\cite{gar88,sha89,sas92,tak97}.

The general framework in the present paper includes both continuous
and Ising synaptic couplings ${\jv}$.  Since the case of principal
interest here is Parisi's CRSB ansatz, in the quantitative numerical
evaluation beyond capacity we will focus on the example of the
continuous, spherical, couplings.  Whether or not a continuous RSB
ansatz will be necessary for more general Ising networks than the
McCulloch-Pitts model, e.g., in multi-layer Ising perceptrons, remains
to be seen \cite{coc96,bar91,bar92}.

\subsection{Training, error measures, and retrieval}
\label{intro-retrieval}

We recall that the patterns to be stored are prescribed as pairs ${\bf
  S}^\mu,\,\xi^\mu$, $\mu=1,...,M$ and the McCulloch-Pitts neuron
(\ref{1}) is required to reproduce $\xi^\mu$ in response to ${\bf
  S}^\mu$. Next we define the so-called local stability parameters
\begin{equation}
\Delta^\mu = \xi^\mu  |{\jv}|^{-1/2} \sum_{k=1}^N J_k\, S_k^\mu \ ,
\label{4}
\end{equation}
where the normalization factor $|{\jv}|^{-1/2}$ guarantees a
sensible behavior in the thermodynamical limit $N\to\infty$ if the
patters ${\bf S^\mu}$ are normalized to a length of the order
$N^{1/2}$.  Introducing an error measure on a pattern as
$V(\Delta^\mu)$, the $V(y)$ called a ``potential'', one is lead to the
Hamiltonian 
\begin{equation}
{\cal H} = \sum_{\mu=1}^M V(\Delta^\mu)
\label{5}.
\end{equation}
Minimizing this Hamiltonian in the space of couplings $\bf J$ is the
task of training.  In particular, maximizing the number of correctly
stored patterns in (\ref{1}) is equivalent to minimizing (\ref{5} if
one chooses the potential $V(y)=\Theta(-y)$, where $\Theta(y)$ denotes
the Heaviside step-function.  Training in $\bf J$ space contrasts with
the neuron (spin) dynamics of the Little-Hopfield model, aimed at
retrieving stored patterns. 

If more than plain memorization of the classifications $\xi^\mu$ of
the training examples ${\sv}^\mu$ is required, then other choices of
$V(y)$ may be advantageous. For instance,
\begin{equation}
V(y) = \Theta (\kappa -y)\, (\kappa -y)^b
\label{6}
\end{equation}
with positive $\kappa$ and $b$ tries to impose, upon minimization in
(\ref{5}), the conditions $\Delta^\mu\geq\kappa$ on all the local
stabilities (\ref{4}).  For the step function potential, $b=0$, the
number of violations of $\Delta^\mu\geq\kappa$ is minimized, but those
$\Delta^\mu$ which violate the condition may take values arbitrarily
far below $\kappa$.  For a softer potential with $b>0$, a compromise
must be made between minimizing the number of violations and of the
``cost'' $(\kappa-\Delta^\mu)^b$ of the committed error.  In any case,
the qualitative effect of positive $\kappa$ after minimization in
(\ref{5}) is that inputs ${\sv}$ in (\ref{1}) close but not
identical to one of the stored patterns ${\sv}^\mu$ can still be
associated with the correct output $\xi^\mu$.  For load parameters
(\ref{3}) below the critical capacity, $\alpha <\alpha_c$, one will
typically choose the largest possible $\kappa$-value admitting a zero
training error in (\ref{5}) and thus $\Delta^\mu \geq\kappa$ for all
patterns.  This maximal $\kappa_{max} (\alpha)$ as a function of the
load parameter $\alpha$ has been calculated by Gardner and Derrida in
\cite{gd88}.  Note that $\kappa_{max}(\alpha)$ is the same for any $b$
and that $\kappa_{max}(\alpha_c)=0$.  Beyond the critical capacity,
$\alpha > \alpha_c$, not all the training patterns can be stored
anyway, thus sacrificing some additional ones by choosing $\kappa > 0$
may still be desirable to create a finite basin of attraction in the
retrieval dynamics for patterns for which $\Delta^\mu\geq\kappa$ can
be achieved \cite{gd88}.  Attractors of the dynamics (\ref{2}) not
corresponding to one of the stored patterns, spurious states,
represent failure of memorization.  We also mention that training the
$\bf J$ couplings for each neuron separately leads to lifting the
symmetry of $J_{ij}$-s in the original Little-Hopfield model. That
leads to the loss of the equilibrium being described by a Hamiltonian,
and to a dynamics exhibiting more complex time series than convergence
to a fixed point \cite{scs88}.

The above concept of training corresponds to $T=0$ dynamics in $\bf J$
space, and can be complemented by a stochastic element to represent
positive temperatures.  The main focus of the present paper is
describing the final equilibrium states of such dynamics.

A further motivation for studying potentials $V(y)$ even more general
than in (\ref{6}) is the fact that a discrete time version of the
gradient descent dynamics of $\bf J$ in the corresponding energy
landscape (\ref{5}) reproduces several well-known learning algorithms
\cite{gg91}.  For instance, the potential (\ref{6}) with $b=1$ induces
a dynamics very similar to the perceptron algorithm of Rosenblatt
\cite{ros62} and later Gardner \cite{eg88}.  Beyond capacity, when the
convergence of such algorithms to a state with minimal positive error
is not proven, there is only an intuitive ground for using such
algorithms, and obviously modifications are necessary
\cite{wen95a,wen95b}.

Next we turn to the retrieval behavior of the Little-Hopfield
associative memory network dynamics (\ref{2}), characterized by the
time dependent overlaps
\begin{equation}
m^\mu(t) = \frac{1}{N}\sum_{k=1}^N S_k^\mu\, S_k(t)
\label{7}
\end{equation}
of the processed pattern ${\sv}(t)$ with the stored patterns (fixed
point attractors) ${\bf S^\mu}$. An input pattern ${\sv} = {\bf
S}(0)$ is associated under the dynamics (\ref{2}) with the stored
pattern ${\sv}^\mu$ if $m^\mu(t)$ evolves towards $1$ in the course
of time, while $m^\nu(t)\to 0$ for all other patterns $\nu\not =\mu$.
The smallest value of $m^\mu(0)$ which still leads to a successful
retrieval, i.e., $m^\nu(t)\to\delta_{\mu\nu}$, is a measure for the
basin of attraction of the stored pattern ${\sv}^\mu$.

In the thermodynamical limit $N\to \infty$ the following result for the
first time step of the evolution in (\ref{2}) has been derived in
\cite{kep88,gar89b}
\begin{equation}
m^\mu(1) =
\int\rho(\Delta)\,\mbox{erf}\left(\frac{m^\mu(0)\,\Delta}{
\sqrt{2[1-m^\mu(0)^2]}}\right)\, d\Delta \ ,
\label{8}
\end{equation}
where $\mbox{erf} (x)=2\pi^{-1/2}\int_0^x e^{-y^2} dy$ and
$\rho(\Delta)$ is the distribution of the local stabilities from
(\ref{4}), defined as
\begin{equation}
 \rho (\Delta) = \frac{1}{N}\sum_{\mu =1}^M \delta(\Delta - \Delta^\mu
 ) \ . 
\label{9}
\end{equation}
In general, $\rho(\Delta)$ depends on the algorithm by which the
vector of synaptic couplings in (\ref{4}) has been computed.  It has
been assumed that all the McCulloch-Pitts units in (\ref{2}) have been
independently trained according to the same algorithm, thus in the
thermodynamical limit $\rho(\Delta)$ will be the same function for all
of them.  We mention that with the Hebb rule this condition of
independence does not hold, thus (\ref{8}) is not valid for the
original version of the Little-Hopfield model.  In the case, when
${\jv}$ has been obtained by minimizing a Hamiltonian function of
the general form (\ref{5}), the resulting distribution of overlaps
$\rho(\Delta)$ will be one of the most important quantities of our
present work.
 
When (\ref{5}) is minimized with the maximal $\kappa$-value in
(\ref{6}) admitting an error-free storage of all training patterns,
i.e., $\kappa=\kappa_{max}(\alpha)$, Kepler and Abbott \cite{kep88}
have observed numerically that retrieval is successful if and only if
\begin{equation}
m^\mu(1)>[1+m^\mu(0)]/2\ .
\label{10}
\end{equation}
In the thermodynamical limit this seems to be exact, or a very good
approximation, at least for sufficiently small load parameters
$\alpha$ such that $\kappa_{max}(\alpha)\geq 0.6$ \cite{kep88}.

In general, the further time evolution of $m^\mu(t)$ becomes
increasingly more complicated than the first time step (\ref{8}).
Analytical approximations as well as numerical studies for various
specific learning rules for the synaptic couplings ${\jv}$
(including the Hebb rule) have been elaborated in
\cite{hor89,col93,gar87,for88,kra88,hen91}.  For randomly dilute
networks such that the fraction of non-zero synaptic couplings
$J_{ik}$ in (\ref{2}) tends to zero like $N^{-1}\ln N $ in the
thermodynamical limit, it has been shown in \cite{gar89b} that the
same dynamics for $m^\mu(t)$ as in (\ref{8}) remains valid for
arbitrary times $t$, provided the initial condition ${\sv}(0)$ has
an appreciable overlap with only one of the stored patterns ${\bf
  S}^\mu$.  Further interesting explorations along these lines can be
found in Refs.\cite{rie88,der87,ami90}

A question of particular interest for our present study has been
addressed by Griniasty and Gutfreund \cite{gg91}, namely whether it
may be an advantage with respect to the retrieval properties to
increase $\kappa$ in (\ref{6}) beyond the threshold
$\kappa_{max}(\alpha)$ of error-free storage in the minimization of
(\ref{5}).  For randomly dilute networks they demonstrated
analytically that this is indeed the case provided
$\alpha<\alpha_c(\kappa=0)=2$, but that the critical storage capacity
$\alpha_c=2$ itself cannot be surpassed by this trick.  For
$\alpha<\alpha_c(\kappa=0)$, the effect of choosing $\kappa >
\kappa_{max}(\alpha)$, with $b=1$ in (\ref{6}), is twofold.  On the
one hand, the patterns ${\sv}^\mu$ themselves are no longer
attractors but converge under the dynamics (\ref{2}) towards nearby
fixed points.  On the other hand, the basins of attraction of these
fixed points steadily grow as $\kappa$ exceeds $\kappa_{max}(\alpha)$
and rather soon reach the ``full basin scenario'', i.e. every input
pattern ${\sv} = {\sv}(0)$ with a finite initial overlap
$m^\mu(0)>0$ will converge towards the same attractor as ${\sv}^\mu$
does.  We remark that these conclusions in \cite{gg91} are based on a
RS ansatz which is not rigorously valid \cite{bou94,der94} for the
potential (\ref{6}) with $\kappa>\kappa_{max}(\alpha)$ and $0\leq
b\leq 1$.  The CRSB scheme of our present work may be needed for an
exact treatment, though the quantitative corrections are not expected
to be large.

\subsection{Multi-layer perceptrons}
\label{intro-multi-layer}

As far as practical applications to real problems are concerned,
multi-layer perceptrons are the most important networks tractable
within a statistical mechanical approach.  They have great
computational abilities and at the same time are not prohibitively
complicated due to the absence of feedback effects.  Still, the very
property that these architectures are able to implement nontrivial
tasks of practical interest makes their theoretical analysis
difficult.  Qualitatively, the flexibility of multi-layer perceptrons
is due to the fact that the individual McCulloch-Pitts units within
each layer can share the effort to produce the correct output.  On the
one hand, this ``division of labor'' gives rise to intricate
anti-correlations between their activities
\cite{bar92,eng92,eng96b,mal97,mal99}.  On the other hand, for not too
small set of training examples, it brings along a spontaneous breaking
of their permutation symmetry, possibly superimposed in addition by a
spontaneous breaking of replica symmetry \cite{saa94}.  Note that
permutation symmetry of units in a layer is understood in the average
sense, for a given training set of patterns there is generically no
such symmetry. We will not present here a systematic discussion of the
ongoing research on these topics but rather highlight two particular
aspects of specific interest from the viewpoint of the simple
perceptron analysis in our present paper.  One being the capacity of
multi-layer networks, and the other one the possibility to mimic
multi-layer structures with a single unit with a non-monotonic
transfer function. For a more detailed overview, especially regarding
learning algorithms and generalization properties, we refer to
\cite{hkp91,wat93} and for the present state-of-the-art to
\cite{mal97,sch95,wes98,win97,ama98} and further references therein.

The storage capacity of multi-layer perceptrons has been analyzed
within a statistical mechanical approach for the first time in
\cite{bar92,eng92,bar90} for spherical and in \cite{bar91,bar92} for
Ising perceptrons, addressing the simplest case with one adaptive
input layer (first layer) and one hidden layer (second layer).  The
latter is governed by a pre-wired Boolean function, mostly either a
so-called committee or a parity machine.  For an Ising parity machine
with non-overlapping receptive fields\ie tree architecture, $1$-RSB
seems to be exact \cite{bar91}.  For fully connected machines with
spherical synaptic weights \cite{bar92,eng92,bar90,kwo97}, the
assumption of RS cannot be upheld since it yields results incompatible
with the rigorous bound of Mitchison and Durbin \cite{mit89} (see also
\cite{bar92,wat93}), based on a generalization of Cover's line of
reasoning \cite{cov65}.  While an improved $1$-RSB ansatz respects these
limits, the necessity of additional steps of RSB in order to draw
reliable quantitative conclusions remains unclear \cite{bar92,kwo97}.

A first alternative approach \cite{saa94} suggests to break the
permutation symmetry of the hidden units explicitly prior to the
actual replica calculations, but the resulting equations are
approximations and difficult to solve for large networks.  A second
alternative method is the cavity approach, elaborated on a level
equivalent to an RS ansatz in \cite{gri93,won95}.  A most promising
new roadway seems to be the multifractal analysis of the space of
synaptic couplings by Monasson and co-workers
\cite{mok94,mon95,mon96,coc96}.  One of the most remarkable findings
of these and subsequent works \cite{urb97,xio97,xio98} is that an RS
ansatz in this approach yields results very close but not identical to
those of a $1$-RSB ansatz in the standard treatment along the lines of
Gardner and Derrida \cite{bar91,bar92,eng92,bar90,kwo97}.

For our present study the salient observation is \cite{mal97} that the
increased power of multi-layer networks in comparison with the simple
perceptron stems from the possibility that the single McCulloch-Pitts
units may all operate in the region beyond their individual storage
capacity, while the network as a whole is still below its maximal
storage capacity, the reason being that via the division of labor, the
errors of one unit may be rectified by another
one, or made up for collectively.  Specifically, results for a simple
perceptron beyond its storage capacity have been utilized for the
exploration of multi-layer networks in \cite{wes98,pri94}.

As a generalization of the simple perceptron with
input-output-relation (\ref{1}), the following setup was introduced in
\cite{kob91}
\begin{eqnarray}
& & \xi = V_\gamma(y)\ ,\ \ y = |{\jv}|^{-1/2} \sum_{k=1}^N J_k S_k
\label{r1}\\
& & V_\gamma(y) = \mbox{sign}\left( [y-\gamma]\, y\,
  [y+\gamma]\right)\ . 
\label{r2}
\end{eqnarray}
Like in (\ref{4}), the scaling by $|{\jv}|^{-1/2}$ guarantees a
sensible thermodynamical limit $N\to\infty$ and, unless indicated
otherwise, we will focus on the case of a spherical constraint ${\bf
  J}^2=N$. The potential (\ref{r2}) outputs $+1$ if $y > \gamma$ or
$\gamma<y<0$ and $-1$ otherwise ($V_\gamma(-y)=-V_\gamma(y)$), hence
the name ``reversed-wedge perceptron'' for the input-output relation
(\ref{r1}) was coined in \cite{wat92}. Without loss of generality, one
can focus on non-negative parameters $\gamma$, reproducing the simple
perceptron (\ref{1}) for $\gamma = 0$, and its equivalent reversed
counterpart for $\gamma\to\infty$.  In \cite{kob91,def93,nis93} the
reversed-wedge perceptron (\ref{r1}) was studied as a generalization
of the simple perceptron with an increased storage capacity as main
result. As revealed in \cite{bmz93}, the assumption of RS, on which
those first works are based, ceases to fulfill the AT stability
condition before the limit of capacity is reached, and an improved
$1$-RSB calculation modifies the storage capacity by more than a
factor of $2$ for $\gamma$-values of the order of one. It has been
conjectured \cite{mok94}, that a consistent treatment of the problem
is only possible by means of the general Parisi RSB framework. 
The storage problem in (\ref{r1}) is equivalent to a minimization
of the cost function (\ref{5}) with the potential from (\ref{r2}), so
the problem becomes a special case of the theory presented later in
this paper. 

In \cite{bmz93} it was observed that by rewriting (\ref{r1}, \ref{r2})
as
\begin{equation}
\xi = \prod_{j=1}^3 \mbox{sign} (y-\theta_j )
\label{r3}
\end{equation}
with $\theta_j = (j-2)\gamma$, the reversed-wedge perceptron may also
be looked upon as a toy model of a multi-layer perceptron. To see
this, we first note that each factor of the form $\mbox{sign}
(y-\theta )$ in (\ref{r3}) is a generalization of the simple
perceptron (\ref{1}) with a ``firing threshold'' $\theta$ as new
feature. Such a threshold has a well founded biological basis but has
been omitted from many a theoretical study \cite{hkp91}.  A systematic
exploration of perceptrons with a threshold by way of a replica
analysis has been undertaken in \cite{wes97}.  Returning to (\ref{r3})
we see that this input-output relation represents a special kind of a
two-layer perceptron with three McCulloch-Pitts units, endowed with
different thresholds but identical synaptic weights ${\jv}$ in the
first layer, and a non-adaptive second layer, pre-wired according to a
so-called parity machine.  Besides the two-layer architecture the
toy model suggests, and the occurrence of RSB before the maximal
storage capacity is reached, several further features of the
reversed-wedge perceptron (\ref{r3}) have been found to qualitatively
agree with characteristic properties of real multi-layer networks
\cite{eng94,bmz93,rei96b}.

As shown in \cite{mok94,bmz93}, the version space is partitioned into
an exponentially large number of disconnected components for any
positive load parameter $\alpha$.  Nevertheless, up to a certain
finite $\alpha$-value, the RS solution appears to be correct
\cite{mok94}.  This observation invalidated the hitherto widespread
belief that unbroken RS signals a connected (possibly convex) version
space and that RSB is tantamount to the breaking of ergodicity.  The
quite subtle point is that, in the thermodynamical limit $N\to\infty$,
each of the exponentially many disconnected components of the version
space, has an infinitezimal contribution to full volume thus
validating the RS result, while beyond a certain $\alpha$ they become
small in number, but each having a significant relative contribution
to the volume of version space, thus causing RSB. 

In all previously mentioned cases, RSB was intimately connected with
frustration and a rugged energy landscape with nearly degenerate
minima. In the case of the reversed-wedge perceptron, below its
maximal storage capacity, there is no frustration.  Then all
constraints in the form of input-output relations (\ref{r1}) are
satisfied for vectors ${\jv}$ belonging to the version space.  The
local minima may now be identified with the exponentially many
disconnected domains of the version space, each having exactly the
same energy and being completely flat with bottom level at zero energy
(muffin-tin shape).  In place of a spontaneous one may now rather
speak of an induced breaking of ergodicity, that can be attributed to
the non-monotonicity of the potential, and may be the reason why RS
remains applicable for smaller $\alpha$-s.  Due to the absence of
frustration it may come as a surprise that, for sufficiently large but
below capacity $\alpha$ values, Parisi's scheme, including the
ultrametric organization of the ergodic components. is apparently
still applicable.  A very similar situation arises for potentials in
(\ref{6}) with negative $\kappa$-values \cite{gd88,wat93} and for
certain unsupervised learning scenarios \cite{mie95,loo95}, involving
potentials of the form
\begin{equation}
V(y) = \Theta (\kappa - |y| )\ ,
\label{r4}
\end{equation}
in the regime below the respective critical capacity value of the load
parameter $\alpha$.  Beyond the critical $\alpha$-value, in all cases
frustration sets in, where it is natural to expect Parisi's RSB
scenario.

Various generalizations of the reversed-wedge perceptron have been
explored, two of which we find particularly interesting.  In
\cite{bol96} the case of more than three discontinuities in (\ref{r2})
has been considered.  As the number of discontinuities increases, the
maximal storage capacity is found to increase and also the
consideration of RSB effects becomes more and more important for
quantitatively reliable results.  The reversed wedge Ising perceptron
with $\gamma = (2\ln 2)^{1/2}$ in (\ref{r2}) was demonstrated in
\cite{bex95} to saturate the information theoretical upper bound for
the maximal storage capacity, a fact which has found its natural
physical explanation by means of a multifractal analysis of the
version space in \cite{bex97}.  The concomitant vanishing of the
Edwards-Anderson order parameter has, like in the high temperature
regime of the SK model \cite{aiz87}, the consequence that the annealed
approximation coincides with the RS solution \cite{bex95}.


\eject

\section{Statistical Mechanics of Pattern Storage}
\label{sec-thermodyn}

We now set the stage for the detailed study of the single neuron by
introducing the model and reviewing basic statistical mechanical
notions.  With the exception of Sections \ref{ssec-hess} and
\ref{ssec-sym-PPDE}, the style of presentation is meant to be
self-contained henceforth.  Some overlaps with Section \ref{history}
are the consequences.

\subsection{The model}
\label{ssec-model}

We consider the McCulloch-Pitts model neuron \cite{hkp91},
\bml  \label{defneu}
\begin{eqnarray}
   \label{defneu1}
   \xi & = & \text{sign}(h),  \\
   \label{defneu2}
   h & =  & N^{-1/2} \sum\nolimits_{k=1}^{N} J_k S_k,  
\end{eqnarray}
\eml where $\jv$ is the vector of synaptic couplings, $\sv$ the
input and $\xi$ the response.  The normalization was chosen so that
$h$ is typically of $O(1)$ when $N\rightarrow \infty$.  Patterns to be
stored are prescribed as pairs 
\begin{equation}
\{ {\sv}^\mu ,\xi ^\mu\}_{\mu
  =1}^{M} 
\label{examples}  
\end{equation}
such that the neuron is required to generate ${\xi^\mu}$ in
response to ${\sv}^\mu$.  Given the ensemble of patterns, the local
stability parameter 
\begin{eqnarray}
   \label{def-delta}
\Delta^\mu = h^\mu \xi^\mu
\end{eqnarray}
obeys some distribution $\rho(\Delta)$ \cite{gg91}.  The $\mu$-th
pattern is stored by the neuron, if the actual response signal from
Eq.\ (\ref{defneu}) equals the desired output $\xi^\mu$\ie $\Delta^\mu
>0$.  The number of patterns $M$ is generically of order $N$, so
\begin{equation} \label{def-alpha}
\alpha = M/N
\end{equation}
is an intensive parameter.  For the sake of simplicity,
we generate the $S_k^\mu$-s independently from a normal distribution,
and take $\xi^\mu = \pm 1$ equally likely. The corresponding
probability density will be denoted by $P\left(\{ {\bf
S}^\mu,\xi^\mu\}\right)$.

Since the output $\xi$ in (\ref{defneu}) is invariant, if $\jv$ is
multiplied by a factor, it is useful to eliminate this degree of
freedom by the spherical constraint $\left|{\jv}\right| = \sqrt{N}$.
In general, a prior distribution $w({\jv})$, not necessarily
normalized, expresses our initial knowledge about the synapses.  The
spherical constraint, which we choose to normalize, corresponds to
\bml
  \label{spher-prior}
\begin{eqnarray}
  \label{spher-prior-a}
  w({\jv})& = &  C_N \ \delta\left(N-\left|{\jv}\right|^2\right) \\
  \label{spher-prior-b}
 C_N & = & N\, \Gamma\left(\frac{N}{2}\right)
  (N\pi)^{-\frac{N}{2}}
\end{eqnarray}
\eml
Another generic type of prior distribution is when it prescribes
independent, identical constrains for the synapses as
\begin{equation}
  \label{gen-prior}
w({\jv}) = \prod_{k=1}^N w_0(J_k) ,
\end{equation}
{\em e.\ g}., binary, or Ising, synapses have 
\begin{equation}
  \label{bin-prior}
w_0(J) = \delta(J-1) + \delta(J+1).
\end{equation}
This prior distribution is not normalized.  Its scale is conveniently
set by requiring that $J_k^2$ averages to unity, that is $\int w_0(J)
J^2 \,{\rm d}J = \int w_0(J)\,{\rm d}J$, whence $N^{-1}\sum_{k=1}^N
J_k^2$ goes to $1$ for large $N$.  ``Soft spins'' are generated by
smooth, multiple-peaked $w_0(J)$-s.

Our main goal is to find those $\jv$-s that store the prescribed
patterns\ie are compatible with the patterns. The problem can be
reformulated as an optimization task with a suitable cost function\ie
Hamiltonian, to be minimized.  A convenient choice here is the sum of
errors committed on the patterns
\begin{equation}
  \label{hamiltonian} {\cal H} = \sum_{\mu =1}^M V(\Delta^\mu),
\end{equation}
where the potential $V(\Delta^\mu)$ measures the error on the $\mu$-th
pattern ${\sv}^\mu ,\xi ^\mu$.  A natural error measure $V(y)$ is
zero for arguments larger than a given threshold $\kappa$ and
monotonically decreases elsewhere \cite{gg91}.  Storage as defined
above corresponds to $\kappa = 0$, while a $\kappa >0$ means a
stricter requirement on the local stability $\Delta$ and ensures a
finite basin of attraction for a memorized pattern during retrieval.
The Hamiltonian (\ref{hamiltonian}) defines through gradient descent a
dynamics in the space of couplings $\jv$.  Specifically,
\begin{equation}
  \label{gen-v}
  V(y)=(\kappa -y )^b\,\theta (\kappa -y)
\end{equation}
corresponds to the perceptron and adatron rules for $b=1,2$,
respectively, where $\theta(y)$ is the Heaviside function, see
\cite{gg91} and references therein.  There is no such dynamics in the
case $b=0$, but because of its prominent static meaning -- the
Hamiltonian counts the incorrectly stored patterns -- we will consider
that in concrete calculations.  Furthermore, the $\theta(y)$ function
can be approximated by a smooth one that does have an associated
dynamics.  Thus our present study of thermal equilibrium with the
Hamiltonian involving (\ref{gen-v}) with $b=0$ can be thought of as
the average asymptotics of such a dynamics.  A non-gradient-descent
algorithm, designed to minimize the Hamiltonian with $b=0$, will be
discussed in Section \ref{ssec-simul}.

\subsection{Thermodynamics}
\label{ssec-thermodyn}

The Hamiltonian (\ref{hamiltonian}) gives rise to a statistical
mechanical system \cite{eg88,gar89b} resembling models of spin glasses
with infinite-range interactions \cite{sgrev87,sgrev91}.  A microstate
is a specific setting of the synaptic weight vector $\jv$, quenched
disorder is due to the randomly generated patterns, and a positive
temperature $T=\beta^{-1}$ has the effect of introducing
tolerance to error of storage. (We use the convention of setting
Boltzmann's constant to unity.)  The partition function assumes the
form
\cite{eg88,gar89b}
\begin{equation}
  \label{partfunc}
   Z = \int \rd^N\!  J\ w({\jv}) \ \exp \left(-\beta \sum_{\mu =1}^M
V(\Delta^\mu)\right). 
\end{equation}
Integration is over the entire real axis if not denoted otherwise.
For large $N$ we expect self-averaging \cite{sgrev87,sgrev91}, that
is, for a given instance of the quenched disorder the extensive
thermodynamical quantities are assumed to approach their quenched
average.  This leads us to the thermal statics of the system, where
the question of breaking of ergodicity on some time scales is not
dealt with.  The replica method \cite{sgrev87,sgrev91} starts with our
writing the mean free energy per coupling as
\begin{equation}
  \label{fe}
  f = - \lim_{N\to\infty}\frac{\left<\ln Z\right>_{qu}}{N\beta}
  =\lim_{N\to\infty} \lim_{n\rightarrow 0}
  \frac{1-\left<Z^n\right>_{\rm qu}}{nN\beta} ,
\end{equation}
where 
\begin{equation}
  \label{qu-average}
 \left<\dots \right>_{\mathrm qu} = \int P\left(\{ {\sv}^\mu , 
 \xi^\mu\}\right)  \dots \prod_{\mu =1}^M \rd\xi^\mu\, \rd^NS^\mu
\end{equation}
stands for the quenched average over patterns.  In order to carry on
with calculations, it is common practice \cite{hkp91,sgrev87,sgrev91}
to interchange the limits $n\to 0$ and $N\to\infty$.  In what follows
we accept the reversal of limits based on numerous examples wherein
the consequent results were verified by other analytic methods or
numerical simulations, see for example Refs.\ \cite{hkp91,sgrev87}.

Introducing the thermal average as 
\begin{equation}
  \label{th-average}
\left<\dots \right>_{\mathrm th} = Z^{-1} \int  \exp \left(-\beta
  \sum_{\mu =1}^M V(\Delta^\mu)\right) \dots w({\jv})\, \rd^N\! J,
\end{equation}
one naturally obtains the mean error per pattern as
\begin{equation} 
  \label{ener-def}
  \varepsilon = \left< \left< V(\Delta) \right>_{\mathrm th} \right>_{\mathrm qu}.
\end{equation}
From the free energy this derives as
\begin{equation} 
  \label{ener}
  \varepsilon =
\frac{1}{\alpha}\frac{\partial\beta f}{\partial\beta},
\end{equation}
  it is thus the analog of the thermodynamical energy.  The mean
  entropy per synapse, or, the specific entropy,
\begin{equation}
  \label{entr}
  s = \beta(\alpha\varepsilon - f)
\end{equation}
is a measure of the volume in coupling space associated with a given
mean error. 

A case of special significance is when at $T=0$ the mean error is
zero\ie storage is perfect. Then the partition function becomes the
integral $\Omega$ of the prior distribution over the version space,
$\Omega = \left. Z \right|_{T=0}$.  If the prior distribution is
uniform over the version space, as it is in the case of spherical
normalization, then $\Omega$ is proportional to its volume.  The zero
temperature entropy measures this volume as $\left. s
\right|_{T=0}=N^{-1} \ln\Omega$.

\subsection{Spherical and independently distributed synapses}
\label{ssec-various-constraints}

Before proceeding, we warn the reader that quantities of different
types -- variables, functions, functionals -- may be denoted by the
same symbol, the difference possibly shown by the type of the
argument.  An example is the free energy and the replica free energy,
as shown below.  Such practice will be limited to cases when there is
little chance for confusion.

In the case of the spherical constraint (\ref{spher-prior}), the free
energy per synapse was first proposed in \cite{eg88,gar89b} for the
special error measure $V(y)=\theta(\kappa -y)$.  The replica symmetric
free energy was given by \cite{gg91} for a general error measure
$V(y)$.  For a general $V(y)$ without the assumption of replica
symmetry the free energy reads as \bml
\label{fe-spher1}
  \begin{eqnarray}
  \label{fe-spher1a}
  f & = & \lim_{n\rightarrow 0} \frac{1}{n}~ {\min_{\sf Q}}
  \ f({\sf  Q}) \\
  \label{fe-spher1b} 
  f({\sf Q}) & = & f_\rs({\sf Q}) + \alpha\, f_\re({\sf Q}) ,\\
  \label{fe-spher1c}
  f_\rs({\sf Q}) & = & - (2\beta)^{-1} \ln\text{det}{\bf\sf Q}, \\
  \label{fe-spher1d}
  f_\re({\sf Q}) & = & - \frac{1}{\beta}\ln \int \frac{\rd^n\! x \ \rd^n\!
    y}{(2\pi)^n} \exp\left( -\beta
    \sum\nolimits_{a=1}^n V(y_a) + i\xv \yv - \frac{1}{2} 
  \xv {\sf Q}\xv\right),
\end{eqnarray}
\eml 
see Appendix \ref{app-repl-fe} for derivation.  Note that the minimum
condition should be understood as the extremum in ${\sf Q}$ and
non-negativity of the eigenvalues of the Hessian in terms of the
matrix elements of $\sf Q$.  Marginal linear stability is allowed, and
as we shall see later, will occur in some phases.  Once the $\sf Q$
matrix is appropriately parametrized then in terms of those parameters
the extremum becomes a maximum for $n<1$.  So the minimum condition
above is meant before such a parametrization is applied.  These
considerations deal with the consequences of our having interchanged
the limits $N\to\infty$ and $n\to 0$ \cite{sgrev87}.

The entropic term $f_\rs$, for which we used the concise form from
\cite{gt90}, is specific to the spherical model.  On the other hand,
the energy-term $f_\re$, first displayed in \cite{mez93}, is independent
of the prior constraint for the synapses.  The $n \times n$ matrix
$\sf Q$ has been introduced through the constraint \cite{eg88,gar89b}
\begin{equation}
  \label{syn-overlap}
  \left[{\sf Q}\right]_{ab} \equiv q_{ab}
  = \frac{1}{N} \sum_{k=1}^N J_{ak}J_{bk}, 
\end{equation}
{\em i.\ e.}, $\sf Q$ is the matrix of the overlaps of the synaptic
couplings, is symmetric and positive semidefinite, with uniform
diagonal elements $q_{aa}\equiv q_\rD =1$ and $-1 \leq q_{ab}\leq 1$.
Here the indices $a, b=1,\dots n$ are so called replica indices; a
quantity with label $a$ belongs to the $a$-th factor in the power
$Z^n$ of the partition function $Z$.  Any quantity carrying a replica
index is intimately related to the replica method, and its
observability needs to be clarified extra.  Only the off diagonals,
$q_{ab}$ with $a\neq b$, entail minimum conditions.  Let us introduce
the mean of some function $A (\xv, \yv)$ as
\begin{equation}
  \label{mean-e} \lav A (\xv, \yv) \rav _\re = e^{n\beta
   f_\re({\sf Q})} \int \frac{\rd^n\! x \ \rd^n\! y}{(2\pi)^n} A (\xv,
   \yv) \exp\left( -\beta \sum\nolimits_{a=1}^n V(y_a)+ i\xv \yv -
   \frac{1}{2} \xv {\sf Q}\xv\right), 
\end{equation}
where the prefactor ensures $\lav 1\rav_e=1$, and the subscript refers
to the fact that the expectation value is associated with the energy
term. Then, using
\begin{equation}
  \label{der-lndet}
  \frac{\ptl \ln\text{det}{\sf Q}}{\ptl q_{ab}} 
  = \left[{\sf Q}^{-1}\right]_{ab} \equiv q^{-1}_{ab},
\end{equation} 
we obtain
\begin{equation}
  \label{stat-q}
q^{-1}_{ab} = \alpha \lav x_a x_b \rav _\re,  \; a\neq b
\end{equation} 
as the extremum condition in terms of $q_{ab}$.

If the prior distribution is like (\ref{gen-prior}) then
\bml
  \label{fe-gen}
  \begin{eqnarray}
    \label{fe-gena}
  f & = & \lim_{n\to 0} \frac{1}{n}\  
  {\ba{c}\text{\footnotesize ~} \\ \text{min}\\ 
    \text{\footnotesize $\sf Q$ }\ea} 
  {\ba{c}\text{\footnotesize ~} \\ \text{extr}\\ 
    \text{\footnotesize $\sf \hat{Q}$}\ea} 
  f({\sf Q,\hat{Q} }),  \\
    \label{fe-genb} 
    f({\sf Q,  \hat{Q}}) & = & f_{i}({\sf Q, \hat{Q}}) 
    + \hat{f}_{\rs}({\sf \hat{Q}}) + \alpha\, f_\re({\sf Q}) ,\\
  \label{fe-genc}
  f_{\mathrm i}({\sf Q, \hat{Q}}) & = & \frac{\beta}{2} \text{Tr}{\sf Q} 
  {\sf \hat{Q}} , \\
  \label{fe-gend}
  \hat{f}_\rs({\sf \hat{Q}}) & = & - \frac{1}{\beta} \ln \int e^{
  \frac{1}{2}\beta^2 {\jvi} {\sf \hat{Q}} {\jvi}} 
  \prod_{a=1}^n w_0(J_a) \rd J_a 
\end{eqnarray}
\eml and $f_\re({\sf Q})$ is given by (\ref{fe-spher1d}).  The special
case of Ising synapses (\ref{bin-prior}) gives the free energy in
\cite{gd88,km89}. Besides $\sf Q$ there is now another symmetric
auxiliary matrix, $\sf \hat{Q}$, whose diagonals are
$\hat{q}_{aa}\equiv \hat{q}_\rD =0$.  The derivation of the above free
energy is given in Appendix \ref{app-repl-fe}.  We emphasize that the
type of extremum in ${\sf \hat{Q}}$ is not restricted to minimum, see
the argument below Eq.\ (\ref{repl-partfunc4}).

The interaction term $f_{\mathrm i}$ together with the entropic
term, here $\hat{f}_\rs$, at extremum corresponds to the entropic term
(\ref{fe-spher1c}) in the spherical case.  If we introduce the mean
associated with the entropic term here as
\begin{equation}
    \label{mean-s} \lav A ({\jv}) \rav _\rs = e^{n\beta
    \hat{f}_\rs({\sf \hat{Q}})} \int A ({\jv})\, e^{
    \frac{1}{2} \beta^2 {\jvi}  {\sf \hat{Q}} {\jvi}} \prod_{a=1}^n
    w_0(J_a) \rd J_a 
\end{equation} 
then the stationarity condition in terms of a $\hat{q}_{ab}$ reads as
\begin{equation}
  \label{stat-q-gen}
q_{ab} = \lav J_a J_b \rav _\rs, 
\end{equation} 
and that by $q_{ab}$ gives
\begin{equation}
  \label{stat-qhat-gen}
  \beta^2 \hat{q}_{ab} = - \alpha \lav x_a x_b \rav _\re.
\end{equation}
For the diagonals are not varied, the above equations should hold only
for $a\neq b$.  Note that in the limit $n\to 0$ the normalization
coefficients in (\ref{mean-e},\ref{mean-s}) each become $1$, so for
most purposes those formulae can be understood as if the coefficients
were absent.

\subsection{Neural stabilities, errors, and overlaps}
\label{ssec-neural-stability}

The probability distribution of the neural stability parameter
$\Delta$ associated with stored patterns is given as \cite{gg91,mez93}
\begin{equation}
  \label{def-dist-stab}
  \rho(\Delta) = \left< \left< \delta \left( \Delta -h^1\xi^1\right)
      \right>_{\mathrm th}\right>_{\mathrm qu},
\end{equation}
where the formula for a $h^\mu$ in (\ref{defneu}) is understood.  Due
to permutation symmetry among patterns there is no loss of generality
in our selecting the first pattern in the definition above.  The above
definition is obviously independent of the replica method.  This
however, can be used to calculate the distribution of stabilities.
Replacing $Z^{-1}$ by $Z^{n-1}$ in the thermal average in
(\ref{def-dist-stab}), keeping in mind that in the end the $n\to 0$
limit should be taken, we recognize that (\ref{def-dist-stab}) is
technically a little modified version of the partition function
integral.  The calculation is in analogy with the derivation of
(\ref{fe-spher1d}), the latter shown in Appendix \ref{app-repl-fe},
and we end up with
\begin{equation}
  \label{repl-dist-stab}
   \rho(\Delta) = \lav \delta \left( \Delta - y_1 \right)
   \rav _\re \big| _{n=0} ,
\end{equation}
where any replica index other than $1$ could equally be chosen.  Thus
the average of an arbitrary function $U(y_a)$, $a$ arbitrary but
fixed, can be written in the form  
\begin{equation} 
  \label{any-fcn-dist}
  \lav U \left( y_a \right) \rav _\re \big| _{n=0} 
  = \int \rd y \ \rho(y)\, U(y),
\end{equation}
where $\Delta$ as integration variable was replaced by $y$.  An
instructive formula for the mean error is obtained in terms of the
distribution of neural stabilities as
\begin{equation} 
  \label{ener-dist}
  \varepsilon = \lav V \left( y_1 \right) \rav _\re \big| _{n=0} 
  = \int \rd y \ \rho(y)\, V(y).
\end{equation}
The first equality comes about from the definition (\ref{ener}), the
energy term (\ref{fe-spher1}), and the notation (\ref{mean-e}), while
the second follows from (\ref{any-fcn-dist}).  In case of replica
symmetry this expression goes over to the mean error displayed in
\cite{gg91}.  In fact, (\ref{any-fcn-dist}) allows us to use error
measures that are not related to the thermodynamic energy.  One can
take a $V(y)$ in the Hamiltonian and define the observable error by
another $U(y)$ measure.  This was done, without assuming replica
symmetry, with $U(y)=\theta(\kappa-y)$ in \cite{mez93}.  Eq.\ 
(\ref{ener-dist}) holds for both constraints (\ref{spher-prior}) and
(\ref{gen-prior}).  In the second case this is due to the fact that
$\beta^2$ can be absorbed into $\hat{q}_{ab}$, consequently $\beta$
times (\ref{fe-genc}) and (\ref{fe-gend}) both become $\beta$
independent.  Thus only $f_e({\sf Q})$ enters the thermodynamical
formula (\ref{ener}) for the mean error, yielding (\ref{ener-dist}).

The overlaps $q_{ab}$ emerged from the replica theory as auxiliary
variables, with no prescription how to measure them.  In analogy to
spin glasses, where the Edwards--Anderson overlap of spins $q_{EA}$
has been defined independently of replicas \cite{ea75}, one can
introduce an Edwards--Anderson overlap of synapses
\begin{equation}
  \label{ea-def}
  q_{\mathrm EA} =  \left< \frac{1}{N} \sum_{k=1}^N \left< J_k\right>_{\rm th}^2 
  \right>_{\rm qu}. 
\end{equation}
If the summation produces a self-averaging quantity, then the quenched
average can be omitted from the definition, but eventually the same
formula holds.  Replacing the $Z^{-2}$ that appears in the thermal
averages by $Z^{n-2}$, we can again apply the replica method.
Attaching the $a=1$ and $a=2$ replica indices to the synaptic
couplings in (\ref{ea-def}), and carrying out a calculation analogous
to that in the case of the local stability distribution, we have
\begin{equation}
  \label{ea}
  q_{\mathrm EA} = q_{12}.
\end{equation}
This is valid for both the spherical and the independently distributed
synapses.  The ambiguity in this expression is obvious: there was some
arbitrariness in selecting the $1$st and $2$nd replica indices for the
the synaptic couplings in (\ref{ea-def}) and labeling the other $n-2$
replicas starting from $a=3$.  In the terminology of replica theory of
spin glasses, the result (\ref{ea}) is in fact the overlap within one
pure thermodynamical state \cite{sgrev87}.  A detailed study of the
probability distributions of overlaps in multiple thermodynamical
states (see \cite{mpstv84} on the SK model) for the neuron is beyond
our present scope.  Nevertheless, in the analysis of correlation
functions the consequences of ultrametricity as described in Ref.
\cite{mv85} will be recovered.  Actually, for special error measures
complex structures can arise in the neuron even without spontaneous
replica symmetry breaking \cite{bmz93,mok94}.  In summary, based on
the manifold of analogies with spin glasses that we expound later in
the paper, we expect that several aspects of the physical
interpretation of the replica theory for spin glasses carry over to
the storage problem of the neuron, even when in the latter replica
symmetry is spontaneously broken.

A moral of this subsection is that the replica method enables us not
only the calculation of extensive quantities, like the free energy,
but also the evaluation of local quantities.  Technically, this is due
to the fact that replicas are useful in taking the average of, besides
the logarithm, also inverse powers of the partition function.


\eject
\section{The Parisi Solution} 
\label{sec-PA-Gen}

\subsection{Finite replica symmetry breaking}
\label{ssec-finite-rsb}

\subsubsection{Recursive evaluation of the free energy term} 
\label{sssec-free-ener-term}

Below we resolve the ``hard'' terms in the free energies, namely,
expressions (\ref{fe-spher1d}) and (\ref{fe-gend}). The derivation
follows the spirit of Parisi's as described concisely for the SK model
in Ref.\ \cite{par80a}.  The added aim here is to present the Parisi
solution in a comprehensive, self-contained manner.  Later we will be
rewarded for this approach, because the calculation of expectation
values shall follow straightforwardly.
  
Our main concern here is 
\begin{equation}
    \varphi[\varPhi(y),{\sf Q}] = \frac{1}{n} \ln \int \frac{\rd^n\! x \,
 \rd^n\! y}{(2\pi)^n} \exp\left(\sum\nolimits_{a=1}^n \varPhi (y_a) +
 i\xv \yv  - \frac{1}{2} \xv {\sf Q}\xv \right) .  
\label{gen-parisi}
\end{equation} 
Whereas this formula would look simpler, if the Fourier transform of
$e^{\varPhi(y)}$ were used, we keep the above notation, because it is the
function $\varPhi(y)$ that will explicitly appear in the final
evaluation.  Both Eqs.\ (\ref{fe-spher1d}) and (\ref{fe-gend}) are of
this type.  Eq.\ (\ref{fe-spher1d}) corresponds to
\begin{equation} 
  \label{e-phi} 
   -\beta f_e({\sf Q}) =  n \varphi[-\beta V(y),{\sf Q}] ,
\end{equation} 
and (\ref{fe-gend}) is obtained by 
\begin{equation} 
  \label{s-phi} 
   - \beta \hat{f}_s({\sf \hat{Q}}) =  n \varphi \left[
    \ln \int \rd x ~w_0(x)~ e^{-\beta yx} , {\sf \hat{Q}} \right] .
\end{equation} 
Note that there are no {\em a priori}\/ bounds for $\hat{q}_{ab}$ and
the diagonal elements $\hat{q}_{aa}$ vanish.  Furthermore, the
function $w_0(x)$ is assumed to cut off sufficiently fast so that
(\ref{s-phi}) exists.  Later in this paper, however, when an integral
expression is displayed and we do not discuss divergence that means we
assume the conditions of finiteness hold.  This does not mean that
under other conditions the expression could not diverge.

We will call (\ref{gen-parisi}) the ``free energy term'', it is
ubiquitous also in long range interaction spin glass models.  We just
have seen that the free energy of the neuron with arbitrary,
independently distributed synapses contains two additive terms of the
type of Eq.\ (\ref{gen-parisi}). We will see later that the spherical
entropic term is also of this type, so the free energy of the
spherical neuron is the sum of two terms of the type
(\ref{gen-parisi}).  Most of the mathematical parts of this paper are
centered about the evaluation of expression (\ref{gen-parisi}).
 
Due to the absence of inherent topology in infinite range spin glass
models, the replica approach led there to a single-site effective free
energy.  For such problems the ansatz of Parisi's turned out to be a
very successful mathematical framework, presently the stepping stone
to the field theory of the spin glass transition (see references in
\cite{dtk98}).  Since the present neuron problem is {\em a priori}\/
single-site, it is reasonable to search for the ${\sf Q}$ minimizing
(\ref{fe-spher1a}) by using Parisi's hierarchical assumption, which
reads as
\begin{equation} 
  \label{pa} 
   {\sf Q} = \sum_{r=0}^{R+1} \left( q_r-q_{r-1}\right)  {\sf U}_{m_r}
\otimes  {\sf I}_{n/m_r},
\end{equation}
where the subscript $k$ to a matrix marks that it is $k\times k$,
${\sf I}_{k}$ is the unit matrix and ${\sf U}_{k}$ has all elements
equal unity, furthermore, 
\bml\label{ieq}
\begin{eqnarray}
  \label{ieq0}
  q_{-1} & = & 0, \;q_{R+1} = q_\rD ,\\
  \label{ieq2}
  m_{R+1} & = & 1 \leq m_R \leq m_{R-1} \dots \leq m_1 \leq m_0 = n,
\end{eqnarray}
\eml where the integer $m_r$ is a divisor of $m_{r-1}$.  In the case
of the $\sf Q$ of Section \ref{sec-thermodyn} there is a presumed
ordering
\begin{equation}
  \label{ieq1}
  q_{-1}  = 0 \leq q_0  \leq q_1 \dots \leq q_R  \leq q_{R+1} =1
\end{equation}
In theory, $q_r<0$ are also possible, but in our numerical
explorations of examples such $q_r$-s did not appear, so we shall
consider the restriction to nonnegative $q$-s part of the ansatz.  For
$\sf \hat{Q}$ of Section \ref{sec-thermodyn} the assumption is
\begin{equation}
  \label{ieq3}
  \hat{q}_{-1} = 0 \leq \hat{q}_0  \leq\hat{q}_1 \dots \leq
  \hat{q}_R, \;\hat{q}_{R+1} =0 .
\end{equation}
These represent the $R$-step replica symmetry breaking scheme
($R$-RSB). At this stage we do not prescribe the ordering of $q_r$-s
and allow uniform diagonals $q_{aa}\equiv q_\rD $ of any magnitude.

The quadratic form in (\ref{gen-parisi}) is then
\begin{equation}
  \label{quad}
  \xv {\sf Q}\xv = \sum_{r=0}^{R+1}
\left( q_r-q_{r-1}\right) \sum_{j_r=1}^{n/m_r}\left( \sum_{a=m_r(j_r
    -1)+1}^{j_r m_r} x_a\right)^2.
\end{equation}
The $\varphi[\varPhi(y),{\sf Q}]$ of (\ref{gen-parisi}) should thus be replaced by
$\varphi[\varPhi(y),\qv,\mv)]$, where the parameters in (\ref{pa}) are
considered to be the elements of the vectors in the argument.  By
using the notation
\begin{equation}
 \label{dz}
\rD z = \frac{\rd z~ e^{-\hf z^2}}{\sqrt{2\pi}} 
\end{equation}
and the identity 
\begin{equation}
\label{gauss-meas}
e^{-\hf Ax^2}=\int \rD z~ e^{-izx\sqrt{A}}
\end{equation}
we obtain  
\begin{eqnarray}
 e^{n\varphi[\varPhi(y),\mbox{\scriptsize\boldmath$q$},\mbox{\scriptsize\boldmath$m$}]} 
  & = & \int \left[ 
  \prod_{r=0}^{R+1} \prod_{j_r=1}^{n/m_r} \rD z_{j_r}^{(r)} \right]
  \frac{\rd^n\! x \ \rd^n\! y}{(2\pi)^{n}} \nn \\ \label{temp1} & & \times
  \exp\left(-i\sum_{r=0}^{R+1} \sqrt{q_r-q_{r-1}}
  \sum_{j_r=1}^{n/m_r}z_{j_r}^{(r)} \sum_{a=m_r(j_r -1)+1}^{j_r m_r}
  x_a \right. \nn \\ && \qquad + \left. i\sum_{a=1}^nx_ay_a +
  \sum_{a=1}^n \varPhi(y_a)\right).
\end{eqnarray}
Appendix \ref{app-sec-deriv-r-rsb} shows that the above expression
equals Eq.\ (\ref{rrsb}).  

The limit $m_0=n\rightarrow 0$ violates the ordering in (\ref{ieq2}).
In fact, experience in spin glasses \cite{sgrev87,sgrev91} and in
$R$-RSB, $R=1,2$ calculations in neural networks (see
\cite{mez93,ws96}) suggests that $m_r$-s get less than $1$ and the
ordering in (\ref{ieq2}) is to be reversed.  This can be understood by
our introducing
\begin{equation}
\label{x-r}
x_r = \frac{n-m_r}{n-1}
\end{equation}
for arbitrary $n$ and using the $x_r$-s for parametrization instead of
the $m_r$-s.  The new parameter $x_r$ should not be confounded with
the integration variable $x_a$ in Eq.\ (\ref{gen-parisi}).  For
integer $n$ and $m_r$-s satisfying (\ref{ieq2}) we have the ordering 
\begin{equation}
  \label{ieq2b} x_{R+1} = 1 \geq x_R \geq x_{R-1} \dots \geq x_1 \geq
   x_0 = 0. 
\end{equation}
Keeping the $x_r$-s fixed as $n\to 0$ defines the $n$-dependence of
the $m_r$-s, and for $n=0$ formally we get $x_r=m_r$.  This explains
the aforementioned practice to treat the $m_r$-s as real numbers in
$[0,1]$ with ordering reversed \wrt (\ref{ieq2}). 

Eq.\ (\ref{rrsb}) becomes for $n\to 0$, in terms of the $x_r$-s,
\begin{eqnarray}
 \label{phirrsb} \varphi[\varPhi(y),\qv,\xv] \big| _{n=0}
 &=& \frac{1}{x_1}\int \rD z_0 \ln \int \rD z_1 \nn \\ && \times \left[ \int
 \rD z_2 \dots \left[ \int \rD z_{R+1}\exp \varPhi\left(\sum_{r=0}^{R+1} z_r
 \sqrt{q_r-q_{r-1}} \right)\right] ^\frac{x_R}{x_{R+1}} \dots
 \right]^\frac{x_{1}}{x_{2}}.
\end{eqnarray} 
This is the general formula for $R$-RSB. Expression (\ref{phirrsb})
can be written in form of an iteration for decreasing $r$-s as
\bml
  \label{rec-psi}
\begin{eqnarray}
  \label{rec-psi1} \psi_{r-1}(y) & = & \int \rD z~
  \psi_r\left(y+z\sqrt{q_r-q_{r-1}}\right) ^ \frac{x_r}{x_{r+1}}, \\
  \label{rec-psi2} \psi_R(y) & = & \int \rD z~ e^{\varPhi\left(y+z
  \sqrt{q_{R+1}-q_{R}}\right)},
\end{eqnarray}
\eml or, we can set $x_{R+2}=1$ and put the initial condition as
\begin{equation}
  \label{rec-alt-init}
    \psi_{R+1}(y) =  e^{\varPhi(y)}.
\end{equation}
In the iterated function we omitted to mark the functional dependence
on $\varPhi(y)$ and $\qv, \xv$.  If a $q_r-q_{r-1}<0$ then the square
root is imaginary.  Since the Gauss measure of integrations suppresses
odd powers in a Taylor expansion of the integrand, the result, if the
integrals exist, will be real. The case of non-monotonic $q_r$
sequence will be briefly discussed in the end of this section.  Then
\begin{equation}
  \label{ferecg}
 \varphi [\varPhi(y), \qv,\xv]\big| _{n=0} = \frac{1}{x_1}
  \int \rD z~
  \ln \psi_0\left(z\sqrt{q_0}\right).
\end{equation}
 
Note that an iteration like (\ref{rec-psi}) can be also understood,
before the $n\to 0$ limit is taken, directly on Eq.\ (\ref{rrsb})
where $m_r/m_{r+1}$ is integer.  Then formally $\varphi [\varPhi(y),
\qv,\xv] = n^{-1} \ln\psi_{-1}(0)$.  Hence for $n\to 0$ we recover
(\ref {ferecg}).  It is, however, an advantage that we can first take
$n\to 0$ then define the recursion (\ref{rec-psi}) with fractional
powers.  Indeed, while dealing with the consequences of the recursion,
the replica limit $n\to 0$ is implied and we do not have to return to
the question of that limit again.

It is instructive to introduce 
\begin{equation}
  \label{conv}
  \varphi_r(y) = \frac{\ln\psi_r(y)}{x_{r+1}},
\end{equation}
lending itself to the recursion
\bml
  \label{recf}
\begin{eqnarray}
  \label{recf1} \varphi _{r-1}(y) & = & \frac{1}{x_r} \ln \int \rD z\,
  e^{x_r \varphi_r(y +z\sqrt{q_r-q_{r-1}})} \\ \label{recf2}
  \varphi_{R+1}(y) & = & \varPhi(y) ,
\end{eqnarray}
\eml
and yielding 
\begin{equation}
  \label{ferecf}
 \varphi [\varPhi(y), \qv,\xv] \big| _{n=0} = 
 \int \rD z~ \varphi_0\left(z\sqrt{q_0}\right) .
\end{equation}
 
\subsubsection{Parisi's PDE} 
\label{sssec-ppde}
 
The above recursions can be viewed as diffusion processes in the
presence of ``kicks''.  Let us introduce here Parisi's order parameter
function (OPF) as
\begin{equation}
  \label{xq}
 x(q)  = \sum_{i=0}^R (x_{i+1}-x_i)\, \theta(q-q_i),
\end{equation}
defined on the interval $[0,1]$, where (\ref{ieq1}) and (\ref{ieq2b})
are understood.  With the standard notation
\begin{equation}
  \label{q+} f(q^{+0}) = \lim_{\epsilon\to 0} f(q+\epsilon),
\end{equation}
we have obviously
\begin{equation}
  \label{x_qr} x(q_r^{+0}) = x_{r+1}, \qquad x(q_r^{-0}) = x_{r},
\end{equation}
and we may set
\begin{equation}
  \label{x_qr2} x(q_r) = x_{r+1}.
\end{equation}

\begin{figure}[h!]
\begin{center}
\setlength{\unitlength}{0.00087489in}
\begingroup\makeatletter\ifx\SetFigFont\undefined%
\gdef\SetFigFont#1#2#3#4#5{%
  \reset@font\fontsize{#1}{#2pt}%
  \fontfamily{#3}\fontseries{#4}\fontshape{#5}%
  \selectfont}%
\fi\endgroup%
{\renewcommand{\dashlinestretch}{30}
{\small%
\begin{picture}(4523,3775)(0,-10)
\put(1767,2623){\ellipse{90}{90}}
\put(1767,1813){\ellipse{90}{90}}
\path(1767,3298)(3522,3298)
\path(1767,2623)(327,2623)
\path(12,418)(4017,418)
\path(3897.000,388.000)(4017.000,418.000)(3897.000,448.000)
\path(1767,508)(1767,373)
\dashline{60.000}(1767,418)(1767,3748)
\path(1927.468,2087.948)(1808.000,2120.000)(1898.330,2035.498)
\path(1808,2120)(2213,1895)
\path(1227,1273)(1722,1768)
\path(1658.360,1661.934)(1722.000,1768.000)(1615.934,1704.360)
\thicklines
\path(1767,2173)(1769,2173)(1772,2174)
	(1778,2175)(1788,2178)(1802,2181)
	(1820,2185)(1843,2189)(1870,2195)
	(1902,2202)(1937,2210)(1976,2218)
	(2017,2227)(2061,2236)(2106,2246)
	(2152,2255)(2198,2265)(2243,2274)
	(2288,2284)(2331,2292)(2372,2301)
	(2412,2309)(2450,2316)(2486,2323)
	(2520,2329)(2552,2335)(2583,2340)
	(2612,2345)(2640,2349)(2667,2353)
	(2698,2357)(2729,2361)(2760,2364)
	(2792,2368)(2825,2370)(2860,2373)
	(2895,2376)(2933,2378)(2971,2380)
	(3011,2383)(3051,2385)(3092,2386)
	(3132,2388)(3171,2390)(3209,2391)
	(3244,2393)(3276,2394)(3305,2395)
	(3329,2396)(3348,2397)(3364,2397)
	(3374,2398)(3381,2398)(3385,2398)(3387,2398)
\path(1767,1808)(1765,1808)(1761,1808)
	(1754,1808)(1742,1807)(1726,1807)
	(1705,1806)(1678,1805)(1647,1804)
	(1612,1803)(1572,1801)(1529,1800)
	(1484,1798)(1438,1796)(1390,1794)
	(1343,1792)(1296,1790)(1251,1788)
	(1207,1786)(1166,1784)(1126,1781)
	(1089,1779)(1055,1777)(1022,1774)
	(992,1772)(964,1769)(937,1766)
	(912,1763)(881,1759)(850,1754)
	(820,1748)(789,1742)(758,1736)
	(726,1728)(693,1720)(659,1711)
	(624,1701)(589,1692)(554,1682)
	(521,1672)(489,1663)(461,1655)
	(436,1647)(415,1641)(398,1636)
	(386,1632)(378,1630)(374,1629)(372,1628)
\path(374,793)(376,793)(380,793)
	(387,794)(399,794)(415,795)
	(436,796)(463,797)(494,799)
	(529,801)(569,803)(612,805)
	(657,807)(703,810)(751,812)
	(798,814)(845,817)(890,819)
	(934,821)(975,824)(1015,826)
	(1052,828)(1086,830)(1119,831)
	(1149,833)(1177,835)(1204,836)
	(1229,838)(1260,840)(1291,842)
	(1321,844)(1352,847)(1383,849)
	(1415,852)(1448,854)(1482,857)
	(1517,860)(1552,863)(1587,867)
	(1620,869)(1652,872)(1680,875)
	(1705,877)(1726,879)(1743,881)
	(1755,882)(1763,882)(1767,883)(1769,883)
\path(1775,879)(1777,879)(1780,880)
	(1787,882)(1798,885)(1813,889)
	(1832,894)(1856,900)(1886,907)
	(1919,916)(1957,925)(1998,936)
	(2043,947)(2089,959)(2137,970)
	(2186,983)(2235,995)(2283,1007)
	(2330,1018)(2376,1029)(2419,1040)
	(2461,1049)(2500,1059)(2538,1067)
	(2573,1075)(2605,1082)(2636,1088)
	(2666,1094)(2693,1099)(2720,1104)
	(2752,1109)(2784,1114)(2816,1118)
	(2848,1122)(2881,1125)(2915,1128)
	(2949,1131)(2985,1134)(3022,1136)
	(3060,1138)(3097,1140)(3135,1141)
	(3171,1143)(3205,1144)(3237,1145)
	(3265,1146)(3290,1147)(3310,1148)
	(3326,1148)(3337,1149)(3344,1149)
	(3348,1149)(3350,1149)
\put(3837,103){\makebox(0,0)[lb]{\smash{{{\SetFigFont{10}{14.4}{\rmdefault}{\mddefault}{\updefault}$q$}}}}}
\put(1632,58){\makebox(0,0)[lb]{\smash{{{\SetFigFont{10}{14.4}{\rmdefault}{\mddefault}{\updefault}$q_r$}}}}}
\put(372,1228){\makebox(0,0)[lb]{\smash{{{\SetFigFont{10}{14.4}{\rmdefault}{\mddefault}{\updefault}$\psi(q_r^{-0},y)$}}}}}
\put(2352,2443){\makebox(0,0)[lb]{\smash{{{\SetFigFont{10}{14.4}{\rmdefault}{\mddefault}{\updefault}$\psi(q,y)$}}}}}
\put(552,2758){\makebox(0,0)[lb]{\smash{{{\SetFigFont{10}{14.4}{\rmdefault}{\mddefault}{\updefault}$x(q)=x_{r}$}}}}}
\put(2037,3433){\makebox(0,0)[lb]{\smash{{{\SetFigFont{10}{14.4}{\rmdefault}{\mddefault}{\updefault}$x(q)=x_{r+1}$}}}}}
\put(2262,1768){\makebox(0,0)[lb]{\smash{{{\SetFigFont{10}{14.4}{\rmdefault}{\mddefault}{\updefault}$\psi(q_r^{+0},y)=\psi_r(y)$}}}}}
\put(2352,913){\makebox(0,0)[lb]{\smash{{{\SetFigFont{10}{14.4}{\rmdefault}{\mddefault}{\updefault}$\varphi(q,y)$}}}}}
\end{picture}
}
}
\end{center}
\vskip0.2in
\caption{Schematic behavior  of $x(q)$, $\psi(q,y)$, and
  $\varphi(q,y)$ at a discontinuity point $q_r$. A fixed $y$ is
  assumed. The function $\varphi(q,y)$ is continuous in $q$ but has a
  discontinuous derivative.  The two limits of $\psi(q_r,y)$ are
  related through Eq.\ (\ref{psi}). The circles are placed where
  the function value is not taken as the limit. }
\label{fig-discontinuity}
\end{figure}
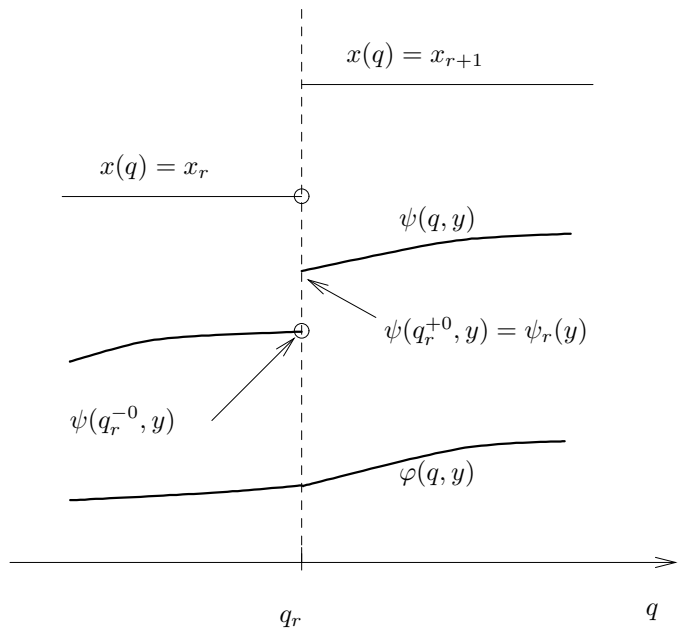

Next we introduce the field $\psi(q,y)$ such that at $q_r$ it has the
discontinuity \bml \label{psi}
\begin{eqnarray}
  \label{psiplus}
 \psi(q_r^{+0},y) & = & \psi_r(y) ,\\
  \label{psiminus}
 \psi(q_r^{-0},y) & = & \psi_r(y)^\frac{x(q_r^{-0})}{x(q_r^{+0})} .
\end{eqnarray}
\eml 
In other words, 
\begin{equation}
  \label{psi_q}  \psi(q,y)^\frac{1}{x(q)} 
\end{equation}
is continuous in $q$.  We may set at the discontinuity 
\begin{equation}
  \label{psi_q-discontinuity}
\psi(q_r,y) = \psi_r(y).
\end{equation}
A graphic reminder to the way $x(q)$ and $\psi(q,y)$ are defined at
the discontinuity is shown on FIG. \ref{fig-discontinuity}\@.  Note
that $r$ was converted to $q$ differently for $x_r$ and $\psi_r$, {\it
  cf.\ } Eqs.\ (\ref{x_qr2}) and (\ref{psi_q-discontinuity}).  All
fields appearing below follow the convention
(\ref{psiplus},\ref{psi_q-discontinuity})).

\vskip0.5in
\vskip0.5in

In the interval $(q_{r-1},q_r)$ we define the $\psi(q,y)$ based on
(\ref{rec-psi1}) as
\begin{equation}  
\label{psi-cont}
  \psi(q,y) = \int \rD z~ \psi\left(q_r^{-0},y+z\sqrt{q_r-q}\right),
\end{equation}
ensuring that (\ref{psiplus}) holds for $r\to r-1$.  Relation
(\ref{psi-cont}) says that the $\psi(q,y)$ evolves in the open
interval from $q_{r}$ to $q_{r-1}$ by the linear diffusion equation
\begin{equation}  
\label{psi-lin-diff}
    \ptl _q \psi = -\hf \ptl _y ^2 \psi.
\end{equation}

Near the discontinuity of $x(q)$ another differential equation can be
derived.  Let us differentiate Eq.\ (\ref{psi_q}) by $q$ as
\begin{equation}  
\label{dif-psi_q} 
    \ptl _q \psi^\frac{1}{x} =
    \frac{1}{x}\psi^\frac{1-x}{x} \left( \ptl _q \psi -
    \frac{\dot{x}}{x}\psi \, \ln\psi \right).
\end{equation} 
Since $\psi(q,y)^\frac{1}{x(q)}$ is continuous in $q$ while
$\psi(q,y)$ and $x(q)$ are not, the two singular derivatives on the
\rhs must cancel in leading order.  Hence we obtain
\begin{equation}  
\label{psi-kick} 
    \ptl _q \psi = \frac{\dot{x}}{x}\psi \, \ln\psi
\end{equation} 
in an infinitezimal neighborhood of $q_r$. The above derivation is
apparently unfounded, because at a discontinuity the rules of
differentiation used in (\ref{dif-psi_q}) loose meaning.  However,
considering (\ref{psi-kick}) at a fixed $y$ as an ordinary
differential equation separable in $q$ helps us through the
discontinuity, and we obtain
\begin{equation}  
\label{psi-discontinuity3}
 \int_{\psi(q_r^{-0},y)}^{\psi(q_r^{+0},y)}
 \frac{\rd\psi}{\psi \ln\psi} =  \int_{x(q_r^{-0})}^{x(q_r^{+0})}
 \frac{\rd x}{x}.
\end{equation}
The integrals yield
\begin{equation}  
 \label{psi-discontinuity2}
  \ln \ln \psi(q,y) \Big|_{q_r^{-0}}^{q_r^{+0}}
 =\ln x(q)\Big|_{q_r^{-0}}^{q_r^{+0}},
\end{equation}
whence by exponentiating twice we recover the continuity condition for
(\ref{psi_q}).  In conclusion, for a discontinuous $x(q)$ equation
(\ref{psi-kick}) can indeed be interpreted as the differential form of
the prescription that (\ref{psi_q}) is continuous in $q$.

Concatenation of (\ref{psi-lin-diff}) and (\ref{psi-kick}) gives, with
regard to the initial condition (\ref{rec-alt-init}), the PDE
\bml 
 \label{psi-pde}
 \begin{eqnarray}
  \label{psi-pde1}
    \ptl _q \psi & = & -\hf \ptl _y ^2 \psi + \frac{\dot x }{x}  \psi
    \ln \psi, \\
  \label{psi-pde2}
  \psi(1,y) & = & e^{\varPhi(y)}.
\end{eqnarray}
\eml Indeed, at a $q_r$ the $\dot{x}(q)$ is singular, so the second
term on the \rhs dominates and we recover (\ref{psi-kick}),
whereas within an interval $\dot{x}(q)\equiv 0$ and thus
(\ref{psi-lin-diff}) holds.
 
The transformation analogous to (\ref{conv}) is 
\begin{equation} 
  \label{psi-to-phi} 
   \psi(q,y) = e^{\varphi(q,y) x(q)} ,
\end{equation} 
and gives rise to \bml 
  \label{ppde} 
\begin{eqnarray} 
  \label{ppde1} 
   \ptl _q \varphi &  = & -\hf \ptl _y ^2 \varphi 
   - \hf x \left(\ptl _y \varphi\right)^2 ,\\
  \label{ppde2} 
    \varphi(1,y) & = & \varPhi(y). 
\end{eqnarray} 
\eml It follows that when $x(q)$ has a finite discontinuity then the
field $\varphi(q,y)$ is continuous in $q$, as on FIG.
\ref{fig-discontinuity}.\@  This is in accordance with the condition
that formula (\ref{psiminus}) is continuous.
 
The PDE (\ref{ppde1}) can be rewritten via the transformation $q=q(x)$
to one evolving in $x$, a PDE first proposed by Parisi with a special
initial condition for the SK model \cite{sgrev87,sgrev91}.  In this
paper we refer to (\ref{ppde}) and its equivalents as Parisi's PDE,
PPDE for short. 
 
When $x(q)\equiv \text{const.}$, differentiation of the PPDE
(\ref{ppde}) in terms of $y$ gives the Burgers equation for the field
$\ptl _y\varphi$.  Then the derivative of Eq.\ (\ref{psi-to-phi}) by
$y$ corresponds to the Cole-Hopf transformation formula
\cite{hopf50,cole51}, which converts the Burgers equation into the PDE
for linear diffusion\footnote{E. Ott has kindly called our attention
  to the Cole-Hopf transformation.}, here (\ref{psi-pde}) with
$\dot{x}\equiv 0$.  If $x(q)$ is not a constant, (\ref{psi-to-phi})
connects two nonlinear PDE-s.  We shall refer to (\ref{psi-to-phi}) as
Cole-Hopf transformation.
 
In the case of a discontinuous initial condition $\varPhi(y)$ the
Cole-Hopf transformation (\ref{psi-to-phi}) connects two discontinuous
functions at $q=1$, while generically diffusion smoothens the
discontinuity for $q<1$.  Even if we succeed in defining the PDE-s for
non-differentiable initial conditions, the equivalence of
(\ref{psi-pde}) and (\ref{ppde}) is doubtful.  In case of ambiguity
precedence is taken by the PDE (\ref{psi-pde}), that directly follows
from the recursion (\ref{rec-psi}).  The question of discontinuity in
the initial condition will be discussed later.
 
Our main focus is the term (\ref{ferecf}), now also a functional of
$x(q)$ 
\begin{equation}
  \label{phi-from-pde}
   \varphi[\varPhi(y),x(q)] = \int \rD z~ \varphi(q_0,z\sqrt{q_0}) ,
\end{equation}
where $n=0$ is implied. Note that $x(q)\equiv 0$ in the interval
$(0,q_0)$, so (\ref{ppde1}) becomes the PDE for linear diffusion,
whose solution at $q=0$ is given by the \rhs of (\ref{phi-from-pde}),
thus
\begin{equation} 
  \label{fee-final}
   \varphi[\varPhi(y),x(q)] =\varphi (0,0) .
\end{equation}
 
In the above PDE-s $q$ is a time-like variable evolving from 1 to 0.
In the context of the PDE-s we will refer to $q$ as time, and ordinary
derivative by $q$ will be denoted by a dot.  The above PDE-s can be
considered as nonlinear diffusion equations in reverse time
direction.  
 
Next we study the case of $\hat{{\sf Q}}$ with Parisi elements obeying
(\ref{ieq3}).  Then the PDE obtained for the field
$\hat{\psi}(\hat{q},y)$ by continuation contains the function
$\hat{x}(\hat{q})$.  We obtain
 \bml
 \label{psihat-pde}
 \begin{eqnarray}
  \label{psihat-pde1}
    \ptl _{\hat{q}} \hat{\psi} & = & -\hf \ptl _y ^2 \hat{\psi} + 
    \frac{\dot {\hat{x}}}{\hat{x}}  \hat{\psi} \ln \hat{\psi}, \\
  \label{psihat-pde2}
  \hat{\psi}(\hat{q}_R,y) & = & \int \rD z\, e^{\varPhi\left(y+iz
  \sqrt{\hat{q}_R} \right)} ,
\end{eqnarray}
\eml where $\hat{\psi}(\hat{q}_R,y)$ is real due to the symmetry of
the $Dz$ measure.  Alternatively, with the Cole-Hopf transformation
(\ref{psi-to-phi}), we have \bml
 \label{phihat-pde}
 \begin{eqnarray}
  \label{phihat-pde1}
    \ptl _{\hat{q}} \hat{\varphi} & = & -\hf \ptl _y ^2 \hat{\varphi} 
    - \hf \hat{x} \left( \ptl _y \hat{\varphi} \right) ^2 , \\
  \label{phihat-pde2}
  \hat{\varphi}(\hat{q}_R,y) & = & \ln \int \rD z\, e^{\varPhi\left(y+iz
  \sqrt{\hat{q}_R} \right)}.
\end{eqnarray}
\eml 
The existence of the integral is a sensitive question here, because
the imaginary term in argument expresses the fact that $\exp\varPhi$ is
evolved by backward diffusion. The meaningfulness of the above
initial condition should be checked case-by-case.  Then the sought
term is
\begin{equation} 
  \label{feehat-final}
   \left. \varphi[\varPhi(y),{\sf \hat{Q}}] \right| _{n=0} =
   \varphi[\varPhi(y),\hat{x}(\hat{q})] = \hat{\varphi} (0,0) .
\end{equation}
In contrast to the PDE-s associated with the matrix $\sf Q$ of
naturally bounded elements, where the time span of the evolution is
the unit interval, in the case of $\sf \hat{Q}$ the PDE-s' evolution
interval is not fixed {\em a priori}.  Now $\hat{q}$ goes from
$\hat{q}_R$ to $0$, where $\hat{q}_R$ itself is a thermodynamical
variable subject to extremization.

Finally we emphasize that the recursive technique may be able to treat
non-monotonic $q_r$ sequences.  Indeed, if $q_r<q_{r-1}$ then an
imaginary term would multiply $z$ in the integrand on the \rhs of
(\ref{rec-psi1}), but the \lhs would have a real function.  If the
integrals involved exist then there is no obstacle to extend the
theory to non-monotonic $q_r$-s.  Such a case did not, however, arise
in our explorations.  As we shall see in Section \ref{ssec-meaning-xq},
the OPF $x(q)$ is a probability measure, a property that
non-monotonicity would contradict.  On the other hand, $\sf \hat{Q}$
can be considered as associated with a non-monotonic $\hat{q}_r$
sequence.  Its diagonals vanish, $\hat{q}_{aa}\equiv \hat{q}_{R+1}=0$,
and so the step from $\hat{q}_R$ to $\hat{q}_{R+1}=0$ goes against the
trend of the otherwise supposedly monotonic increasing $\hat{q}_r$
sequence, $r=0,\dots,R$.  Accordingly, an imaginary factor of $z$
appears on the \rhs of (\ref{phihat-pde2}), and the recursion is as
meaningful as it was in the case of a monotonic $q_r$ sequence.

The generalization of the picture above is straightforward to an
order parameter with more components, when the structure of the free
energy term remains essentially the same.  We briefly discuss this
case in Appendix \ref{app-high-dim}.


\subsection{Finite and continuous replica symmetry breaking} 
 \label{ssec-PDEs}

\subsubsection{The continuous limit} 
\label{sssec-continuousRSB}

If the minimum of the free energy is found at an OPF given in
(\ref{xq}) with $R=\infty$, then the $q_r$-s accumulate
infinitezimally densely in some region.  If this happens in an
interval, the OPF $x(q)$ is expected to increase there strictly
monotonically, given its physical interpretation as mean probability
distribution of the overlaps, as discussed in Section
\ref{ssec-meaning-xq}.  Within that interval the recursions go over to
the PDE-s of Section \ref{sssec-ppde}, which can be obtained in this
case in the spirit of the approach of Ref.\ \cite{dup80}, as discussed
in Appendix \ref{app-sec-duplantier}.  In other regions in $q$, where
the $x(q)$ remains a step function, the recursions discussed in
Section \ref{sssec-free-ener-term} can be used, but the PDE-s are also
still valid, as described in Section \ref{sssec-ppde}.  In either
case, the PDE-s are applicable independently of whether the minimizing
OPF is continuous or step-like.  

In physical systems so far, including spin glass and neural network
models, out of finite $R$-s only $R=0$ and $R=1$ RSB phases were found
thermodynamically stable.  The significance of $2\leq R<\infty$ RSB
seems to be in approximating the $R=\infty$ case.  Generically, both
finite and infinite $R$ states are characterized by the border values 
\bml
\label{q-extremities}
\begin{eqnarray}
\label{border-q0}
    q_{(0)} &=& \left\{ \begin{array} {ll} q_0 & \text{if}\qquad R <
         \infty\\ \lim_{R\to\infty} q_0 \quad& \text{if}\qquad R = \infty
         \end{array} \right., \\
\label{border-q1} q_{(1)} & =& \left\{ \begin{array}
         {ll} q_R & \text{if}\qquad R < \infty\\ \lim_{R\to\infty} q_R
         \quad& \text{if}\qquad R = \infty \end{array} \right.,
\end{eqnarray}
\eml
where $q_{(0)}\geq 0$ and $q_{(1)}\leq 1$.  These are delimiters of
the trivial plateaus of the OPF $x(q)$ as
\begin{equation}
\label{q-plateaus}
    x(q) \equiv \left\{ \begin{array} {ll} 0\quad & \text{if}\quad
        0\leq q < q_{(0)} \\ 1 & \text{if}\quad 1\geq q\geq q_{(1)}
        \end{array} \right..
\end{equation}
The border values (\ref{q-extremities}) apply to both the finite and
infinite $R$ cases, the difference remaining in the shape of the OPF
$x(q)$ within the interval $(q_{(0)},q_{(1)})$.  Here we assumed
$q_{R+1} = q_\rD = 1$, this makes the $q=1$ value special and we will
use that in the general discussion. 

When $R=\infty$ a typical situation is when extremization of the free
energy yields
\begin{equation}
\label{xq-parisi-phase}
    x(q) = \left\{ \begin{array} {ll} 0 & \text{if}\quad 0 \leq q <
         q_{(0)}, \\ x_\rc(q), \quad x_{(0)} \leq x_\rc(q) < x_{(1)}, \,
         0<\dot{x}_c(q)<\infty \quad & \text{if}\quad q_{(0)} \leq q<
         q_{(1)}, \\ 1 & \text{if}\quad q_{(1)} \leq q \leq 1.
         \end{array} \right.
\end{equation}
In words, the OPF has a strictly increasing, continuous segment
$x_\rc(q)$ between the border values (\ref{q-extremities}).  Here
$x_{(1)} = x(q^{-0}_{(1)})$. 

The case with an OPF having a smooth, strictly increasing, segment
$x_\rc(q)$ will be referred to as continuous RSB (CRSB).  Obviously,
CRSB always implies $R\to\infty$.  In principle, then the OPF may be
more complicated than (\ref{xq-parisi-phase})\eg there may be
nontrivial ($x\neq 0,1$) plateaus and several $x_\rc(q)$ segments
separated by them.  So far, however, no system was found whose replica
solution involved more than one strictly increasing segments $x_\rc(q)$,
separated by a plateau or a discontinuity.

In what follows we will use the term ``continuation'', when we
understand the $n\to 0$ limit, the usage of $x(q)$ based on Eqs.\
(\ref{x-r}, \ref{xq}), as well as we give allowance for but do not
necessarily imply CRSB.

If the OPF in question is $\hat{x}(\hat{q})$, defined analogously to
(\ref{xq}) with the parameters $\{\hat{q}_r,\hat{x}_r\}$, continuation
goes along similar lines.

\subsubsection{Derivatives of Parisi's PDE} 
 \label{sssec-deriv-PDEs}

The iterations derived in Section \ref{sssec-free-ener-term} only
describe finite $R$-RSB, including the $R=0$ replica symmetric case,
while the PDE-s incorporate both finite and continuous RSB.  We
therefore study the PDE-s.

For later purposes it is worth summarizing some PDE-s related to the 
PPDE (\ref{ppde}) and its Cole-Hopf transformed Eq.\
(\ref{psi-pde}). The field 
\begin{equation}
  \label{def-mu}
  \mu(q,y)=\ptl _y\varphi(q,y),
\end{equation}
satisfies the PDE 
\bml
\label{deriv1-ppde}
\begin{eqnarray}
\label{deriv1-ppde-evolv}
   \ptl _q \mu &  = & -\hf \ptl _y ^2  \mu -  x \mu \ptl _y \mu,\\
  \label{deriv1-ppde-init}
  \mu(1,y) & = & \varPhi^\prime(y),
\end{eqnarray}
\eml obtained from the PPDE by differentiation in terms of $y$.  One
more differentiation introduces
\begin{equation}
  \label{def-kappa}
  \kappa(q,y)=\ptl _y\mu(q,y), 
\end{equation}
which evolves according to
\bml
\label{deriv2-ppde}
\begin{eqnarray}
\label{deriv2-ppde-evolv}
   \ptl _q \kappa &  = & -\hf \ptl _y ^2  \kappa -  
   x \left(\kappa^2 +\mu\ptl _y\kappa  \right),\\
  \label{deriv2-ppde-init}
  \kappa(1,y) & = & \varPhi^{\prime\prime} (y).
\end{eqnarray}
\eml Note that while the PPDE (\ref{ppde}) and Eq.\ 
(\ref{deriv1-ppde}) are self contained equations, in principle
solvable for the respective fields, (\ref{deriv2-ppde}) is not such
and should rather be considered as a relation between the fields
$\mu(q,y)$ and $\kappa(q,y)$.

The Cole-Hopf transformation for the first derivative $\mu(q,y)$ can
be conveniently defined as the field $\mu(q,y)\psi(q,y)$.  This can be
further differentiated to produce the Cole-Hopf transformed field for
$\kappa(q,y)$.  The PDE-s for the transformed fields each reduce to
the linear diffusion equation along plateaus of $x(q)$.

\subsubsection{Linearized PDE-s and their adjoints} 
 \label{sssec-lin-PDEs}
 
As we shall see, in the calculation of expectation values linear PDE-s
associated with the above equations play an important role.  A
perturbation $\varphi(q,y)+\epsilon \vartheta(q,y)$, around a known
solution $\varphi(q,y)$ of the PPDE, itself satisfies the PPDE to
$O(\epsilon)$ if
\begin{equation} 
\label{theta-pde} 
\ptl _q \vartheta = -\hf\ptl _y^2\vartheta -x\mu\ptl _y\vartheta.
\end{equation} 
This equation is satisfied by $\mu(q,y)$ with initial condition
(\ref{deriv1-ppde-init}).  The field
\begin{equation}
  \label{def-eta}
  \eta(q,y)=\ptl _y\vartheta(q,y), 
\end{equation}
then evolves according to
\begin{equation}
  \label{eta-pde}
   \ptl _q \eta   = -\hf\ptl _y^2\eta -x\ptl _y(\mu\eta),
\end{equation}
obviously satisfied by $\kappa(q,y)$ if the initial condition is
specified by (\ref{deriv2-ppde-init}).

The field $P(q,y)$ adjoint to $\vartheta(q,y)$ and crucial in the
computation of expectation values can be introduced by the requirement
that
\begin{equation}
  \label{theta-p}
   \int \rd y \, P(q,y)\,\vartheta(q,y)
\end{equation}
be independent of $q$.  Differentiating by $q$, using Eq.\
(\ref{theta-pde}), and partially integrating with the assumption that
$P(q,y)$ decays sufficiently fast for large $|y|$, we wind up with
the PDE
\begin{equation}
  \label{spde} \ptl _q P = \hf\ptl _y^2P -x\ptl _y(\mu P).
\end{equation}
Here the time $q$ evolves in forward direction, from $0$ to $1$.  The
equivalent of the field $P(q,y)$, evolving from the initial condition
in our notation
\begin{equation}
  \label{spde-init}
   P(0,y)=\delta(y),
\end{equation}
was introduced by Sompolinsky in a dynamical context for the SK model
\cite{som81c}.  In this case the average (\ref{theta-p}) assumes the
alternative forms
\begin{equation}
  \label{theta-p2}
   \int \rd y \, P(q,y)\,\vartheta(q,y) \equiv \int \rd y \,
   P(1,y)\,\vartheta(1,y) =  \vartheta(0,0). 
\end{equation}
Eq.\ (\ref{spde}) is in fact a Fokker-Planck equation with
$x(q)\,\mu(q,y)$ as drift. The initial condition (\ref{spde-init}) is
normalized to $1$ and localized to the origin.  Hence follows the
conservation of the norm
\begin{equation}
  \label{norm-of-p} \int \rd y \, P(q,y) \equiv 1,
\end{equation}
and the non-negativity of the field $P(q,y)$.  Thus $P(q,y)$ can be
interpreted as a $q$-time-dependent probability density.  We will
refer to the initial value problem (\ref{spde},\ref{spde-init}), which
determines Sompolinsky's probability field $P(q,y)$, as Sompolinsky's
PDE (SPDE) hereafter.  

Analogously, the field $S(q,y)$ adjoint to $\eta(q,y)$ satisfies
\begin{equation}
  \label{s-pde}
   \ptl _q S   = \hf\ptl _y^2S -x\mu \ptl _y S,
\end{equation}
that renders
\begin{equation}
  \label{eta-s}
  \int \rd y \, S(q,y)\,\eta(q,y)
\end{equation}
constant in $q$.  Obviously $\ptl_ yS$ satisfies the SPDE
(\ref{spde}). 
 
The Cole-Hopf transformation can be extended to $\vartheta(q,y)$.
This is done by the recipe that in the intervals with $\dot{x}\equiv0$
the new field exhibits pure diffusion.  Suppose that $\psi(q,y)$
satisfies (\ref{psi-pde}), then let
\begin{equation}
  \label{def-nu}
  \nu(q,y) = \vartheta(q,y)\psi(q,y), 
\end{equation}
whence
\begin{equation}
  \label{nu-pde}
   \ptl _q \nu   = -\hf\ptl _y^2\nu + \frac{\dot{x}}{x} \nu \ln\psi.
\end{equation}
Similarly, the analog of the Cole-Hopf transformation for the field
$P(q,y)$ adjoint to $\vartheta(q,y)$ is
\begin{equation}
   \label{def-t} 
  T(q,y)=P(q,y)/\psi(q,y),
\end{equation} 
satisfying
\begin{equation} 
  \label{t-pde} 
  \ptl _q T   = \hf\ptl _y^2T - \frac{\dot{x}}{x} T \ln\psi.
\end{equation} 
If $\dot{x}=0$ then the PDE-s (\ref{nu-pde},\ref{t-pde}) indeed reduce
to the equation for pure diffusion.  Based on that the $\vartheta$ and
$P$ fields can be evaluated along plateaus of $x(q)$
straightforwardly.

\subsubsection{Green functions} 
 \label{sssec-gf} 
  
The PDE-s previously considered were of the form
\begin{equation} 
  \label{L-q} \ptl _q X(q,y) = \hat{\cal L}(q,y,\ptl _y)\, X(q,y) +
  h(q,y),
\end{equation} 
where the unknown field is $X(q,y)$ and the time $q$ may evolve in
either increasing or decreasing direction.  The differential operator
$\hat{\cal L}(q,y,\ptl _y)$ is possibly nonlinear in $X$, may be $q$-
and $y$-dependent, and contains partial derivatives by $y$.  For
vanishing argument $X=0$ the operator gives zero, $\hat{\cal L} \,
0=0$.  We included the additive term $h(q,y)$ for the sake of
generality, it was absent from the PDE-s we encountered so far.
 
In what follows we shall introduce Green functions (GF-s) for linear
as well as nonlinear PDE-s.  Suppose that $X(q,y)$ is the unique
solution of a PDE like (\ref{L-q}) with some initial condition.  The
GF associated with the PDE for the field $X(q,y)$ is defined as
\begin{equation} 
   \label{gf-X} {\cal G}_{X}(q_1,y_1;q_2,y_2) = \frac{\delta
   X(q_1,y_1)}{\delta X(q_2,y_2)}.
\end{equation}
This may be viewed as the response of the solution $X$ at $q_1$ to an
infinitezimal change of the initial condition at $q_2$.  The above
definition yields a retarded GF, that is, if the PDE for $X$ evolves
towards increasing (decreasing) $q$ then the GF vanishes for $q_1 <
q_2$ ($q_1 > q_2$).  Obviously 
\begin{equation} 
   \label{gf-reduces} {\cal G}_X(q,y_1;q,y_2) = \delta(y_1-y_2).
\end{equation}
The chain rule for the functional derivative in
(\ref{gf-X}) can be expressed as
\begin{equation} 
   \label{gf-evol} {\cal G}_{X}(q_1,y_1;q_2,y_2) = \int \rd y\, {\cal
   G}_{X}(q_1,y_1;q,y) \, {\cal 
   G}_{X}(q,y;q_2,y_2), 
\end{equation} 
where $q$ is in the interval delimited by $q_1$ and $q_2$.  This is
just the customary composition rule for GF-s.  In terms of the adjoint
property, (\ref{gf-evol}) means that the adjoint field to the GF in
its fore variables is the same GF in its hind variables. The PDE-s the
GF satisfies in its fore and hind variables are, therefore, each
other's adjoint equations.
 
The definition (\ref{gf-X}) applies both to linear and nonlinear PDE-s
(\ref{L-q}).  It is the specialty of the linear PDE that ${\cal
G}_X(q_1,y_1;q_2,y_2)$ satisfies the same PDE in the variables $q_1,
y_1$ with additive term $h(q,y) = \pm \delta(q_1-q_2)\,
\delta(y_1-y_2)$, where the sign is $+$ if the time $q$ in the PDE
(\ref{L-q}) increases and $-$ if it decreases. Then the solution can
be given in terms of the GF-s in the usual form
\begin{equation}  
  \label{gf-evolve} X(q_1,y_1) = \int \rd y_2 \,{\cal
  G}_X(q_1,y_1;q_2,y_2)\, X(q_2,y_2) + \int \rd y_2 \int_{q_2}^{q_1}
  \rd q  \,{\cal G}_X(q_1,y_1;q,y_2)\, h(q,y_2). 
\end{equation} 
If the PDE for $X$ is nonlinear then ${\cal G}_X(q_1,y_1;q_2,y_2)$ is
the GF for the PDE that is obtained from the aforementioned PDE by
linearization as performed at the beginning of Section
\ref{sssec-lin-PDEs}.  In short, the GF of a nonlinear PDE is the GF
of its linearized version.  Note that for a nonlinear PDE the GF is
associated with a solution $X$ of it, for that solution usually enters
some coefficients in the linearized PDE the GF satisfies.
 
Suppose now that the differential operator in (\ref{L-q}) is
$\hat{\cal L}(q,\ptl _y)$\ie it is translation invariant in $y$.  Such
is the case for the PPDE (\ref{ppde}) and its derivatives.  Then it is
easy to see that
\begin{equation} 
  \label{gf-Y-def} Y(q,y)=\ptl _y X(q,y)
\end{equation}
will obey the PDE that is the linearization of the PDE for
$X$.  Therefore 
\begin{equation}  
  \label{gf-Y-evolve} Y(q_1,y_1) = \int \rd y_2 \,{\cal
  G}_X(q_1,y_1;q_2,y_2)\, Y(q_2,y_2) + \int \rd y_2 \int_{q_2}^{q_1}
  \rd q  \,{\cal G}_X(q_1,y_1;q,y_2)\, \ptl _{y_2} h(q,y_2).
\end{equation} 
If the PDE for $X$ is nonlinear then Eq.\ (\ref{gf-evolve}) does not but
Eq.\ (\ref{gf-Y-evolve}) does hold.  The latter, however, is merely an
identity and should not be considered as the solution producing $Y$
from an initial condition, because in order to calculate ${\cal G}_X$
the knowledge of $X$ and thus that of $Y$ is necessary.
 
A prominent role will be played by the GF for the field $\varphi(q,y)$
from the PPDE (\ref{ppde}), that is, by ${\cal G}_{\varphi} (q_1,y_1
;q_2,y_2)$.  This GF was introduced and studied in Refs.\
\cite{dl83,sd84}.  Our first observation here is based on the fact
that the linearization of the PPDE yielded Eq.\ (\ref{theta-pde}) and
the linearization of the derivative of the PPDE, Eq.\
(\ref{deriv1-ppde}), produced Eq.\ (\ref{eta-pde}).  Therefore the
respective GF-s are identical,
\begin{eqnarray}  
  \label{gf-varphi} {\cal G}_{\varphi}(q_1,y_1;q_2,y_2) & = & {\cal
   G}_{\vartheta}(q_1,y_1;q_2,y_2),\\ \label{gf-mu} {\cal
   G}_{\mu}(q_1,y_1;q_2,y_2) & = & {\cal G}_{\eta}(q_1,y_1;q_2,y_2).
\end{eqnarray} 
Given the initial condition (\ref{spde-init}) of the SPDE, its 
solution is 
\begin{equation} 
  \label{gf-p} 
  P(q,y) = {\cal G}_P(q,y;0,0) 
\end{equation}
The GF-s ${\cal G}_P$ and ${\cal G}_{\varphi}$ were discussed for the
SK model in Ref.\ \cite{sd84}. Considering the constancy of
(\ref{theta-p}) and (\ref{eta-s}) we have 
\begin{eqnarray} 
  \label{gf-theta-p} {\cal G}_{\varphi}(q_1,y_1;q_2,y_2) & = & {\cal
   G}_P(q_2,y_2;q_1,y_1),\\ \label{gf-eta-s} {\cal
   G}_{\mu}(q_1,y_1;q_2,y_2) & = & {\cal G}_S(q_2,y_2;q_1,y_1).
\end{eqnarray} 
An identity between derivatives of GF-s can be obtained from Eqs.\
(\ref{def-eta}) and (\ref{gf-evolve}) as
\begin{equation} 
  \label{gf-theta-eta}
  \ptl _{y_{2}}{\cal G}_{\varphi}(q_2,y_2;q_1,y_1) = - \ptl _{y_{1}} {\cal
    G}_{\mu}(q_2,y_2;q_1,y_1).
\end{equation}
Because of their central significance, we display the equations the GF
of the field $\varphi$ satisfies.  In its fore set of arguments the
${\cal G}_{\varphi} (q_1,y_1 ;q_2,y_2)$ satisfies
\begin{equation}
  \label{gf-pde-forw} \ptl _{q_{1}} {\cal G}_{\varphi} = -\hf\ptl
   _{y_{1}}^2 {\cal G}_{\varphi} -x(q_1)\,\mu(q_1,y_1)\, \ptl _{y_1} {\cal
   G}_{\varphi} - \delta(q_1-q_2)\delta(y_1-y_2),
\end{equation}
where the differential operator on the \rhs is the same as on the \rhs
of (\ref{theta-pde}).  In the hind set, with regard to the identity
(\ref{gf-theta-p}) and the SPDE (\ref{spde}), we obtain a PDE 
\begin{equation}
  \label{gf-pde-backw} \ptl _{q_{2}} {\cal G}_{\varphi} = \hf\ptl
   _{y_{2}}^2 {\cal G}_{\varphi} -x(q_2)\, \ptl _{y_2} \left(
   \mu(q_2,y_2)\, {\cal G}_{\varphi}\right) +
   \delta(q_1-q_2)\delta(y_1-y_2)
\end{equation} 
whose \rhs contains the same differential operator as on \rhs of the
SPDE.  The norm in the second $y_2$ argument is conserved as
\begin{equation}
  \label{gf-norm} \int {\cal G}_{\varphi}(q_1,y_1;q_2,y_2) \, \rd y_2
  \equiv 1  
\end{equation} 
for $q_1\leq q_2$. 

Equation (\ref{gf-evolve}) shows how a particular solution of the
linear PDE with a source can be expressed by means of the GF.  For
example, suppose that the source field $h(q,y)$ is added to the
linearized PPDE as
\begin{equation}
  \label{theta-with-source-pde} \ptl _q \vartheta = -\hf\ptl
   _y^2\vartheta -x\mu\ptl _y\vartheta + h
\end{equation}
and an initial condition $\vartheta(q_1,y)$ is set for some $0<q_1\leq
1$.  Then we have the solution for $0\leq q\leq q_1$ in the form
\begin{equation}
 \label{theta-with-source-solve} \vartheta(q,y) = \int \rd y_1\,
 {\cal G}_{\varphi}(q,y;q_1,y_1)\, \vartheta(q_1,y_1) - \int_q^{q_1}
 \rd q_2 \int \rd y_2\, {\cal G}_{\varphi}(q,y;q_2,y_2) h(q_2,y_2).
\end{equation}
The derivative field (\ref{def-mu}) satisfies (\ref{deriv1-ppde}),
thus it also satisfies the above PDE (\ref{theta-with-source-pde})
with zero source, whence
\begin{equation}
 \label{mu-evolve} \mu(q,y) = \int \rd y_1\, {\cal
 G}_{\varphi}(q,y;1,y_1)\, \varPhi^\prime (y_1).
\end{equation} 
Derivation of $\mu$ gives $\kappa$ as from (\ref{def-kappa}) which
satisfies the PDE (\ref{deriv2-ppde}).  Its solution can be expressed
in terms of the GF associated to $\mu$ as
\begin{equation}
 \label{kappa-evolve} \kappa(q,y) = \int \rd y_1\, {\cal
 G}_{\mu}(q,y;1,y_1)\, \varPhi^\pp (y_1).
\end{equation} 
Note that the relation (\ref{gf-theta-eta}) is necessary to maintain
(\ref{def-kappa}).

So far we considered the GF-s of $\varphi$ and its derivative fields.
It is also instructive to see their relation to the GF of the field
$\psi$.  Starting from the definition (\ref{gf-X}) of the GF and using
the Cole-Hopf formula (\ref{psi-to-phi}) we get
\begin{equation}
 \label{gf-psi} {\cal G}_{\psi}(q_1,y_1;q_2,y_2) = \frac{x(q_1)\,
 \psi(q_1,y_1)}{x(q_2)\, \psi(q_2,y_2)} {\cal
 G}_{\varphi}(q_1,y_1;q_2,y_2).
\end{equation} 
From the PDE-s (\ref{gf-pde-forw},\ref{gf-pde-backw}) for ${\cal
G}_{\varphi}$ we have for ${\cal G}_{\psi} (q_1,y_1 ;q_2,y_2)$
\bml  \label{gf-psi-pde} 
\begin{eqnarray}
  \label{gf-psi-pde-forw} \ptl _{q_{1}} {\cal G}_{\psi} &=& -\hf\ptl
   _{y_{1}}^2 {\cal G}_{\psi} + \frac{\dot{x}(q_1)}{x(q_1)}\,
   \left(\ln\psi(q_1,y_1) +1\right)\, {\cal G}_{\psi} -
   \delta(q_1-q_2)\delta(y_1-y_2), \\ \label{gf-psi-pde-backw} \ptl
   _{q_{2}} {\cal G}_{\psi} &=& \hf\ptl _{y_{2}}^2 {\cal G}_{\psi} -
   \frac{\dot{x}(q_1)}{x(q_1)}\, \left(\ln\psi(q_1,y_1) +1\right)\,
   {\cal G}_{\psi} + \delta(q_1-q_2)\delta(y_1-y_2).
\end{eqnarray} 
\eml
Equation (\ref{gf-psi-pde-forw}) could also be obtained by
linearization of the PDE (\ref{psi-pde1}), while
(\ref{gf-psi-pde-backw}) is its adjoint.  These PDE-s are particularly
useful if $\dot{x}(q)= 0$, because then they reduce to pure diffusion.
 One can view relation (\ref{gf-psi}) as the
translation of the Cole-Hopf transformation (\ref{psi-to-phi}) onto
the GF-s.  We again see the advantage of keeping track of a Cole-Hopf
transformed pair like ${\cal G}_{\psi}$ and ${\cal G}_{\varphi}$,
because ${\cal G}_{\psi}$ is simple for plateaus in $x(q)$ and ${\cal
G}_{\varphi}$ is useful when $\dot{x}(q)>0$, especially at jumps.

Notation in subsequent sections can be shortened by the introduction
of what we shall call vertex functions 
\begin{eqnarray}
  \label{vertex-ttt}
   \Gamma_{\varphi\varphi\varphi} \left( q;\{q_i,y_i\}_{i=1}^3 \right) 
    &=& \int \rd y\, {\cal G}_{\varphi}(q_1,y_1;q,y)
    {\cal G}_{\varphi}(q,y;q_2,y_2) \, {\cal G}_{\varphi}(q,y;q_3,y_3), \\
  \label{vertex-tee}
   \Gamma_{\varphi\mu\mu} \left( q;\{q_i,y_i\}_{i=1}^3 \right) 
   &= & \int \rd y\, {\cal G}_{\varphi}(q_1,y_1;q,y) 
   {\cal G}_{\mu}(q,y;q_2,y_2) \, {\cal G}_{\mu}(q,y;q_3,y_3).   
\end{eqnarray}
The ordering $q_1\leq q\leq q_2$, $q_1\leq q\leq q_3$ is understood.
The vertex functions satisfy the appropriate linear PDE in each pair
$q_i,y_i$, furthermore, if $q$ coincides with say $q_j$ then the
vertex functions reduce to the product of the other two GF-s with
$q_i, i\neq j$.  
  
As shown for $q_1< q< q_2$, $q_1<q < q_3$ in Appendix
\ref{app-sec-GF-ident}, we have the useful identity
\begin{equation}
  \label{gf-diff-vertex}
  \ptl _q \Gamma_{\varphi\varphi\varphi} = \ptl _{y_2} \ptl _{y_3}
  \Gamma_{\varphi\mu\mu}.
\end{equation}
A notable consequence of that is obtained from the fact that
$\mu(q,y)$ of Eq.\ (\ref{def-mu}) and $\kappa(q,y)$ of Eq.\ 
(\ref{deriv2-ppde})) are evolved by ${\cal G}_\varphi$ and ${\cal
  G}_\mu$, respectively, as it follows from Eq.\ (\ref{gf-Y-evolve}).
Therefore, multiplication of (\ref{gf-diff-vertex}) by the initial
conditions $\varPhi^\prime(y_i) = \mu(1,y_i)$, for $i=2,3$, and
integration by those $y_i$-s gives for $q_1<q$
\begin{equation}
  \label{gf-mu-kappa} \ptl _q \int \rd y \, {\cal
  G}_{\varphi}(q_1,y_1;q,y)\, \mu(q,y)^2 = \int \rd y \, {\cal
  G}_{\varphi}(q_1,y_1;q,y)\, \kappa(q,y)^2.
\end{equation}
The mathematical properties of the PDE-s will acquire physical meaning
in subsequent chapters where thermodynamical properties are studied. 

\subsubsection{Evolution along plateaus} 
 \label{sssec-triv-plats}

Here we collect the few obvious formulas describing the evolution of
some fields along the trivial $x=0$ and $x=1$ plateaus, and give the
GF for $\varphi$ for any plateau.

Let us consider firstly the $x=0$ plateau\ie the region $0\leq q <
q_{(0)}$.  We recall the Cole-Hopf formula (\ref{psi-to-phi}) for the
field $\psi(q,y)$ to obtain
\begin{equation}  
 \label{psi-for-x=0} \psi(q,y) \equiv 1.
\end{equation}
The field $\varphi(q,y)$ obeys the PPDE (\ref{ppde}), thus is purely
diffusive for $x=0$ as
\begin{equation}  
 \label{phi-for-x=0} \varphi(q,y) = \int \rD z\,
 \varphi(q_{(0)},y+z\sqrt{q_{(0)}-q}).
\end{equation} 
Due to continuity of $\varphi$ in $q$ this also holds for $q=q_{(0)}$.
The probability field $P(q,y)$ from the SPDE
(\ref{spde},\ref{spde-init}) is the Gaussian function
\begin{equation}  
  \label{P-for-x=0} P(q,y) = G(y,q),
\end{equation} 
where the notation 
\begin{equation}
  \label{gauss}
   G(x,\sigma) = \frac{1}{\sqrt{2\pi\sigma}} \exp \left(-\frac{x^2}{2
   \sigma} \right)
\end{equation}
was used.

In the region $q_{(1)} \leq q \leq 1$ is the $x=1$ plateau, we have 
\bml
\label{psi-phi-for-x=1}
\begin{eqnarray}  
 \label{psi-for-x=1} \psi(q,y) &=& \int \rD z\,
 \exp\varPhi(y+z\sqrt{1-q}), 
 \\ \label{phi-for-x=1} \varphi(q,y) &=& \ln\psi(q,y).
\end{eqnarray} 
\eml 
The time-dependent probability field $P(q,y)$ is best evaluated along
plateaus by its own version, (\ref{def-t}), of the Cole-Hopf
transformation.  The transformed field $T(q,y)$ obeys (\ref{t-pde}),
so it reduces to pure diffusion along a plateau.  Thus, assuming the
knowledge of $P\left( q_{(1)},y\right)$ and having the $\varphi(q,y)$
from (\ref{psi-phi-for-x=1}) we get
\begin{equation}  
 \label{p-for-x=1} P(q,y) = e^{\varphi(q,y)} \int \rD z\,
 e^{-\varphi\left(q_{(1)},y+z\sqrt{q-q_{(1)}}\right)} P\left(
 q_{(1)},y+z\sqrt{q-q_{(1)}}\right).
\end{equation} 

The GF for the field $\varphi$, ${\cal G}_{\varphi}$, will be given on
any plateau. Suppose that $\dot{x}(q)\equiv 0$ in the closed interval
$[q_1,q_2]$.  Then from (\ref{gf-psi-pde}), for a positive plateau
value $x$, ${\cal G}_{\psi}$ is a Gaussian function.  Then ${\cal
G}_{\varphi}$ becomes from (\ref{gf-psi})
\begin{equation}  
 \label{gf-phi-gauss} {\cal G}_{\varphi}(q_1,y_1;q_2,y_2) =
 e^{x(\varphi(q_2,y_2)-\varphi(q_1,y_1))} \, G(y_2-y_1,q_2-q_1),
\end{equation} 
where the notation (\ref{gauss}) has been used.  The GF remains to be
determined on the trivial plateau $x=0$, that is obtained from say
(\ref{gf-pde-forw}) as
\begin{equation}   
 \label{gf-phi-gauss-x=0} {\cal G}_{\varphi}(q_1,y_2;q_2,y_2) =
 G(y_2-y_1,q_2-q_1). 
\end{equation} 
This is the same as we would get from (\ref{gf-phi-gauss}) by
substituting $x=0$.

\subsubsection{Discontinuous initial conditions}  
 \label{sssec-dic}  
  
If the initial condition $\varPhi(y)$ of the PPDE (\ref{ppde}) is
discontinuous then special care is necessary near $q=1$.  While
strictly speaking the PPDE is defined only for initial conditions
twice differentiable by $y$, one may expect that for practical
purposes a much less strict condition suffices.  For instance, in the
textbook example of pure diffusion any function whose convolution with
the Gaussian GF gives a finite result, can be accepted as initial
condition irrespective of its differentiability.  The physical picture
is that diffusion smoothens steps and spikes and brings the solution
into a differentiable form within an infinitezimal amount of time.
  
The problem with the PPDE for discontinuous initial condition lays
deeper.  It can be traced back to the fact that the Cole-Hopf
transformation no longer connects the two PDE-s (\ref{psi-pde}) and
(\ref{ppde}).  Even if by means of the Dirac delta we accept
differentiation through a discontinuity, the derivatives of $\psi(1,y)
= \exp\varPhi(y)$ and $\varphi(1,y) = \varPhi(y)$ are not related by the
chain rule, namely
\begin{equation} 
  \label{no-chain} \varPhi^\prime(y) \,e^{\varPhi(y)} \neq \left(
  e^{\varPhi(y)}\right)^\prime. 
\end{equation} 
This can be seen easily by taking for example the step function
\begin{equation} 
  \label{Phi=step}
\varPhi(y) = \alpha\, \theta(y). 
\end{equation} 
Then  
\begin{equation} 
  \label{disc-deriv} e^{\varPhi(y)} = 1+(e^\alpha-1)\, \theta(y),
\end{equation} 
and the inequality (\ref{no-chain}) now takes the form 
\begin{equation}  
  \label{no-chain-xmp} \alpha\, \delta(y) \left[ 
  1+(e^\alpha-1)\,\theta(y)\right]\neq (e^\alpha-1)\, \delta(y). 
\end{equation}  
Equality could only be restored if $\theta(y=0)$ were chosen
$\alpha$-dependent, an artifact we do not accept.  However, the
derivation of the PPDE (\ref{ppde}) from the PDE (\ref{psi-pde}) is
invalid if the chain rule cannot be applied.
 
The difficulty can be circumvented by our using the explicit 
expressions (\ref{psi-phi-for-x=1}) for the fields $\psi(q,y), \, 
\varphi(q,y)$ in the interval $[q_{(1)},1]$.  Obviously, even if there
is a discontinuity -- a finite step -- in $\varPhi(y)$, the $\psi(q,y)$
and thus $\varphi(q,y)$ will become smooth for $q<1$.  For instance
for (\ref{Phi=step}), using the notation
\begin{equation}  
\label{def-H-x}  
H(x) = \int_x^\infty \rD z = \frac{1}{2}\left[ 1 -
\text{erf}\left(x/\sqrt{2}\right) \right],
\end{equation} 
we have
\begin{equation}   
  \label{smoothened-psi} \psi(q,y) = e^\alpha + (1-e^\alpha)\, H\left(
  y/\sqrt{1-q}\right) \qquad \text{for} \quad q_{(1)}\leq q \leq 1.
\end{equation}  
This is an analytic function for $q\neq 1$ and becomes
(\ref{disc-deriv}) for $q\to 1$.  Then $\varphi(q,y)$ is obtained in
$[q_{(1)},1]$ by (\ref{phi-for-x=1}), also analytic for $q\neq 1$, and
$\varphi(1,y)$ becomes indeed (\ref{Phi=step}).  The above formulas
extend down to $q_{(1)}$.  Interestingly, as we shall see later, in
the limit of the ground state $T\to 0$, we have $q_{(1)}\to 1$, but
the discontinuity of the fields equally disappears at $q_{(1)}$,
although analyticity will not hold.

Thus we have the fields for $q<1$, the only problem remains that we
cannot say that $\varphi(q,y)$ satisfies the PPDE (\ref{ppde}) at
$q=1$, because of the inequality (\ref{no-chain}). 
 
The difference in nature between the $\psi$ and $\varphi$ functions
for $q_{(1)}\leq q \leq 1$ can be illustrated by the following.  The
singularity of the PDE-s can be tamed by our considering the fields as
integral kernels.  Let us take an analytic function $a(y)$ such that
itself and its derivatives decay sufficiently fast for large arguments
and consider
\begin{equation} 
  \label{A-psi-q} A_{\psi}(q) = \int \rd y\,a(y)\, \psi(q,y).
\end{equation}  
Starting from (\ref{psi-for-x=1}), changing the integration
variable as $y\to y-z\sqrt{1-q}$, and formally expanding in terms of
$\sqrt{1-q}$ we get
\begin{equation} 
  \label{A-psi-q-series} A_{\psi}(q) = \sum_{k=0}^\infty
  \frac{(1-q)^k}{2^k k!}  \int \rd y\, a^{[2k]} (y) \, e^{\varPhi(y)},
\end{equation}  
where we used $\int \rD z\, z^{2k+1} = 0$, $\int \rD z\, z^{2k} =
(2k-1)!!$, 
and the notation
\begin{equation}
  \label{shorthand-diff} f^{[k]}(x) = \frac{\rd^k~}{\rd x^k} f(x).
\end{equation} 
On the other hand, a similar procedure can be carried out for
\begin{equation}  
  \label{A-phi-q} A_{\varphi}(q) = \int \rd y\,a(y)\, \varphi(q,y),
\end{equation}  
a case we illustrate on (\ref{Phi=step}).  From (\ref{smoothened-psi})
we have
\begin{equation}  
  \label{smoothened-phi} \varphi(q,y) = \ln \left[ e^\alpha +
  (1-e^\alpha)\, H\left( y/\sqrt{1-q}\right) \right] = \varphi\left(
  y/\sqrt{1-q}\right),
\end{equation} 
where the last equality defines the single-argument function
$\varphi(z)$.  Then 
\begin{equation} 
  \label{A-phi-q-2} A_{\varphi}(q) = A_{\varphi}(1) + \int \rd y\, a(y)
  \, \left( \varphi\left( y/\sqrt{1-q} \right) - \alpha\,\theta(y)
  \right),
\end{equation}
where $ A_{\varphi}(1)$ was added to and subtracted from the \rhs.
Changing the integration variable as $y\to y\sqrt{1-q}$ and formally
expanding by $\sqrt{1-q}$ we get 
\begin{equation} 
  \label{A-phi-q-series} A_{\varphi}(q) = A_{\varphi}(1) +
 \sum_{k=0}^\infty \frac{(1-q)^{\frac{k+1}{2}}}{k!} a^{[k]}(0) \, \int
 \rd y\, y^k\, \left( \varphi(y) - \alpha\,\theta(y) \right).
\end{equation} 
Thus in leading order we have from 
(\ref{A-psi-q-series},\ref{A-phi-q-series})  
\bml\label{A-asymp} 
\begin{eqnarray} 
\label{A-psi-asymp} 
 A_{\psi}(q) - A_{\psi}(1) &\propto& 1-q \\ A_{\varphi}(q) - 
 A_{\varphi}(1) &\propto& \sqrt{1-q}.  \label{A-phi-asymp} 
\end{eqnarray} \eml  
So, considering the fields as integral operators in the case of
non-differentiable initial conditions, we see from Eq.\
(\ref{A-asymp}) that $\psi$ does but $\varphi$ does not have a
finite derivative by $q$ at $q=1$.  This explains why we could
maintain the PDE for $\psi$ while the PPDE had to be given up in $q=1$
with a non-differentiable initial condition.
 
If the PPDE (\ref{ppde}) is ill defined for $q=1$ then so may be the
PDE-s for the derivative fields, the linearized PDE-s, and the PDE-s
for the GF-s, as discussed in sections \ref{sssec-deriv-PDEs},
\ref {sssec-lin-PDEs}, and \ref{sssec-gf}.  We settle the ambiguity by
redefining the derivative field $\mu(q,y)$ as
\begin{equation} 
  \label{def-mu-disc} \mu(q,y) \, x(q) \, \psi(q,y) = \ptl _y
  \psi(q,y),
\end{equation} 
so in $[q_{(1)},1]$, where $x(q)\equiv 1$
\begin{equation} 
  \label{def-mu-disc-x=1} \mu(q,y) \, \psi(q,y) = \int \rD z\, \ptl _y
  e^{\varPhi\left(y+z\sqrt{1-q}\right)}.
\end{equation} 
For a smooth $\varPhi(y)$ one recovers the original definition
(\ref{def-mu}) for any $q$.  If, however, $\varPhi(y)$ is discontinuous
then, due to the inequality (\ref{no-chain}), the new formula
(\ref{def-mu-disc-x=1}) will, in general, differ from (\ref{def-mu})
at $q=1$.  The $\mu(q,y)$ from (\ref{def-mu-disc-x=1}) satisfies in
$[q_{(1)},1]$
\bml
\label{mu-pde-disc}
\begin{eqnarray}
 \label{mu-pde-disc-1} \ptl _q\mu &=& - \hf \ptl _y^2\mu - \mu\, \ptl
  _y \mu,\\ \mu(1,y) \, e^{\varPhi(y)} &=& \left(
  e^{\varPhi(y)}\right)^\prime.
\label{mu-pde-disc-init} 
\end{eqnarray} 
\eml The specialty here is that the derivation of (\ref{mu-pde-disc})
could be done without the now invalid chain rule.  The above PDE
coincides with (\ref{deriv1-ppde-evolv}) at $x=1$, with an initial
condition that may be different from (\ref{deriv1-ppde-init}).
 
In a similar spirit it can be shown that the $\mu(q,y)$ redefined
above enters the PDE-s (\ref{gf-pde-forw},\ref{gf-pde-backw}) for the
GF ${\cal G}_{\varphi}$, provided the latter is introduced by our
first giving ${\cal G}_{\psi}$ via (\ref{gf-X}) then defining ${\cal
G}_{\varphi}$ via (\ref{gf-psi}).  Note that the GF ${\cal
G}_{\varphi}$ is given in the interval $[q_{(1)},1]$ by
(\ref{gf-phi-gauss}) with $x=1$, a smooth function in the
$y$-arguments if both $q$ arguments are less than $1$.
 
The continuous framework, with PDE-s, was meant to be a practical
reformulation of the iteration (\ref{rec-psi}).  Real use of it is in
the $R\to \infty$ limit, when it allows more liberty in
parametrization of a finite approximation than just the taking of a
large but finite $R$.  In case of ambiguity, however, the iteration
takes precedence.  That argument helped us to refine our formalism of
PDE-s for discontinuous initial conditions.

In what follows we will use the short notation made possible by the
PDE formalism as if we were dealing with a continuous initial
condition $\varPhi(y)$.  However, if $\varPhi(y)$ is discontinuous
then the PPDE must not be applied at $q=1$, rather
(\ref{psi-phi-for-x=1}) yields the field $\varphi(q,y)$ in
$[q_{(1)},1]$.  So although then the PPDE is not true at $q=1$, we
keep it and understand it as the above recipe.  The derivative of the
PPDE can be upheld with the above definition of the derivative field
$\mu$ as can the PDE-s for the GF ${\cal G}_{\varphi}$.  In concrete
computations on a discontinuous initial condition we shall see that
this takes care of most of the problem.


\eject

\section{Correlations and Thermodynamical Stability}
\label{sec-corr-stab}

\subsection{Expectation values}
\label{ssec-expect}

\subsubsection{Replica averages}
\label{sssec-replica-av}

In this section we evaluate important special cases of the generalized
averages (\ref{mean-e}) and (\ref{mean-s}) within Parisi's ansatz.  In
what follows, generically the knowledge of $\sf Q$, or equivalently,
in the $n\to 0$ limit, that of $x(q)$ will be assumed.  Practically,
all fields introduced above as solutions of various PDE-s, for given
$x(q)$, will be considered as known and expectation values expressed
in terms of those fields.  The pioneering works in this subject are
that of de Almeida and Lage \cite{dl83} and of M\'ezard and Virasoro
\cite{mv85}, who evaluated the average magnetization and its low-order
moments in the SK model.  What follows in Section \ref{ssec-expect} can
be viewed as the generalization of the mechanism these authors
uncovered. 

We shall call the variable $y$ in (\ref{gen-parisi}) ``local field''.
In the SK model $y$ corresponds to the local magnetic field, for the
neuron it is the local stability parameter, and it is useful have a
name for it even in the present framework.

The generic formula comprising (\ref{mean-e}) and (\ref{mean-s}) is 
\begin{equation}      
  \label{repl-aver-def} \lav A(\xv ,\yv) \rav = \int
  \frac{\rd ^n\! x \ \rd ^n\!  y} {(2\pi)^n} \, A(\xv ,\yv)\,
  \exp\left( \sum\nolimits_{a=1}^n \varPhi(y_a)+ i\xv \yv - \case{1}{2}
  \xv  {\sf Q}\xv \right).
\end{equation}  
The normalizing coefficient, analogous to the prefactors in Eqs.\
(\ref{mean-e}) and (\ref{mean-s}), is not included here, since in the
limit $n\to 0$ it becomes unity.  We shall automatically disregard
such factors henceforth, furthermore, we will take $n\to 0$ silently
whenever appropriate.  Dependence on $\varPhi$ and $\sf Q$ is not marked
on the \lhs.

The quantity (\ref{repl-aver-def}) will be called the replica average
of the function $A(\xv ,\yv)$.  Such formulas emerge in most
cases when we set out to evaluate thermodynamical quantities in or
near equilibrium.

\subsubsection{Average of a function of a single local field}
\label{sssec-single-y}

A case of import is when the quantity to be averaged depends only on
the local field $y_a$ of a single replica.  Such is the form of the
distribution of stabilities given in Eq.\ (\ref{repl-dist-stab}) and
the energy (\ref{ener-dist}).  Due to the fact that $y_a$ and $x_a$
are each other's Fourier transformed variables, the expectation values
of replicated $x$-s, like in Eqs.\
(\ref{stat-q},\ref{stat-q-gen},\ref{stat-qhat-gen}) are related to the
averages of products of functions of local fields $y_a$-s.  The latter
can be straightforwardly understood once the case of a function of one
$y_a$ argument is clarified.  Thus we firstly focus on
\begin{equation}      
  \label{y-aver} C_{A} = \lav A(y_1) \rav.
\end{equation}  
There is no loss of generality in choosing the first replica, $a=1$,
because RSB only affects groups of two or more replicas.  Within
Parisi's ansatz (\ref{pa}) the $C_A$ evaluates to a formula like the
\rhs of (\ref{temp1}) with the difference that here $A(y_1)$ is
inserted into the integrand.  In analogy with (\ref{temp2}) we obtain
\begin{equation}
   C_{A} = \int \left[ \prod_{r=0}^{R+1}\prod_{j_r=1}^{n/m_r}
   \rD z_{j_r}^{(r)}\right] A \left(\sum_{r=0}^{R+1} z_1^{(r)}
   \sqrt{q_r-q_{r-1}}\right) \, \prod_{a=1}^{n}
   \exp\varPhi\left(\sum_{r=0}^{R+1} z_{j_r(a)}^{(r)}
   \sqrt{q_r-q_{r-1}}\right). \label{dist_temp2}
\end{equation}
We used the definition of $j_r(a)$ from Eq.\ (\ref{j_r-def}).  In the
argument of $A$ the $j_r(1)=1$ label was inserted for the $z^{(r)}$-s.
After a reasoning similar to that followed in Section
\ref{sssec-free-ener-term} again expressing the integer $m_r$ by the
real $x_r$ from (\ref{x-r}) and taking $n\to 0$, we arrive at the
recursion
\bml
  \label{rec-theta}
\begin{eqnarray}
  \label{rec-theta1} \vartheta_{r-1}(y) ~ \psi_{r-1}(y) & = & \int \rD z~
  \vartheta_r\left(y+z\sqrt{q_r-q_{r-1}}\right) \psi_r\left( y +
  z\sqrt{q_r-q_{r-1}}\right) ^ \frac{x_r}{x_{r+1}} , \\
  \label{rec-theta2} \vartheta_{R+1}(y) & = & A(y) ,
\end{eqnarray}
\eml while the iteration of $\psi_r(y)$ is defined by Eqs.\ 
(\ref{rec-psi},\ref{rec-alt-init}).  The final average is obtained at
$r=0$ as
\begin{equation}
\label{theta-aver2}
   C_{A} = \int \rD z~ \vartheta_0\left(z\sqrt{q_0}\right).
\end{equation}
Using the identity (\ref{diffop}) we are lead to the operator form
\begin{equation}
  \label{rec-theta-psi-op}
    \vartheta_{r-1}(y) \psi_{r-1}(y) = e^{\frac{1}{2}(q_r-q_{r-1}) 
      \frac{d^2}{dy^2}}\, \vartheta_r(y)~ \psi_r(y) ^
      \frac{x_r}{x_{r+1}} ,
\end{equation}
whence by continuation it is easy to derive the PDE
\begin{equation}
  \label{theta-psi-pde}
   \ptl _q \left( \vartheta \psi \right)  = -\hf \ptl _y ^2  \left( 
     \vartheta \psi  \right) + \frac{\dot x}{x} \vartheta \psi \ln \psi .
\end{equation} 
In the spirit of Section \ref{sssec-ppde} it is straightforward to show
that this equation holds also for finite $R$-RSB as well.  Then at
discontinuities of $x(q)$ the singular second term on the \rhs is
absorbed by the requirement that $\psi(q,y)^{1/x(q)}$ is continuous in
$q$.  The initial condition for $\psi(q,y)$ was previously given in
(\ref{psi-pde2}) and that for $\vartheta(q,y)$ is set by
(\ref{rec-theta2}) as
\begin{equation}
  \label{theta-pde-init}
   \vartheta (1,y) = A(y).
\end{equation}
In Eq.\ (\ref{theta-psi-pde}) we recognize the PDE (\ref{nu-pde}) for
the field (\ref{def-nu}).  Now we again have a product like
(\ref{def-nu}), so the field $\vartheta(q,y)$ here also satisfies the
PDE (\ref{theta-pde}).  Thus the sought average (\ref{theta-aver2})
can be written as
\begin{equation}
  \label{theta-aver3}
  C_{A} = \int \rD z~ \vartheta\left(q_{(0)},z\sqrt{q_{(0)}}\right),
\end{equation}
a functional of $\varPhi(y)$ and $x(q)$, where the definition of
$q_{(0)}$ by (\ref{border-q0}) was used.  A practical expression for
the above average involves the adjoint field $P(q,y)$, obeying the PDE
(\ref{spde}) and rendering the formula (\ref{theta-p}) independent of
$q$.  Let us recall the abbreviation for the Gaussian (\ref{gauss}),
then (\ref{theta-aver3}) is of the form of (\ref{theta-p}) at
$q=q_{(0)}$ if 
\begin{equation} 
  \label{P-init1}
   P(q_{(0)},y) = G(y,q_{(0)}).
\end{equation} 
Given the purely diffusive evolution in the interval $(0,q_{(0)})$,
this condition means that $P(0,y)$ is localized at $y=0$\ie $P(q,y)$
satisfies the SPDE (\ref{spde},\ref{spde-init}), whence we can write
the expectation value in the form (\ref{theta-p2}) as
\begin{equation}
  \label{theta-final-aver} C_{A} = \int\, \rd y\, P(1,y)\,A(y).
\end{equation}
This is the main result of this section.  Here the initial condition
(\ref{theta-pde-init}) was used, which is just the function we
intended to average.  This expression reveals that $P(1,y)$ is the
probability distribution of the quantity $y$, or, for a general $q$,
$P(q,y)$ is the distribution at an intermediate stage of evolution.  

Note that in \cite{our-paper-00} we gave a shorter derivation for
(\ref{theta-final-aver}), which avoided the use of the recursion
(\ref{rec-theta}).  The reason for our going the longer way here is
that it straightforwardly generalizes to the case of higher order
correlation functions.  

\subsubsection{Correlations of functions of local fields }
\label{sssec-multiple-y}

The expectation value of a product of functions each depending on a
single local field variable reads as
\begin{equation}
  \label{repl-corr-functions} C_{AB\dots Z}(a,b,\dots ,z) =\lav
  A(y_a)\,B(y_b)\dots Z(y_z) \rav
\end{equation}
This will be called replica correlation function, or correlator, of
the functions $A,\, B,\dots Z$ of respective local fields $y_a,\,
y_b,\dots y_z$.  Its ``order'' is the number of different local fields
it contains.  The natural generalization of the observations in the
previous section allows us to construct formulas for the above
correlation function.  This will be undertaken in the present and the
following two sections.
 
Let us first consider the second order local field correlator
\begin{equation}
  C_{AB}(a,b) = \label{repl-corr-functions-AB} \lav
  A(y_a)\,B(y_b)\rav.
\end{equation}
The Parisi ansatz allows us to parametrize $C_{AB}$ by the $q$
variable, rather than the replica indices $a$ and $b$, remnants of the
$n\times n$ matrix character of $\sf Q$.  This goes as follows.
Fixing the replica indices $a$ and $b$ we obtain two iterations like
(\ref{rec-theta}), with respective initial conditions $A(y)$ and
$B(y)$ at $q=1$.  The iterated functions we denote by $\vartheta_A$
and $\vartheta_B$, respectively.  The iterations evolve until they
reach an index $r(a,b)$ specified by the property that for
$r<r(a,b)$-s, all $j_r$ indices coincide, $j_r(a)=j_r(b)$.  Here we
used the definition of the labels $j_r(a)$ from Eq.\ 
(\ref{j_r-def})\ie if $j_r=1,\dots,n/m_r$ are the labels of ``boxes''
of replicas that contain $m_r$ replicas then $j_r(a)$ is the ``serial
number'' of the box containing the $a$-th replica.  The $r(a,b)$ marks
the largest $r$ index for which the replicas $a$ and $b$ fall into the
same box.  Obviously, since for decreasing $r$ the box size $m_r$
increases, for any given $r\leq r(a,b)$ the said replicas will fall
into the same box of size $m_r$.  The $r(a,b)$ will be referred to
hereafter as merger index, and is a given function of $a$ and $b$ for
a given set of $m_r$-s of Eq.\ (\ref{ieq2}).

The hierarchical organization of the replicas implies the following
property.  Consider three different replica indices $a,\, b,\,
\text{and}\, c$, then either all three merger indices coincide as 
$r(a,b) = r(a,c) = r(b,c)$, or two merger index coincide and the third
one is smaller\eg $r(a,c) = r(b,c) > r(a,b)$.  This is characteristic
for tree-like structures, for example, a maternal genealogical scheme.

The merger index allows us to relabel the matrix elements (\ref{ieq1})
in the Parisi ansatz as
\begin{equation}
  \label{merger}
  q_{r(a,b)}=q_{ab}. 
\end{equation} 
This we can consider as the definition of $r(a,b)$, provided that
giving $q_r$ uniquely determines $r$, that is, in (\ref{ieq1}) strict
inequalities hold.  At the juncture $r=r(a,b)$ the two aforementioned
iterations, so far each obeying (\ref{rec-theta1}), merge into one,
such that the product of the two ``incoming'' $\vartheta_A$ and
$\vartheta_B$ fields at $r=r(a,b)$ give the initial condition for the
one ``outgoing'' iteration, denoted by $\vartheta_{AB}$.  That is, for
$r<r(a,b)$, again the iteration (\ref{rec-theta1}) is to be used for
$\vartheta_{AB,r}(y)$ such that at $r=r(a,b)$ it satisfies the initial
condition
\begin{equation}
  \label{interm-init} \vartheta_{AB,r(a,b)}(y) =
  \vartheta_{A,r(a,b)}(y)\,\vartheta_{B,r(a,b)}(y).
\end{equation} 
Such merging of $\vartheta$ fields to produce an initial condition for
further evolution will turn out to be ubiquitous whenever correlators
are computed.  After changing from the discrete $r$ index to the $q$
time variable, we obtain the expectation value in a form similar to
(\ref{theta-aver3}) as
\begin{equation}
  \label{AB-corr-1} C_{AB}(q_{r(a,b)}) = \int \rD z~
     \vartheta_{AB}\left(q_{(0)},z\sqrt{q_{(0)}}\right).
\end{equation}
Here we switched notation and denote the dependence on the initial
$a,b$ replica indices through $q_{r(a,b)}$.  Equivalently, replacing
$q_{r(a,b)}$ by $q$, we get
\begin{equation}
  \label{AB-corr-2}
  C_{AB}(q) = \int\, \rd y\, P(q,y)\,
  \vartheta_{AB} (q,y) = \int\, \rd y\, P(q,y)\,
  \vartheta_{A} (q,y)\, \vartheta_{B} (q,y)
\end{equation}
Here only such $q$ is meaningful that equals a $q_r$ in the $R$-RSB
ansatz, or, is a limit of a $q_r$ if $R\to\infty$.  However, this
expression can be understood, at least formally, for all $q$-s in the
interval $[0,1]$.

\subsubsection{Replica correlations in terms of Green functions} 
\label{sssec-x-corr-gf}
 
It is instructive to redisplay the formulas for $C_A$ and $C_{AB}(q)$
in terms of GF-s.  Their natural generalization will yield the GF
technique and the graphical representation for general correlation
functions.

The time evolution of the $\vartheta$ field can be expressed by means
of the GF.  Based on the relation between $P(q,y)$ and the GF given by
(\ref{gf-p}) we can write
\begin{equation}
  \label{A-corr-GF}
C_{A} = \int \, \rd y \, {\cal G}_\varphi (0,0;1,y) A(y).
\end{equation} 
Correlators can be conveniently represented by graphs.  On the obvious
case of $C_A$, see Fig.\ \ref{c-A}, we can illustrate the graph rules.

\vbox{
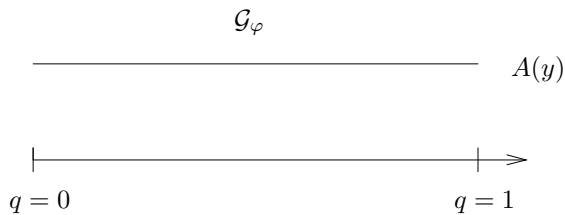
\begin{figure}[h!]
\begin{center}
\setlength{\unitlength}{0.00083333in}
\begingroup\makeatletter\ifx\SetFigFont\undefined%
\gdef\SetFigFont#1#2#3#4#5{%
  \reset@font\fontsize{#1}{#2pt}%
  \fontfamily{#3}\fontseries{#4}\fontshape{#5}%
  \selectfont}%
\fi\endgroup%
{\renewcommand{\dashlinestretch}{30}
{\small%
\begin{picture}(3623,1353)(0,-10)
\path(150,945)(2925,945)
\path(150,345)(3225,345)
\path(3105.000,315.000)(3225.000,345.000)(3105.000,375.000)
\put(0,45){\makebox(0,0)[lb]{\smash{{{\SetFigFont{10}{12.0}{\familydefault}{\mddefault}{\updefault}$q=0$}}}}}
\put(2775,45){\makebox(0,0)[lb]{\smash{{{\SetFigFont{10}{12.0}{\familydefault}{\mddefault}{\updefault}$q=1$}}}}}
\put(3300,870){\makebox(0,0)[b]{\smash{{{\SetFigFont{10}{14.4}{\rmdefault}{\mddefault}{\updefault}$A(y)$}}}}}
\put(1500,1170){\makebox(0,0)[b]{\smash{{{\SetFigFont{10}{14.4}{\rmdefault}{\mddefault}{\updefault}${\cal G}_\varphi$}}}}}
\path(150,420)(150,345)
\path(150,420)(150,270)
\path(2925,420)(2925,270)
\end{picture}
}
}
\vskip0.2in
\caption{Graphical representation of Eq.\ (\protect{\ref{A-corr-GF}})
  for $C_A$.  The line corresponds to the GF associated with the field
  $\varphi$.  Its two $q$-coordinates are taken at the endpoints of
  the line and the two $y$-coordinates are integrated over.  At $q=1$
  the function included in the integrand is displayed.  At $q=0$ the
  Dirac delta $\delta(y)$, understood in the integrand and forcing the
  zero $y$-argument in (\protect{\ref{A-corr-GF}}), is not indicated,
  because it is present for all correlators.\label{c-A} }
\end{center}
\end{figure}}
 
We symbolize the GF ${\cal G}_\varphi (q_0,y_0;q_1,y_1)$ by a line
stretching between $q_0$ and $q_1$.  Over the $y$-s appropriate
integrations will be understood.  If $q_0=0$ the corresponding $y_0$
is set to zero\ie integration is done after multiplication by a Dirac
delta.  For this is always the case in our examples, we do not put any
marks at $q=0$.  A weight function under the integral at $q=1$, like
$A(y)$ in (\ref{A-corr-GF}), should be marked at the right end of the
line.  In sum, $C_{A}$ is a single line between $q=0$ and $q=1$,
labeled by $A(y)$ at $q=1$.
 
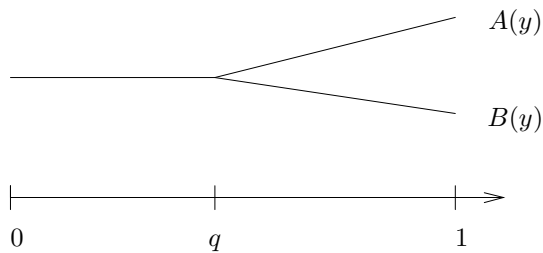
\begin{figure}[h!]
\begin{center}
\setlength{\unitlength}{0.00083333in}
\begingroup\makeatletter\ifx\SetFigFont\undefined%
\gdef\SetFigFont#1#2#3#4#5{%
  \reset@font\fontsize{#1}{#2pt}%
  \fontfamily{#3}\fontseries{#4}\fontshape{#5}%
  \selectfont}%
\fi\endgroup%
{\renewcommand{\dashlinestretch}{30}
{\small%
\begin{picture}(3485,1596)(0,-10)
\path(1287,438)(1287,288)
\path(12,1113)(1287,1113)
\path(1287,1113)(2787,1488)
\path(1287,1113)(2787,888)
\put(1287,63){\makebox(0,0)[b]{\smash{{{\SetFigFont{10}{14.4}{\rmdefault}{\mddefault}{\updefault}$q$}}}}}
\put(3162,813){\makebox(0,0)[b]{\smash{{{\SetFigFont{10}{14.4}{\rmdefault}{\mddefault}{\updefault}$B(y)$}}}}}
\put(3162,1413){\makebox(0,0)[b]{\smash{{{\SetFigFont{10}{14.4}{\rmdefault}{\mddefault}{\updefault}$A(y)$}}}}}
\path(12,363)(2787,363)
\path(12,438)(12,363)
\path(12,438)(12,288)
\path(2787,438)(2787,288)
\path(2787,363)(3087,363)
\path(2967.000,333.000)(3087.000,363.000)(2967.000,393.000)
\put(2787,63){\makebox(0,0)[lb]{\smash{{{\SetFigFont{10}{12.0}{\rmdefault}{\mddefault}{\updefault}1}}}}}
\put(12,63){\makebox(0,0)[lb]{\smash{{{\SetFigFont{10}{12.0}{\rmdefault}{\mddefault}{\updefault}0}}}}}
\end{picture}
}
}
\end{center}
\caption{The correlation function $C_{AB}(q)$.\hspace{0.5in}~}
\label{c-AB}
\end{figure}

As to the second order correlator (\ref{AB-corr-2}), based on Eqs.\
(\ref{theta-with-source-pde},\ref{theta-with-source-solve}) we can
write $\vartheta_{A}$ and $\vartheta_{B}$ in terms of the GF and
obtain
\begin{equation}
  \label{AB-corr-GF} C_{AB}(q) = \int \, \rd y\, \rd y_1\, \rd y_2 \,
  {\cal 
G}_\varphi (0,0;q,y)\, {\cal G_\varphi} (q,y;1,y_1)\, A(y_1)\, {\cal
G}_\varphi (q,y;1,y_2)\, B(y_2).
\end{equation} 
Its graphic representation is given in Fig.\ \ref{c-AB}, it consists
of a single vertex.

\begin{figure}[h!] 
\begin{center} 
\setlength{\unitlength}{0.00083333in}
\begingroup\makeatletter\ifx\SetFigFont\undefined%
\gdef\SetFigFont#1#2#3#4#5{%
  \reset@font\fontsize{#1}{#2pt}%
  \fontfamily{#3}\fontseries{#4}\fontshape{#5}%
  \selectfont}%
\fi\endgroup%
{\renewcommand{\dashlinestretch}{30}
{\small%
\begin{picture}(3560,2208)(0,-10)
\path(1212,375)(1212,225)
\path(2037,375)(2037,225)
\path(1212,1425)(2037,1650)
\path(2767,1504)(2017,1654)(2767,2029)
\put(1212,525){\makebox(0,0)[b]{\smash{{{\SetFigFont{10}{14.4}{\rmdefault}{\mddefault}{\updefault}$q_1$}}}}}
\put(2037,525){\makebox(0,0)[b]{\smash{{{\SetFigFont{10}{14.4}{\rmdefault}{\mddefault}{\updefault}$q_2$}}}}}
\put(3237,675){\makebox(0,0)[b]{\smash{{{\SetFigFont{10}{14.4}{\rmdefault}{\mddefault}{\updefault}$C(y)$}}}}}
\put(3237,1500){\makebox(0,0)[b]{\smash{{{\SetFigFont{10}{14.4}{\rmdefault}{\mddefault}{\updefault}$B(y)$}}}}}
\put(3237,2025){\makebox(0,0)[b]{\smash{{{\SetFigFont{10}{14.4}{\rmdefault}{\mddefault}{\updefault}$A(y)$}}}}}
\path(1212,1425)(2787,750)
\path(12,1425)(1212,1425)
\path(12,300)(2787,300)
\path(12,375)(12,300)
\path(12,375)(12,225)
\path(2787,375)(2787,225)
\path(2787,300)(3087,300)
\path(2967.000,270.000)(3087.000,300.000)(2967.000,330.000)
\put(2787,0){\makebox(0,0)[lb]{\smash{{{\SetFigFont{10}{12.0}{\rmdefault}{\mddefault}{\updefault}1}}}}}
\put(12,0){\makebox(0,0)[lb]{\smash{{{\SetFigFont{10}{12.0}{\rmdefault}{\mddefault}{\updefault}0}}}}}
\end{picture}
}
} 
\end{center} 
\caption{The correlation function $C_{ABC}(q_1,q_2)$.\hspace{0.5in}~} 
\label{c-ABC}  
\end{figure}
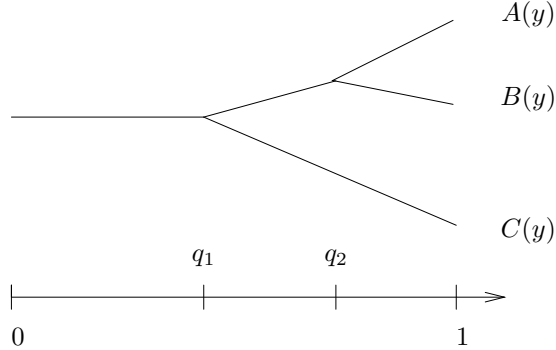

The third order correlator $C_{ABC}(a,b,c)$, see
(\ref{repl-corr-functions}) for notation, can be analogously
calculated.  We can assume without restricting generality that $r(a,b)
\geq r(a,c) = r(b,c)$, and use the notation $q_1 = q_{r(a,c)} \leq q_2 =
q_{r(a,b)}$.  The $q_i$-s, $i=1,2$, used here should not be confounded
with the $q_r$-s of (\ref{ieq1}) from the $R$-RSB scheme.  In this
case the two iterations (\ref{rec-theta1}) with respective initial
conditions $A(y)$ and $B(y)$ merge at $r(a,b)$.  Switching to
parametrization by $q$ means that the PDE (\ref{theta-pde}) rather
than the iteration (\ref{rec-theta1}) is to be considered.  Thus
(\ref{theta-pde}) should now be used in two copies, one with initial
condition $\vartheta_A(1,y) = A(y)$ and the other with
$\vartheta_B(1,y) = B(y)$.  They merge at $q_2$.  That means, the
``incoming'' fields multiply to yield a new initial condition
$\vartheta_{AB}(q_2,y) = \vartheta_{A}(q_2,y) \vartheta_{B}(q_2,y)$,
like in (\ref{interm-init}), and hence for $q_1
\leq q \leq q_2$ the field $\vartheta_{AB}(q,y)$ obeys the PDE
(\ref{theta-pde}).  In $q_1$ another merger takes place with the
incoming field $\vartheta_C(q,y)$.  This started from the initial
condition $\vartheta_C(1,y) = C(y)$ and has evolved according to
(\ref{theta-pde}) until $q=q_1$.  Here the product of the two incoming
fields $\vartheta_{ABC}(q_1,y) = \vartheta_{AB}(q_1,y)
\vartheta_{C}(q_1,y)$ becomes the initial condition at $q=q_1$ for the
final stretch of evolution by (\ref{theta-pde}) down to $q=0$.  The
resulting correlator is easy to formulate in terms of GF-s.  Indeed,
(\ref{theta-with-source-solve}) with $h\equiv 0$ gives the solution of
the PDE (\ref{theta-pde}) starting from an arbitrary initial
condition, specified at an arbitrary time.  Hence 
\begin{eqnarray} 
 \label{ABC-corr-GF} C_{ABC}(q_1,q_2) &=& \lav A (y_{a})\, B(y_{b})\,
C(y_c)\rav \nn \\ &=& \int \rd y_1\,d{y_2}\,dy_3\,dy_4\,dy_5\,
P(q_1,y_1)\, {\cal G}_\varphi(q_1,y_1;1,y_3)\, C(y_3) \nn \\ && \times
\, {\cal G}_\varphi(q_1,y_1;{q_2},{y_2})\, {\cal G}_\varphi(q_2,y_2;
1,y_4)\, A(y_4)\,{\cal G}_\varphi(q_2,y_2; 1,y_5) \, B(y_5).
\end{eqnarray}
The corresponding graph is on Fig.\ \ref{c-ABC}, it has two vertices.
The special case $r(a,b) = r(a,c) = r(b,c)$ corresponds to $q_1=q_2$.
Then we wind up with a single vertex of altogether four legs, and
accordingly, the ${\cal G}_\varphi(q_1,y_1;{q_2},{y_2})$ in
(\ref{ABC-corr-GF}) should be replaced by $\delta(y_1-y_2)$.

A general correlator of local fields $y$ can be graphically
represented starting out of the full ultrametric tree
\cite{sgrev87}.  This can be visualized as a tree with $R+1$
generations of branchings and at the $r$-th generation having
uniformly the connectivity $m_r/m_{r+1}$.  The $R+1$-th generation has
$n$ branches, to the end of each a ``leaf'' can be pinned.  The leaves
are labeled by the replica index $a=1,\dots,n$.  Between $r=0$ and
$r=1$ is the ``trunk''.  For a -- possibly large -- integer number of
replicas $n$ this is a well defined graph.  If $n\to 0$ then the
$m_r$-s cannot be held integers and possibly the $q_r$-s densely fill
an interval, thus the full tree looses graphical meaning.  On the
other hand, the graphs representing replica correlators can be
understood as subtrees of the full tree for integer $n$, and
remarkably, they remain meaningful even after continuation.

On Figs.\ \ref{c-A}, \ref{c-AB}, \ref{c-ABC} we illustrated the first
three simplest local field correlations by graphs.  There a branch
connecting vertices of time coordinates say $q_1$ and $q_2 > q_1$ was
associated with ${\cal G}_{\varphi} (q_1,y_1;q_2,y_2)$, with implied
integrations over the local field coordinates.  This feature holds
also for higher order correlations.  Similarly to the case explained
in Section \ref{sssec-single-y}, then again the iteration
(\ref{rec-theta}), or, equivalently, the PDE (\ref{theta-pde})
emerges.  Given an interval $(q_1,q_2)$ the initial condition for a
field $\vartheta$ is set at the upper border $q_2$, then $\vartheta$
undergoes evolution by the linearized PPDE (\ref{theta-pde}), and the
result is the solution at $q_1$.  Since ${\cal G}_{\varphi}
(q_1,y_1;q_2,y_2)$ is the GF that produces the solution of
(\ref{theta-pde}) out of a given initial condition, it is natural to
associate the GF with the branch of a graph linking $q_1$ with $q_2$.
Since the GF is in fact an integral kernel, integration is to be
performed over variables $y_1$ and $y_2$ at the endpoints of the
branch.  This automatically yields the merging of incoming fields
$\vartheta$ at a vertex to form a new initial condition, as
exemplified (before continuation) for the second order local field
correlator in Eq.\ (\ref{interm-init}).  Indeed, the local field $y$
associated with a vertex at $q$ of altogether three legs is the fore
$y$ argument of two incoming GF-s and the hind $y$ argument of one
outgoing GF, so the latter evolves the product of the incoming
$\vartheta$ fields towards decreasing times starting from $q$.

The graph rules for the general local field correlator $C_{AB\dots
Z}(a,b\dots z)$, defined by (\ref{repl-corr-functions}), can be
summarized as follows.  Draw continuous lines starting out from the
leaves corresponding to the replica indices $a,b\dots z$ along
branches until the trunk is reached.  Lines will merge occasionally,
and in the end all lines meet at the trunk.  The merging points are
specified by the merger indices $r(a,b)\dots$, or equivalently, by the
$q_{r(a,b)}\dots$ values from (\ref{merger}) for each pair of the
replica indices we started with.  Obviously, not all such $q$-s for
different replica index pairs from the set $a,b\dots z$ need to be
different, in the extreme case all such $q$-s may be equal.  The graph
thus obtained is, from the topological viewpoint, uniquely determined
by the given set of replica indices of a correlator.  Then the
explicit dependence on the replica indices $a,b,\dots z$ is no longer
kept, instead they appear through merger indices $r(a,b),\dots$, or,
equivalently, $q_{r(a,b)},\dots$.  This allows us to take the $n\to 0$
limit.  In the end, the correlator becomes a function of all
$q_{r(a,b)},\dots$-s that can be formed from the replica indices
$a,b,\dots z$ of (\ref{repl-corr-functions}).  Now that each branch
merging has a given time $q$ value, it is useful to include the
coordinate axis of $q$ with a graph.
 
The calculation of a correlator implies evolution by the PDE
(\ref{theta-pde}), first with different $y$ variables along the
respective branches, from the leaves towards the trunk.  The functions
$A(y), B(y),\dots Z(y)$ are the initial conditions of this evolution
until the first respective merging points.  Whenever branches meet,
say at a $q_{i}$, the fields $\vartheta_{(1)}(q_{i},y)$,
$\vartheta_{(2)}(q_{i},y)$, etc., associated with the different
incoming lines multiply, all having a common $y$ local field.  Thus is
created a new initial condition for further evolution by
(\ref{theta-pde}), from $q_{i}$ onward to decreasing $q$-s.  At the
last juncture, say $q_1$, the $y$-integral of the product of the
incoming fields weighted with $P(q_1,y)$ yields the correlator in
question.  Obviously, the branches that connect merging points can be
associated with the GF ${\cal G}_\varphi$ of the PDE
(\ref{theta-pde}). It follows that at a merging point of two branches
the $y$-integral gives the vertex function
$\Gamma_{\varphi\varphi\varphi}$ of (\ref{vertex-ttt}).
 
It should be noted that the correlator $C_{AB\dots Z}(a,b\dots z)$ is
now expressed as an integral expression, where the product
$A(y_a)\,B(y_b)\dots Z(y_z)$ appears in the integrand.  Thus an
average of the more general form
\begin{equation}
  \label{gen-aver-loc-field} \lav A(y_a,y_b,\dots,y_z) \rav
\end{equation} 
is obtained by our replacing $A(y_a)\,B(y_b)\dots Z(y_z)$ by
$A(y_a,y_b,\dots,y_z)$ in that expression.  Then we loose the picture
of $\vartheta$ fields independently evolving from $q=1$ by the PDE
(\ref{theta-pde}) and then merging for some smaller $q$-s, because the
function $A(y_a,y_b,\dots,y_z)$ couples the $\vartheta$ fields at the
outset $q=1$.  In what follows we will not encounter averages
(\ref{gen-aver-loc-field}) of non-factorizable functions.

In summary, a given correlation function is represented by a tree,
that is a finite subtree of the full ultrametric tree.  Leaves are
associated with initial conditions of the evolution by
(\ref{theta-pde}).  Branches directed from larger to decreasing $q$
correspond to the GF ${\cal G}_{\varphi}$.  Each vertex, including the
leaves and the bottom of the trunk, has a $q,y$ pair associated with
it.  At the leaves $q=q_{R+1}=1$, and there is integration over $y$-s
in each vertex. At $q=0$ simply $y=0$ should be substituted into the
final formula, so the GF of the trunk becomes just Sompolinsky's field
$P$ due to (\ref{gf-p}).  The intermediate $q$-s will be the
independent variables by those we characterize the correlation
function.  Thus a tree uniquely defines an integral expression,
furthermore, topologically identical trees correspond to the same type
of integral.  Of course, two topologically identical trees can have
different functions associated with their respective leaves, and then
the two integrals will evaluate to different results. 
  
Elementary combinatorics gives the number $N(K)$ of topologically
different trees of $K$ leaves in terms of a recursion.  Denoting the
integer part of $z$ by $[z]$ we have
\bml
\label{n-k}
\begin{eqnarray}
\label{n-k-init}
N(1) &=& 1, \\
\label{n-k-rec}
N(K) &=& \sum_{k=1}^{[(K-1)/2]} N(k)\, N(K-k) + \left(
\left[\frac{K}{2}\right] - \frac{K-1}{2}\right)\,
N\left(\frac{K}{2}\right) \,\left(N\left(\frac{K}{2}\right) + 1
\right).
\end{eqnarray}
\eml 
The basis of this recursion is the fact that in a tree with $K$ leaves
two subtrees meet at the trunk, one having $k$ and the other $K-k$
number of leaves.  The sum is interpreted as zero for $K=2$.  The
second term on the \rhs contributes only for $K$ even, it gives the
number of trees that are composed out of two subtrees both having
$K/2$ leaves.  Some terms generated by the above recursion are
$N(2)=1$, $N(3)=1$, $N(4)=2$, $N(5)=3$, $N(6)=6$, $N(7)=11$,
$N(8)=23$.  For $K=1,2,3$ we have $N(K)=1$, in accordance with our
previous finding that in each of those cases there is only one graph,
see Figs.\ \ref{c-A}, \ref{c-AB}, \ref{c-ABC}.
 
In deriving (\ref{n-k}) we assumed that vertices have altogether three
legs.  In that case the number of vertices is $K-1$.  If $q$-s
coincide because branches shrink to a point then the number of
vertices decreases and vertices with more than three legs arise.  The
corresponding integral expressions are consistent with the graph rules
laid done before.  Indeed, a branch of zero length is associated with
the GF as in (\ref{gf-reduces})\ie gives rise to a Dirac delta
equating the local fields at its two endpoints, wherefore each
remaining branch still represents a GF and the vertex with more than
three legs will still have a single $y$ variable to be integrated
over.
  
\subsubsection{Replica correlations of $x$-s}   
\label{sssec-x-corr} 
  
Derivatives by $q_{ab}$ of the archetypical expression
(\ref{gen-parisi}) play an important role in determining
thermodynamical properties.  Let us introduce the expectation values
(\ref{repl-aver-def}) of products of $x_a$-s as
\begin{equation}   
  \label{repl-corr} C_x^{(k)}(a_1,\dots,a_k) = (-i)^k \,\lav 
  x_{a_1}x_{a_2}\dots x_{a_k} \rav. 
\end{equation} 
The $(-i)^k$ is factorized for later convenience.  This is the
correlation function of order $k$ of the variables $x_{a_j}$.
Correlators of even, $2k$, order are related to the derivatives of
(\ref{gen-parisi}) by the matrix elements $q_{ab}$ as
\begin{equation}   
  \label{repl-corr-deriv-by-q} C_x^{(2k)}(a_1,\dots,a_{2k}) = \frac{
  \,\ptl ^k e^{n\, \varphi [\varPhi(y),{\sf Q}]}} {\ptl q_{a_1a_2}\dots
  \ptl q_{a_{2k-1}a_{2k}}}.
\end{equation} 
Second order correlators enter the stationarity conditions
(\ref{stat-q}, \ref{stat-q-gen}, \ref{stat-qhat-gen}), and fourth
order ones appear in studies of thermodynamical stability, as we shall
see it later.

By partial integration (\ref{repl-corr}) can be brought to the form of
the average of products of various derivatives of $\varPhi(y_a)$ as
\begin{equation}  
 \label{phideriv-corr} C_x^{(k)}(a_1,\dots,a_k) = \int \frac{\rd^n\! x \
   \rd ^n\! y} {(2\pi)^n} ~e^{i\xvs \yvs - \case{1}{2} \xvs  {\sf
   Q}\xvs } \,\ptl _{y_{a_1}}\, \ptl _{y_{a_2}}\dots \ptl _{y_{a_k}}\,
   \exp\left(\sum\nolimits_{a=1}^n \varPhi(y_a) \right),
\end{equation} 
where coinciding replica indices give rise to higher derivatives.  In
the special case when all $a_j$ indices are different we have
\begin{equation} 
  \label{phiprime-corr} C_x^{(k)}(a_1,\dots,a_k) = \lav
  \varPhi^\prime(y_{a_1})\, \varPhi^\prime(y_{a_2}) \dots
  \,\varPhi^\prime(y_{a_k}) \rav. 
\end{equation} 

Note that in the case of a discontinuous $\varPhi(y)$ we may not use the
chain rule of differentiation, therefore in (\ref{phideriv-corr}) the
derivatives should act directly on the exponential.  Then, in the
spirit of Section \ref{sssec-dic}, we can conclude that $\mu(1,y)$ as
defined in (\ref{mu-pde-disc-init}) should be used in lieu of
$\varPhi^\prime (y)$, so the field $\mu(q,y)$ defined in
(\ref{def-mu-disc}) evolves from $q=1$ down until the first merging
point in its way (the first vertex to be met when coming from a leaf
at $q=1$).  In the following general treatment we assume a smooth
$\varPhi(y)$, with the note that the adaptation of the results to
discontinuous ones is straightforward.
 
Expression (\ref{phiprime-corr}) is of the form
(\ref{repl-corr-functions}), so 
\begin{equation} 
  \label{phiprime-corr2} C_x^{(k)}(a_1,\dots,a_k) =
  C_{\varPhi^\prime\dots {\varPhi^\prime}}(a_1,\dots,a_k).
\end{equation} 
We review some low order correlators below. 
 
\subsubsection{One- and two-replica correlators of $x$-s}
\label{sssec-upto-xx} 
  
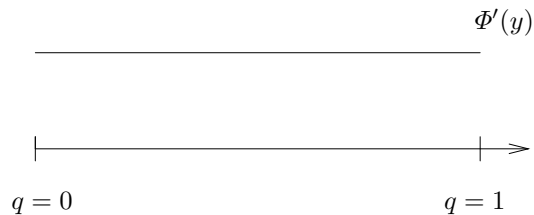
\begin{figure}[h!] 
\begin{center} 
\setlength{\unitlength}{0.00083333in}
\begingroup\makeatletter\ifx\SetFigFont\undefined%
\gdef\SetFigFont#1#2#3#4#5{%
  \reset@font\fontsize{#1}{#2pt}%
  \fontfamily{#3}\fontseries{#4}\fontshape{#5}%
  \selectfont}%
\fi\endgroup%
{\renewcommand{\dashlinestretch}{30}
{\small%
\begin{picture}(3844,1353)(0,-10)
\path(150,1020)(2925,1020)
\put(0,45){\makebox(0,0)[lb]{\smash{{{\SetFigFont{10}{12.0}{\familydefault}{\mddefault}{\updefault}$q=0$}}}}}
\put(2700,45){\makebox(0,0)[lb]{\smash{{{\SetFigFont{10}{12.0}{\familydefault}{\mddefault}{\updefault}$q=1$}}}}}
\put(3075,1170){\makebox(0,0)[b]{\smash{{{\SetFigFont{10}{14.4}{\rmdefault}{\mddefault}{\updefault}$\varPhi^\prime (y)$}}}}}
\path(150,420)(2925,420)
\path(150,495)(150,420)
\path(150,495)(150,345)
\path(2925,495)(2925,345)
\path(2925,420)(3225,420)
\path(3105.000,390.000)(3225.000,420.000)(3105.000,450.000)
\end{picture}
}
}
\caption{ The graph of $C^{(1)}_x$ is a single line.\hspace{0.5in}~}
\label{c-x} 
\end{center} 
\end{figure}

The simplest case of replica correlation function of $x$-s is the
average of a single $x$. Eq.\ (\ref{phiprime-corr2}) for $k=1$ becomes
independent of the single replica index and gives a formula of the
type (\ref{P-init1}) as  
\begin{equation} 
  \label{x-aver} C_x^{(1)} = C_{\varPhi^\prime} = \int \, \rd y \,
     P(1,y)\, \varPhi^\prime(y). 
\end{equation} 
Comparison of (\ref{deriv1-ppde}) and (\ref{theta-pde}) shows that
with the present initial condition $\vartheta(q,y)=\mu(q,y)$. Thus,
recalling that $P(1,y)={\cal G}_\varphi (0,0;1,y)$, we get
alternatively  
\begin{equation}  
  \label{x-aver2} C_x^{(1)} = \mu(0,0).
\end{equation} 
This is shown on Fig.\ \ref{c-x} graphically, it is a special case of
Fig.\ \ref{c-A}. 

\begin{figure}[h!] 
\begin{center} 
\setlength{\unitlength}{0.00083333in}
\begingroup\makeatletter\ifx\SetFigFont\undefined%
\gdef\SetFigFont#1#2#3#4#5{%
  \reset@font\fontsize{#1}{#2pt}%
  \fontfamily{#3}\fontseries{#4}\fontshape{#5}%
  \selectfont}%
\fi\endgroup%
{\renewcommand{\dashlinestretch}{30}
{\small%
\begin{picture}(7210,1656)(0,-10)
\path(12,1152)(935,1152)
\path(935,1152)(2289,1336)
\path(935,1152)(2289,844)
\path(996,414)(996,291)
\path(4287,921)(6537,921)
\put(3581,1029){\makebox(0,0)[b]{\smash{{{\SetFigFont{10}{12.0}{\rmdefault}{\mddefault}{\updefault}$+\theta(q-1^{-0})\times$}}}}}
\put(996,45){\makebox(0,0)[b]{\smash{{{\SetFigFont{10}{12.0}{\rmdefault}{\mddefault}{\updefault}$q$}}}}}
\put(2562,621){\makebox(0,0)[b]{\smash{{{\SetFigFont{10}{12.0}{\rmdefault}{\mddefault}{\updefault}$\varPhi^\prime(y)$}}}}}
\put(2562,1521){\makebox(0,0)[b]{\smash{{{\SetFigFont{10}{12.0}{\rmdefault}{\mddefault}{\updefault}$\varPhi^\prime(y)$}}}}}
\put(6387,1071){\makebox(0,0)[b]{\smash{{{\SetFigFont{10}{12.0}{\rmdefault}{\mddefault}{\updefault}$\varPhi^{\prime\prime}(y)$}}}}}
\path(12,352)(2289,352)
\path(12,414)(12,352)
\path(12,414)(12,291)
\path(2289,414)(2289,291)
\path(2289,352)(2535,352)
\path(2436.570,327.390)(2535.000,352.000)(2436.570,376.610)
\path(4312,414)(4312,291)
\path(6527,352)(6773,352)
\path(6674.570,327.390)(6773.000,352.000)(6674.570,376.610)
\path(6527,414)(6527,291)
\path(4312,352)(6588,352)
\put(2289,106){\makebox(0,0)[lb]{\smash{{{\SetFigFont{9}{10.8}{\rmdefault}{\mddefault}{\updefault}1}}}}}
\put(12,106){\makebox(0,0)[lb]{\smash{{{\SetFigFont{9}{10.8}{\rmdefault}{\mddefault}{\updefault}0}}}}}
\put(6527,106){\makebox(0,0)[lb]{\smash{{{\SetFigFont{9}{10.8}{\rmdefault}{\mddefault}{\updefault}1}}}}}
\put(4312,106){\makebox(0,0)[lb]{\smash{{{\SetFigFont{9}{10.8}{\rmdefault}{\mddefault}{\updefault}0}}}}}
\end{picture}
}
}
\end{center} 
\caption{The correlation function $C^{(2)}_x(q)$.\hspace{0.5in}~}
\label{c-x2} 
\end{figure}
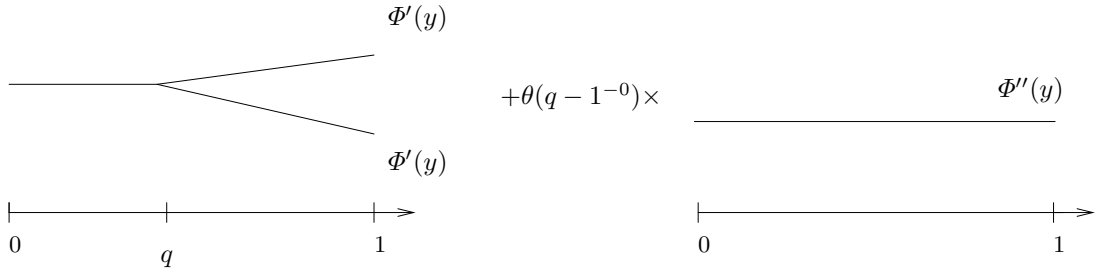

Let us now turn to the correlator of two $x_a$-s as defined in
(\ref{repl-corr}). If the replica indices are different then
(\ref{phiprime-corr}) applies; that should be complemented to allow
for coinciding indices as
\begin{equation}
  \label{xx-corr} C_x^{(2)}(a,b) = C_{\varPhi^\prime\varPhi^\prime}
   (a,b) + \delta_{ab}\, C_{\varPhi^{\prime\prime}}.
\end{equation}
This function depends on the replica indices through the overlap $q$
at the merger $q=q_{r(a,b)}$.  The first term on the \rhs is a special
case of the correlation function $C_{AB}(q)$ given in Eq.\ (\ref
{AB-corr-GF}) with $A(y)=B(y)=\varPhi^\prime(y)$.  Note, however, that
the $\mu$ field satisfying (\ref{deriv1-ppde}) is in fact the
$\vartheta$ of (\ref{theta-pde}) starting from the initial condition
$\varPhi^\prime(y)$. Therefore the two instances of convolution of the GF
with $\varPhi^\prime(y)$ give $\mu(q,y)$ in (\ref {AB-corr-GF}) and we
get 
\begin{equation}
  \label{xx-corr-first}
  \int\, \rd y\, P(q_{r(a,b)},y)\, \mu(q_{r(a,b)},y)^2
\end{equation}
for the first term on the \rhs of Eq. (\ref{xx-corr}). The second term
there is of the type studied in Section \ref{sssec-single-y}.  Note that
the initial condition is by (\ref{deriv2-ppde-init}) just
$\kappa(1,y)$.  Furthermore, $r(a,a)=R+1$ and $q_{aa}=q_{r(a,a)}=1$.
In summary, for the $q$-dependent two-replica correlation function we
obtain
\begin{equation} 
\label{def-corr2}  
     C_x^{(2)}(q) = \left\{ \begin{array} {ll} \int\, \rd y\, P(q,y)\,
        \mu(q,y)^2 & \text{~~if~} q<1 \\ \int\, \rd y\, P(1,y)\,
        \left[\mu(1,y)^2 + \,\kappa(1,y)\right] & \text{~~if~} q=1,
        \end{array} \right. 
\end{equation} 
having omitted the subscript $r(a,b)$ from $q$.  Note that the second
term on the \rhs of (\ref{xx-corr}) contributes at $q=1$.  The above
formula can be abbreviated as
\begin{equation}  
  \label{corr2} C_x^{(2)}(q)= \int\, \rd y\, P(q,y)\, \left[\mu(q,y)^2 +
  \theta(q-1^{-0})\,\kappa(1,y)\right], 
\end{equation} 
where the second term is nonzero only if $q=1$.  We will use the  
shorter notation with the Heaviside function in similar cases 
hereafter.  Fig.\ \ref{c-x2} summarizes the result graphically. 

As it was emphasized earlier, the correlator is meaningful for $q$
arguments at the stationary $q_{ab}$-s, or at their limits for $n\to
0$.  For $q$-s where $\dot x(q)\equiv 0$ the extension of the
correlators is not unique.  For instance, we can write any
$q_{(1)}<q<1$ (for finite $R$-RSB, $q_{(1)}=q_R$, and for continuation
see Section \ref{sssec-continuousRSB}) in lieu of $1^{-0}$ in the
argument of the Heaviside function in (\ref{corr2}).  Note that the
two-replica correlation function, like the fields obeying the PPDE and
the PDE-s described in Sections \ref{sssec-deriv-PDEs} and
\ref{sssec-lin-PDEs}, does not have a plateau in $(q_{(1)},1)$.  In
summary, expression (\ref{corr2}) is the two-replica correlation
function for both the finite $R$-RSB and $R\to\infty$, at arguments
$q$ where $\dot{x}(q)\neq 0$. 
 
\subsubsection{Four-replica correlators} 
\label{sssec-x-4corr}
 
The native form of the four-replica average is by (\ref{phideriv-corr})
\begin{eqnarray}
  C^{(4)}_x(a,b,c,d) & = & C_{ \varPhi^\prime \varPhi^\prime \varPhi^\prime
  \varPhi^\prime} (a,b,c,d) + \left[ \delta_{ab}\, C_{ \varPhi^{\pp}
  \varPhi^\prime \varPhi^\prime} (a,c,d) + \text{ 5 comb's}\right] \nn \\ &
  & + \left[ \delta_{ab}\delta_{cd} C_{\varPhi^{\pp} \varPhi^{\pp}}(a,c) +
  \text{ 2 comb's} \right] + \left[ \delta_{abc} C_{\varPhi^{\pp\prime}
  \varPhi^{\prime}} (a,d) + \text{ 3 comb's} \right] \nn \\
  \label{4corr-native} & & + \delta_{abcd} C_{\varPhi^{[4]}} .
\end{eqnarray}
Here ``comb's'' stands for combinations, then we used the shorthand
notation that a $\delta_{ab\dots c}=1$ only if all $a$, $b\dots c$
indices are equal, else $\delta_{ab\dots c}=0$, furthermore, the
abbreviation (\ref{shorthand-diff}) is understood.

In order to simplify notation, we switch to using $q_i$ for the
parametrization of expectation values.  The $q_i$-s should not be
confounded with the $q_r$ values introduced in (\ref{ieq1}) for the
$R$-RSB scheme.

\begin{figure}[h!]
\begin{center}
\setlength{\unitlength}{0.00083333in}
\begingroup\makeatletter\ifx\SetFigFont\undefined%
\gdef\SetFigFont#1#2#3#4#5{%
  \reset@font\fontsize{#1}{#2pt}%
  \fontfamily{#3}\fontseries{#4}\fontshape{#5}%
  \selectfont}%
\fi\endgroup%
{\renewcommand{\dashlinestretch}{30}
{\small%
\begin{picture}(6999,2745)(0,-10)
\path(1287,225)(1287,375)
\path(1512,375)(1512,225)
\path(1962,375)(1962,225)
\path(4512,375)(4512,225)
\path(5037,375)(5037,225)
\path(6012,375)(6012,225)
\put(12,0){\makebox(0,0)[lb]{\smash{{{\SetFigFont{10}{12.0}{\familydefault}{\mddefault}{\updefault}0}}}}}
\put(2712,0){\makebox(0,0)[lb]{\smash{{{\SetFigFont{10}{12.0}{\familydefault}{\mddefault}{\updefault}1}}}}}
\put(3987,0){\makebox(0,0)[lb]{\smash{{{\SetFigFont{10}{12.0}{\familydefault}{\mddefault}{\updefault}0}}}}}
\put(6687,0){\makebox(0,0)[lb]{\smash{{{\SetFigFont{10}{12.0}{\familydefault}{\mddefault}{\updefault}1}}}}}
\put(1887,525){\makebox(0,0)[lb]{\smash{{{\SetFigFont{10}{14.4}{\familydefault}{\mddefault}{\updefault}$q_3$}}}}}
\put(1437,525){\makebox(0,0)[lb]{\smash{{{\SetFigFont{10}{14.4}{\familydefault}{\mddefault}{\updefault}$q_2$}}}}}
\put(1212,525){\makebox(0,0)[lb]{\smash{{{\SetFigFont{10}{14.4}{\familydefault}{\mddefault}{\updefault}$q_1$}}}}}
\put(5937,525){\makebox(0,0)[lb]{\smash{{{\SetFigFont{10}{14.4}{\familydefault}{\mddefault}{\updefault}$q_3$}}}}}
\put(4962,525){\makebox(0,0)[lb]{\smash{{{\SetFigFont{10}{14.4}{\familydefault}{\mddefault}{\updefault}$q_2$}}}}}
\put(4437,525){\makebox(0,0)[lb]{\smash{{{\SetFigFont{10}{14.4}{\familydefault}{\mddefault}{\updefault}$q_1$}}}}}
\path(12,1500)(1287,1500)
\path(1287,1500)(1962,2475)
\path(1287,1500)(1512,1200)
\path(12,375)(12,225)
\path(1512,1200)(2712,975)
\path(1512,1200)(2712,525)
\path(1962,2475)(2712,2700)
\path(1962,2475)(2712,2175)
\path(12,300)(3012,300)
\path(2892.000,270.000)(3012.000,300.000)(2892.000,330.000)
\path(2712,375)(2712,225)
\path(3987,1200)(4512,1200)
\path(4512,1200)(5037,1800)
\path(5037,1800)(6012,2325)
\path(6012,2325)(6687,2100)
\path(6012,2325)(6687,2475)
\path(5037,1800)(6687,1350)
\path(4512,1200)(6687,600)
\path(3987,300)(6987,300)
\path(6867.000,270.000)(6987.000,300.000)(6867.000,330.000)
\path(3987,375)(3987,225)
\path(6687,375)(6687,225)
\put(312,2475){\makebox(0,0)[lb]{\smash{{{\SetFigFont{10}{14.4}{\familydefault}{\mddefault}{\updefault}(a)}}}}}
\put(3987,2550){\makebox(0,0)[lb]{\smash{{{\SetFigFont{10}{14.4}{\rmdefault}{\mddefault}{\updefault}(b)}}}}}
\end{picture}
}
}
\end{center}
 \caption{The correlation functions (a) $C^{(4,1)}_x(q_1,q_2,q_3)$,
 (b) $C^{(4,2)}_x(q_1,q_2,q_3)$, when all $q_i<1$ and are different
 from each other.  The $\varPhi^\prime(y)$ functions at the tip of the
 branches at $q=1$ are understood but not marked.  }
\label{c-x4}
\end{figure}
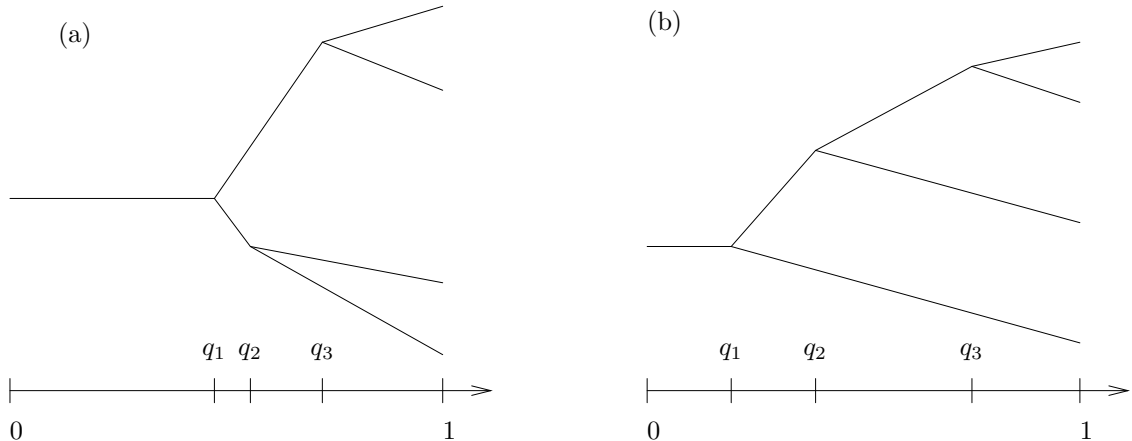

There are only two essentially different correlation functions,
because two topologically different trees with four leaves can be
drawn.  Indeed, $N(4)=2$, {\em c.f.} Eq.\ (\ref{n-k}).  The graphs are
shown on Fig.\ \ref{c-x4}.  They correspond to the first term on the
\rhs of Eq.\ (\ref{4corr-native}) and thus represent the case when all
replica indices are different.  Taking into account coinciding indices
is somewhat involved both analytically and graphically, we give below
only the formulas.  

The graph in Fig.\ \ref{c-x4}a corresponds to
\begin{equation}
  \label{4corr1}
  C^{(4,1)}_x(q_1,q_2,q_3) =  \int\, \rd y P(q_1,y)\,
   \Xi\left(q_1,y;q_2\right)\Xi\left(q_1,y;q_3\right)
   + \,\theta(q_1-1^{-0}) \int\, \rd y\, P(1,y)\,
   \varPhi^{[4]}(y),
\end{equation}
where
\begin{equation}
\label{def-2corr-aux} 
 \Xi\left(q_1,y_1;q_2\right) =  \int\, \rd y_2\,  {\cal
 G}_\varphi (q_1,y_1;q_2,y_2) \left[\mu(q_2,y_2)^2 + \theta(q_2-1^{-0})
  \varPhi^{\pp}(y_2)\right]. 
\end{equation}
Note that $\Xi\left(q_1,y_1;q_2\right)$ can be considered as a
generalized two-replica correlation with extra $q_1,y_1$ dependence,
because $\Xi(0,0;q) = -C_x^{(2)}(q)$.  The inequalities
\begin{equation}
  \label{corr1-q-ineq}
 q_1\leq q_2\leq 1, \;\;  q_1\leq q_3\leq 1
\end{equation}
are understood, so the last term on the \rhs of (\ref{4corr1}) is
nonzero only, if $q_i=1$, $i=1,2,3$.

The topologically asymmetric tree of Fig.\ \ref{c-x4}b is associated with
\begin{eqnarray}
  C^{(4,2)}_x\left(q_1,q_2,q_3\right) & = &\int\, \rd y_1\, \rd y_2\,
   P(q_1,y_1)\, \mu(q_1,y_1)\, {\cal
   G}_\varphi\left(q_1,y_1;q_2,y_2\right)
\mu(q_2,y_2)\, \Xi\left(q_2,y_2;q_3\right) \nn \\
 \label{4corr2}  
  & & +   \theta(q_2-1^{-0}) \int\, \rd y_1\, \rd y_2\, P(q_1,y_1)\,
   \mu(q_1,y_1)\, {\cal G}_\varphi\left(q_1,y_1;1,y_2\right) \, 
 \varPhi^{\prime\prime\prime} (y_2),
\end{eqnarray} 
where we assume 
\begin{equation}
  \label{corr2-q-ineq}
 q_1\leq q_2\leq q_3\leq 1
\end{equation}
but also require $q_1<1$, because the case $q_1=1$ has been
settled by Eq.\ (\ref{4corr1}). 

In conclusion, given the GF for the linear PDE (\ref{theta-pde}),
correlation functions can be calculated in principle.  Interestingly,
the GF for a Fokker-Planck equation also assumes the role here as the
traditional field theoretical GF\@.  Note that this is an instance
where a mean-field property transpires: the graphs to be calculated
are all trees.  It should be added that here the tree structure is the
direct consequence of ultrametricity \cite{sgrev87}, and may carry
over to non-mean-field-like systems with ultrametricity \cite{par99}.
That simple form of graphs is {\em a priori} far from obvious, since
there are techniques for long range interaction systems where diagrams
with loops are present \cite{sgrev91}.  In hindsight we can say that
by using the GF of a Fokker-Planck equation with a nontrivial drift
term, we implicitly performed a summation of infinitely many graphs of
earlier approaches.


\subsection{Variations of the Parisi term} 
\label{ssec-var}

The variation of the free energy term by the OPF $x(q)$ is necessary
in order to formulate later stationarity conditions, and second order
variations yield the matrix of stability against fluctuation of the
OPF.  In this chapter only the mathematical properties are
investigated, physical significance will be elucidated later.

\subsubsection{First variation}
\label{sssec-var1}

The main result of Section \ref{sec-PA-Gen} is that the ubiquitous term
(\ref{gen-parisi}) boils down within the Parisi ansatz to
(\ref{fee-final})\ie
\begin{equation}
  \label{fe-to-vary} \lim_{n\to 0} \varphi[\varPhi(y),{\sf Q}] =
 \varphi[\varPhi(y),x(q)] =\varphi (0,0) .
\end{equation}
In order to determine the variation of $\varphi (0,0)$ in terms of
$x(q)$ we introduce small variations as $x\to x+\delta x$ and $\varphi
\to {\varphi}+\delta {\varphi}$ and require that the varied quantities
also satisfy the PPDE (\ref{ppde1}) with the same initial condition
(\ref{ppde2}) for ${\varphi}+\delta {\varphi}$. Linearization of the
PPDE in the variations gives
\bml
  \label{ppde-var-ext}
\begin{eqnarray}
  \label{ppde-var-ext1}
   \ptl _q \delta {\varphi} &  = & -\hf \ptl _y ^2 \delta {\varphi}
   - x \mu\, \ptl _y \delta {\varphi} - \hf \mu^2 \delta x , \\
  \label{ppde-var-ext2}
    \delta {\varphi}(1,y) & = & 0,
\end{eqnarray}
\eml 
where $\mu(q,y)=\ptl _y{\varphi}(q,y)$ satisfies the PDE
(\ref{deriv1-ppde}).  Eq.\ (\ref{ppde-var-ext}) is an inhomogeneous,
linear PDE for $\delta {\varphi}(q,y)$, given $x(q)$, $\delta x(q)$,
and $\mu(q,y)$.  Note that this is of the form of the linearized PPDE
with source (\ref{theta-with-source-pde}).  Its solution is given in
(\ref{theta-with-source-solve}), whence
\begin{equation}
  \label{ppde-var-sol-ext} \delta {\varphi}(q_1,y_1) = \hf
  \int_{q_1}^1\rd q_2 \int\rd y_2 \, {\cal
  G}_{\varphi}(q_1,y_1;q_2,y_2) \, 
  \mu(q_2,y_2)^2\, \delta x(q_2),
\end{equation}
whence
\begin{equation}
  \label{var-phi-ext} \frac{\delta {\varphi}(q_1,y_1)}{\delta x(q_2)}
  = \hf \theta (q_2-q_1) \int\rd y_2 \, {\cal
  G}_{\varphi}(q_1,y_1;q_2,y_2) \, \mu(q_2,y_2)^2.
\end{equation}
Thus the variation of the term (\ref{fe-to-vary}) is
\begin{equation}
  \label{var-fe-ext} \frac{\delta {\varphi}(0,0)}{\delta x(q)} = \hf
  \int\rd y \, {\cal G}_{\varphi}(0,0;q,y) \, \mu(q,y)^2 = \hf \int\rd y
  \, P(q,y) \, \mu(q,y)^2.
\end{equation}
Here we used the identity (\ref{gf-p}) between the GF and the field
$P(q,y)$.  It is interesting that the above formula is in fact
proportional to the two-replica correlation of Eq.\ (\ref{corr2})
\begin{equation}
 \label{var1-corr} \frac{\delta {\varphi}(0,0)}{\delta x(q)} = \hf
 C_x^{(2)}(q)
\end{equation}
for $q<1$.  Since the correlation function can also be obtained by
differentiation in terms of $q_{ab}$, we have by Eqs.\
(\ref{gen-parisi}, \ref{def-corr2}, \ref{var-fe-ext}), for $q<1$
\begin{equation}
\label{def-deriv-q} 
 \lim_{n\to 0}\left. \frac{\ptl n \varphi[\varPhi(y),{\sf Q}]}{\ptl
 q_{ab}} \right|_{q_{ab}=q} = \frac{\delta {\varphi}(0,0)}{\delta
 x(q)}.
\end{equation}
This relation tells us that if a free energy is the sum of terms
(\ref{gen-parisi}) then the two stationarity conditions, one obtained
by differentiation in terms of the matrix elements $q_{ab}=q$ and the
other by variation in terms of $x(q)$, are equivalent.  Such is the SK
model, the spherical neuron, and the neuron with arbitrary,
independent synapses.

In the case of a discrete $R$-RSB scheme (\ref{ieq}) variation by
$x(q)$ is made with the assumption of a plateau\ie $x(q)\equiv x$,
$0<x<1$, in an interval $I$.  Then the role of the variation will be
taken over by the derivative in terms of the plateau value $x$ and of
the endpoints $q_1$ and $q_2$.  It is straightforward to show that
\begin{equation}
\label{var-fe-m} 
  \frac{\ptl {\varphi}(0,0)}{\ptl x} = \hf \int_I C_x^{(2)}(q)\rd q
\end{equation}
results.  Since the fields $P$ and $\mu$ are purely diffusive in $I$,
the $q$-integral is Gaussian.  On the other hand, the derivatives in
terms of the endpoints are $\hf C_x^{(2)}$ at the endpoints due to
Eqs.\ (\ref{def-deriv-q}, \ref{var1-corr}).  If we work with an ansatz
for the OPF that has both $\dot{x}(q)>0$ and $x(q)\equiv x$, $0<x<1$,
segments, then (\ref{var-fe-ext}) should be used in an interval where
$\dot{x}(q)>0$ and (\ref{var-fe-m}) along a plateau.  If
$\dot{x}(q)>0$ at isolated points, like in a finite $R$-RSB scheme at
jumps, differentiation in terms of the location of that points results
in (\ref{var-fe-ext}) at that points.

\subsubsection{Second variation}
\label{sssec-var2}
 
The stability of a thermodynamic state against fluctuations in the
space of the OPF $x(q)$, the so called longitudinal fluctuations, can
be studied through the second variation of the free energy term
(\ref{fe-to-vary}).  We will present here briefly the way the
longitudinal Hessian can be calculated.
 
In order to determine the variation of the first derivative
(\ref{var-fe-ext}) we should vary the fields $\mu$ and $P$.  For $\mu$
we obtain by definition 
\begin{equation}
  \label{var-m} \frac{\delta {\mu}(q_1,y_1)}{\delta x(q_2)} = \ptl
  _{y_1} \frac{\delta \varphi(q_1,y_1)}{\delta x(q_2)} = \hf \theta
  (q_2-q_1) \int\rd y_2 \, \ptl _{y_1} {\cal
  G}_{\varphi}(q_1,y_1;q_2,y_2) \, \mu(q_2,y_2)^2.
\end{equation}
In order to calculate the variation of the field $P$ we need to vary
the SPDE (\ref{spde}).  This yields
\bml
  \label{spde-var-ext}
\begin{eqnarray}
  \label{spde-var-ext1} \ptl _q \delta {P} & = & \hf \ptl _y ^2 \delta
   {P} - x\, \ptl _y \left(\mu \delta {P}\right) - x\, \ptl _y
   \left(P \delta \mu \right) - \delta x\, \ptl _y \left(\mu
   {P}\right), \\ \label{spde-var-ext2} \delta {P}(0,y) & = & 0,
\end{eqnarray}
\eml 
This can be solved by using the fact that the GF for the SPDE is the
reverse of ${\cal G}_\varphi$. Thus
\begin{eqnarray}
  \label{spde-var-sol-ext} \delta P(q_1,y_1) & = &- \int_0^{q_1}\rd q_2
  \int\rd y_2 \, {\cal G}_{\varphi}(q_2,y_2;q_1,y_1) \nn \\ && \times \,
  \left\{ x(q_2) \, \ptl _{y_2} \left(P(q_2,y_2)\, \delta \mu(q_2,y_2)
  \right) + \ptl _{y_2} \left(P(q_2,y_2)\, \mu(q_2,y_2)\right) \, \delta
  x(q_2) \right\}.
\end{eqnarray}
Hence the variation of $P(q_1,y_1)$ by $x(q_2,y_2)$ is straightforward
to obtain, where also Eq.\ (\ref{var-m}) should be used.  

The above preliminaries allow us to express the second variation of
the free energy functional.  Varying (\ref{var-fe-ext}) gives
\begin{equation} 
  \label{var2-fe-ext} \frac{\delta^2 {\varphi}(0,0)}{\delta
  x(q_1)\delta x(q_2)} = \hf \int\rd y_1 \, \frac{\delta
  P(q_1,y_1)}{\delta x(q_2)} \, \mu(q,y_1)^2 + \int\rd y_1 \, P(q_1,y_1)
  \mu(q_1,y_1) \frac{\delta \mu(q_1,y_1)}{\delta x(q_2)}.
\end{equation} 
Substitution of the variation of $P(q_1,y_1)$ and of $\mu(q_1,y_1)$
yields after some manipulations
\begin{eqnarray} 
  \label{var2-fe-ext-final} \frac{\delta^2 {\varphi}(0,0)}{\delta
  x(q_1)\delta x(q_2)} & = & \hf \int\rd y_1\rd y_2 \, \ptl _{y_1} {\cal
  G}_\varphi (q_{\text{min}}, y_1; q_{\text{max}}, y_2) \,
  P(q_{\text{min}}, y_1)\, \mu(q_{\text{min}},y_1)\,
  \mu(q_{\text{max}},y_2)^2 \nn \\ & & + \case{1}{4}
  \int_0^{q_{\text{min}}}\rd q_3\, x(q_3)\, \int\rd y_1\rd y_2\rd y_3 \,
  P(q_3,y_3) \nn \\ & & ~~~\times \,\ptl _{y_3} {\cal G}_\varphi (q_3,
  y_3; q_1, y_1) \,\ptl _{y_3} {\cal G}_\varphi (q_3, y_3; q_2, y_2)
  \, \mu(q_1,y_1)^2 \mu(q_2,y_2)^2,
\end{eqnarray}
where 
\bml \label{q-min-max}
\begin{eqnarray} 
\label{q-min}
q_{\text{min}} &=& \min(q_1,q_2), \\
\label{q-max}
q_{\text{max}} &=& \max(q_1,q_2). 
\end{eqnarray}
\eml
Note the symmetry of (\ref{var2-fe-ext-final}) \wrt the interchange of
$q_1$ and $q_2$.  If we have the extremizing $x(q)$ as well as the GF
${\cal G}_\varphi$, the latter yielding by (\ref{mu-evolve}) the field
$\mu$, then Eq.\ (\ref{var2-fe-ext-final}) is an explicit expression
for the second functional derivative.


\subsection{The Hessian matrix} 
\label{ssec-hess}

There are results in the literature on the algebraic properties of
ultrametric matrices that can be straightforwardly applied to the
present problem.  As we shall see below, this amounts to finding, in
the state described by a general OPF $x(q)$, an explicit expression
for the eigenvalues of the Hessian in the so called replicon sector,
deemed to be ``dangerous'' from the viewpoint of thermodynamical
stability.

\subsubsection{Ultrametric matrices} 
\label{sssec-hess-def}

The Hessian, or, stability matrix of the free energy term
(\ref{gen-parisi}) is
\begin{equation}
  \label{def-hess} M _{ab,cd} = \frac{\ptl ^2 n \varphi[\varPhi(y),{\sf
 Q}]}{\ptl q_{ab}\, \ptl q_{cd}}.
\end{equation}
If the replica correlations of $x_a$-s as in
(\ref{repl-corr-deriv-by-q}) are thought as moments then
(\ref{def-hess}) is analogous to a cumulant, and can obviously be
expressed as
\begin{equation}
  \label{hess-corr}
 M _{ab,cd} = \lav x_ax_bx_c x_d \rav -  
 \lav x_ax_b \rav \lav x_cx_d \rav = C_x^{(4)}(a,b,c,d) -
 C_x^{(2)}(a,b)\, C_x^{(2)}(c,d).
\end{equation}
The transposition symmetry of the matrix $\sf Q$ was understood in the
above definition.  The Hessian (\ref{def-hess}) becomes a so called
ultrametric matrix \cite{tdk94} once the $R$-RSB form (\ref{pa}) for
$\sf Q$ is substituted.  Note that while constructing the stability
matrix we did not differentiate in terms of the indices $x_r$.
Indeed, one produces the Hessian before the hierarchical form
for $\sf Q$ is substituted, and at that stage the parameters of the
$R$-RSB scheme do not appear.

We can now comfortably apply the results of the elaborate study by
Temesv\'ari, De Dominicis, and Kondor \cite{tdk94} about ultrametric
matrices.  Such matrices have four replica indices and are in essence
defined by the property that they exhibit the same symmetries \wrt the
interchange of indices as the Hessian (\ref{def-hess}) with a Parisi
$\sf Q$ matrix substituted in it.  The theory was originally
formulated for finite $R$-RSB \cite{tdk94}, but, as we shall see,
continuation of the formulas comes naturally.  Firstly we should
clarify notation.  Let us remind the reader to the merger index
$r(a,b)$ defined in the $R$-RSB ansatz by Eq.\ (\ref{merger}) in Section
\ref{sssec-x-corr}.  The  $r(a,b)$ was denoted by $a\cap b$ in  Ref.\ 
\cite{tdk94}.  According to the convention of \cite{tdk94}, the
elements of the ultrametric matrix $\sf M$ can be characterized in a
symmetric way by four merger indices, among them three independent.
Redundancy is the price payed for a symmetric definition.  The new
indices are
\bml \label{merg}
\begin{eqnarray}
   \label{merg2a}
  r_0 & = & r(a,b), \\
  \label{merg2b}
  r_1 & = & r(c,d), \\
  \label{merg2c}
  r_2 & = & \max [ r(a,c),  r(a,d)], \\
  \label{merg2d}
  r_3 & = & \max [ r(b,c),  r(b,d)],
\end{eqnarray}
\eml 
whence
\begin{equation}
  \label{stab-m} 
         M^{r_0,r_1}_{r_2,r_3} \equiv M_{ab, cd}
\end{equation}
is just a relabeling of the Hessian matrix elements.  

According to \cite{tdk94} one can distinguish among three main
invariant subspaces --- sectors --- of the space of $\sf Q$ matrices.
Here we give a loosely worded brief account of the decomposition,
emphasizing also the physical picture that transpires from comparison
with earlier results on the SK model.

The longitudinal sector is spanned by Parisi matrices with the same
set of $m_r$, or, equivalently, $x_r$ (its relation to the $m_r$ is
given by (\ref{x-r})), indices as the matrix $\sf Q$ had that was
substituted into (\ref{def-hess}). In the general case (without
restrictions like the fixing of the diagonal elements) this space has
$R+1$ dimensions. The projection of the Hessian onto the longitudinal
sector is a $R+1\times R+1$ matrix, whose diagonalization cannot be
performed based solely on its utrametric symmetry, but should be done
differently for different free energy terms $\varphi[\varPhi(y),{\sf
Q}]$.  The longitudinal Hessian in the $R\to\infty$ limit is related
to the Hessian of the functional $\varphi[\varPhi(y),x(q)]$ (see Section
\ref{sssec-var2}).  This is demonstrated by the variational
stability analysis of the SK model, within the continuous RSB scheme,
near the spin glass transition, as performed in Ref.\ \cite{tam80}.
The eigenvalue equation obtained by variation was recovered by taking
the $R\to\infty$ limit of the eigenvalue problem within the
longitudinal sector of the Hessian (\ref{def-hess}).  The longitudinal
subspace can be considered as the generalization of a deviation from
the RS solution that equally has RS structure\ie the longitudinal
eigenvector of de Almeida and Thouless (AT) \cite{at78}.

The second sector has been called anomalous in Ref.\ \cite{tdk94}.  It
may be viewed as the generalization of the second family of AT
eigenvectors.  The ultrametric symmetry allowed the transformation of
the Hessian restricted to this invariant subspace into a
quasi-diagonal form of $n-1$ pieces of $R+1\times R+1$ matrices
\cite{tdk94}.  Some of these submatrices are identical, there are only
$R$ different of them in the generic case.  Again, the diagonalization
of these submatrices is a task to be performed on a case-by-case
basis.  To our knowledge no such study has been performed for $R>1$.

The third is the so called replicon sector.  Here the ultrametric
symmetry made it possible to fully diagonalize the Hessian, resulting
in an explicit expression for the replicon eigenvalues in terms of
Hessian matrix elements \cite{tdk94}. The replicon modes, the elements
of this subspace, are the generalization of the eigenvectors of de
Almeida and Thouless that destabilized the RS solution of the SK
model.  In other words, these can be thought as responsible for
replica symmetry breaking.  In the stability analysis by Whyte and
Sherrington \cite{ws96} on the $1$-RSB solution of the storage problem
of the spherical neuron (by Ref.\ \cite{mez93}) it was equally the
replicon eigenvalue that caused thermodynamical instability.  Note
that the replicon modes were also termed as ergodons by Nieuwenhuizen
\cite{nie95,nie95b}, due to their role in the breakdown of ergodicity
in an RSB phase.

\subsubsection{Replicons} 
\label{sssec-replicons}

The replicon sector has special physical significance, since
instability there in known cases signaled the need for higher order
$R$-RSB.  The replicon eigenvalues of an ultrametric matrix can be
written as \cite{tdk94}
\begin{equation}
  \label{repl} \lambda^{r_1}_{r_2,r_3} = \sum_{s=r_2}^{R}
  \sum_{t=r_3}^{R} m_{s+1} m_{t+1} \left( M^{r_1,r_1}_{s+1,t+1} -
  M^{r_1,r_1}_{s+1,t} - M^{r_1,r_1}_{s,t+1} + M^{r_1,r_1}_{s,t}
  \right),
\end{equation}
where $0\leq r_1\leq R$ and $r_1\leq r_2,r_3 \leq R$.  The $r_i$-s are
no longer attached to replica labels as they had been in Eqs.\
(\ref{merg}).  This discrete expression lends itself to continuation,
when one uses parametrization by $q_{r_i}$ to relabel as
\begin{equation}
  \label{m-cont}
  M\left(q_{r_1},q_{r_2},q_{r_3}\right) \equiv M^{r_1,r_1}_{r_2,r_3}.
\end{equation}
Here the inequalities (\ref{corr1-q-ineq}) are implied. Using the
simpler notation of $q_i$-s for parameterization we get  for the
replicon eigenvalues
\begin{equation}
  \label{repl-cont} \lambda\left(q_1,q_2,q_3\right) =
  \int_{q_2^{+0}}^1\rd \bar{q}_2 \int_{q_3^{+0}}^1\rd \bar{q}_3
  \, x(\bar{q}_2) \, x(\bar{q}_3)\, \ptl _{\bar{q}_2} \ptl _{\bar{q}_3}
  M\left(q_1,\bar{q}_2,\bar{q}_3\right).
\end{equation}
Comparison with the sum above shows that the inequalities $q_1 \leq
q_2, q_3\leq q_{(1)}(=q_R)$ need to hold, and, of course, the
eigenvalue is defined only in those $q_i$-s where ${\dot x}(q_i)\neq
0$.  Expression (\ref{repl-cont}) is unambiguous even though the
correlation functions and so the integrand are ill defined over
intervals where $x(\bar{q}_i)$ has a plateau.  In such an interval the
integrand becomes a derivative and we define the quadrature as the
difference between values at the endpoints of the interval.  Eq.\ 
(\ref{repl-cont}) is equivalent to a formula expressed in terms of the
variable $x$ that was quoted in \cite{dtk98}.  We call the reader's
attention also to the fact that the continuation of the sum
(\ref{repl}) implies that in case of ambiguity the right-hand-side
limit in $q$ of the partial derivatives are to be used.  This
distinction is generically of no import in regions where $\infty
>\dot{x}(q)>0$, but is necessary to be made at steps, where the left
and right limits are different.  The lower integration limits in
(\ref{repl-cont}) carry the superscript $+0$ for this reason.  In
order to simplify notation, hereafter we often omit the mark $+0$ but
understand it tacitly wherever necessary. 

\begin{figure}[h!] 
\begin{center} 
\setlength{\unitlength}{0.00083333in}
\begingroup\makeatletter\ifx\SetFigFont\undefined%
\gdef\SetFigFont#1#2#3#4#5{%
  \reset@font\fontsize{#1}{#2pt}%
  \fontfamily{#3}\fontseries{#4}\fontshape{#5}%
  \selectfont}%
\fi\endgroup%
{\renewcommand{\dashlinestretch}{30}
{\small%
\begin{picture}(3869,2013)(0,-10)
\path(762,513)(762,363)
\path(1512,513)(1512,363)
\path(2037,513)(2037,363)
\path(12,1338)(762,1338)
\path(762,1338)(1512,1638)
\path(762,1338)(2037,963)
\dashline{60.000}(2037,963)(2787,963)
\dashline{60.000}(1512,1638)(2787,1638)
\put(762,63){\makebox(0,0)[b]{\smash{{{\SetFigFont{10}{14.4}{\rmdefault}{\mddefault}{\updefault}$q_1$}}}}}
\put(1437,63){\makebox(0,0)[b]{\smash{{{\SetFigFont{10}{14.4}{\rmdefault}{\mddefault}{\updefault}$q_2$}}}}}
\put(2037,63){\makebox(0,0)[b]{\smash{{{\SetFigFont{10}{14.4}{\rmdefault}{\mddefault}{\updefault}$q_3$}}}}}
\put(3162,888){\makebox(0,0)[b]{\smash{{{\SetFigFont{10}{14.4}{\rmdefault}{\mddefault}{\updefault}$\varPhi^{\pp}(y)$}}}}}
\put(3162,1563){\makebox(0,0)[b]{\smash{{{\SetFigFont{10}{14.4}{\rmdefault}{\mddefault}{\updefault}$\varPhi^{\pp}(y)$}}}}}
\put(1062,1563){\makebox(0,0)[b]{\smash{{{\SetFigFont{10}{14.4}{\rmdefault}{\mddefault}{\updefault}${\cal G}_\varphi$}}}}}
\put(2112,1788){\makebox(0,0)[b]{\smash{{{\SetFigFont{10}{14.4}{\rmdefault}{\mddefault}{\updefault}${\cal G}_\mu$}}}}}
\path(12,438)(2787,438)
\path(12,513)(12,438)
\path(12,513)(12,363)
\path(2787,513)(2787,363)
\path(2787,438)(3087,438)
\path(2967.000,408.000)(3087.000,438.000)(2967.000,468.000)
\put(2787,138){\makebox(0,0)[lb]{\smash{{{\SetFigFont{10}{12.0}{\rmdefault}{\mddefault}{\updefault}1}}}}}
\put(12,138){\makebox(0,0)[lb]{\smash{{{\SetFigFont{10}{12.0}{\rmdefault}{\mddefault}{\updefault}0}}}}}
\end{picture}
}
} 
\end{center} 
\caption{The replicon eigenvalue in terms of GF-s.  The full line is
  ${\cal G}_\varphi$ as before, the dashed line represents ${\cal
    G}_\mu$, the GF for the PDE (\ref{deriv1-ppde-evolv}).}
\label{lambda}  
\end{figure}
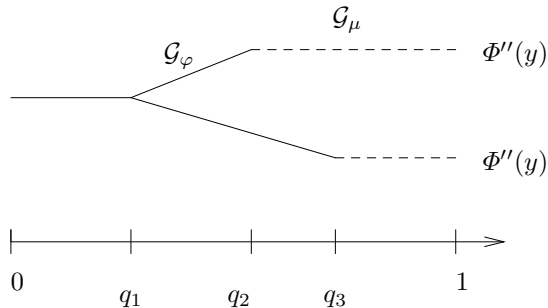

Next we use the expression of the Hessian through correlators as given
by (\ref{hess-corr}).  After inspection of how the discrete labeling
was converted to continuous parametrization we get
\begin{equation}
  \label{hess-corr-q}
  M\left(q_1,q_2,q_3\right) =  C^{(4,1)}_x \left(q_1,q_2,q_3\right) -
  C^{(2)}_x \left(q_1\right)^2,
\end{equation}
where the fourth order correlator defined in (\ref{4corr1}) appears.
Hence the replicon spectrum is
\begin{equation}
  \label{repl-spectr-by-corr} \lambda\left(q_1,q_2,q_3\right) =
  \int_{q_2}^1\rd \bar{q}_2 \int_{q_3}^1\rd \bar{q}_3 \, x(\bar{q}_2) \,
  x(\bar{q}_3) \, \ptl _{\bar{q}_2} \ptl _{\bar{q}_3} C^{(4,1)}_x
  \left(q_1,\bar{q}_2,\bar{q}_3\right).
\end{equation}  
From expression (\ref{4corr1}) for the correlator we obtain
\begin{equation}
\label{repl-spectr}
 \lambda\left(q_1,q_2,q_3\right) =  \int\rd y\, P\left(q_1,y\right)\,
  \Lambda\left(q_1,y;q_2\right)\,
  \Lambda\left(q_1,y;q_3\right),
\end{equation}  
where by definition
\begin{equation}
\label{def-Lambda}
  \Lambda\left( q_1,y_1;q_2 \right) = \int_{q_2}^1
 \rd \bar{q}_2 \, x(\bar{q}_2)\, \ptl _{\bar{q}_2}
  \Xi \left( q_1,y_1; \bar{q}_2 \right).
\end{equation}  
Using Eq.\ (\ref{def-2corr-aux}) for $\Xi$ and the identity
(\ref{gf-mu-kappa}) then substituting for the product $x(q)\,
\kappa(q,y)^2$ the other terms in Eq.\ (\ref{deriv2-ppde-evolv}), next
performing partial integration and noting that ${\cal G}_\varphi$
satisfies in its hind variables the SPDE (\ref{spde}), we obtain
\begin{equation}
\label{Lambda}
  \Lambda\left( q_1,y_1;q_2 \right) = \int\,\rd y_2\, {\cal G}
  _\varphi (q_1,y_1;q_2,y_2) \, \kappa(q_2,y_2).
\end{equation}  
The replicon spectrum can be expressed equivalently by the vertex
function (\ref{vertex-ttt}) as
\begin{equation}
\label{repl-final}
\lambda\left(q_1,q_2,q_3\right) = \int\,\rd y_2\,\rd y_3\,
\Gamma_{\varphi\varphi\varphi} \left( q_1;0,0;q_2,y_2;q_3,y_3
\right) \kappa \left(q_2,y_2 \right)\, \kappa \left(q_3,y_3
\right). 
\end{equation}
This formula can be graphically represented, if we recall that the
field $\kappa$ is produced by the GF ${\cal G}_\mu$ for the PDE
(\ref{deriv1-ppde-evolv}) by (\ref{kappa-evolve}). Let us mark ${\cal
G}_\mu$ with a dashed line, then we have the graph on Fig.\
\ref{lambda}. 

Here we reemphasize that the solution of the relevant PDE-s, in
particular, the field $\varphi(q,y)$ with its derivatives and the GF-s
are assumed to be known, so the correlation functions and the
replicon spectrum are considered as resolved if they are expressed in
terms of the above fields and GF-s.

\subsubsection{A Ward-Takahashi identity} 
\label{sssec-corr-wt}

Recent results indicate the existence of an infinite series of
identities among derivatives of a function of $\sf Q$, such as a free
energy term, provided this term exhibits permutation symmetry in
replica indices and the derivatives are considered with a Parisi
matrix substituted as argument \cite{dtk98,tkd98}.  An equivalent
source of the same identities is a ``gauge'' invariance, namely, the
property that the free energy term looses its dependence on the
specific $m_r$ and $q_r$ values and winds up depending only on $x(q)$
in the $n\to 0$ limit \cite{tkd98}.  These relations can be considered
as analogous to the Ward-Takahashi identities (WTI-s), arising in field
theory for a thermodynamical phase wherein a continuous symmetry is
spontaneously broken \cite{iz9x}.  The continuous symmetry that is
held responsible for the WTI-s is the replica permutation symmetry in
the $n\to 0$ limit, together with the appearance of an interval in $q$
where $x(q)$ is continuous and strictly increasing \cite{dtk98,tkd98}.
In our case, the free energy term (\ref{gen-parisi}) is of the
aforementioned type, so we expect the WTI-s to hold.

Interestingly, the lowest order nontrivial WTI can be easily obtained
based on the results expounded in the previous section.  Let us
consider the replicon eigenvalues (\ref{repl-final}) in the case of
coinciding arguments
\begin{equation} 
  \label{wt-limit-q}
  q = q_1 = q_2 = q_3 \leq q_{(1)}.
\end{equation}
The behavior of the vertex function $\Gamma_{\varphi\varphi\varphi}$ for
coinciding $q$-arguments can be easily deduced from the requirement
that the GF-s become Dirac-deltas for coinciding times.  Then the
replicon eigenvalue assumes the form
\begin{equation} 
\label{wt-lambda} 
\lambda(q,q,q) = \int\,\rd y\, P(q,y)\, \kappa(q,y)^2 = \lambda(q).
\end{equation} 
This is precisely the r.\ h.\ s.\ of the identity (\ref{gf-mu-kappa})
at $q_1=0$, $y_1=0$, while on the l.\ h.\ s.\ of same we discover the
2nd order correlator (\ref{corr2}) for $q<1$, therefore
\begin{equation} 
\label{wt} 
\lambda(q) = \dot{C}^{(2)}_x(q).
\end{equation} 
Strictly, this formula should be taken only at $q$-s where the
correlation function is defined\ie $q$-s that are limits of some
$q_r$-s in the Parisi scheme (\ref{ieq1}).  Nevertheless, we find that
it holds with the smooth continuation of (\ref{corr2}) and
({\ref{wt-lambda}) for any $0\leq q< 1$, the more so remarkable
because the replicon eigenvalues were not defined for arguments
larger than $q_{(1)}$. 
  
Our present derivation yields just one identity out of a set of
infinitely many, but its advantage is that it uses analytic forms, and
it is brief due to our prior knowledge about the properties of the
relevant PDE-s.  Note that the WTI (\ref{wt}) was obtained for a
mathematical abstraction, the formula (\ref{gen-parisi}), but will
gain physical significance once we return to thermodynamics in
Sections \ref{ssec-spher-neur} and \ref{ssec-indep-synaps}.


\eject

\section{Interpretation and Special Properties}
\label{sec-special-properties}
\subsection{Physical meaning of $x(q)$ }
\label{ssec-meaning-xq}

In relation to spin glasses it has been shown that the OPF $x(q)$ is
the average probability that the overlap of two spin configurations
from two different pure (macro)states is smaller than $q$ \cite{p83}.
Furthermore, this property was found to naturally hold for
combinatorial optimization problems that can be mapped to various spin
glass models \cite{sgrev87}.  Similar feature follows from Parisi's
ansatz for $\sf Q$ in the present neuron model evidently, but because
of its significance we briefly give the derivation. Several further
consequences of the hierarchical form of $\sf Q$, as discussed in
\cite{sgrev87}, also carry over to the neuron in the case of RSB.

Firstly let us consider the expression (\ref{quad}), where we replace
$x_i$ by $1$ and $q_{ab}$ by some function of the elements
$F(q_{ab})$.  We obtain, using $m_0=n\to 0$,
\begin{equation}
  \label{f-q-indent1}
  \frac{1}{n} \sum_{a\neq b} F(q_{ab}) = - F(1) +
  \sum_{r=0}^{R+1} \left[  F(q_{r}) -  F(q_{r-1}) \right]  m_r,
\end{equation}
whence, by continuation in the sense of Section \ref{sssec-continuousRSB}
\begin{equation}
  \label{f-q-indent2}
  \left. \frac{1}{n} \sum_{a\neq b} F(q_{ab}) \right| _{n=0} = 
  - F(q_{(1)}) + \int_0^{q_{(1)}}\rd q ~ \dot{F}(q) ~ x(q) 
  =  - \int_0^1\rd q ~ \dot{x}(q) ~ F(q).
\end{equation} 
Here the assumption that only nonnegative $q$-s are relevant and
$q_{R+1}=q_\rD =1$ was used.

A density for the off-diagonal matrix elements of $\sf Q$ can be
obtained by substituting the Dirac delta for $F(q)$ as
\begin{equation}
  \label{distrib-q}
  \left. \frac{2}{n(n-1)} \sum_{a< b} \delta(q-q_{ab}) 
  \right| _{n=0} =   \int_0^1\rd \bar{q} ~ 
  \dot{x}(\bar{q}) ~\delta(q-\bar{q}) = \dot{x}(q).
\end{equation}
Finally, using the notation $\langle \dots \rangle_{(n)}$ for thermal
average with $n$ replicated partition functions, also averaged over
the patterns, the mean probability density of overlaps $P(q)$ is, by
the definition of $q_{ab}$,
\begin{equation}
  \label{distrib-J-overlap1}
  P(q) = \left. \frac{2}{n(n-1)} \sum_{a< b} \left\langle \delta ( q - N^{-1} 
    \jv_a \cdot \jv_b) \right\rangle_{(n)}~ \right| _{n=0} 
  = \left. \frac{2}{n(n-1)} \sum_{a< b}  \left\langle \delta(q-q_{ab})  
    \right \rangle_{(n)} ~ \right| _{n=0}.
\end{equation}
 Since the quantity to be averaged on the r.\ h.\ s.\ does not depend
exponentially on $N$, the saddle point known from the free energy
calculation does not move.  The average $\langle \dots \rangle_{(n)}$
can be thus obtained by simple substitution of the saddle point value
in the Dirac deltas, that is, the $\langle \dots \rangle_{(n)}$ sign
can be removed and we obtain (\ref{distrib-q}), that is,
\begin{equation}
  \label{distrib-J-overlap2}
  P(q) =  \dot{x}(q).
\end{equation}
The $P(q)$ considered here is not to be confounded with the
probability field $P(q,y)$ of Section \ref{sssec-lin-PDEs}.  This
interpretation of $x(q)$ indeed restricts the physically relevant
space to monotonous functions.  Further consequence that should be
born in mind is that $q$-s where $P(q)=0$ have vanishing relative
weight in the thermodynamical limit.  So any quantity depending on
$q$ carries direct physical meaning only for $q$-s where $\dot
x(q)> 0$.  This reservation will hereafter be understood.  The
significance of the $x(q)$ (or $q(x)$) order parameter in long range
interaction systems extend to the finite range problems.  Indeed, the
``mean field'' $q(x)$ plays a role also in the field theory of spin
glasses as discussed in Ref.\ \cite {dtk98}.

It should be emphasized that the distribution $P_{\sf S}(q)$ of
overlaps for a given instance of the patterns ${S_k^\mu}$, is not
self-averaging.  So the quenched average included in $\langle \dots
\rangle_{(n)}$ and hence in the definition of $P(q)$ leads to loss of
information about the distribution of the random variable $q$.


\subsection{Diagonalization of a Parisi matrix } 
\label{ssec-P-matrix}

Since spectral properties of Parisi matrices (\ref{pa}) play an
essential role in our framework, here we briefly review known results
about them (see\eg Refs.\ \cite{cri92,mez91}).  Only the case
$q_{aa}=q_\rD =1$ will be considered here, extension to any diagonals is
straightforward. The eigenvalue problem is
\begin{equation} 
  \label{pa-eigen} 
  {\sf Q} {\bf v}^{(r)} = D^{(r)} {\bf v}^{(r)},
\end{equation} 
where $r$ labels the eigenvalues and -vectors.  The simplest
eigenvector belongs to $r=0$ and has uniform elements, say $ {\bf 
v}^{(0)} = (1,1,\dots ,1)$.  The $r=1$ subspace is spanned by
vectors, orthogonal to ${\bf v}^{(0)}$, that are uniform over boxes of
the first generation, each having $m_1$ number of elements.  An
example is $v_a^{(1)} = 1$ if $a=\ell_1 m_1 +1, \dots ,\ell_1 (m_1
+1)$, $v_a^{(1)} = -1$ if $a=\ell_2 m_1 +1, \dots ,\ell_2 (m_1 +1)$,
with $\ell_1, \ell_2 < n/m_1$, integers, and $v_a^{(1)}= 0$ for other
$a$-s. For a general $r$, the eigenvectors are uniform over boxes of
size $m_r$ and orthogonal to all eigenvectors of lower indices,
yielding the eigenvalues
\begin{equation} 
  \label{pa-spectrum}
  D^{(r)} = \sum  _{p=r} ^ {R+1} m_p \left( q_p - q_{p-1} \right).
\end{equation} 
The dimension of the space of vectors uniform in boxes of size $m_r$
is $n/m_r$, this space is spanned by all eigenvectors of index not
larger than $r$.  Given the fact that the $r=0$ eigenvalue is
non-degenerate, it follows that the degeneracy of the $r$-th, $r>0$,
eigenvalue is 
\begin{equation} 
  \label{pa-multiplicity}
\mu_r=n(m_r^{-1} - m_{r-1}^{-1}).
\end{equation} 
Continuation of (\ref{pa-spectrum}) in the sense of Section
\ref{sssec-continuousRSB} results in eigenvalues indexed by $q$ as
\begin{equation}
  \label{pa-cont-spectrum}
  D(q) = \int_q^1 \rd\bar{q} ~ x(\bar{q}) .
\end{equation}
In the case of finite $R$-RSB, comparison with (\ref{pa-spectrum})
gives
\begin{equation}
  \label{pa-spectr-equiv}
  D(q_r) =  D^{(r+1)}, 
\end{equation}
thus formula (\ref{pa-cont-spectrum}) incorporates both the $R$-RSB
case and the one when $x(q)$ is made up of plateaus and curved
segments.  According to the conclusions of Section
\ref{ssec-meaning-xq}, whereas the function $D(q)$ is defined for all
$0\leq q\leq 1$, it gives eigenvalues only for $q$-s where $\dot x(q)>
0$.  In particular, after continuation and with the notation of
Section \ref{sssec-continuousRSB}, $x(q)\equiv 1$ in the interval
$[q_{(1)},1]$, so we have there from (\ref{pa-cont-spectrum})
\begin{equation}
 \label{linear-Dq} D(q)=1-q.
\end{equation}
While $D(q_{(1)})$ is an
eigenvalue, $D(q_{(1)})=1-q_{(1)}=D^{R+1}$, the $D(q)$ from Eq.\
(\ref{linear-Dq}) has not the meaning of eigenvalue for $q>q_{(1)}$.

The above results allow us to calculate the trace of a matrix function
$F({\sf Q})$   
\begin{equation}
  \label{pa-trace-func}
  {\mbox Tr} F({\sf Q}) = \sum_{r=0}^{R+1}\mu_r F(D^{(r)}) 
  =  n \sum_{r=0}^{R}\frac{1}{m_r}\left[ F(D^{(r)}) - F(D^{(r+1)})
  \right] + n F(D^{(R+1)}). 
\end{equation}
In the continuation process we obtain
\begin{eqnarray}
\lim_{n\to 0} \frac{1}{n}{\text Tr}
  F({\sf Q}) &=& \int_0^{q_{(1)}} \rd q \, F^\prime (D(q)) +
  F(D(q_{(1)})) \nn \\ &=& \int_0^1 \rd q \, \left[
  F^\prime(D(q))-F^\prime (1-q)\right]+F(1) \nn \\ &=& \int_0^1 \rd q \,
  F^\prime(D(q)) + F(0).
  \label{pa-trace-func-cont} 
\end{eqnarray}
Note that depending on $F(q)$ not all alternative forms may be
meaningful\eg if $F(x)=\ln(1-x)$ or $F(x)=\ln x$ then the second or
the third expression is ill defined, respectively.  The explicit
dependence on $q_{(1)}$ was eliminated from the second and third
formulas.  These expressions stay valid also for finite $R$-RSB\@.  A
special case is the calculation of the determinant for
(\ref{fe-spher1c})
\begin{equation}
  \label{pa-lndet}
   \lim_{n\to 0} \frac{1}{n} \ln\text{det}{\sf Q} 
   =  \lim_{n\to 0} \frac{1}{n}{\text Tr} \ln({\sf Q}) 
   = \int_0^1  \rd q \, \left[ \frac{1}{D(q)} - \frac{1}{1-q}\right],
\end{equation} 
where the second formula from (\ref{pa-trace-func-cont}) was used.

Since in the stationarity relation (\ref{stat-q}) the inverse of a
Parisi matrix appears, we will calculate that herewith.  Because of
the fact that that the diagonalizing transformation depends only on
the $m_r$-s, but not on the $q_r$-s, the inverse of a Parisi matrix is
a Parisi matrix with the same $\{m_r\}$ set.  Thus also the elements
of the inverse matrix depend only on the merger index $r(a,b)$
introduced in (\ref{merger}).  It is convenient to parametrize them
also by $q$ as
\begin{equation}
  \label{inverse-q-explicit} [{\sf Q}^{-1}]_{ab} \equiv
  q^{(-1)}(q_{r(a,b)}).
\end{equation}
This defines a function $q^{(-1)}(q)$ by continuation, that has
plateaus within $(q_{r-1},q_r)$ in the $R$-RSB scheme.  Equivalently,
the inverse matrix can be represented by the inverse of $q^{(-1)}(q)$,
the function $x^{(-1)}(q)$ (not to be confounded with the inverse of
$q(x)$ that is $x(q)$). The two characteristics are related through
\begin{equation}
  \label{inverse-xq-xq}
  x^{(-1)}(q^{(-1)}(q)) \equiv x(q). 
\end{equation}
This expresses the fact that in a finite $R$-RSB the set of $x_r$
indices is the same for $\sf Q$ and ${\sf Q}^{-1}$. The spectra are in
reciprocal relation, for $q\leq q_{(1)}$
\begin{equation}
  \label{pa-spectr-inverse}
  D^{(-1)}(q^{(-1)}(q)) =  \frac{1}{D(q)}, 
\end{equation}
whence by differentiation, using (\ref{pa-cont-spectrum}) on each
side, and requiring $q^{(-1)}(0)=0$, we arrive at
\begin{equation}
  \label{pa-xq-inverse}
  q^{(-1)}(q) =  - \int_0^q \frac{\rd\bar{q}}{D(\bar{q})^2}. 
\end{equation}
This leaves the diagonal elements $(q^{-1})_{R+1}=q^{(-1)}(1)$ of
${\sf Q}^{-1}$ undetermined, that is obtained from the reciprocal
relation of the respective eigenvalues of index $R+1$, yielding
\begin{equation}
  \label{Rplus1-inverse}
  q^{(-1)}(1)=\frac{1}{1-q_{(1)}}-\int_0^{q_{(1)}} 
  \frac{\rd\bar{q}}{D(\bar{q})^2}. 
\end{equation}
An attempt to continuation of $q^{(-1)}(q)$ between $q_{(1)}$ and $1$
shows that $q^{(-1)}(q)$ is non-monotonic.  Again, relations
(\ref{pa-xq-inverse},\ref{Rplus1-inverse}) equally hold for the
discrete $R$-RSB case, as well as when $x(q)$ has both plateaus and
curved segments, with the usual reservation that (\ref{pa-xq-inverse})
relates matrix elements only when $\dot x(q)>0$.


\subsection{Symmetries of Parisi's PDE} 
\label{ssec-sym-PPDE}

A systematic procedure of identifying all continuous symmetries of a
PDE is the so called prolongation method \cite{prolo86}.  The knowledge
of a continuous symmetry group allows one to generate out of a given
solution a family of other solutions.

Via the prolongation method we find by construction that there are
altogether three one-parameter transformations leaving the PPDE
(\ref{ppde}) invariant.  The action of these symmetries on a solution
$\varphi(q,y)$ can be given as a one-parameter family
$\varphi(s,q,y)$, with $\varphi(0,q,y)=\varphi(q,y)$.  These
one-parameter families are \bml
  \label{sym-ppde}
\begin{eqnarray}
  \label{sym-ppde1}
  \varphi_1(s,q,y) & = & \varphi(q,y+s), \\
  \label{sym-ppde2}
  \varphi_2(s,q,y) & = & \varphi(q,y) + s, \\
  \label{sym-ppde3}
  \varphi_3(s,q,y) & = & \varphi\left(q,y-D(q)\,s\right) 
  -ys+\hf D(q)s^2 ,
\end{eqnarray}
\eml where $D(q)$ is defined by (\ref{pa-cont-spectrum}). The fact
that the above families are solutions of the PPDE (\ref{ppde}),
provided $ \varphi(q,y)$ is also a solution, can also be shown by
substitution.  The additional statement, namely, that there are no
more continuous symmetries, follows from the construction of the
prolongation method that we cannot undertake to describe here.

Eq.\ (\ref{sym-ppde1}) represents translation in $y$, while
(\ref{sym-ppde2}) is a shift of the field $\varphi$ by a constant,
these symmetries are obvious.  The third one, (\ref{sym-ppde3}), is
less so, it is a shift of the origin in $y$ and of the field $\varphi$
and a 'tilting' of the field $\varphi$ in $y$.

The symmetry transformation equally changes the initial condition.  As
a forward reference we note that, in the case of the energy term for
the storage problem of a single neuron, the PPDE (\ref{ppde-ener}) has
the error measure potential $V(y)$ as initial condition.  The constant
shift and the 'tilting' in $y$ changes $V(y)$ such that it no longer
satisfies the properties of $V(y)$ outlined in Section
\ref{ssec-model}.  Thus uncovering the above symmetries is of little
help in finding solutions to the PPDE in the neuron problem at hand.
However, given the relevance of the Parisi solution to a vast class of
disordered systems, we considered the symmetries worth presenting.


\subsection{Spherical entropic term: a solvable case of Parisi's PDE}
\label{ssec-sher-solving-P-equ}

While most of the relevant quantities related to the spherical
entropic term are straightforward to calculate, from the technical
viewpoint they represent a solvable example of Parisi's framework,
suitable for an exercise.

Note that the general distribution (\ref{gen-prior}) does not include
the overall spherical normalization (\ref{spher-prior}), so the
results on independent synapses do not carry over.  Nevertheless, we
can cast (\ref{fe-spher1c}) into the general form (\ref{gen-parisi})
with the association
\begin{equation}
  \label{spher-assoc}
  e^{\varPhi_\rs (y)} = \sqrt{2\pi}\, \delta(y).
\end{equation}
The subscript $s$ signals that we are dealing with the entropic term
of the free energy.  For we need to regularize the Dirac-delta, we use
a Gaussian with small variance $\sigma$. With the notation
(\ref{gauss}) we have
\begin{equation}
  \label{spher-phi} \varPhi_{\rs ,\sigma}(y) = \ln \left( \sqrt{2\pi}\,
  G(y,\sigma)\right) = - \frac{1}{2} \left[ \frac{y^2}{\sigma}
  +\ln\sigma \right].
\end{equation} 
Thus 
\begin{equation} 
  \label{spher-by-parisi}
  f_\rs ({\sf Q}) = - (2\beta)^{-1} \ln\text{det}{\bf\sf Q} 
  = \lim_{\sigma\to 0} \beta^{-1}n
  \varphi[\varPhi_{\rs ,\sigma}(y),{\sf Q}].
\end{equation} 
We keep $\sigma$ finite while performing continuation\ie the limits
$n\to 0$ and $\sigma\to 0$ will be interchanged.  Then we need to
solve the PDE (\ref{ppde1}) with initial condition
\begin{equation}  
  \label{spher-by-parisi-init}
  \varphi_{\rs ,\sigma}(1,y) =\varPhi_{\rs ,\sigma}(y).
\end{equation} 
This can be done by our assuming that $\varphi_{\rs ,\sigma}(q,y)$ is a
quadratic polynomial in $y$.  With the notation of
(\ref{pa-cont-spectrum}) and $D_\sigma(q)= \sigma + D(q)$, the
solution is
\begin{equation} 
  \label{shper-by-parisi-solution} \varphi_{\rs ,\sigma}(q,y) = -
  \frac{1}{2} \left[ \frac{y^2}{D_\sigma(q)} + \ln\sigma + \int_q^1
  \frac{\rd\bar{q}}{D_\sigma(\bar{q})} \right].
\end{equation} 
Hence we obtain 
\begin{equation} 
  \label{spher-by-parisi-final} \lim_{n\to 0} \frac{1}{n} f_\rs ({\sf Q})
  = f_\rs [x(q)] = \lim_{\sigma\to 0} \beta^{-1} \varphi_{\rs ,\sigma}(0,0)
  = \lim_{\sigma\to 0} - (2\beta)^{-1} \left[ \int_0^1
  \frac{\rd q}{D_\sigma(q)} + \ln\sigma \right].
\end{equation} 
The rightmost expression is, apart from a prefactor, equivalent to
(\ref{pa-lndet}).  Either of them thus gives Eq.\ (\ref{fe-spher1c}).

As it has been described in Section \ref{sec-corr-stab}, expectation
values are calculated by using ${\cal G}_\varphi$.  That is by Eq.\
(\ref{gf-varphi}) the GF of Eq.\ (\ref{theta-pde}), the linear PDE for
the field $\vartheta$.  Given the field 
\begin{equation} 
  \label{entr-mu} \mu_{\rs,\sigma}(q,y)=-\frac{y}{D_{\sigma}(q)},
\end{equation} 
from the definition (\ref{def-mu}) and from
(\ref{shper-by-parisi-solution}), the GF is found to be Gaussian
\bml 
\label{spher-gf}
\begin{eqnarray}
  {\cal G}_{\rs ,\varphi}(q_2,y_2;q_1,y_1) &=& G( A,B), \\ A &=& y_2
  \frac{D_\sigma (q_1)}{D_\sigma(q_2)} - y_1, \\ B &=& D_\sigma(q_1)^2
  \left[E_\sigma(q_1)-E_\sigma(q_2)\right],\\ \label{def-eq}
  E_\sigma(q) &=& \int_0^q \frac{\rd\bar{q}}{D_\sigma(\bar{q})^2}.
\end{eqnarray}
\eml
Note that we omitted the subscript $\sigma$ from the GF.
Sompolinsky's time-dependent density is by (\ref{gf-p})
\begin{equation}
  \label{spher-p} P_{\rs }(q,y) = G\left( y , D_\sigma(q)^2
  E_\sigma(q) \right).
\end{equation}
The generalized two-replica correlator (\ref{def-2corr-aux}) is thus
\begin{equation}
  \label{spher-gen-corr2}
  \Xi_\rs (q_1,y_1;q_2) = E_\sigma(q_2) -  E_\sigma(q_1) +
  \frac{y_1^2}{D_\sigma(q_1)^2} - \sigma^{-1}\theta(q_2 - 1^{-0}),
\end{equation}
whence the correlation function (\ref{corr2}) is
\begin{equation}
  \label{spher-corr2} C^{(2)}_{\rs,x}(q) = \Xi_\rs (0,0;q) = E_\sigma(q) -
  \sigma^{-1} \theta(q-1^{-0}).
\end{equation}
The regularizing parameter $\sigma$ can be taken zero at many a place
in the above formulae, an exception being the correlator at $q=1$,
where the integration should be performed first in $(q_{(1)},1)$ to
get the finite result $C^{(2)}_{\rs ,x}(1) = E(q_{(1)}) -
(1-q_{(1)})^{-1}$. We evaluate the first of the two 4-replica
correlators from (\ref{4corr1}) as
\begin{equation}
  \label{spher-4corr1} C^{(4,1)}_{\rs ,x}\left(q_1,q_2,q_3\right) = 2\,
  E_\sigma\left(q_1\right)^2 + \left[ E_\sigma\left(q_2\right)\, -
  \sigma^{-1}\, \theta\left(q_2-1^{-0}\right) \right] \left[
  E_\sigma\left(q_3\right) - \sigma^{-1}\,
  \theta\left(q_2-1^{-0}\right) \right].
\end{equation}
The replicon eigenvalue from (\ref{repl-spectr-by-corr}) is as
\begin{equation}
  \label{spher-repl}
  \lambda_\rs \left(q_1,q_2,q_3\right) = \left[
  D_\sigma\left(q_2\right) \,  D_\sigma\left(q_3\right) \right] ^{-1},
\end{equation}
independent of $q_1$. Note that the maximal argument allowed in
the eigenvalue is $q_{(1)}$, if this is smaller than 1 then the
regularization parameter $\sigma$ can be omitted.  The one WTI
(\ref{wt}) can be checked directly by comparing Eqs.\ 
(\ref{spher-corr2},\ref{spher-repl}).

In this example the PPDE could be solved in closed form.  The question
obviously arises, under what conditions can the solution be obtained
analytically.  It is easy to see that if the initial condition at
$q=1$ is quadratic in $y$ then, for arbitrary $x(q)$, the solution can
be explicitly given as a quadratic function in $y$, with $q$-dependent
coefficients.  Other analytic solutions we did not find for a general
$x(q)$, but of course for special $x(q)$-s, like step functions, the
PPDE can be solved in closed form.  

\subsection{Small field expansion}
\label{ssec-small-potential}

The case of an overall small function $\varPhi(y)$ in (\ref{gen-parisi})
is of interest because, on the one hand, in the neuron problem it
corresponds to the high temperature limit, and on the other, it will
yield the usual energy term in several of the infinite range
interaction spin glass models.  The latter feature stresses the
generality of the framework discussed in this paper.  We can apply
straightforward perturbation expansion by introducing a small
parameter $\epsilon$ and writing
\begin{equation}
  \label{small-pot-phi}
  \varphi[\epsilon\varPhi(y),{\sf Q}] = \epsilon\, \varphi_1[\varPhi(y),{\sf
    Q}] + \epsilon^2 \, \varphi_2[\varPhi(y),{\sf Q}] + O(\epsilon^3),
\end{equation}
where we took into account that the $O(\epsilon^0)$ term vanishes.
The linear term is 
\begin{eqnarray}
  \label{small-pot-phi1a} \varphi_1[\varPhi(y),{\sf Q}] & = & \frac{1}{n}
  \int \! \frac{\rd^n\! x \,\rd^n\! y}{(2\pi)^n} e^{i\xvs\yvs -
  \case{1}{2} \xvs {\sf Q}\xvs} \sum_{a=1}^n \varPhi(y_a) =
  \frac{1}{n} \sum_{a=1}^n \int\! \frac{\rd x\,\rd y}{2\pi} \, \varPhi(y)
  e^{ixy - \case{1}{2} x^2 q_{aa}} \nn \\ & =&
  \frac{1}{n}\sum_{a=1}^n\int\! \rD z\,
  \varPhi\left(z\sqrt{q_{aa}}\right), 
\end{eqnarray}
which, for $q_{aa}\equiv 1$, gives
\begin{equation}
  \label{small-pot-phi1b}
  \varphi_1[\varPhi(y),{\sf Q}]= \int\! \rD z\, \varPhi(z).
\end{equation}
In $O(\epsilon^2)$ we obtain 
\begin{eqnarray}
  \label{small-pot-phi2a} \varphi_2[\varPhi(y),{\sf Q}] & = &
   \frac{1}{2n} \int \! \frac{\rd^n\! x \,\rd^n\! y}{(2\pi)^n} e^{i\xvs
   \yvs - \case{1}{2} \xvs {\sf Q}\xvs} \sum_{a,b=1}^n
   \varPhi(y_a)\varPhi(y_b) - \frac{n}{2} \varphi_1^2 \nn \\ &=&
   \frac{1}{2n}\sum_{a,b=1}^n\int\!
   \frac{\rd x_1\,\rd x_2\,\rd y_1\,\rd y_2}{(2\pi)^2} \,
   \varPhi(y_1)\varPhi(y_2) 
   e^{i\xvs \yvs - \case{1}{2} ( x_1^2 q_{aa} + x_2^2 q_{bb} + 2 x_1
   x_2 q_{ab})} - \frac{n}{2} \varphi_1^2.
\end{eqnarray}
In the last expression $i\xv\yv$ is shorthand for
$i(x_1y_1+x_2y_2)$.  When $n\to 0$ the term $\hf n \varphi_1^2$
vanishes.

In the generic case of $q_{aa}=q_\rD =1$ we obtain after elementary
manipulations
\begin{equation}
  \label{small-pot-phi2b}
   \varphi_2[\varPhi(y),{\sf Q}] = \frac{1}{2n} \sum_{a,b=1}^n
   \varPsi(q_{ab}), 
\end{equation}
where 
\bml \label{small-pot-psi-def}
\begin{eqnarray}
   \label{small-pot-psi}
   \varPsi(q) & = & \int \! \rD z_1 \rD z_2 \, \varPhi(\nv_1 \zv) \,
   \varPhi(\nv_2 \zv), \\
  \label{small-pot-n-cond}
  |\nv_1| & = & |\nv_2| =1, \;\; \nv_1 \nv_2 = q.
\end{eqnarray}
\eml
Here $\nv_1 \zv $, etc.\ denote scalar products of
two-dimensional vectors. In the continuous limit
\begin{equation}
  \label{small-pot-phi2c}
     \varphi_2[\varPhi(y),x(q)] = \frac{1}{2} \int_0^1 \! \rd q\,
     x(q)\, \dot{\varPsi}(q) 
\end{equation}
results, where (\ref{f-q-indent2}) with $q_{aa}\equiv 1$ was used,
thus the term (\ref{small-pot-phi}) is hereby resolved up to
$O(\epsilon^2)$.

For the derivatives of $\varPsi(q)$, with the notation
(\ref{shorthand-diff}), we obtain the suggestive formula
\begin{equation}
  \label{small-pot-psi-deriv}
  \varPsi^{[k]}(q) = \int \! \rD y_1 \rD y_2 \, \varPhi^{[k]} (\nv_1
   \yv) \,    \varPhi^{[k]} (\nv_2 \yv),
\end{equation}
together with the condition (\ref{small-pot-n-cond}). This yields a
simple relation between appropriate expansion coefficients of the
functions $\varPhi(y)$ and $\varPsi(q)$. Namely, if
\begin{equation}
  \label{psi-taylor}
  \varPsi(q) = \sum_{k=0}^{\infty} \, \varPsi_k \, q^k
\end{equation}
then applying (\ref{small-pot-psi-deriv}) with
(\ref{small-pot-n-cond}) at $q=0$ we get
\begin{equation}
  \label{psi-taylor-coeff}
  \varPsi_k = \frac{1}{k!}\varPsi^{[k]}(0) = \frac{1}{k!} \left[
    \int\! \rD y\, 
    \varPhi^{[k]} (y) \right]^2.
\end{equation}
On the other hand, assuming that $\varPhi(y)$ is not diverging too fast
for large $|y|$, we have
\begin{equation}
  \label{phi-hermite1}
  \int\! \rD y \, \varPhi^{[k]} (y) = (-1)^k \int\!  \rD y\, \varPhi (y)
  e^{\frac{1}{2}y^2}\, 
  \frac{\rd^k~}{\rd y^k} e^{-\frac{1}{2}y^2}  = 
  \frac{(-1)^k}{2^{\frac{k}{2}}}\,  \int\! \rD y\, \varPhi(y) H_k\left(
  \frac{y}{\sqrt{2}}\right),
\end{equation}
where $H_k(y)$ is the $k$-th Hermite polynomial.  Hence, given the
Hermite expansion of $\varPhi(y)$ as 
\begin{equation}
  \label{phi-hermite2} \varPhi(y) = \sum_{k=0}^\infty\, \varPhi_k\,
  H_k\left(\frac{y}{\sqrt{2}}\right),
\end{equation}
then using the orthogonality
\begin{equation}
  \label{hermite-orthog}
  \int \! \rD y\,  H_k\left(\frac{y}{\sqrt{2}}\right)\,  H_{l} 
  \left(\frac{y}{\sqrt{2}}\right) = 2^k\, k!\, \delta_{kl}
\end{equation}
we have for the Taylor coefficients of $\varPsi(q)$ 
\begin{equation}
  \label{phi-psi-coeff}
  \varPsi_k = k!\, 2^k\, \varPhi_k^2.
\end{equation}
In conclusion, for a given analytic $\varPsi(q)$, with nonnegative Taylor
coefficients, we can thus construct a $\varPhi(y)$ that reproduces
$\varPsi(q)$ through the expression (\ref{small-pot-psi}). The
correspondence between $\varPsi(q)$ and $\varPhi(y)$ is not one-to-one,
because all $\varPhi(y)$-s with Hermite coefficients $\pm\varPhi_k$ will
yield the same $\varPsi(q)$.

The expression (\ref{small-pot-phi2b}) is the ubiquitous form for the
energy term in various SK-type models, 
\begin{equation}
  \label{gen-energ}
   f_\re ({\sf Q}) = A \sum_{a,b=1}^n \varPsi(q_{ab}), 
\end{equation}
where $A$ is a prefactor depending of the model.  In particular, we
have in the cases of the SK spin glass \cite{sgrev87}, the $p$-spin
interaction \cite{gm84}, and Nieuwenhuizen's multi-$p$-spin
interaction model \cite{nie95}, for $\varPsi(q)$ the functions $q^2$,
$q^p$, and $f(q)$, respectively.  (The corresponding formula with
$\varPsi(q)=q^2$, in the SK replica free energy, is the second term in
Eq.\ (\ref{sk4}).)  The multi-$p$-spin interaction model, which
incorporates the Ising and $p$-spin as special cases, has the $p$-spin
component entering through the characteristic exchange constant $J_p$
and leads to
\begin{equation} 
  \label{nieuw}
  \varPsi(q) = \sum_{p=1}^{\infty} \, \frac{q^p}{p}\, J_p^2,
\end{equation}
whence 
\begin{equation}
  \label{nieuw-our}
  \varPhi(y) = \sum_{p=1}^{\infty} \, \frac{J_p}{p\sqrt{2^p(p-1)!}}\,
    H_p\left( \frac{y}{\sqrt{2}} \right).
\end{equation}
Given $\varPhi(y)$, 
\begin{equation}
  \label{gen-energ-our} f_\re ({\sf Q}) = n\,A
  \left. \frac{\rd^2~}{\rd\epsilon^2} \varphi[\epsilon\varPhi(y),{\sf
  Q}]\right|_{\epsilon =0}
\end{equation}
relates the spin glass energy term to the general framework expounded
in this chapter.  Due to the fact, that the Taylor coefficients of
$\varPsi(q)$ in a multi-$p$-spin interaction system are necessarily
non-negative, a given $\varPhi(q)$ uniquely determines the corresponding
multi-$p$-spin interaction model. 

Whereas for the evaluation of the free energy term
(\ref{small-pot-phi}) the usage of PDE-s could be avoided, we invoke
the auxiliary $q$- and $y$-dependent fields for the calculation of
expectation values.  Since now the initial condition of the PPDE
(\ref{ppde}) and thus the solution of it, $\varphi(q,y)$, is of
$O(\epsilon)$, in lowest order the nonlinear term in (\ref{ppde}) can
be omitted.  Writing $\varphi(q,y)\approx \epsilon\varphi_1(q,y)$, and
using similar notation for the derivative fields $\mu(q,y)$ and
$\kappa(q,y)$, we obtain linear diffusion equations for the fields
$\varphi_1(q,y)$, $\mu_1(q,y)$, and $\kappa_1(q,y)$.  Hence
\begin{eqnarray}
  \label{small-pot-varphi}
  \varphi_1(q,y) & = & \int\! \rd y_1\, G(y-y_1,1-q)\, \varPhi(y_1),
  \\
  \label{small-pot-mu}
  \mu_1(q,y) & = & \int\! \rd y_1\, G(y-y_1,1-q)\, \varPhi^{\prime}(y_1),
  \\
  \label{small-pot-kappa}
  \kappa_1(q,y) & = & \int\! \rd y_1\, G(y-y_1,1-q)\, \varPhi^{\prime\prime}(y_1).
\end{eqnarray}
The GF (\ref{gf-varphi}) is in leading order a Gaussian
\begin{equation}
  \label{small-pot-gf} {\cal G} _{\varphi} (q_2,y_2;q_1,y_1) =
  G(y_1-y_2,q_1-q_2) + O(\epsilon),
\end{equation}
and 
\begin{equation} 
  \label{small-pot-P} 
  P(q,y) = {\cal G} _{\varphi} (0,0;q,y) = G(y,q) + O(\epsilon), 
\end{equation} 
Thus the two-replica correlator (we only treat here the $q<1$ case) is
by Eq.\ (\ref{corr2}) in leading order 
\begin{eqnarray} 
  \label{small-pot-corr2} C^{(2)}_x(q) &=&  \epsilon^2 \int\!
  \left[\prod_{i=1}^3 \rd y_i\right] G(y_1,q) \, G(y_1-y_2,1-q)\,
  \varPhi^{\prime}(y_2)\, G(y_1-y_3,1-q)\, \varPhi^{\prime}(y_3)\nn \\
  &=&  
  \epsilon^2 \dot{\varPsi}(q).
\end{eqnarray} 
From the non-negativity of the Taylor coefficients of $\varPsi(q)$, see
Eq.\ (\ref{nieuw}), it follows that $ C_x^{(2)}(q)\geq 0$.  The replicon
spectrum of (\ref{repl-spectr}) can also be evaluated by our noting
that (\ref{Lambda}) is now
\begin{equation} 
  \label{small-pot-L} 
  \Lambda(q_1,y_1;q_2) = \epsilon \kappa_1(q_1,y_1) + O(\epsilon^2),
\end{equation} 
independent of $q_2$, whence in leading order
\begin{eqnarray}    
  \label{small-pot-repl-spectr} \lambda\left( q_1, q_2, q_3\right) &=&
  \epsilon^2 \int\!  \left[\prod_{i=1}^3 \rd y_i\right] G(y_1,q_1)\,
  G(y_1-y_2,1-q_1)\, \varPhi^{\pp}(y_2) \nn \\ && \times \,
  G(y_1-y_3,1-q_1)\, \varPhi^{\pp}(y_3) \nn \\ &= &\epsilon^2
  \ddot{\varPsi}(q_1).
\end{eqnarray}  
Due to the non-negativity of the Taylor coefficients in Eq.\
(\ref{nieuw}) we have $\lambda\left( q_1, q_2, q_3\right) \geq 0$.
Comparison with (\ref{small-pot-corr2}) shows immediately that the WTI
(\ref{wt}) is satisfied.  The eigenvalues associated with the SK-type
energy term (\ref{gen-energ}) are obtained, based on
(\ref{gen-energ-our}), as $2A\ddot{\varPsi}(q_1)$.


\eject
\section{The Neuron: Spherical Synapses }
\label{ssec-spher-neur}

Having worked out the technical tools in the previous sections, we are
now in the position to apply them for the storage problem of the
McCulloch-Pitts neuron.

\subsection{General results}
\label{ssec-gen-err-measure}

\subsubsection{Free energy and stationarity condition}
\label{ssec-fe-stati}

The free energy (\ref{fe-spher1}) can be resolved based on the results
of Sections \ref{sec-PA-Gen}, \ref{sec-corr-stab}, and
\ref{sec-special-properties} with the substitution
\begin{equation}
  \label{phi-ener}
  \varPhi(y) = - \beta V(y). 
\end{equation}
The specific formula for the free energy is one of our main results,
so however elementary the above substitution is, we collect the
relevant expressions below.  Introducing the field
\begin{equation}
  \label{fe-field-ener}
  f(q,y) = - \beta^{-1} \varphi(q,y), 
\end{equation}
we obtain, from Eqs.\ (\ref{e-phi}) and (\ref{fee-final}), the energy
contribution to the free energy term (\ref{fe-spher1d}), as a functional
of the OPF $x(q)$
\begin{equation}
  \label{fee-ener-parisi}
  f_\re [x(q)] = \lim_{n\to 0} \frac{1}{n}  f_\re ({\sf Q}) = - \beta^{-1}
   \varphi[-\beta V(y),{\sf Q}] \big| _{n=0} = f(0,0),
\end{equation}
where, from (\ref{ppde}), the  $f(q,y)$ is the solution of 
\bml
  \label{ppde-ener}
\begin{eqnarray}
  \label{ppde-ener1}
   \ptl _q f &  = & -\hf \ptl _y ^2 f
   + \hf \beta x \left(\ptl _y f \right)^2 , \\
  \label{ppde-ener2}
    f(1,y) & = & V(y).
\end{eqnarray}
\eml 
The analog of the function $\mu(q,y)$ of Eq.\ (\ref{def-mu}),
useful for the calculation of replica correlators, is now
\begin{equation}
  \label{intro-m}
  m(q,y) = \ptl_y f(q,y) = -\beta^{-1} \mu(q,y)
\end{equation}
and from (\ref{deriv1-ppde}) we get
\bml
\label{deriv-ppde-ener}
\begin{eqnarray}
  \label{deriv-ppde-ener-evolv}
   \ptl _q m &  = & -\hf \ptl _y ^2  m + \beta x m \ptl _y m , \\
  \label{deriv-ppde-ener-init}
  m(1,y) & = &V^\prime(y).
\end{eqnarray}
\eml
By introducing
\begin{equation}
  \label{intro-khi}
  \chi(q,y) = \ptl _y^2 f(q,y) = - \beta^{-1} \kappa(q,y)
\end{equation}
we obtain
\bml
\label{deriv2-ppde-ener}
\begin{eqnarray}
  \label{deriv2-ppde-ener-evolv}
   \ptl _q \chi &  = & -\hf \ptl _y ^2  \chi + \beta x (m \ptl _y \chi
   + \chi^2), \\
  \label{deriv2-ppde-ener-init}
  \chi(1,y) & = &V^\pp (y) .\end{eqnarray}
\eml
The $q$-dependent probability density $P(q,y)$, satisfying the SPDE
(\ref{spde},\ref{spde-init})  now obeys
\bml
\label{spde-ener}
\begin{eqnarray}
  \label{spde-evolv-ener}
 \ptl _q P  & = & \hf \ptl _y ^2 P  + \beta x \ptl _y (P m) , \\
  \label{spde-init-ener}
  P(0,y) & = &\delta(y) .
\end{eqnarray} 
\eml The entropic term (\ref{fe-spher1c}) has  essentially been
calculated through the formula (\ref{pa-lndet}), whence we have
\begin{equation}
  \label{fes-spher-parisi}
  f_\rs[x(q)] =  \lim_{n\to 0} \frac{1}{n}  f_\rs({\sf Q}) 
  = - \frac{1}{2\beta} \int_0^1  
  \rd q\, \left[ \frac{1}{D(q)} - \frac{1}{1-q}\right] .
\end{equation} 
The specific form of the stationarity condition (\ref{stat-q})
immediately follows from (\ref{pa-xq-inverse}) and
(\ref{corr2}) as
\begin{equation}
  \label{stat-cond}
  \int_0^q \frac{\rd \bar{q}}{D(\bar{q})^2} = \alpha \beta^2 \int \rd y~
  P(q,y)~ m(q,y)^2 .   
\end{equation}
This equation holds at isolated $q_r$-s in an $R$-RSB scheme, and does
so identically in an interval where $\dot{x}(q)>0$.  The question of
plateaus with value in $(0,1)$ will be efficiently treated by the
variational formalism of Section \ref{ssec-variation}.  The stationary OPF
$x(q)$ should be substituted into (\ref{fee-ener-parisi}) and
(\ref{fes-spher-parisi}), which by (\ref{fe-spher1b}) sum up to the
value of the thermodynamical free energy
\begin{equation}
  \label{fe-spher-xq-final}
  f = f_\rs[x(q)] + \alpha f_\re[x(q)],
\end{equation}
whose ingredients we redisplay as 
\bml
\label{fe-functional-spher-ext} 
\begin{eqnarray}
  \label{fe-functional-spher-s-ext}
  f_\rs[x(q)] & = & - \frac{1}{2\beta} \int_0^1  
  \rd q ~ \left[ \frac{1}{D(q)} - \frac{1}{1-q}\right], \\ 
  \label{fe-functional-spher-e-ext}
  f_\re[x(q)] & = & f(0,0).
\end{eqnarray}
\eml 
The distribution of local stabilities, introduced in
(\ref{repl-dist-stab}), is an expectation value of a type previously
 calculated.  Using Eq. (\ref{theta-final-aver}) we have the simple
result
\begin{equation}
  \label{dist-stab-spher}  
  \rho(\Delta) = \lav \delta(y_1-\Delta) \rav = \int \rd y ~ P(1,y)~
  \delta(y-\Delta) = P(1,\Delta).  
\end{equation}
The energy can be directly obtained from this distribution by Eq.\ 
(\ref{ener-dist}) as
\begin{equation} 
  \label{ener-dist-final}
  \varepsilon = \lav V \left( y_1 \right) \rav 
  = \int \rd y\, P(1,y)\, V(y).
\end{equation}
The average of any function of $\Delta$ is, in general, the function's
average over the distribution $P(1,y)$.  This shows the physical
meaning of the auxiliary variable $y$ within the Parisi framework: it
is the stability parameter, extended to any intermediary stage $q$,
obeying a distribution $P(q,y)$, that becomes at $q=1$ the physically
observable distribution $P(1,y)$.  The entropy is by (\ref{entr},
\ref{fe-spher-xq-final}, \ref{fe-functional-spher-ext})
\begin{equation} 
  \label{spher-entr}
  s = \frac{1}{2} \int_0^1  
  \rd q ~ \left[ \frac{1}{D(q)} - \frac{1}{1-q}\right] 
   + \alpha\beta \left[ \int \rd y\, P(1,y)\, V(y)- f(0,0)\right].
\end{equation}
Given the monotonicity of the OPF, the Edwards-Anderson order
parameter (\ref{ea}) can be cast as
\begin{equation} 
  \label{q_ea-neuron} q_{\mathrm EA} = \max_{x<1} q(x).
\end{equation}
This is the maximal $q$ that has non-vanishing probability,
$P(q)\equiv {\dot x}(q)>0$, in the notation of Section
\ref{sssec-continuousRSB} we have $q_{\mathrm EA}=q_{(1)}$.

In summary, as we demonstrated it in Section \ref{ssec-meaning-xq},
$x(q)=\int_0^q \rd \bar{q}~ P(\bar{q})$, where $P(q)$ is the
probability density of the overlap $q$ between two synaptic
configurations. Thus $x(q)$ is monotonous and invertible with inverse
$q(x)$, allowance given for plateaus and isolated discontinuities in
these functions.  The conclusion of the present section is that the
equilibrium properties of the neuron model are determined by the
stationary shape of $x(q)$, or its inverse $q(x)$, thus they play the
role of order parameter function, in close analogy to spin glasses
\cite{sgrev87,sgrev91}.

\subsubsection{Variational principle: the PPDE as external constraint}
\label{ssec-variation-ext}

In Section \ref{ssec-fe-stati} we have given specific forms for the free
energy and stationarity conditions of Section \ref{sec-thermodyn}, for
the case of the Parisi ansatz.  Those formulas were originally
expressed in terms of the $\sf Q$ matrix, while Section
\ref{ssec-fe-stati} has the field $f(q,y)$, obeying the PPDE.  It is
natural to ask, what happens if we express the free energy in terms of
$x(q)$, and look for its extremum by varying $x(q)$.  This is
reversing the order of the original recipe, when the stationarity
condition in terms of the elements of $\sf Q$ was taken first and the
resulting formula, Eq.\ (\ref{stat-q}), expressed in terms of $x(q)$.
The equivalence of these two procedures has been seen in the cases
$R=0,1$ for spin glasses (see {\em e.\ g.}  \cite {cri92}) and the
neuron \cite{eg88,gg91,mez93}.  It is our observation that the
equivalence carries over to the continuous Parisi ansatz.  The proof
is in principle given by Eq.\ (\ref{def-deriv-q}), an identity which
tells us that the variation by $x(q)$ is proportional to the
two-replica correlation, obtained by differentiation by $q_{ab}$.  We
will, however, not leave the matter there and give a self-contained
presentation of the variational theory.

We shall consider two approaches.  In this section the PPDE will be
maintained as external constraint, while in the next one it will be
included by a multiplier field into the functional to be extremized.
The variational formulation opens the way to alternative methods to
find stationarity states.  Indeed, given the variational free energy
to be extremized, we are no longer bound to the stationarity
prescription (\ref{stat-cond}) for finding the extremum, rather we can
choose any suitable procedure that is capable to locate the extremum
of a functional.

The free energy is then
\begin{equation}
  \label{fe-spher-var-ext}
  f =  {\ba{c}\mbox{\footnotesize ~} \\ \mbox{max}\\ 
    \mbox{\footnotesize{\em x(q)}}\ea}
  f\left[ x(q) \right], 
\end{equation}
with the free energy functional
\begin{equation}
  \label{fe-functional-spher-def-ext} 
  f[x(q)] =  f_\rs[x(q)] + \alpha f_\re[x(q)] 
\end{equation}
as defined by (\ref{fe-functional-spher-ext}).  The maximization in
terms of $x(q)$ is a transfiguration of the original minimization by
the matrix elements of $\sf Q$ due to the $n\to 0$ limit.  The PPDE
(\ref{ppde-ener1}) is understood as external constraint, and in what
follows its solution, and in fact the solutions of the related PDE-s
of Sections \ref{sssec-deriv-PDEs}, \ref{sssec-lin-PDEs}, as well as the
GF-s of Section \ref {sssec-gf}, are assumed to be known.

Variation of the free energy gives, following the result of Section
\ref{sssec-var1}, the sum of the two-replica correlations for the
entropic and the energy term.  In fact, in the special case of the
spherical entropic term, the functional derivative of
(\ref{fe-functional-spher-s-ext}) can be straightforwardly calculated.
This gives, of course, the same result as that obtained from the
correlator (\ref{spher-corr2}).  Concerning the energy term, we write
the correlator (\ref{corr2}) with the notation (\ref{intro-m}).  We
recall that the entropic term was related to the generic free energy
term by (\ref{spher-by-parisi-final}) while the energy term (\ref
{fee-ener-parisi}) also involved a minus sign.  Finally, applying
(\ref{var1-corr}) to both the entropic and the energy term, we get for
$q<1$ 
\bml
\label{var-fe-ext-b}  
\begin{eqnarray} 
  \frac{\delta f[x(q)]}{\delta x(q)} & \equiv & \frac{\beta}{2}
  F(q,[x(q)]) = \frac{1}{2\beta} \left( C^{(2)}_{s,x}(q) - \alpha\,
  C^{(2)}_{e,x}(q)\right), \label{var-fe-ext1} \\ F(q,[x(q)]) & = &
  \int_0^q \frac{\rd \bar{q}}{\beta^2D(\bar{q})^2} - \alpha \int \rd y\,
  P(q,y)\,m(q,y)^2 \label{var-fe-ext2}.
\end{eqnarray} 
\eml Note that we never displayed the functional dependence on the OPF
in the correlators, but are doing so in the functional derivative for
clarity.  Furthermore, in the second correlator in (\ref{var-fe-ext1})
the subscript $e$ signals that it comes from the energy term
(\ref{fe-functional-spher-e-ext}), nevertheless, we omit that
subscript from the related fields $P(q,y)$ and $\mu(q,y)$, introduced
in the previous section.

When $x(q)$ can be freely varied, the stationarity condition is
\begin{equation} 
  \label{var-fe-stat-b} 
  F (q,[x(q)]) = 0, 
\end{equation} 
thus (\ref{stat-cond}) is recovered.  In the case of stationarity for
a discrete $R$-RSB scheme (\ref{ieq}), the vanishing of
(\ref{var-fe-ext}) at each $q_r, r=0,\dots,R$ is required.  This,
however, gives only $R+1$ equations, insufficient for the
determination of all $x_r$-s and $q_r$-s.  If variation by $x(q)$ is
made with the assumption that $x(q)\equiv x$ in an interval $I$, where
$0<x<1$, that is, there is a nontrivial plateau in $I$, then from
(\ref{var-fe-m}) follows
\begin{equation} 
  \label{var-fe-m-b} 
  \int_I \rd q\, F (q,[x(q)]) = 0 
\end{equation} 
as stationarity condition.  Thus (\ref{var-fe-m-b}) should hold in
each interval $(x_r,x_{r+1}),\, r=0,\dots,R-1$, within an $R$-RSB
scheme.  If the stationary OPF has both $\dot{x}(q)>0$ and $x(q)\equiv
x\neq 0,1$ parts in some intervals, then these imply the usage of
(\ref{var-fe-stat-b}) and (\ref{var-fe-m-b}) in the respective
intervals of $q$, and (\ref{var-fe-stat-b}) at the jumps between
plateaus.  We will see that such a phase, characterized by an $x(q)$
concatenated from a plateau -- with a nontrivial plateau value -- and
a strictly increasing segment, does arise in the neuron.

\subsubsection{Variational principle: inclusion of the PPDE}
\label{ssec-variation}

Sommers and Dupond \cite{sd84} introduced a variational formalism for
the Ising spin glass by including the PPDE into the free energy
functional with a Lagrange multiplier field.  The latter turned out to
be the field satisfying the SPDE, and it could be interpreted as the
probability density of the local magnetic field. The free energy
functional also depended on and needed to be varied by the an
auxiliary function $\Delta(x)$. The latter function turned out not to
bring new degrees of freedom in play because of an additional relation
between $q(x)$ and $\Delta(x)$.  In contrast to former studies of the
SK \cite{sd84} and Little-Hopfield \cite{tok94} models, we did not
find it necessary to introduce an additional function, the analog of
$\Delta(x)$.  The reason for that is, we surmise, that we had chosen
$x(q)$ as order parameter function.  That has an immediate physical
meaning, as demonstrated in Section \ref{ssec-meaning-xq}, thus no
allowance remained for the ``gauge'' invariance, inherent in the
traditional approach \cite{sd84}.  Moreover, the continued spectrum
(\ref{pa-cont-spectrum}) of the $\sf Q$ matrix turned out to be
proportional to the auxiliary function $\Delta(x(q))$ of Ref.\
\cite{sd84} (the cited authors also found this relation), so
introducing the latter as an independent field to be varied does not
lead to technical simplification.  It should be emphasized that the
``gauge'' invariance appeared to be of limited significance only when
the stationarity criterion was studied. It is, however, of import as
to fluctuations of $\sf Q$ violating Parisi's ansatz and is the source
of the WTI-s \cite{tkd98}.

Following Sommers and Dupond, we shall use the condition
(\ref{ppde-ener}) in the free energy functional as a constraint.
Forcing the PDE (\ref{ppde-ener1}) gives rise to a Lagrange multiplier
field $P(q,y)$, while the initial condition (\ref{ppde-ener2}) should
be set separately. The result is
\begin{equation}
  \label{fe-spher-var}
  f =  {\ba{c}\mbox{\footnotesize ~} \\ \mbox{max}\\ 
    \mbox{\footnotesize{\em x(q)}}\ea}
  {\ba{c}\mbox{ \footnotesize ~} \\ \mbox{extr}\\
    \mbox{\footnotesize{\em f(q,y),P(q,y)}} \ea}\!\!\! 
  f\left[ x(q),f(q,y),P(q,y) \right], 
\end{equation}
with, on the r.\ h.\ s., the functional 
\bml
\label{fe-functional-spher} 
\begin{eqnarray}
  \label{fe-functional-spher-def} 
  f[\dots] & = & f_\rs[\dots] +
    \alpha (f_\re[\dots] + f_a^{(1)}[\dots] + f_a^{(2)}[\dots]), \\ 
  \label{fe-functional-spher-s}
  f_\rs[\dots] & = & - \frac{1}{2\beta} \int_0^1  
  \rd q ~ \left[ \frac{1}{D(q)} - \frac{1}{1-q}\right], \\ 
  \label{fe-functional-spher-e}
  f_\re[\dots] & = & f(0,0),   \\ 
  \label{fe-functional-spher-a1}
  f_a^{(1)}[\dots] & = & \int_0^1\!\!  \rd q\ \int \rd y ~ P(q,y)
  \left[ \ptl _q f (q,y) +\case{1}{2} 
    \ptl _y^2 f(q,y) - \case{1}{2}\beta x(q) \left( \ptl _y  
    f(q,y) \right)^2 \right], \\
  \label{fe-functional-spher-a2}
  f_a^{(2)}[\dots] & = &\int \rd y~ P(1,y)~ \left[ V(y)-f(1,y) \right]. 
\end{eqnarray}
\eml The functional dependence on appropriate arguments is marked by
$[\dots]$.  

There is no physical restriction on the type of extremum in terms of
the auxiliary fields $f(q,y)$ and $P(q,y)$, we keep, therefore, the
more general ``extr'' condition.  The auxiliary functional
$f_a^{(1)}[\dots]$ enforces the PDE (\ref{ppde-ener1}).  The form of
$f_a^{(2)}[\dots]$ can be understood if we impose the initial
condition on the PDE by adding the term
\begin{equation}
  \label{ppde-init-by-delta}
   \delta(q-1) \left[ V(y)-f(q,y) \right] 
\end{equation}
to the l.\ h.\ s.\ of (\ref{ppde-ener1}). The ambiguity of the Dirac
delta centered at $q=1$ can again be taken care of by our using
$\delta(q-1^{-0})$ whenever necessary. Note that the sign of
expression (\ref{ppde-init-by-delta}) matters, it is the above choice
that forces the right initial condition no matter what $f(1,y)$ was
before. This feature can be shown by considering an infinitezimal
decrement in $q$ from $1$ in the PDE complemented by
(\ref{ppde-init-by-delta}).  The two auxiliary terms
(\ref{fe-functional-spher-a1}) and (\ref{fe-functional-spher-a2}) can
be concatenated and variation by $P(q,y)$ gives the PDE
(\ref{ppde-ener}) for $f(q,y)$, initial condition included.  For the
sake of clarity we keep (\ref{fe-functional-spher-a2}) specifying the
initial condition separate.  The terms
(\ref{fe-functional-spher-s},\ref{fe-functional-spher-e}) are
identical to (\ref{fes-spher-parisi},\ref{fee-ener-parisi}),
respectively.

Given the constraint on $f(q,y)$ by the Lagrange term, one should vary
$f(q,y)$ independently, yielding the PDE (\ref{spde-ener}) for
$P(q,y)$ including the initial condition, with the notation
(\ref{intro-m}).  Variation by $x(q)$ can then be done while $f(q,y)$
and $P(q,y)$ are kept fixed, and we find that
\begin{equation}
 \label{var-fe-incl-xq} \frac{\delta\, f\left[
 x(q),f(q,y),P(q,y)\right]}{\delta\, x(q)}
\end{equation}
is equal to (\ref{var-fe-ext-b}).  It should not cause confusion that
the free energy functional $f[\dots]$ and the auxiliary field
$f(q,y)$ have the same symbol, because the argument tells the
difference.  The variational free energy with the PPDE included as
constraint was one of our main results in Ref.\
\cite{our-paper}.

It should be emphasized that while the variational formalism is very
useful for the description of the equilibrium properties, it does not
account for such fluctuations of the matrix elements of $\sf Q$ that
cannot be captured by the OPF $x(q)$.  Thus in order to study
thermodynamical stability we need to resort to the more general
framework of Section \ref{sec-corr-stab}.

\subsubsection{On thermodynamical stability} \label{ssec-stability}

Based on the formulas derived in Section \ref{sec-corr-stab}, we can give
an explicit expression for the replicon spectrum in terms of
$q,y$-dependent fields.  We will only treat explicitly the spherical
neuron, generalization for arbitrary independent synapses is, in
principle, straightforward.

The free energy, as function of the $\sf Q$ matrix, is the sum of the
entropic and the energy terms.  Due to the fact that both undergo the
same scheme of spontaneous RSB, their Hessians can be simultaneously
quasi-diagonalized, based solely on the ultrametric symmetry of the
Hessian (see Section \ref{sssec-hess-def}).  This results in the
longitudinal-anomalous sector $(R+1)\times(R+1)$ matrices in the
diagonals, and in the replicon sector the replicon eigenvalues as
diagonal elements.  Hence a replicon eigenvalue of the complete
Hessian is the sum of the two eigenvalues, one from the entropic term
and one from the energy term.  We do not deal with the
longitudinal-anomalous sector in the general case, mostly because
complete diagonalization there depends on the specific system under
consideration. 

The entropic eigenvalue has been calculated in Section
\ref{ssec-sher-solving-P-equ}} as (\ref{spher-repl}), so we have
\begin{equation}
    \lambda_\rs(q_1,q_2,q_3)= [D(q_2)\, D(q_3)]^{-1} .
\label{repl-neur-spher-entr}
\end{equation}
In order to get the contribution from the energy term, we introduce
the GF for the field $f$.  Using the fact that $f$ and $\varphi$ are
proportional we obtain
\begin{equation}
  \label{gf-f} {\cal G}(q_1,y_1;q_2,y_2) = \frac{\delta
f(q_1,y_1)}{\delta f(q_2,y_2)} = \frac{\delta \varphi
(q_1,y_1)}{\delta \varphi (q_2,y_2)} = {\cal
G}_{\varphi}(q_1,y_1;q_2,y_2).
\end{equation}
Note that on the \lhs we omitted the subscript $f$ that we consider
the default.  Hence by (\ref{vertex-ttt}) we obtain the vertex
function $\Gamma$.  Then the eigenvalue from the energy term is given
by (\ref{repl-final}) with the substitution $\kappa(q,y) = -\beta
\chi(q,y)$, where $\chi(q,y)$ satisfies the PDE
(\ref{deriv2-ppde-ener-evolv}), yielding
\begin{equation}
    \lambda_\re(q_1,q_2,q_3)= - \beta^2 \int\, \rd y_2\, \rd y_3\,
    \Gamma \left(
    q_1;0,0;q_2,y_2;q_3,y_3 \right) \chi \left(q_2,y_2 \right)\, \chi
    \left(q_3,y_3
\right).  
\label{repl-neur-spher-ener}
\end{equation}
The final formula for the replicon spectrum is thus
\begin{equation}
    \lambda(q_1,q_2,q_3)= \lambda_\rs(q_1,q_2,q_3) + \alpha
    \lambda_\re(q_1,q_2,q_3).
\label{repl-neur-spher}
\end{equation}
Note that here the solutions of the relevant PDE-s were assumed to be
known.

The WTI discussed in Section \ref{sssec-corr-wt} implies the existence of
zero modes.  Indeed, using the fact that the functional derivative
(\ref{var-fe-ext-b}) is made up of two-replica correlators, by
(\ref{wt}) we have
\begin{equation}
  \label{zero-mode}
  \lambda(q,q,q) = \beta^2\, \dot{F}(q,[\dots]) = \lambda(q) 
\end{equation}
as the WTI for the spherical neuron.  Here the dot means derivative in
terms of the explicit $q$-dependence.  But stationarity for strictly
increasing segments of $x(q)$ means the vanishing of the r.\ h.\ s.,
so the eigenvalue for such $q$-s is zero.  Note that in an $R$-RSB
scheme stationarity at the $q_r$-s does not imply the vanishing of
(\ref{zero-mode}).  Based on the interpretation, quoted in Section 
\ref{sssec-corr-wt}, of the WTI as a consequence of spontaneously
broken permutation symmetry, the zero modes found here can be
considered as Goldstone modes of the symmetry broken phase.

In order to decide about thermodynamical, linear, stability of a
stationary $x(q)$, the analysis of the full replicon spectrum is
necessary.

 \subsubsection{Main types of the OPF} \label{ssec-types}

It has been the experience in the study of various long range
interaction disordered systems that only a few main types for the OPF
$x(q)$ satisfy the stationarity condition and are at least marginally
stable at the same time \cite{sgrev91}. Below we review those
that appear in the storage problem.

The $R$-RSB ansatz (see Eq.\ (\ref{xq}) proved to describe
thermodynamical equilibrium for $R=0$ and $R=1$ in several different
systems in some parameter range.  The former is the $RS$, the latter
the $1$-RSB state.  Interestingly, we have not found any examples in
the literature when $R$-RSB with $R>1$ would have described
thermodynamical equilibrium.  In the storage problem Whyte and
Sherrington \cite{ws96} have shown that at $T=0$ all finite $R$-RSB
solutions are unstable.  In fact, the eigenvalue causing instability
is $-\infty$, a typical $T=0$ phenomenon, also observed for such
eigenvalues in the SK model.

As to CRSB states, the shape of the OPF that corresponds to the phase
discovered by Parisi in the SK model is displayed in
(\ref{xq-parisi-phase}).  In the nomenclature of \cite{nieuw97} this
is the SG-I state.

Another type of phase also arises in the storage problem, namely, a
concatenation of a $1$-RSB plateau and a strictly increasing segment
of the OPF. This has the form
\begin{equation}
\label{xq-sg-iv-def}
    x(q) = \left\{ \begin{array}{ll} 1 & \text{if~} 
        q_{(1)} \leq q \leq 1 \\ 
        x_\rc(q) & \text{if~} q_1 \leq q <
        q_{(1)} \\ 
        x_1 & \text{if~}  q_{(0)} < q \leq q_1 \\
        0  & \text{if~}  0\leq q < q_{(0)}, \end{array}
        \right. 
\end{equation}
Such OPF has been observed in spin glasses with spins of more than two
states, like the Potts model \cite{gks85}.  This type of continuous
OPF with a plateau has been termed SG-IV in Ref.\ \cite{nieuw97}.


\subsubsection{Stationarity and its consequences}
\label{ssec-conseq}

The stationarity conditions displayed in Section
\ref{ssec-variation-ext} can be cast in more useful forms.  First of
all, note that also the entropic term is of the generic form
(\ref{gen-parisi}), as shown in Eq.\ (\ref{spher-by-parisi}) of
Section \ref{ssec-sher-solving-P-equ}.  We will thus formulate
stationarity in terms of the correlators in (\ref{var-fe-ext2}).  The
fields for the energy term will not be labeled, while the fields
belonging to the entropic term and treated in Section
\ref{ssec-sher-solving-P-equ} will carry now the subscript $s$ like
$\varphi_\rs $, $\mu_\rs $, $\kappa_\rs $, and $P_\rs $.

The stationarity conditions for the regions of positive $P(q)$ can be
cast into an equation that holds for all $q$-s as
\begin{equation}
\label{stationarity-everywhere}
        \int_0^q \rd {\bar q}\, {\dot x}({\bar q})\, 
         F[q,x(q)] \equiv 0.
\end{equation}
where $F$ was defined in Eq.\ (\ref{var-fe-ext-b}).  Indeed, $F=0$
must hold unless $P(q)=\dot{x}(q)=0$.  The lower limit of integration
can be safely chosen to be zero.  Alternatively, we have a combined
stationarity condition that contains the requirements about plateaus
and smooth $x_\rc(q)$ segments, but can be imposed only at $q$-s where
$P(q)>0$, namely
\begin{equation}
\label{stationarity-at-qr}
        \int_0^q \rd {\bar q}\, x({\bar q})\, F[q,x(q)] \equiv 0.
\end{equation}

Next we summarize a few identities that follow from the PDE-s of Section
\ref{ssec-fe-stati} for the fields in the energy term
\bml
\label{further-relations}
\begin{eqnarray}
\label{further-relations1}
    \frac{\rd }{\rd q} \int \rd y\, P(q,y)\, f(q,y) &=& - \frac{1}{2}
    \beta x(q) 
    \int \rd y\, P(q,y)\, m(q,y)^2 ,\\
\label{further-relations2}
    \frac{\rd }{\rd q} \int \rd y\, P(q,y)\, m(q,y) &=& 0 ,\\
\label{further-relations3}
     \frac{\rd }{\rd q} \int \rd y\, P(q,y)\, \chi(q,y) &=& \beta x(q)
     \int \rd y\, 
     P(q,y)\, \chi(q,y)^2 ,\\
\label{further-relations4}
     \frac{\rd }{\rd q} \int \rd y\, P(q,y)\, m(q,y)^2 &=& \int \rd y\,
     P(q,y)\, \chi(q,y)^2.
\end{eqnarray}
\eml 
Similar identities among the fields $\varphi_\rs $, $\mu_\rs $,
$\kappa_\rs $, and $P_\rs $ of the spherical entropic term can be
naturally obtained, when the factors $-\beta$ are erased as well.

Note that the PPDE (\ref{ppde-ener}) was used in deriving
(\ref{further-relations1}).  According to what has been said in
Section \ref{sssec-dic}, for a discontinuous potential $V(y)$ the PPDE
is invalid at $q=1$, so (\ref{further-relations1}) holds only for
$q$-s where $f(q,y)$ is smooth in $y$, generically for $q<1$.  Let us
consider the integral of (\ref{further-relations1})
\begin{equation}
\label{further-relations1b}
  \int \rd y\, P(q,y)\, f(q,y) - f(0,0) = - \frac{1}{2} \beta \int_0^q
    x(q) \int \rd y\, P(q,y)\, m(q,y)^2.
\end{equation}
Suppose that the PPDE (\ref{ppde-ener1}) holds for any $q<1$,
furthermore, that both sides are continuous in $q$ at $q=1$, a
condition that is met if $\beta$ is finite.  Due to the first
assumption (\ref{further-relations1b}) holds for any $q<1$, and due to
continuity it does so also at $q=1$.
 
First we consider (\ref{stationarity-everywhere}).  After partially
integrating it, and recalling (\ref{var-fe-ext-b}) where $F$ was a sum
of two correlation functions, we can use the relations
(\ref{further-relations3}, \ref{further-relations4}) to express the
$x(q){\dot F}$ term as a derivative in $q$.  The result is
\begin{eqnarray}
\label{stationarity-everywhere2}
        \beta\int_0^q \rd {\bar q}\, {\dot x}({\bar q}) F[{\bar q},x(q)]
         &=& \beta x({q})\, F[q,x(q)] \nn \\ && - \left. \int \rd y
         \left[\beta^{-1}P_\rs (q,y)\, \kappa_\rs (q,y) + \alpha
         \, P(q,y)\, \chi(q,y) \right] \right|_0^q \nn \\ &=& 0.
\end{eqnarray}
The subscript $s$ refers to the fields related to the entropic term,
discussed in Section \ref{ssec-sher-solving-P-equ}.  It follows from
Eq.\ (\ref{entr-mu}) that
\begin{equation}
\label{p-kappa}
          \kappa_\rs (q,y) = \ptl _y \mu_\rs (q,,y) = -\frac{1}{D(q)},
\end{equation}
hence
\begin{equation}
\label{stationarity-everywhere3}
        \beta x(q)\,F[q,x(q)] + \frac{1}{\beta D(q)} - \alpha \int \rd y
        P(q,y)\chi(q,y) = \frac{1}{\beta D(0)} - \alpha \,\chi(0,0),
\end{equation}
where we took into account that $P(0,y)=\delta(y)$.  Obviously, for
$P(q)=\dot{x}(q)>0$ the first term on the \lhs vanishes by
(\ref{var-fe-stat-b}), thus the rest is constant for such $q$-s.  Note
that this constant is nontrivial, in contrast to some spin models
\cite{sd84,tok94}, where the analogous constant vanishes.

The form (\ref{stationarity-at-qr}) immediately suggests the use of
(\ref{further-relations1b}) and yields for any $q$ with $P(q)>0$ the
following form for the free energy
\begin{equation}
  f = \beta^{-1} \varphi_\rs (0,0) + \alpha \, f(0,0) = \beta^{-1} \int
  \rd y\, P_\rs (q,y)\, \varphi_\rs (q,y) + \alpha \int \rd y\,
  P(q,y)\, f(q,y). 
\label{stationarity-at-qr2}
\end{equation}
This equation remains true in the interval $[0,q_{(0)}]$, where
$x(q)\equiv 0$.  So there the fields $P$ and $f$ are mutually adjoint.
Substituting $q=0$ we get the standard expression
(\ref{fe-spher-xq-final}).  From Section
\ref{ssec-sher-solving-P-equ}, we can recalculate the spherical
contribution
\begin{equation}
 \beta^{-1}\int \rd y\, P_\rs (q,y)\, \varphi_\rs (q,y) =
 \frac{1}{2\beta}\int_0^q \rd \bar{q}\, \frac{D(\bar{q}) -
 D(q)}{D(\bar{q})^2} + \beta^{-1} \varphi_\rs (0,0),
\label{spher-ident1}
\end{equation}
that can be substituted into (\ref{stationarity-at-qr2}) to yield a
more useful formula.  Note that (\ref{stationarity-at-qr2}) is not an
alternative form for the free energy functional, rather an expression
for the free energy at stationarity.

We mention that differentiations of (\ref{stat-cond}) yield further
stationarity conditions, valid only in intervals, but not at isolated
points, where $P(q)>0$.  We display the first one
\begin{equation}
 {\dot F}[q,x(q)] = \frac{1}{\beta^2 D^2(q)} - \alpha \int \rd y\,
 P(q,y)\, \chi(q,y)^2 = 0,
\label{diff-stat-cond}
\end{equation}
where (\ref{further-relations4}) was used.  The same formula, without
becoming zero, is useful in $R$-RSB schemes.  By Eq.\ 
(\ref{zero-mode}) it represents a replicon eigenvalue with coinciding
$q$ arguments 
\begin{equation}
  \label{eq:neur-lambda}
  \lambda(q) =  \frac{1}{D^2(q)} - \alpha \,\beta^2 \int \rd y\,
 P(q,y)\, \chi(q,y)^2.
\end{equation}
For $R=0$ (RS solution) this is at $q=q_0$ the AT eigenvalue, for $R=1$
it gives at $q=q_1$ the typically most dangerous eigenvalue,
responsible for the destabilization of the $1$=RSB state.

\subsubsection{The entropy} \label{ssec-entropy}
 
Based on the identity (\ref{further-relations1b}) the entropy
(\ref{spher-entr}) can be cast into the alternative form
\begin{equation} 
  \label{spher-entr-altern} s = \frac{1}{2} \int_0^1 \rd q \; \left[
  \frac{1}{D(q)} - \frac{1}{1-q}\right] - \hf\,\alpha\beta^2\, \int_0^1
  \rd q\, x(q)\, \int \rd y\, P(q,y)\, m(q,y)^2.
\end{equation}
This is valid when (\ref{further-relations1b}) can be extended to
$q=1$, for example at finite temperatures. 

It is useful here to separate the interval for $q$-integration into
$(0,q_{(1)})$ and $(q_{(1)},1)$.  Consider the first term
$-\beta\,f_\rs [x(q)]$ on the \rhs of (\ref{spher-entr-altern})
\begin{eqnarray}
-\beta\,f_\rs [x(q)] &=&   \int_0^{q_{(1)}} \rd q \,
\frac{1}{D(q)} + \ln(1-q_{(1)}) \nn \\
&=& \int_0^{q_{(1)}} \rd q \, x(q)\, \int_0^q
\frac{\rd \bar{q}}{D(\bar{q})^2} + (1-q_{(1)})\, \int_0^{q_{(1)}}
\frac{\rd {q}}{D({q})^2} + \ln(1-q_{(1)})\nn \\
&=& \alpha\beta^2\, \int_0^{q_{(1)}} \rd q,\, x(q)
\int \rd y\, P(q,y)\, m(q,y)^2 \nn \\  &&+  
(1-q_{(1)})\, \alpha\beta^2\, \int \rd y\, P(q,y)\, m(q,y)^2.
\label{separate-integr-fs}
\end{eqnarray}
The last relation comes from the alternative stationarity condition
(\ref{stationarity-at-qr}), which can indeed be applied, because for
the upper limit of integration $q_{(1)}$, the Edwards-Anderson order
parameter, we have $P(q_{(1)})>0$. Substitution into
(\ref{spher-entr-altern}) leads to cancellation, thus
\begin{eqnarray} 
s &=& \frac{1}{2} \ln(1-q_{(1)}) + \frac{ \alpha\beta^2}{2}\,
(1-q_{(1)}) \int \rd y\, P(q_{(1)},y)\, m(q_{(1)},y)^2  \nn \\
    && - \frac{ \alpha\beta^2}{2}\, \int_{q_{(1)}}^1 \rd q   \int \rd
    y\, P(q,y)\, m(q,y)^2 
\label{spher-entr-altern2} 
\end{eqnarray}
where $x(q)\equiv 1$ was used in the third term.  It follows from
(\ref{further-relations4}) that $\int \rd y\, P(q,y)\, m(q,y)^2$, in
general, strictly increases in $q$.  But for an increasing function
$h(q)$ 
\begin{equation}
  \label{eq:inequality-for-hq}
   \int_{q_{(1)}}^1 \rd q\, h(q) >(1-q_{(1)})\, h(q_{(1)}),
\end{equation}
therefore in (\ref{spher-entr-altern2}) the second and third term
together is generally negative. The first term is obviously negative,
thus so is the entropy.  The above formulation is useful, besides for
the consistency check of negativity of the entropy, because the
constituent functions are needed only in $[q_{(1)},1]$. There the only
nontrivial ingredient is $P(q,y)$, for $m(q,y)$ is explicitly given
by a Gaussian integral over the known $m(1,y)$. 

\subsubsection{The high temperature limit} \label{ssec-hT}

At high temperatures, if the relative number of examples $\alpha$ is
appropriately rescaled, the neuron exhibits nontrivial thermodynamical
properties\footnote{As it has been pointed out to the author by M.
  Opper, the limit studied here is equivalent to the thermodynamics of
  an $N$-dimensional vector in a Gaussian random, quenched, potential,
  with variance characterized by the function $W(q)$ of
  (\ref{hT-w-def}).}.  This should be contrasted with the fully
connected SK-type spin glasses, which are paramagnetic in the high-T
limit.

For $\beta\to 0$ the energy term (\ref{e-phi}) can be expanded in
terms of the potential and the results of Section 
\ref{ssec-small-potential} apply.  We identify in Eq.\ 
(\ref{small-pot-phi}) $\epsilon$ with $\beta$ and $\Phi(y)$ with
$-V(y)$.  In analogy with the definition (\ref{small-pot-psi-def}) we
introduce
\bml \label{hT-w-def}
\begin{eqnarray}
   \label{hT-w}
  W(q) & = & \int \! \rD z_1 \rD z_2 \, V({\bf n}_1 {\bf z}) \,
   V({\bf n}_2 {\bf z}), \\
  \label{hT-n-cond}
  |{\bf n}_1| & = & |{\bf n}_1| =1, \;\; {\bf
   n}_1 {\bf n}_2 = q.
\end{eqnarray}
\eml The energy term in the free energy functional is expanded as
\begin{equation}
  \label{hT-fe-ener}
  f_\re [x(q)] =  f_{e0} + \beta \, f_{e1}[x(q)] + O(\beta^2),
\end{equation}
where from Eqs.\ (\ref{e-phi}), (\ref{small-pot-phi}), and
(\ref{small-pot-phi1b}) we have
\begin{equation}
  \label{hT-f0}
   f_{e0} = \int \rD z\, V(z) \equiv  \sqrt{W(0)},
\end{equation}
which does not depend on the OPF $x(q)$, and
\begin{equation}
  \label{hT-f1}
  f_{e1}[x(q)] = - \frac{1}{2}\int_0^1 \rd q\, x(q)\, \dot{W}(q).
\end{equation}
The relative number of examples $\alpha$ should be scaled so that
\begin{equation}
  \label{hT-gamma}
  \gamma = \alpha \beta^2
\end{equation}
remains finite. Large $\alpha$-s will counterbalance the homogenizing
effect of high temperatures.  The full free energy functional is thus
singular in the small $\beta$ limit such that
\begin{equation}
  \label{hT-fe}
  \beta f[x(q)] = \beta^{-1}\phi_0 + \phi_1[x(q)] +  O(\beta),    
\end{equation}
where 
\bml
\label{hT-phi}
\begin{eqnarray}
  \phi_0 & = & \gamma\, \sqrt{W(0)}, \label{hT-phi0} \\
  \phi_1[x(q)] & =& - \frac{1}{2} \int_0^1 \rd q\, \left[
  \frac{1}{D(q)}\! - \!\frac{1}{1-q} + \gamma\, x(q)\, \dot{W}(q)
  \right].
\label{hT-phi1}
\end{eqnarray}
\eml 
The entropic contribution was inserted from Eq.\ 
(\ref{fes-spher-parisi}).  The $\beta^{-1}\phi_0$ is singular for
$\beta\to 0$ but is independent of the OPF $x(q)$, thus it does not
lead to meaningful thermodynamics.  The important feature here is that
the third term in (\ref{hT-phi1}) is linear in $x(q)$, because
expansion in $\beta$ is equivalent to expansion in $x(q)$ in the PPDE
(\ref{ppde-ener1}).

The term $\phi_1[x(q)]$ is equivalent to the free energy functional of
Nieuwenhuizen's spherical multi-$p$-spin interaction spin glass, a
most general SK-type spherical system, incorporating the spherical SK
and the more general $p$-spin interaction models \cite{nie95}.  Note
that the above result can be obtained also by solving the PPDE
perturbatively in $\beta$, a longer calculation.

Variation of the $O(\beta)$ term of (\ref{hT-fe}) gives
\begin{equation}
  2 \frac{\delta\phi_1[x(q)]}{\delta x(q)}
     = \int_0^q \frac{\rd \bar{q}}{D(\bar{q})^{2}} 
        - \gamma \dot{W}(q) \equiv F_0(q,[x(q)]),
   \label{hT-func-deriv}
\end{equation}
the leading term in formula (\ref{var-fe-ext2}) for $\beta\to 0$.  In
intervals with $\dot{x}(q)>0$ the stationarity condition $
F_0(q,[x(q)])=0$ can be explicitly solved for the segment $x_\rc (q)$
of the OPF to give
\begin{equation}
  \label{hT-xq}
  x_\rc(q) = \frac{\overdots{W}(q)}{2 \gamma^{1/2} \ddot{W}(q)^{3/2}}.
\end{equation}
If at a $q_r$ the stationary OPF exhibits a step then
$F_0(q_r,[x(q)])=0$ holds, and for a plateau $x(q)\equiv x$ of value
$0<x<1$ in the interval $I$ the condition (\ref{var-fe-m}) should be
applied. 

The replicon eigenvalues can be easily calculated by our adding the
contributions from the entropic term (\ref{spher-repl}) 
\begin{equation}
    \lambda_{\rs 0}(q_1,q_2,q_3) = [D(q_2)\, D(q_3)]^{-1} 
\label{hT-repl-entr}
\end{equation}
and the result from the expansion (\ref{small-pot-repl-spectr})
applied to the energy term,
\begin{equation}
    \alpha \lambda_{\re 0}(q_1,q_2,q_3) = - \gamma \ddot{W}(q_1), 
\label{hT-repl-ener}
\end{equation}
Thus in leading order $\lambda_e$ depends only on one $q$-variable.
Adding them up gives for the replicon spectrum
\begin{equation}
    \lambda_0(q_1,q_2,q_3) = \lambda_{\rs 0}(q_1,q_2,q_3) + \alpha
    \lambda_{\re 0}(q_1,q_2,q_3).
\label{hT-repl}
\end{equation}
This the leading term in (\ref{repl-neur-spher}). If $q_1$ falls into
an interval where $\dot x(q)>0$ then by (7.56) we have $\gamma\ddot
W(q_1) = 1/D(q_1)^2$.  Since $q_1 \leq q_2,q_3\leq q_{(1)}$ and $D(q)$
monotonically decreases, it is easy to see that
$\lambda_0(q_1,q_2,q_3) \geq 0$.  So these replicons are never
linearly unstable.  Such a general statement cannot be made if $q_1$
is a discontinuity point between two plateaus of $x(q)$.  Longitudinal
stability -- we have not discussed this question in the general case
-- can also be checked explicitly in the high-T limit.  Indeed, the
second variation of the free energy functional gives
\begin{equation}
    \frac{\delta^2\phi_1[x(q)]}{\delta
    x(q_1)\delta x(q_2)} =  - \int_0^{\min(q_1,q_2)} 
    \frac{\rd \bar{q}}{D(\bar{q})^3}.  
\label{hT-long}
\end{equation}
This is negative definite, as shown in App.\ \ref{app-stab-high-T}, so
the extremum of $f[x(q)]$ is indeed maximum as required in
(\ref{fe-spher-var}).

The main local quantity of interest is the stability $\Delta$
associated with individual patterns.  The distribution of stability
parameters $\rho(\Delta)$ is by (\ref{dist-stab-spher}) determined
through the field $P(q,y)$, so we have to solve the SPDE
(\ref{spde},\ref{spde-init}) for high temperatures. As described in
Appendix \ref{app-sec-highT-pde}, we find
\begin{equation}
\rho(y) = \rho_0(y) + \beta\, \rho_1(y) + O(\beta^2), 
\label{hT-rho-def}
\end{equation}
where
\bml
\label{hT-rho}
\begin{eqnarray}
        \rho_0(y) & = & G(y,1), \label{hT-rho0} \\
  \rho_1(y) & = & \ptl _{y} \Big[ G(y,1) \int_0^1 \rd q\,
  x(q)\, \int \rD z\, V^{\prime}(y
  q+z\sqrt{1-q^2}) \Big].
\label{hT-rho1}
\end{eqnarray}
\eml 
The first correction $\rho_1$ shows the deviation from the Gaussian,
an effect that is expected to be dramatic for low temperatures.  The
error per pattern is by (\ref{ener-dist}) 
\begin{equation}
\varepsilon = \varepsilon_0 + \beta\, \varepsilon_1+ O(\beta^2), 
\label{hT-ener-def}
\end{equation}
where
\bml
\label{hT-ener}
\begin{eqnarray}
        \varepsilon_0 & = & \int \rd y\,\rho_0(y)\, V(y) =
  \sqrt{W(0)}, \label{hT-ener0} \\ 
        \varepsilon_1 & = & \int \rd y\,\rho_1(y)\, V(y)
        = - \int_0^1 \rd q\, x(q)\, \dot{W} (q).
\label{hT-ener1}
\end{eqnarray}
\eml 
The leading term of the entropy is obtained from the definition
(\ref{entr}) as
\begin{equation}
s_0 = \frac{1}{2} \int_0^1 \rd q\, \left[ \frac{1}{D(q)} - \frac{1}{1-q}
    - \gamma\, x(q)\, \dot{W}(q) \right].
\label{hT-entr}
\end{equation}

\subsubsection{Scaling by temperature} \label{ssec-lowT} 

Gardner and Derrida recognized that at $T=0$, for positive stability
threshold $\kappa$ (see Eq.\ (\ref{gen-v}) for the definition), when
the limit of capacity was approached, the RS order parameter $q$
converged to unity \cite{eg88,gd88}. This is a manifestation of the
fact that, at the limit of capacity, the volume of version space,
compatible with the patterns to be stored, no longer diverges
exponentially in $N$ \cite{hkp91}.  In other words, the volume per
synapse goes to zero if $N\to \infty$, accompanied by the divergence
of the entropy to $-\infty$.  Discrete $R$-RSB calculations at $T=0$,
beyond capacity, showed that the $q$-s belonging to any $0<x\leq 1$
were also equal to unity.  At the same time, $q<1$ values were
associated with the variable $\xi=\beta x$, when this was kept finite
in the limit $x\to 0$ and $\beta\to\infty$.  Furthermore, it is
plausible to assume that for $T\to 0$\ie $q_{(1)}\to 1$, a positive
limit of $x_R\to x_{(1)}=x(q_{(1)}^{-0})\leq 1$ exists.

The above observations suggest a natural scaling for the OPF, valid
for any temperatures, but providing a smooth $T\to 0$ limit.  Let us
introduce \bml
\label{scaling}
\begin{eqnarray}
\label{scaling1}
q(t) &=& q_{(1)} - \left( q_{(1)} - q_{(0)}\right) \left( 1 - (1 +
q_{(1)})\,t+ q_{(1)}t^2\right), \quad 0\le t \le 1 , \\
\label{scaling2}
\xi(t) &=& \beta x(q(t))\,\dot{q}(t), \\
\label{scaling3}
\eta&=& \beta \left(1-q_{(1)}\right), \\
\label{scaling4}
\Delta(t) &=& \beta D(q(t)) = \int_t^1 \xi(\bar{t}) \rd \bar{t} + \eta, 
\end{eqnarray}
\eml
where the subscripted $q$-s are defined in (\ref{q-extremities}),
$x(q)$ vanishes for $0\le q<q_{(0)}$ and $x(q)\equiv 1$ for $1\ge
q>q_{(1)}$. The time variable is changed to $t$ via the invertible
function $q(t)$, that was constructed so that $\dot{q}(1)=0$ for
$q_{(1)}=1$. The scaled function $\Delta(t)$ is not to be confounded
with the local stability parameter $\Delta$.  In the $T\to 0$ limit we
expect $\eta$ to be finite and thus the scaled OPF $\xi(t)$ to be
bounded.  Indeed, in the most dangerous point, $t=1$\ie $q=q_{(1)}$,
where $\beta x(q_{(1)})$ may diverge, we have an expectedly finite
\begin{equation} 
\label{xi-at-1}
        \xi(1) = \eta\, x_{(1)}\, \left( q_{(1)} - q_{(0)}\right) <
        \eta.
\end{equation}
It is advisable to use the above scaling even for $T>0$, because the
scaled formulae remain manageable for small temperatures.  In what
follows $q(t)$ may in fact be any monotonic function with boundary
conditions $q(0)=q_{(0)}$ and $q(1)=q_{(1)}$, our taking the simple
(\ref{scaling1}) is just a numerically useful parametrization.

The auxiliary fields now depend on $t$ and $y$, like $f(t,y)$,
$m(t,y)$, etc.  Of course, $f(t,y)$ equals $f(q(t),y)$ and not the
field $f(q,y)$ in the point $t=q$. We will write the arguments in the
way that no ambiguity remains about which function is meant. The PPDE
should be rewritten as
\begin{equation}
  \label{ppde-rescaled} \ptl _t f(t,y) =  -\hf \,\dot{q}(t)\, \ptl
   _y ^2 f(t,y) + \hf\, \xi(t)\, \left(\ptl _y f(t,y) \right)^2,
\end{equation}
whose initial condition at $t=1$ is the former $f(q_{(1)},y)$.  At
this point it is worth displaying the $f(q,y)$ in the interval
$[q_{(1)},1]$.  There $x(q)\equiv 1$, so Eqs.\ (\ref{phi-ener}) and
(\ref{psi-to-phi}) define the Cole-Hopf transformed field, obeying
linear diffusion, that gives
\begin{equation}
   \label{f-field-in-q1-1} f(q,y) = - \frac{1}{\beta} \ln \int \rD z\,
   e^{-\beta V\left(y+z\sqrt{1-q} \right)}, \;\; q_{(1)}\le q\le 1 .
\end{equation}
The initial condition of the PPDE (\ref{ppde-rescaled}) is
\begin{equation} 
\label{ppde-rescaled-init}
   f(t=1,y) = f(q_{(1)},y) ,
\end{equation}
whence for $T\to 0$, after change of integration variable in
(\ref{f-field-in-q1-1}) as $\bar{y}=y+z\sqrt{1-q}$, we get
\begin{equation} 
\label{ppde-rescaled-init-T=0}
   f(t=1,y) = \min_{\bar{y}} \left(V(\bar{y}) +
   \frac{(y-\bar{y})^2}{2\eta} \right).
\end{equation}
The expression on the \rhs first appeared in Ref.\ \cite{gg91} as a
free energy term in the RS approximation. For small $q$-s we have
\begin{equation}
   \label{f-field-in-0-q0} f(q,y) = \int \rD z\, f\left(q_{(0)},
   y+z\sqrt{q_{(0)}-q} \right), \qquad 0\le q\le q_{(0)}.
\end{equation}
The rescaled SPDE reads as
\begin{equation}
  \label{spde-rescaled} 
   \ptl _t P(t,y)  =  \hf\, \dot{q}(t)\, \ptl _y
  ^2 P(t,y) + \xi(t)\, \ptl _y \left(P(t,y)\, m(t,y)\right) .
\end{equation} 
Along the plateau $[0,q_{(0)}]$ we have $x(q)\equiv 0$, so the SPDE
(\ref{spde}) is a linear diffusion equation, and in $[q_{(1)},1]$ the
Cole-Hopf-type transformation (\ref{def-t}) leads to linear diffusion,
whence
\bml
\label{P-in-plateaus}
\begin{eqnarray}
\label{P-in-plateau-0-q0}
P(q,y) & = & G(y,q) , \;\; 0\le q\le q_{(0)} ,\\
\label{P-in-plateau-q1-1}
P(q,y) & = & e^{ - \beta f(q,y)} \int \rD z\, P\left(t=1 ,
y+z\sqrt{q-q_{(1)}} \right) \, e^{\beta f\left(t=1 ,
y+z\sqrt{q-q_{(1)}} \right)}, \;\, q_{(1)}\le q\le 1.
\end{eqnarray}
\eml
Thus the initial condition for $P(t,y)$ in the SPDE is
\begin{equation}
\label{spde-rescaled2} 
        P(t=0,y) = P(q_{(0)},y) = G\left(y,q_{(0)}\right) .
\end{equation} 
The stationarity condition (\ref{var-fe-stat-b}) for $q$-s where
$\dot{x}(q)>0$\ie 
\begin{equation}
\label{xdot>0} 
\dot{\xi}(t)\, \dot{q}(t) - \xi(t)\,\ddot{q}(t) > 0
\end{equation} 
reads now as
\begin{equation}
  \label{stat-cond-scaled} 
  F[t,\xi(t)] \equiv \frac{q_{(0)}}{\Delta(0)^2} + \int_0^t
  \rd \bar{t} \,\frac{\dot{q}(\bar{t})}{\Delta(\bar{t})^2} - \alpha \int
  \rd y~ P(t,y)~ m(t,y)^2 =0.
\end{equation}
Of course, if one has a nontrivial plateau within the $t$-interval
$(0,1)$\ie (\ref{xdot>0}) fails in a subinterval, then
(\ref{stat-cond-scaled}) is invalid in that subinterval and one should
extremize by the parameters of the plateau extra.  In the PDE-s and
the stationarity condition the temperature does not appear explicitly
and allows for a smooth limit in case $T\to 0$.  Assuming that we
solved the above PDE-s, in the scaled variables the free energy
becomes
\bml
\label{free-energy-in-rescaled}
\begin{eqnarray}
\label{free-energy-in-rescaled0}
        f &=& f_\rs  + \alpha f_\re , \\
\label{free-energy-in-rescaled1} f_\rs  &=& -
 \frac{1}{2} \frac{q_{(0)}}{\Delta(0)} - \frac{1}{2} \int_0^1
 \frac{\dot{q}(t)\, \rd t}{\Delta(t)} + \frac{1}{2\beta} \ln
 \frac{\beta}{\eta}, \\ \label{free-energy-in-rescaled2} f_\re  &=& \int
 \rD z\, f(t=0,z\sqrt{q_{(0)}}).
\end{eqnarray}
\eml
The distribution of local stabilities, based on
(\ref{P-in-plateau-q1-1}), is 
\begin{equation}
  \label{local-stability-distr-in-rescaled} \rho(y) = P(q=1,y) = e^{ -
\beta V(y)} \int \rD z\, P\left(t=1 , y+z\sqrt{1-q_{(1)}} \right) \,
e^{\beta f\left(t=1 , y+z\sqrt{1-q_{(1)}} \right)} .
\end{equation}
It is straightforward to show that $\rho(y)$ is normalized if
$P(t=1,y)$ was normalized, and the latter property follows from the
fact that the SPDE preserves the normalization of its initial
condition (\ref{spde-rescaled2}).  The mean error per pattern can be
calculated as
\begin{equation}
  \label{energy-in-rescaled} \varepsilon = \int \rd y\, \rho(y)\, V(y) .
\end{equation}
In practical cases the limit of the local stability distribution for
$T\to 0$ can be calculated by the saddle point method from
(\ref{local-stability-distr-in-rescaled}) and contains only
nonsingular, scaled variables.  

Concerning the entropy, it is obvious from (\ref{spher-entr-altern})
that beyond capacity, at $T=0$, the entropy is $s=-\infty$.  Indeed,
the first term of (\ref{spher-entr-altern2}) goes to $-\infty$ for
$\beta\to\infty$, and the rest being generally negative, see reasoning
in the end of Section \ref{ssec-entropy}, it cannot compensate for the
negative singularity.  This complements the known result for
$\kappa=0$ that when one approaches the capacity from below then the
entropy diverges to $-\infty$ \cite{gd88}.  Together with the fact
that the overlap beyond capacity equals $1$ with probability $1$, this
demonstrates freezing in the ground state.  This is an effect
analogous to the vanishing of the $T=0$ entropy beyond capacity for
the Ising perceptron \cite{km89}, both show that the number of states
with minimal error is subexponential in $N$.

\subsubsection{The RS state and storage below capacity}
\label{ssec-neur-rs}

For a general potential $V(y)$ that vanishes beyond a certain
stability parameter, $y>\kappa$, and is positive below it, the
original results of Gardner \cite{eg87,eg88} describe the storage
problem at $T=0$ below capacity.  The reason for that is that if all
examples are satisfied then the positive part of the error measure
does not matter.

At finite temperatures the potential comes into play, the equations
are easily obtained from what has been said before.  The free energy
is a function of the only variational parameter
$q=q_0=q_R=q_{(0)}=q_{(1)}$ as 
\bml
  \label{eq:neur-fe-rs}
\begin{eqnarray}
  \label{eq:neur-fe-rs1}
  f(q)   &=& f_\rs (q) + \alpha\, f_\re (q), \\
  \label{eq:neur-fe-rs2}
  - \beta f_\rs (q) &=& \frac{1}{2}\left( \ln(1-q) + \frac{q}{1-q}
  \right), \\ 
  - \beta f_\re (q) &=& \int \rD z_1 \ln \int \rD z_2\,e^{-\beta\,
  V(\zeta)},  
  \label{eq:neur-fe-rs3}
  \\
  \zeta &=& z_1\sqrt{q}+ z_2\sqrt{1-q}.
  \label{eq:neur-fe-rs4}
\end{eqnarray}
\eml Stationarity is given by (\ref{var-fe-stat-b}) that now reads as
\begin{equation}
  \label{eq:neur-statio-rs}
  \beta^2 F(q) = \frac{q}{(1-q)^2} - \alpha\, \int \rD z_1\, \left(
  \frac{\ptl}{\ptl \zeta} \ln \int \rD z_2\,e^{-\beta\, V(\zeta)}
  \right)^2 = 0, 
\end{equation}
with the abbreviation (\ref{eq:neur-fe-rs4}), and the AT eigenvalue
from (\ref{eq:neur-lambda}) is
\begin{equation}
  \label{eq:neur-AT-rs}
  \lambda(q) = \frac{1}{(1-q)^2} - \alpha\, \int \rD z_1\, \left(
  \frac{\ptl^2}{\ptl \zeta^2} \ln \int \rD z_2\,e^{-\beta\, V(\zeta)}
  \right)^2. 
\end{equation}
Note that (\ref{zero-mode}) is not in contradiction with
$\lambda(q)\neq\beta^2\, \dot{F}(q)$ here.  This is because in
(\ref{zero-mode}) the derivative is understood by the explicit $q$
dependence of $F$, while $x(q)$ is fixed, but in $F(q)$ both kinds of
arguments are denoted by the same $q$.  Here $\lambda(q)$ is only
meaningful at the stationary $q$.  The probability density of local
stabilities is given by say (\ref{local-stability-distr-in-rescaled}),
now
\begin{equation}
  \label{eq:neur-rho-rs}
  \rho(y) = e^{-\beta\, V(y)} \int \rD z_1\, \frac{
  G(y+z_1\sqrt{1-q},q)}{\int 
  \rD z_2\, e^{-\beta\, V(y+(z_1+z_2)\sqrt{1-q})}}.
\end{equation}

In the ground state ($T=0$) below capacity the positive part of $V(y)$
is suppressed.  This means that for $\beta\to\infty$ only arguments of
$V$ matter that are greater than $\kappa$.  Closer inspection shows
that this reasoning holds only if $q$ does not approach $1$ at the same
time.  For $\beta\to\infty$ the free energy and the energy goes to
zero, but $\beta f$ remains typically finite, 
\begin{equation}
  \label{eq:neur-fe-t0-rs}
  -\beta  f(q) = \frac{1}{2}\ln(1-q) + \frac{q}{2(1-q)} +
  \alpha\,\int \rD z\, \ln H\left( \frac{\kappa - 
  z\sqrt{q}}{\sqrt{1-q}}\right), 
\end{equation}
where (\ref{def-H-x}) was used for the definition of $H(x)$. The
ground state entropy per synapse is now
\begin{equation}
  \label{eq:neur-entr-t0-rs}
   s=-\beta f,
\end{equation}
the subject of the pioneering works \cite{eg87,eg88,gd88}.  In the
units of the prior volume $\int w(\jv)\, \rd ^N\! J=1$, where
(\ref{spher-prior}) gives the prior density, the volume of version
space is $e^{Ns}$.  The stationarity condition
(\ref{eq:neur-statio-rs}) simplifies in the ground state to
\begin{equation}
  \label{eq:neur-statio-t0-rs}
  \frac{q}{1-q} = \frac{\alpha}{2\pi}\, \int \rD z\,
  \frac{\exp\left(-\frac{(\kappa-z\sqrt{q})^2}{1-q}\right)}{H\left(
  \frac{\kappa-z\sqrt{q}}{\sqrt{1-q}}\right)^2},
\end{equation}
and the AT eigenvalue becomes
\begin{equation}
  \label{eq:neur-lambda-t0-rs} 
  \lambda(q) = \frac{1}{(1-q)^2} - \frac{\alpha}{2\pi q}\, \int \rD z\,
  \left(\frac{\ptl}{\ptl
  z}\frac{\exp\left(-\frac{(\kappa-z\sqrt{q})^2}{2(1-q)}\right)}{H\left(    \frac{\kappa-z\sqrt{q}}{\sqrt{1-q}}\right)}\right)^2.
\end{equation}

Numerical evaluation shows that for increasing $\alpha$ the $q$ goes
to $1$ and the entropy decreases towards $-\infty$.  For $q\alt 1$ the
dominant contribution in the above expressions comes from the region
of exponentially small $H$\ie when its argument is large, positive.
That is ensured by $\kappa>z$.  The asymptotics of $H(x)$ for large
$x$ can be found in \cite{as65}
\begin{equation}
  \label{eq:neur-H-asymp} 
  \frac{e^{-\hf x^2}}{H(x)} \approx \sqrt{2\pi}\, x.
\end{equation}
Thus the limit $q\to 1$ is realized, from Eq.\ 
(\ref{eq:neur-statio-t0-rs}), for $\alpha=\alpha_\rc (\kappa)$ satisfying 
\begin{equation}
  \label{eq:neur-ac-t0-rs}
  1 = \alpha_\rc (\kappa)\, \int_{-\infty}^\kappa \rD z\, (\kappa-z)^2,
\end{equation}
that evaluates to the capacity curve
\begin{equation}
 \label{eq:neur-ac-explicit}
 \alpha_\rc (\kappa) = \left[(\kappa^2+1)\,(1-H(\kappa)) + 
\frac{\kappa\, e^{-\kappa^2/2}}{\sqrt{2\,\pi}}\right]^{-1}.
\end{equation}
This gives $\alpha_\rc (0)=2$, the known result of Refs.\ 
\cite{win61,cov65}.  The recovery of that by Gardner \cite{eg87,eg88}
raised much confidence in the statistical mechanical approach combined
with the replica method.  For the $\kappa$-dependent capacity
(\ref{eq:neur-ac-t0-rs}) several sources could be cited, see\eg
\cite{gg91}.  The $\alpha_\rc (\kappa)$ curve is shown on Fig.\ 
\ref{fig:alpha-t0-rs}, by tradition the horizontal axis \hbox{is
  $\alpha$}. 

From (\ref{eq:neur-fe-t0-rs}) one can convince oneself of the negative
divergence of the entropy for $\alpha\to\alpha_\rc$ from below.
Alternatively, the conclusion in the end of Section \ref{ssec-lowT}
about $s=-\infty$ also applies at the limit of capacity.

\vbox{
\begin{figure}[h!]
 \begin{center}
   \psfig{figure=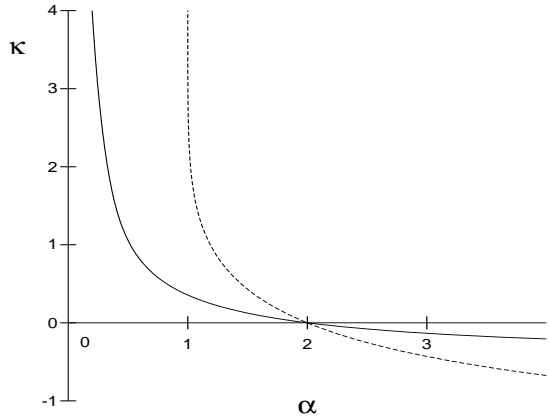,
     width=3in,height=3.5in,angle=0}\vspace*{1cm}
 \end{center}
 \vspace{-0.7in}
 \caption{The limit of capacity $\alpha_\rc $ from Eq.\
   (\protect\ref{eq:neur-ac-explicit}), solid line, and the indicator
   $\alpha^\ast$ of AT stability from
   (\protect\ref{eq:neur-aast-t0-rs}), dashed line.}
 \label{fig:alpha-t0-rs}
\end{figure}
}

A glance at the AT eigenvalue (\ref{eq:neur-lambda-t0-rs}) reveals
that $\lambda$ is typically singular for $q\to 1$.  Decisive is its
sign, that can be determined in that limit by again using the
asymptotics (\ref{eq:neur-H-asymp}) for $\alpha\alt \alpha_\rc
(\kappa)$.  If the latter is inserted into
(\ref{eq:neur-lambda-t0-rs}) before derivation, one immediately sees,
that the amplitude of singularity is
\begin{equation}
  \label{eq:neur-lambda-t0-q1-rs} 
  \lambda(q)\,(1-q)^2 \approx  1 -
  \frac{\alpha_\rc (\kappa)}{\alpha^\ast(\kappa)}, 
\end{equation}
where we have given a name to 
\begin{equation}
  \label{eq:neur-aast-t0-rs}
  \alpha^\ast(\kappa) =  \frac{1}{H(\kappa)},
\end{equation}
and depicted it on Fig.\ \ref{fig:alpha-t0-rs}.  It follows that if
$\alpha_\rc <\alpha^\ast$ ($\alpha_\rc >\alpha^\ast$) then the AT
eigenvalue has a positive (negative) singularity on the capacity line.
Since the RS solution presented here is only valid for $q<1$, below
the capacity line, we can conclude that for $\kappa\geq 0$ this region
is AT-stable.  This RS state ceases to exists at the critical line not
because of AT instability, rather the overlap $q$ reaches the border
$q=1$ of its physical range.  We mention but do not further elaborate
on the property, that for negative $\kappa$-s the RS solution
destabilizes before the capacity is reached.  Thus RSB is necessary
below capacity, like in \cite{bmz93}, but in the present case this
occurs without the need for non-monotonic potentials.  Note that
$\alpha^\ast$ is not the AT stability boundary for the RS solution,
because it was calculated in the limit $q\to 1$.

In sum, for $\kappa\geq 0$ the region below capacity for $T=0$ can be
described by the RS solution.  However, beyond it the particular
form of the potential $V(y)$ affects the behavior, so we specify one
for further studies. 
 

\subsection{The special error measure $\theta ( \kappa - y)$}
\label{ssec-special-error-measure}

Now we apply the framework of Section \ref{ssec-gen-err-measure} to the
error measure (\ref{gen-v}) with $b=0$, that is, to
\begin{equation}
\label{theta-v}
V(y) = \theta ( \kappa - y),
\end{equation}
a much studied case.  This potential does not weigh erroneous patterns
by ``how much'' wrong they are as measured by the local stability
parameter $\Delta$, it simply counts them.  We will often use the
function $H(y)$ as defined by (\ref{def-H-x}).  The initial condition
for the PPDE (\ref{ppde-rescaled}) is obtained by substitution of
(\ref{theta-v}) into Eqs.\ 
(\ref{f-field-in-q1-1},\ref{ppde-rescaled-init})
\begin{equation}
\label{ppde-rescaled-init-Theta}
f(q_{(1)},y) =  - \frac{1}{\beta} \ln \left[ \left(1-e^{-\beta}\right)
H\left(\frac{\kappa -y}{\sqrt{1-q_{(1)}}}\right)+e^{-\beta}\right],
\end{equation}
whence by derivation in terms of $y$ the initial conditions for
$m(q,y)$, etc.\, result.

\subsubsection{The ground state}
\label{ssec-b0-T0}

For $\alpha>\alpha_\rc $ none of the finite $R$-RSB solutions
\cite{gd88,mez93,et93,ws96} are thermodynamically stable \cite{ws96},
thus it is necessary to resort to a CRSB ansatz.  In the time
parameter $t$ the initial conditions at $t=1$ are as follows for
$T=0$.  The initial condition for the field $f(t,y)$ is given by
(\ref{ppde-rescaled-init-T=0}).  The minimum is realized at
\begin{equation}
\label{ppde-min-y-T=0-theta-v}
    \bar{y}_0 = \left\{ \begin{array} {ll} y & \text{~~if~} y \le
        \kappa-\sqrt{2\eta} \\ \kappa  &
        \text{~~if~} \kappa-\sqrt{2\eta} \le y \le \kappa \\ y &
        \text{~~if~} y \ge \kappa , \end{array} \right.
\end{equation}
Hence, at $t=1$, 
\begin{equation}
\label{ppde-rescaled-init-T=0-theta-v}
    f(1,y) = \left\{ \begin{array} {ll} 1 & \text{if~} y \le
        \kappa-\sqrt{2\eta} \\ \frac{(\kappa -y)^2}{2\eta} &
        \text{if~} \kappa-\sqrt{2\eta} \le y \le \kappa \\ 0 &
        \text{if~} y \ge \kappa , \end{array} \right.
\end{equation}
whence by differentiation in terms of $y$ we get the initial
conditions $m(1,y)$ and $\chi(1,y)$ for the evolution in time $t$.
Presuming that $P(t=1,y)$ is known, we calculate the local stability
distribution by applying the saddle point method to Eq.\
(\ref{local-stability-distr-in-rescaled}) yielding
\begin{equation}
\label{local-stability-distr-T=0-theta-v}
    \rho(y) = \left\{ \begin{array} {ll} P(1,y) & \text{if~} y <
        \kappa-\sqrt{2\eta} \\ 0 & \text{if~} \kappa-\sqrt{2\eta} < y
        < \kappa \\ P(1,y) + \delta(y-\kappa)\,
        \int_{\kappa-\sqrt{2\eta}}^{\kappa} \rd \bar{y}\, P(1,\bar{y})
        & \text{if~} y \ge \kappa . \end{array} \right.
\end{equation}
Thus in $\rho(y)$ a gap develops, but normalization is restored by the
$\delta$-peak at $y=\kappa$.  Similar feature was observed in various
approximations, RS and $1$-RSB, in previous works \cite{gg91,mez93},
but there the function appearing in the place of $P(1,y)$ was
explicitly known.  Note the singularities in $y$ in the initial
conditions -- these make numerical calculations more difficult, but do
not alter the fact that for averaged quantities the limit $T\to 0$ is
generically smooth. 

We do not elaborate more on the $T=0$ case, because numerical
evaluation cannot be avoided anyhow.  But, due to the scaling
described in Section \ref{ssec-lowT}, the singularity of the $T\to 0$
limit has been lifted and both $T=0$ and $T>0$ could be treated within
the same numerical framework. 

\subsubsection{The high temperature limit}
\label{ssec-b0-hT} 

As reported in our Letter \cite{our-paper} and discussed for a general
error measure $V(y)$ in Section \ref{ssec-hT}, in the limit when both the
temperature $T$ and relative number of examples $\alpha$ are large,
much can be said by analytic treatment about even the CRSB states.

The general formulae were presented in Section \ref{ssec-hT}, where the
effective free energy to be extremized was given as $\phi_1[x(q)]$ in
Eq.\ (\ref{hT-phi1}).  The error measure under consideration
determines the function $W(q)$ via (\ref{hT-w-def}).  The simplest way
to give $W(q)$ is by
\bml
\label{hT-w-q}
\begin{eqnarray}
\label{hT-w-q1}
 \dot{W}(q) & = &
 \frac{\exp\left(-\frac{\kappa^2}{1+q}\right)}{2\pi\sqrt{1-q^2}}, \\
\label{hT-w-q2}
 W(0) & = & H(-\kappa)^2,
\end{eqnarray}
\eml
whence $W(q)$ and all its derivatives can be calculated. 

The RS ansatz means that $x(q)=\theta(q-q_0)$.  Using Eq.\ 
(\ref{hT-phi1}) we get (the subscript of $q_0$ is omitted) 
\begin{equation}
\label{hT-RS-phi1}
 \phi_1(q) = - \frac{1}{2}\left[ \frac{q}{1-q} + \ln(1-q) +
 \gamma (W(1) - W(q)) \right] 
\end{equation}
and the stationarity condition reads as 
\begin{equation}
\label{hT-RS-stationary}
 \frac{q}{(1-q)^2} = \frac{\gamma
\exp\left(-\frac{\kappa^2}{1+q}\right)}{2\pi\sqrt{1-q^2}}.
\end{equation} 
Local thermodynamical stability is determined from
(\ref{hT-repl-entr}-\ref{hT-repl}), thus the AT
line is given by 
\begin{equation} 
    \lambda_0^{RS}(q) = \lambda_0(q,q,q) =
\frac{1}{(1-q)^2} - \frac{\gamma}{2\pi}
\exp\left(-\frac{\kappa^2}{1+q}\right)
\frac{\kappa^2(1-q)+q(1+q)}{(1+q)^{5/2}(1-q)^{3/2}} = 0.
\label{hT-repl-RS-error-counting} 
\end{equation} 

The $1$-RSB ansatz is equivalent to $x(q)=(1-x_0)\, \theta(q-q_1)+x_0\,
\theta(q-q_0)$ and yields by Eq.\ (\ref{hT-phi1})
\begin{eqnarray} 
 \phi_1(q_0,q_1,x_0) &=& - \frac{1}{2}\Bigg \{
        \frac{q_0}{1-q_1+x_0(q_1-q_0)} - \frac{1-x_0}{x_0}\ln(1-q_1) \
        \nn \\ && + \frac{1}{x_0}\ln[1-q_1+x_0(q_1-q_0)] \nn \\ & & +
        \gamma \left[W(1) - (1-x_0)W(q_1) - x_0 W(q_0)\right] \Bigg
        \}. 
\label{hT-1RSB-phi1} 
\end{eqnarray} 
The leading replicon eigenvalue is given by
(\ref{hT-repl-RS-error-counting}) with $q_1$ substituted, thus the
boundary of local stability is
\begin{equation}
    \lambda_0^{1RSB}(q_1) = \lambda_0(q_1,q_1,q_1) = 0.
\label{hT-repl-1RSB-error-counting} 
\end{equation} 

The classic Parisi phase, or SG-I, is characterized by the OPF
(\ref{xq-parisi-phase}).  There $x_\rc(q)$ is the continuously
increasing part of the OPF, for which we obtain from Eqs.\
(\ref{hT-xq}, \ref{hT-w-q})
\begin{equation}
\label{hT-xqc-error-counting}
x_\rc(q) = \frac{1}{\gamma} \sqrt{\frac{\pi}{2}}\, {\frac {
 \kappa^4(q^2-2q+1)+2\kappa^2(-2q^3+q^2+2q-1)+2q^4+4q^3+3q^2+2q+1 }
 {\left ({q}^{2}- q\,{\kappa}^{2}+q+{\kappa}^{2} \right )^{3/2} \left
 (1-q\right)^{1/4}\left (1+q\right)^{3/4}}}\,e^{{\frac {\kappa^2}{2(1+
 q)}}}.
\end{equation}
The interesting feature is that the OPF has an explicit and
non-perturbative form.  The perturbation is in $\beta$ now, and a
small $\beta$ apparently does not make $x(q)$ degenerate.  We shall
need
\begin{equation}
\label{hT-Dq-error-counting}
    D(q) = \left\{ \begin{array} {ll} 1-q & \text{if~} 
        q_{(1)}\leq q \leq 1 \\ D_\rc (q)\equiv 1-q_{(1)} +
        \int_q^{q_{(1)}}d\bar{q}\,x_\rc(\bar{q}) & \text{if~} q_{(0)}
        \leq q \leq q_{(1)} \\ 
        D_\rc (q_{(0)}) & \text{if~}  0\leq q \leq  q_{(0)} .
        \end{array}
        \right.
\end{equation}
The leading term nontrivial in the free energy, (\ref{hT-phi1}),
depends only on the endpoints of the interval as
\begin{eqnarray}
\label{hT-fe-error-counting}
 \phi_1(q_{(0)},q_{(1)}) & = &-\frac{1}{2} \left[ \frac{q_{(0)}}{D_\rc
 (q_{(0)})} + \int_{q_{(0)}}^{q_{(1)}} \frac{\rd q}{D_\rc (q)} +
 \ln(1-q_{(1)}) \right. \nn \\ && + \left.  \gamma
 \int_{q_{(0)}}^{q_{(1)}}\rd q\,  x_\rc(q) \dot{W}(q) + \gamma W(1) -
 \gamma W(q_{(1)}) \right].  
\end{eqnarray}
The replicon eigenvalues with identical arguments vanish due to the
Ward-Takahashi identity, as described in Section \ref{ssec-stability},
so the SG-I phase is at best marginally stable.  Nonlinear stability
analysis is not available, but believed not to result in instability.

The fourth type of phase found here is a concatenation of a nontrivial
plateau of $x(q)$, like in $1$-RSB, and a continuously increasing
$x_\rc(q)$.  This CRSB spin glass state is also called SG-IV.  The
$x_\rc(q)$ is again given by (\ref{hT-xqc-error-counting}), but extra
variational parameters \wrt the classic Parisi phase (SG-I) should be
introduced: the value $x_0$ of the plateau stretching from $q_{(0)}$ to
a $q_1$, and its upper border $q_1$.  The OPF is given by
(\ref{xq-sg-iv-def}) with $x_\rc(q)$ as in
(\ref{hT-xqc-error-counting}), and
\begin{equation}
\label{hT-Dq-error-counting-sg4}
    D(q) = \left\{ \begin{array} {ll} 1-q &\quad \text{if~} q_{(1)}
        \leq q \leq 1  \\ 
        D_\rc (q)\equiv 1-q_{(1)} +
        \int_q^{q_{(1)}}\rd\bar{q}\,x_\rc(\bar{q}) 
        & \quad \text{if~} q_1 \leq q \leq q_{(1)} \\ 
        x_1\,(q_1-q) + D_\rc (q_1)  & \quad\text{if~}  q_{(0)} \leq q
        \leq q_1 \\        
        D_0= 1-q_{(1)} + D_\rc (q_1) + x_1\,(q_1-q_{(0)})   & \quad
        \text{if~} 0\leq q \leq q_{(0)}. \end{array}
        \right. 
\end{equation}
The resulting free energy can be straightforwardly constructed from
Eq.\ (\ref{hT-phi1}) as 
\begin{eqnarray}
\label{hT-fe-error-counting-sg4}
 \phi_1(q_{(0)},q_1,q_{(1)},x_1) & = &-\frac{1}{2} \left[
 \frac{q_{(0)}}{D_0} + \frac{1}{x_1}\, 
 \ln\left(1+\frac{x_1\,(q_1-q_{(0)})}{D_\rc (q_1)}\right)
 + \int_{q_{1}}^{q_{(1)}} \frac{\rd q}{D_\rc (q)} + \ln(1-q_{(1)})
 \right. \nn \\ && + \left.  \gamma
 x_1\,\left(W(q_1)-W(q_{(0)})\right) \right. \nn \\ && + \left. \gamma
 \int_{q_1}^{q_{(1)}}\rd q\, 
 x_\rc(q) \dot{W}(q) + \gamma W(1) - \gamma W(q_{(1)}) \right].
\end{eqnarray}

The specialty of the high-$T$ limit is that the numerical evaluation
of all spin-glass-like phases involves extremization only in a few
scalars, because the $x_\rc(q)$ is explicitly known.  This has been done
in Ref.\ \cite{our-paper}, the results are demonstrated in the figures
there, which we redisplay for illustration.  On Fig.\ 
\ref{fig-high-t-phd} the phase diagram is shown, with one RS region
and three different types of RSB. If more than one of the ans\"atze
(\ref{hT-RS-phi1}, \ref{hT-1RSB-phi1}, \ref{hT-fe-error-counting},
\ref{hT-fe-error-counting-sg4}) worked, we considered the averaged
equilibrium state the one with the maximal free energy.

\vbox{
\begin{figure}[htb]
  \vspace{1.6in}
\begin{center}
  \epsfig{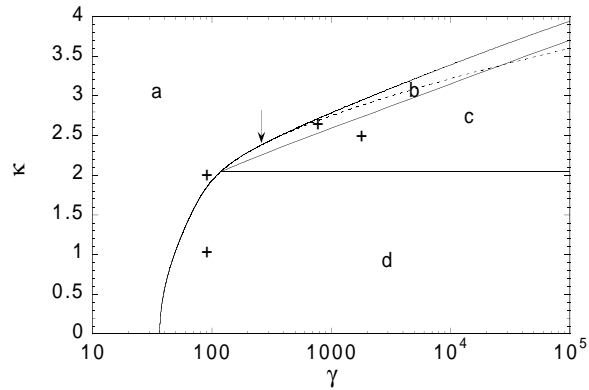}
\vspace{0.1in}
  \caption{Phase diagram for the potential $V(y)=\theta(\kappa -y)$ in
    the $(\gamma ,\kappa)$ plane for high $T$ by numerical
    maximization of the free energy Eq.\ (\protect\ref{hT-phi1}) with
    the ans\"atze described in this section. The full lines separate
    phases with different types of global maxima.  The RS, $1$-RSB,
    SG-IV, and SG-I phases are indicated by {\protect\em a, b, c}, and
    {\protect\em d}, respectively.  The AT curve is the RS phase
    boundary for $\kappa <\kappa_2\simeq 2.38$ and to the right of the
    arrow it analytically continues in the dashed line, no longer a
    phase boundary.  Reprinted from Ref.\ \protect\cite{our-paper}.}
\label{fig-high-t-phd}\end{center}
\end{figure} 
}

It is a plausible conjecture that the RS phase is obtained by analytic
continuation from the phase of perfect storage below capacity
$\alpha<\alpha_\rc (\kappa)$.  So although at high temperatures there
is no phase with zero error, the analog phase is the one with RS
(labeled by a).  Note that at high $T$ we lost the intuitive picture,
valid at $T=0$, that increasing $\kappa$ takes us into the frustrated
phase.  We obtain three RSB phases.  One is $1$-RSB (b), the other the
classic Parisi CRSB (SG-I, labeled by d), the third one is also CRSB,
but with an extra plateau (SG-IV, labeled by c).  The characteristic
shapes for the OPF are shown in Fig.\ \ref{fig-high-t-xq}, note the
plateau in the SG-IV phase (c).
 
\vbox{
\begin{figure}[htb]
  \vspace{0.3in}
\begin{center}
  \epsfig{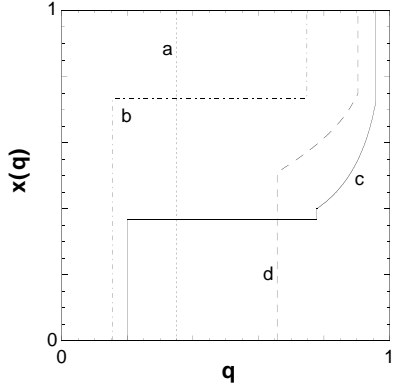}
\vspace{0.1in}
 \caption{ The $x(q)$ function at representative points as marked
   on Fig.\ \protect\ref{fig-high-t-phd} by crosses. Reprinted from
   Ref.\ \protect\cite{our-paper}.}
 \label{fig-high-t-xq} 
\end{center}
\end{figure} 
}
 
Extensive thermodynamical quantities are shown on Fig.\ 
\ref{fig-high-t-entr} for $\kappa=0$.  The entropy is negative and
decreases as it should for increasing $\alpha$\ie increasing $\gamma$.
The mean error per pattern in the high-$T$ limit is $\hf$, and our
approximation tells the correction $\varepsilon_1=T(\hf-\varepsilon)$
from the formula (\ref{hT-ener1}).  For $\gamma\to\infty$ we expect
that even the correction $\varepsilon_1$ vanishes, this is indeed
suggested by the picture.  The transition from RS to CRSB is of third
order from the viewpoint of the free energy\ie its third derivative
exhibits a discontinuity.

\vbox{
\begin{figure}[htb]
  \vspace{1in}
  \begin{center}
  \epsfig{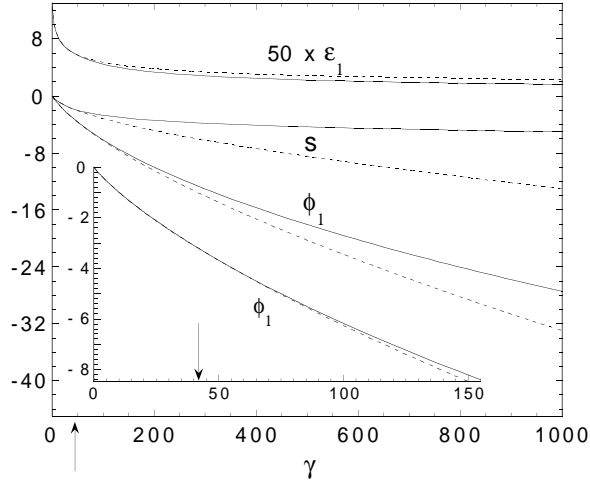}
\vspace{0.1in}
 \caption{The entropy $s$, in leading order $s_0$ from Eq.\ 
   (\protect\ref{hT-entr}), the free energy term $\phi_1$ from Eq.\ 
   (\protect\ref{hT-phi1}), and the enlarged correction 
   $\varepsilon_1$ of the energy (\ref{hT-ener1}) in the high
   $T$ limit, for $\kappa=0$.  The RS---SG-I transition is marked by
   an arrow. The dashed lines correspond to the thermodynamically
   unstable RS state beyond this transition point.   The inset
   demonstrates the smoothness of the transition.  Reprinted from
   Ref.\ \protect\cite{our-paper}. }
\label{fig-high-t-entr} 
\end{center}
\end{figure} 
}
 
If at $T=0$ the region beyond capacity is a single CRSB phase (SG-I),
for $T\to \infty$ the decomposition into three different phases
suggests singular surfaces born at finite $T$-s, whose precise
locations we did not determine. For $\kappa<2$ the transition from RS
to CRSB SG-I is of similar type as at $T=0$, but for $\kappa>2$ we
have a transition from RS into $1$-RSB that does not have a
counterpart at $T=0$.  Nevertheless, the main picture, namely, that
for low $\alpha$-s (translates into low $\gamma$-s here), there is a
normal phase analog to a paramagnet, and for large $\alpha$-s the
system exhibits complex\ie spin-glass-like behavior, is captured in
the high-$T$ limit. 


\subsubsection{Numerical evaluation method for arbitrary
  temperatures$^\ast$  }
\label{ssec-numeric}

As demonstrated in Sect. \ref{ssec-variation-ext}, the maximization of
$f[x(q)]$ with respect to $x(q)$ in (\ref{fe-spher-var-ext}) is done
with the side conditions that $f(q,y)$ satisfies the PPDE
(\ref{ppde-ener}) and $P(q,y)$ the SPDE (\ref{spde-ener}).  We further
recall that $x(q)\equiv 0$ for $q<q_{(0)}$ and $x(q)\equiv 1$ for
$q>q_{(1)}$, and that in the remaining non-trivial regime $q_{(0)}\leq
q\leq q_{(1)}$ it is convenient to use for $x(q)$ the parametrization
by $\xi(t)$ from (\ref{scaling2}).

For an actual numerical implementation, $\xi(t)$ has
to be rewritten once more in the form of some ansatz
with a finite number $N$ of variational parameters 
$v_1,...,v_N$:
\begin{equation}
\xi (t)=\xi(t;v_1,...,v_N )\ .
\label{num1}
\end{equation}
For instance, $v_1,...,v_N$ may be the coefficients of a polynomial
ansatz for $\xi (t)$. Another option would be a piecewise linear
ansatz with $v_n =\xi (t=(n-1)/(N-1))$. Also a finite-step RSB ansatz
with the steps and plateaus parametrized by the $v_n$ is possible.
Given any such parametrization (\ref{num1}), we are left with
maximizing the free energy functional with respect to the $N+2$
variational parameters
\bml
\begin{eqnarray}
\vv  & = & (v_0,v_1,...,v_{N+1}) \label{num2a}\\
v_0 & = & q_{(0)} \label{num2b}\\
v_{N+1} & = & \eta =\beta(1-q_{(1)})\ ,\label{num2c}
\end{eqnarray}
\eml where we have expressed $q_{(1)}$ through $\eta$ according to
(\ref{scaling3}). This maximization of the free energy functional
$f[\vv ]$ has to be performed under the (non-holonomous)
constraints $x(q_{(0)})\geq 0$, $x(q_{(1)})\leq 1$, and $\dot x(q)\geq
0$ for $q_{(0)}\leq q\leq q_{(1)}$, or equivalently, $\xi(0)\geq 0$,
$\xi (1)/\beta \dot q(1)\leq 1$ (cf. (\ref{scaling2})), and $\dot\xi
(t)\dot q(t) - \xi(t)\ddot q(t)\geq 0$ for $1\leq t\leq 1$
(cf. (\ref{xdot>0})).  It is convenient to incorporate these
constraints into an augmented free energy functional $f_\mu (\vv )$
in the form of soft penalty terms: \bml
\begin{eqnarray}
f_\mu (\vv ) & = & f[\vv ]-\mu_0\psi(-\xi(0))
-\mu_t\int_0^1\rd t\, \psi(\xi(t)\ddot q(t) - \dot\xi (t)\dot q(t))\nonumber
\\ 
& &
-\mu_1\psi(\xi(1)/\beta\dot q(1)-1)\label{num3a}\\
\psi(x) & = & x^2\theta(x)/2 \ . \label{num3b}
\end{eqnarray}
\eml
Thus, by successively increasing the coefficients $\mu_0$, $\mu_t$, and
$\mu_1$ in the course of the maximization procedure of $f_\mu(\vv )$,
the respective constraints will be respected more and more rigorously.

Before we proceed, the following points are worth mentioning: (i) Like
in Section \ref{ssec-lowT}, our only assumption on $q(t)$ is that it
should be a monotonically increasing function with $q(0)=q_{(0)}$ and
$q(1)=q_{(1)}$. But for concrete numerical calculations, especially at
low temperatures $T=\beta^{-1}$, the specific choice (\ref{scaling1})
has proven to be particularly appropriate.  In any case, the implicit
dependence of $q(t)$ on the variational parameters $v_0=q_{(0)}$ and
$v_{N+1}=\eta$ should be kept in mind:
\begin{equation}
q(t)=q(t;v_0,v_{N+1})\ .
\label{num4}
\end{equation}
(ii) In our experience, the maximization procedure typically ends not
at the border of the admitted parameter-regime, where the soft
constraints (\ref{num3a}) come into action, but rather in the interior
of this admitted region. However, in the course of the maximization
this border may be visited, and, in the absence of the soft
constraints in (\ref{num3a}), the maximization procedure often goes
out of the admitted region and diverges eventually.  (iii) Strictly
speaking, there are additional constraints on $v_0$ and $v_{N+1}$
associated with the restrictions $0\leq q_{(0)}<q_{(1)}\leq 1$; in our
experience they, however, were never in danger to be violated with the
obvious exception of cases with a stable RS solution.  (iv) As in any
variational ansatz, the necessary number $N$ of parameters depends on
how well the ansatz is adapted to the problem.  In principle, a
polynomial or piecewise linear ansatz (\ref{num1}) with a sufficiently
large number $N$ of parameters can approximate any shape of $x(q)$
arbitrarily well. Whether or not $N$ is sufficiently large in a given
case should follow from the accuracy with which the stationarity
conditions (\ref{var-fe-stat-b}, \ref{var-fe-m-b}) are satisfied.  In
practice, unavoidable numerical inaccuracies make things more
complicated. As has been observed already in Ref. [16] within a 2-RSB
ansatz, in the neighborhood of its maximum the free energy functional
$f_\mu(\vv )$ changes extremely little upon certain
parameter-variations, that is, the energy landscape $f_\mu(\vv )$ is
very ``flat'' in certain directions. In our experience, with
increasing number of parameters $N$ in (\ref{num1}), this problem
becomes worse and worse in that the finite numerical accuracy gives
rise to a spurious ``roughness'' in the already very ``flat'' energy
landscape. As a consequence, any maximization strategy becomes slow or
even fails for too large $N$.  Similarly, the stability conditions are
satisfied very well (in comparison with their numerical uncertainty)
within a fairly large neighborhood of the true maximizing $x(q)$.  As
a consequence, in any specific case, a carefully tailored ansatz with
not too many parameters has to be used and the criterion for
convergence should be that $q_{(0)}$, $\eta$, and $\xi (t)$ change
negligibly upon refining the parametrization (\ref{num1}).

In order to maximize the augmented free energy functional
(\ref{num3a}), a good compromise between robustness against the
spurious numerical fine structure in the energy landscape and speed of
convergence turned out to be a plain steepest descent procedure along
the following lines: given a ``working'' parameter set $\vv $, the
direction of the steepest increase of $f_\mu(\vv )$ is along the
gradient $\partial f_\mu(\vv )/\partial \vv $.  Taking into
account all the implicit dependencies on $\vv $ in (\ref{num1}),
(\ref{num4}) and the expression (\ref{var-fe-ext2}) for the gradient
of the original free energy functional, a straightforward but somewhat
tedious calculation yields for the gradient of $f_\mu(\vv )$ from
(\ref{num3a}) the result
\bml
\begin{eqnarray}
 \frac{\partial f_\mu}{\partial v_0} & = & \int_0^1\frac{\dot
 F(t)\xi(t)}{2\dot q(t)}\frac{\partial q(t)}{\partial q_{(0)}}\, \rd t
 +\frac{M_1}{\dot q(1)}\,\frac{\partial\dot q(1)}{\partial
 q_{(0)}}\nonumber \\ & & -\int_0^1 M(t)\left(\xi(t)\frac{\partial
 \ddot q(t)}{\partial q_{(0)}} -\dot\xi (t)\frac{\partial\dot
 q(t)}{\partial q_{(0)}}\right)\,\rd t \label{num5a} \\ \frac{\partial
 f_\mu}{\partial v_n} & = & \int_0^1\frac{F(t)}{2}\frac{\partial
 \xi(t)}{\partial v_n}\,\rd t
 +\mu_0\,\psi'(-\xi(0))\frac{\partial\xi(0)}{\partial v_n}
 -\frac{M_1}{\xi(1)}\,\frac{\partial\xi(1)}{\partial v_n}\nonumber \\
 & & -\int_0^1 M(t)\left(\frac{\partial \xi(t)}{\partial v_n}\ddot
 q(t) -\frac{\partial \dot\xi (t)}{\partial v_n}\dot q(t) \right)\,\rd t
\label{num5b}
\\ \frac{\partial f_\mu}{\partial v_{N+1}} & = & \frac{F(1)}{2}
-\int_0^1\frac{\dot F(t)\xi(t)}{2\beta \dot q(t)}\frac{\partial
q(t)}{\partial q_{(1)}}\,\rd t -\frac{M_1}{\beta \dot
q(1)}\,\frac{\partial\dot q(1)}{\partial q_{(1)}}\nonumber \\ & &
-\int_0^1 M(t)\left(\dot \xi(t)\frac{\partial \dot q(t)}{\beta
\partial q_{(1)}} -\xi (t)\frac{\partial\ddot q(t)}{\beta \partial
q_{(1)}}\right)\,\rd t \,
\label{num5c}
\end{eqnarray}
\eml
where $1\leq n\leq N$, $\psi'(x)= x\,\theta (x)$, we have
introduced the quantities
\bml
\begin{eqnarray}
M_1 & = & \mu_1 \, \psi'(\xi(1)/\beta\dot q(1) -1)\, 
\xi(1)/\beta\dot q(1),
\label{num6b}
\\
M(t) & = & \mu_t\, \psi'(\xi(t)\ddot q(t)-\dot\xi (t)\dot q(t)),
\label{num6c}
\end{eqnarray}
\eml
and used $F(t)$ to denote the \lhs of Eq.\ (\ref{stat-cond-scaled})
for a given $\xi(t)$ function.

Along the direction $\partial f_\mu(\vv )/\partial\vv $ of
steepest increase, one now searches for the maximum, {\em i.e.}, the
expression $f_\mu(\vv +\lambda\, \partial f_\mu({\bf
v})/\partial\vv )$ has to be maximized with respect to
$\lambda$. This implies the condition
\begin{equation}
J(\lambda_{{\rm max}}) =  0
\label{num7a}
\end{equation}
for the maximizing $\lambda=\lambda_{{\rm max}}$, where
\begin{equation}
J(\lambda) = \frac{\partial f_\mu(\vv +\lambda\, \partial
f_\mu(\vv )/ \partial\vv )}{\partial\vv }\cdot \frac{\partial
f_\mu(\vv )}{\partial\vv }\ .
\label{num7b}
\end{equation}
By updating the parameter set as
\begin{equation}
\vv \mapsto\vv +\lambda_{{\rm max}}\, \partial f_\mu(\vv )/
\partial\vv 
\label{num8}
\end{equation}
one completes one iteration step of the steepest descent procedure.
This iteration scheme is then repeated until $\vv $ does not
appreciably change any more. Note that due to the numerical
inaccuracies it makes little sense to locate the zero from
(\ref{num7a}) very precisely in each iteration step.  Our usual
strategy was based on the assumption that $J(\lambda)$ behaves
approximately linear near its zero at $\lambda=\lambda_{{\rm max}}$.
If $J(\lambda)$ is given at two nearby $\lambda$-values, one then
obtains an approximation for $\lambda_{{\rm max}}$ by linear
interpolation.  One such readily available $J(\lambda)$-value is that
for $\lambda =0$, the second one follows by choosing for $\lambda$ the
approximation for $\lambda_{{\rm max}}$ from the previous iteration
step.

\subsubsection{The CRSB state}
\label{ssec-numeric-results}

In Ref.\ \cite{our-paper-00} we presented some characteristic results,
obtained by the method expounded in the previous section, for the
error measure (\ref{theta-v}).  In a non-exhaustive search we found
that if the RS solution is AT-unstable, at $T=0$ beyond capacity and
also for some low temperatures, only a classic Parisi CRSB state
emerges.  Its OPF is given in \ref{xq-parisi-phase}), and also denoted
as SG-I.  We conjecture that at $T=0$ the region beyond capacity is
such a phase.  Sufficiently high $T$-s, where the $1$-RSB and the CRSB
state with a plateau (SG-IV) would have arisen, as described in Section 
\ref{ssec-b0-hT}, were not reached in our explorations.  

The scaling introduced in Section \ref{ssec-lowT}, and notably the
introduction of the OPF $\xi(q)=\beta\,x(q)$, allows the description
of the CRSB state at any temperatures, at the same time maintaining a
smooth transition to the ground state, $T=0$.  Physically, the fact
that $x(q)\to 0$, at $T=0$, for any $q<1$ means that $q=1$ with
probability one.  Thus freezing sets in, similarly to the ground state
of the SK model \cite{sd84}.  At the same time, the degenerate $x(q)$
is no longer a useful OPF, because the free energy becomes a
functional of rather $\xi(q)$.    

On Fig.\ \ref{fig:opf-low-t} the scaled OPF $\xi(q)=\beta\,x(q)$ is
displayed for various parameters.

\vbox{
\begin{figure}[htb]
 \begin{center}
 \psfig{figure=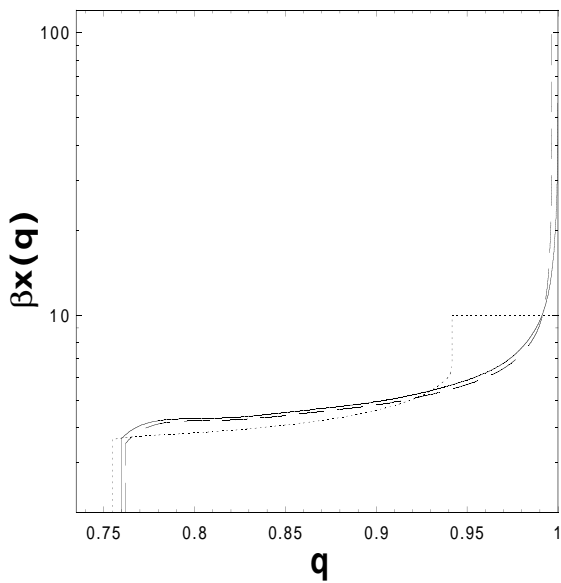,
 width=3in,height=3.5in,angle=0}
 \end{center}
 \caption{Scaled order parameter function $x(q)$ for $\kappa=0$,
   $\alpha=3$ at $T=0$ (solid), $T=0.01$ (dashed), and $T=0.1$
   (dotted). The first discontinuity is at $q_{(0)}$, below the
   function is constantly zero.  The second discontinuity for $T>0$ is
   $q_{(1)}$, which goes to $1$ for $T\to 0$.  Reprinted from Ref.\ 
   \protect\cite{our-paper-00}}
 \label{fig:opf-low-t}
\end{figure}
}

All parameter settings are in the AT-unstable region.  This figure is
the first indication, to our knowledge, of Parisi's CRSB state for low
temperatures in a system that is not a model of long range interaction
spin glasses, or closely related to such as the Little-Hopfield
network.  It is remarkable that the scaling by $\beta$ makes the
continuously increasing segment $\xi_c(q)=\beta\,x_c(q)$ of the OPF
little sensitive to the temperature.  Equally stable is the lower end
$q_{(0)}$ of the $\xi_c(q)$ segment, but the upper end $q_{(1)}$ shows
linear temperature dependence, $1-q_{(1)}\propto T$.  The rightmost
plateau's value is obviously $\xi(1)=\beta$.  

At the same parameter settings as before the local stability density
is displayed on Fig.\ \ref{fig:local-stabilities-low-t}. 

\vbox{
\begin{figure}[htb]
 \begin{center}
   \psfig{figure=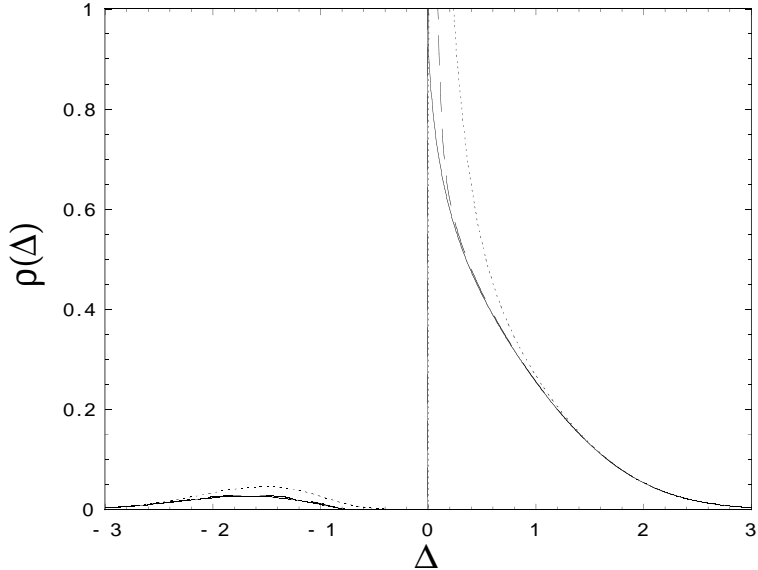,
     width=4in,height=3in,angle=0}
 \end{center}
  \caption{Density of
    local stabilities $\rho(\Delta)$ from theory for $\kappa=0$,
    $\alpha=3$ at $T=0$ (solid), $T=0.01$ (dashed), and $T=0.1$
    (dotted).  Reprinted from Ref.\ \protect\cite{our-paper-00}}
 \label{fig:local-stabilities-low-t}
\end{figure}
}

Since in the method of Section \ref{ssec-numeric} the evaluation of the
probability field $P(q,y)$ by the scaled SPDE (\ref{spde-rescaled}) is
done in every approximant step, we obtain the sought field in the end
by (\ref{local-stability-distr-in-rescaled}).  Not shown is the Dirac
delta peak at $T=0$, this restores normalization to one there.  A gap
exists at $T=0$, with right border $\Delta=\kappa$, in accordance with
(\ref{local-stability-distr-T=0-theta-v}), but the gap immediately
disappears for any positive $T$, as it can be seem from
(\ref{local-stability-distr-in-rescaled}).  At $T=0$ the density
$\rho(\Delta)$ linearly vanishes at the lower edge of the gap.

Comparison between the CRSB solution and earlier RS \cite{gd88},
$1$-RSB \cite{mez93,et93}, and $2$-RSB \cite{ws96} approaches shows
that averaged quantities, like the mean error per pattern do not show
significant differences.  The qualitative behavior of the error, that
it is zero below and is positive beyond capacity at $T=0$, furthermore
that it linearly increases for small $\alpha-\alpha_c$, is reflected
by the previous solutions.  The $1$- and $2$-RSB $\varepsilon(\alpha)$
curves look the same on a resolution of a figure \cite{ws96}.  On the
other hand, the difference is more conspicuous in the distribution of
non-self-averaging quantities.  The OPF $x(q)$ is the averaged
probability measure of the overlap of coupling vectors, and the
definitely continuously increasing part of it in Figs.\ 
\ref{fig:opf-low-t}, \ref{fig-high-t-xq} shows that finite $R$-RSB-s
are qualitatively in error.  Further qualitative difference can be
found in the distribution of local stabilities $\rho(\Delta)$.
Indeed, for finite $R$-RSB the $\rho(\Delta)$ exhibits a discontinuity
at the lower edge of the gap.  The right tendency is shown by the
feature that the size of the discontinuity is smaller in the $1$-RSB
than in the RS solution \cite{mez93}.

\subsubsection{Simulation}
\label{ssec-simul}

In this section we describe the simulation results from
\cite{our-paper-00}.  Wendemuth adapted existing algorithms for below
capacity of the simple perceptron, with potentials of the form
(\ref{gen-v}), to the region beyond it by specially dealing with
patterns with positive stabilities \cite{wen95a}, and performed a
series of simulations \cite{wen95b}.  The most sensitive part of his
work was on the potential with $b=0$, which counts the number of
unstable patterns, an NP-complete problem from the algorithmic
viewpoint \cite{ama91}.  His data showed significant deviation from
the then available best theoretical prediction from the $1$-RSB
calculation of Majer, Engel, and Zippelius \cite{mez93}.  He evaluated
the probability density of local stabilities at $\alpha=1$ and
$\kappa=1$, a point known to be beyond capacity.  Although the shapes
roughly resembled, a gap, and a peak at its right end, were present,
the simulation data gave systematically and discouragingly larger
stabilities than predicted by theory.

Essentially following Wendemuth's algorithm we redid the
simulation in order to see how persistent the deviation is.  The first
step is to generate random patterns (\ref{examples}).  We 
selected numbers with uniform distribution from an interval centered
around zero and in the end normalized them as
\begin{equation}
  \label{eq:simu-norm}
  \sum_{k=1}^N (S^{\mu}_k)^2=N
\end{equation}
The output for the patterns, $\xi^{\mu}$, were taken uniformly $1$,
not restricting generality, for $S^{\mu}_k$ have random signs.  The
algorithm goes in discrete time $t=0,1,\dots$.  We initialized at
$t=0$ the coupling vector according to the Hebb rule
\begin{equation}
\label{initj}
J_k(0) = \text{const.}\,\sum_{\mu=1}^M S_k^{\mu},
\end{equation}
with the constant chosen so that the Eucledian norm was $|\jv(0)|=N$.
At time $t$ the local stabilities
\begin{equation}
\label{locstab}
\Delta^{\mu}(t) = \frac{\jv(t)\cdot \sv ^{\mu}}{|\jv(t)|}
\end{equation}
are computed and among the unstable ones\ie $\Delta^{\mu}(t)<\kappa$,
the one with the largest $\Delta^{\mu}(t)$ is selected.  This is the
least unstable pattern, characterized by the index $\mu_0(t)$.  The
couplings are updated according to the rule of Wendemuth
\cite{wen95a,wen95b}.  We took
\begin{equation}
\label{update1}
\jv(t+1) = \jv(t) + \lambda \left( \sv ^{\mu_{0}(t)} +
\Delta \sv (t)\right),
\end{equation}
where 
\begin{equation}
  \label{eq:-update2}
    \Delta \sv (t) = \left\{ \begin{array}{ll} 0 & \quad \text{if~}
         \Delta^{\mu_{0}(t)}(t)>0 
         \\ 
         \jv(t) \frac{N/|\jvi(t)| -
         \Delta^{\mu_{0}(t)}(t)}{|\jvi(t)| -
         \Delta^{\mu_{0}(t)}(t)} & \quad \text{if~}
         \Delta^{\mu_{0}(t)}(t)<0.  \end{array}
        \right. 
\end{equation}
The $\lambda$ is the gain parameter, chosen in Ref.\ \cite{wen95a} as
$\lambda=N^{-3/2}$.   By trial and error we found that a larger gain
parameter $\lambda=N^{-1}$ did not endanger overall convergence, and
made the final approach for a given pattern, $\Delta^{\mu_0(t)}(t)\to
\kappa$, faster.  The second row in the update rule
(\ref{eq:-update2}) is Wendemuth's term introduced to specially cope
with patterns with negative stability.

At the next time step $t+1$ we again find the least unstable pattern
with index $\mu_0(t+1)$ and update the couplings by the above rule.
The usual course of the algorithm is that the least unstable pattern
is the same, $\mu_0(0)=\mu_0(1)=\dots$, until it becomes stabilized at
say $t_1-1$, whence another pattern is taken for some steps,
$\mu_0(t_1)=\mu_0(t_1+1)=\dots$, again until it becomes stabilized.
In principle, another pattern may become least stable before the one
in question is stabilized, but typically this was not the case.

The above recipe is repeated until a pattern cannot be stabilized in a
reasonable time.  The notion of reasonable time could be quantified,
because the time needed to stabilize a pattern showed a systematic
increase as function of the total number of patterns stabilized
before.  Therefore, it is a good recipe to halt the algorithm, when a
pattern cannot be stabilized within a small multiple of the
extrapolated convergence time.  In test runs, if the last pattern
could not be stabilized within twice the extrapolated convergence
time, it could not within ten times of the same either.  Thus we are
confident that we exploited the possibilities of the update rule
described above.
 
Wendemuth algorithm is based on the argument that one has the highest
chance to stabilize the pattern among all patterns with
$\Delta^\mu<\kappa$ whose $\Delta^\mu$ is closest to $\kappa$.  So
this algorithm may maximize the number of stable patterns, by
successively pushing the stability of the least unstable pattern to
$\kappa$ from below.  A consequence is that the remaining
non-stabilized patterns with $\Delta^\mu<\kappa$ will have relatively
large distance $\kappa-\Delta^\mu$, but the latter quantity does not
enter the present error measure.  Nevertheless, the principle of
stabilizing the least unstable pattern resembles qualitatively the
gradient descent algorithm for differentiable error measures, because
every step is made in the momentarily most promising direction.  The
shortcomings of such algorithms in NP-complete problems is known, and
we cannot be certain that the number of unstable patterns is indeed
minimized.
 
The result of the simulation at the parameter setting
$\alpha=\kappa=1$ is shown on Fig.\ \ref{fig:numeric-data}.  Since
$\kappa>0$, in the final approach $\Delta^{\mu_0(t)}\to\kappa$ for the
momentarily least unstable pattern the stability is positive, so the
second row in the update rule (\ref{eq:-update2}) does not come into
play.

\vbox{
\begin{figure}[htb]
 \begin{center}
    \psfig{figure=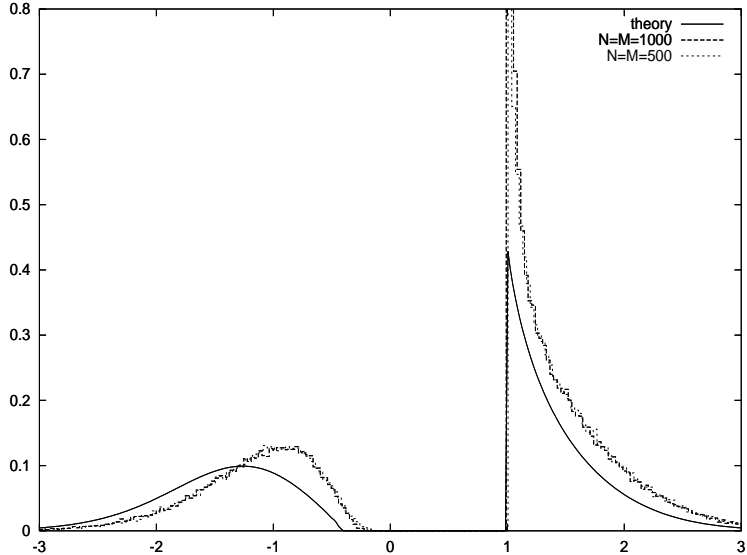,
     width=3in,height=4in,angle=270} \vspace*{0.5cm}
  \end{center}
  \caption{Density of local stabilities $\rho(\Delta)$ at
    $\alpha=\kappa=1$.  The horizontal axis is $\Delta$, the vertical
    one $\rho$.  The theoretical prediction is given by the full line.
    The two empirical densities are normalized histograms, taken with
    $M=N=500$ and $1000$.  Reprinted from Ref.\ 
    \protect\cite{our-paper-00}.}
  \label{fig:numeric-data} 
\end{figure}
}

The full line is the result of numerical extremization of the
variational free energy (\ref{fe-functional-spher-def-ext}) by the
method explained in Section \ref{ssec-numeric}.  We omitted the Dirac
delta peak of the theoretical probability density at $\kappa=1$.  The
dashed lines are the histograms for the local stabilities from
simulation for two sizes, $M=N=500$ and $1000$, with proper
normalization.  We do not enclose the original data of Wendemuth
\cite{wen95b}, but mention that his histogram showed a much larger
systematic error.  To quantify the deviation let us consider the mean
error $\varepsilon$\ie the relative number of misclassified patterns.
Wendemuth's number is $0.21$, the present simulation gives $0.15$,
while theory predicts $0.1358$.  Thus we are still about $10\%$ off
the theoretical value, but it is a remarkable improvement \wrt the
previous deviation of $55\%$.  The size of the gap from simulation is
also within about $10\%$ of the theoretical value.  The simulation
data reproduces, for the larger size $M=N=1000$, the property that the
density $\rho(\Delta)$ linearly vanishes at the lower edge of the gap.
This should be contrasted with the $1$-RSB result in Ref.\ 
\cite{mez93}, where the size of the discontinuity at the lower edge of
the gap is about the third of the height of left peak.  The simulation
clearly favors the CRSB solution.

In summary, the theoretical and simulation data do not match
perfectly, however, given the NP-completeness of the numerical
problem, this does not disprove theory.  We mention that the algorithm
used had the primitive side of being deterministic, furthermore, it
does not have a rigorous mathematical basis for convergence to the
desired state.  There is obviously room for further improvements.


\section{The Neuron: Independently Distributed  Synapses}
\label{ssec-indep-synaps}

\subsection{Free energy and stationarity condition}
\label{ssec-fe-stati-is}

In this paper we focus mostly on the spherical neuron.  Since,
however, the main formulas for the case of prior distribution
(\ref{gen-prior}), where synapses are independent and obey arbitrary
distribution, follow straightforwardly from Section \ref{sec-PA-Gen}, we
now briefly review them. In the course of continuation the limits \bml
\label{limits-entr}
\begin{eqnarray}
  \label{limits1-entr}
   q_0\to q_{(0)}~ &, & ~~~q_R\to q_{(1)}, \\
  \label{limits2-entr}
  \hat{q}_0\to \hat{q}_{(0)}~ & , & ~~~\hat{q}_R\to \hat{q}_{(1)} 
\end{eqnarray}
\eml 
are assumed.  The corresponding free energy (\ref{fe-gen}) can be
characterized by two OPF-s
\bml
  \label{hat-order-param1}
\begin{eqnarray}
  \label{hat-order-param11}
  q(x)~ & , & ~~~q(0) = q_{(0)}~, ~~~q(1) = q_{(1)}~, \\
  \label{hat-order-param12}
  \hat{q}(x)~ & , & ~~~\hat{q}(0) = \hat{q}_{(0)}~, ~~~\hat{q}(1) =
  \hat{q}_{(1)} . 
\end{eqnarray}
\eml
Alternatively, we can take as OPF-s the respective inverses 
\begin{equation}
  \label{hat-order-param2}
  x(q)~, ~~~\hat{x}(\hat{q})
\end{equation}
Then
\begin{equation}
  \label{hat-order-param3}
  \hat{q}= \hat{q}(x(q)),
\end{equation}
or its inverse function
\begin{equation}
  \label{hat-order-param4}
   q=q(\hat{x}(\hat{q}))
\end{equation}
establishes a relation between the overlaps $q$ and $\hat{q}$.

Concerning the $f_\re $ term, Eqs.\ (\ref{phi-ener}-\ref{spde-ener}) from
the spherical case carry over unchanged.  The entropic term
(\ref{fe-gend}) is a transcript of (\ref{feehat-final}) with
(\ref{s-phi}) together with the appropriate equations that produce the
averages. We introduce the field (see Eqs.\ 
(\ref{phihat-pde1},\ref{phihat-pde2})
\begin{equation}
  \label{field-entr}
  \hat{f}(\hat{q},y) =- \beta^{-1} \hat{\varphi}(\hat{q}, y) 
\end{equation}
to get
\begin{equation}
  \label{fes-parisi}
  \hat{f}_\rs [\hat{x}(\hat{q})] = \lim_{n\to 0} \frac{1}{n}
  \hat{f}_\rs ({\sf \hat{Q}}) 
  = -\beta^{-1} \left. \hat{\varphi}\left[\ln \int\rd u 
  ~w_0(u) ~ e^{-\beta uy},{\sf \hat{Q}}\right] \right| _{n=0} 
  =   \hat{f}(0,0),
\end{equation}
where  $\hat{f}(\hat{q},y)$ is the solution of 
\bml
  \label{ppde-entr}
\begin{eqnarray}
  \label{ppde-entr1} \ptl _{\hat{q}}\hat{f} & = & -\hf \ptl _y ^2
   \hat{f} + \hf \beta \hat{x} \left(\ptl _y \hat{f} \right)^2 , \\
   \label{ppde-entr2} \hat{f}(\hat{q}_{(1)},y) & = & - \beta^{-1} \ln
   \int\rD z \int\rd u ~w_0(u)~ e^{-\beta u(y+iz\sqrt{\hat{q}_{(1)}})} .
\end{eqnarray}
\eml
Introducing
\begin{equation}
  \label{intro-m-entr}
  \hat{m}(\hat{q},y) = \ptl _y \hat{f}(\hat{q},y),
\end{equation}
we have
\bml
\label{deriv-ppde-entr}
\begin{eqnarray}
  \label{deriv-ppde-entr-evolv}
   \ptl _{\hat{q}} \hat{m} &  = & -\hf \ptl _y ^2  \hat{m} 
   + \beta \hat{x} \hat{m} \ptl _y \hat{m} , \\
  \label{deriv-ppde-entr-init}
  \hat{m}(\hat{q}_{(1)},y) & = &  \ptl _y  \hat{f}(\hat{q}_{(1)},y).
\end{eqnarray}
\eml
Furthermore, the $\hat{~}$-ed 'susceptibility field' is 
\begin{equation}
  \label{intro-chi-entr}
  \hat{\chi}(\hat{q},y) = \ptl _y \hat{m}(\hat{q},y),
\end{equation}
obeying
\bml
\label{second-deriv-ppde-entr}
\begin{eqnarray}
  \label{second-deriv-ppde-entr-evolv} \ptl _{\hat{q}} \hat{\chi} & = & -\hf
   \ptl _y ^2 \hat{\chi} + \beta \hat{x}( \hat{m} \ptl _y \hat{\chi} +
   \hat{\chi}^2) , \\ \label{second-deriv-ppde-entr-init}
   \hat{\chi}(\hat{q}_{(1)},y) & = & \ptl _y^2 \hat{f}(\hat{q}_{(1)},y).
\end{eqnarray}
\eml
The probability density $\hat{P}(\hat{q},y)$ satisfies a variant of
the SPDE
\bml
\label{spde-entr}
\begin{eqnarray}
  \label{spde-evolv-entr}
 \ptl _{\hat{q}}\hat{P}  & = & \hf \ptl _y ^2\hat{P} 
 + \beta \hat{x} \ptl _y (\hat{P} \hat{m}) , \\
  \label{spde-init-entr}
  \hat{P}(0,y) & = &\delta(y) .
\end{eqnarray} 
\eml The interaction term (\ref{fe-genc}) is simplest if expressed
through the functions (\ref{hat-order-param1})
\begin{equation}
  \label{fei-entr}
  f_i[x(q),\hat{x}(\hat{q})] = -\beta \int_0^1\rd x ~ q(x) ~ \hat{q}(x).
\end{equation}
Since a function is a functional of its inverse, the $f_i[\dots]$ can
be considered as functional of $x(q)$ and $\hat{x}(\hat{q})$.

The stationarity conditions (\ref{stat-q-gen},\ref{stat-qhat-gen}) now
read as 
\bml  
\label{stationarity-soft-spin}
\begin{eqnarray}
  \label{stationarity-soft-spin1} 
    q & = & \int\rd y~ \hat{P}(\hat{q},y)~ 
    \hat{m}(\hat{q},y)^2 ,\\   
  \label{stationarity-soft-spin2} 
    \hat{q} & = & \alpha \int\rd y~ P(q,y)~ m(q,y)^2 ,   
\end{eqnarray}
\eml where the connection between $q$ and $\hat{q}$ is established by
(\ref{hat-order-param3}) or (\ref{hat-order-param4}).  The r.\ h.\ 
sides are respective functionals of $\hat{x}(\hat{q})$ and $x(q)$.
Note that solving these equations involves also finding the starting
point $\hat{q}_{(1)}$, in contrast to the evaluation of the energy
term, where the initial condition is fixed at $q=1$.  Given the
solution for the stationary $x(q)$ and $\hat{x}(\hat{q})$, by
substituting them into the r.\ h.\ s.\ of
\begin{equation}
  \label{fe-soft-spin-xq-final}
  f = \hat{f}_\rs [\hat{x}(\hat{q})] +  f_i[x(q),\hat{x}(\hat{q})] 
  + \alpha f_\re [x(q)]
\end{equation}
we obtain the final result for the mean free energy.

A special case of independently distributed synapses is the clipped
neuron\ie the neuron with discrete synapses. The most studied such
model is the Ising neuron with binary synapses, which has attracted
considerable interest (see \cite{hkp91} and \cite{wes97} for
references).  The prior distribution in the Ising case involves
(\ref{bin-prior}), so the initial conditions for the PDE-s are \bml
\label{ic-I}
\begin{eqnarray}
  \label{ic-I1} \hat{f}(\hat{q}_{(1)},y) & = & - \beta^{-1} \ln
   \text{cosh} \beta y + \hf \beta \hat{q}_{(1)}, \\ \label{ic-I2}
   \hat{m}(\hat{q}_{(1)},y) & = & - \text{tanh} \beta y.
\end{eqnarray}
\eml 
The Ising neurons studied in the literature so far were reminiscent to
the random energy model in that they involved at most $1$-RSB
\cite{derrida80}.  However, only a few choices of the error measure
potential $V(y)$ were considered, and at this stage it cannot be
excluded that the full Parisi scheme becomes of import for other
potentials.

Finally we emphasize that previously studied SK-type spin glasses,
with Ising, or, as a matter of fact, any kind of individual spin
constraint (\ref{gen-prior}), are included in the considerations of
this section.  Formulae equivalent with (\ref{ic-I}) are well known
from the SK model.  This is not surprising, because the entropic term
with the Ising constraint is the same for both the SK spin glass and
the present neuron model.  In the case of other constraints for the
SK-type spin glass the first two terms on the r.\ h.\ s.\ of
(\ref{fe-soft-spin-xq-final}) remain valid.  Concerning the third, the
energy term, let us assume the most general multi-spin interaction of
Nieuwenhuizen's \cite{nie95} resulting in a term
(\ref{gen-energ}). Then indeed, if in (\ref{fe-soft-spin-xq-final})
the $\alpha f_\re $ is replaced by the energy term (\ref{gen-energ-our}),
with the understanding of the correspondence
(\ref{nieuw},\ref{nieuw-our}), one obtains by
(\ref{fe-soft-spin-xq-final}) the full free energy functional of the
spin glass problem.

\subsection{Variational principle}
\label{sssec-vari-indep-syanps}

The results of Section \ref{ssec-fe-stati-is} can also be derived from
a variational principle.  A reasoning similar to what we followed in
the spherical problem yields a free energy functional $f[\dots]$ that
produces the mean free energy as
\begin{equation}
  \label{fe-indep-var} f = {\ba{c}\mbox{\footnotesize ~} \\
  \mbox{max}\\ \mbox{\footnotesize{\em x(q)}}\ea} {\ba{c}\mbox{
  \footnotesize ~} \\ \mbox{extr}\\ \mbox{\footnotesize{\em
  f(q,y),P(q,y)\,}} \ea}\!\!\!  {\ba{c}\mbox{\footnotesize ~} \\
  \mbox{extr}\\ \mbox{\footnotesize{$\,\hat{x}(\hat{q})$}}\ea}
  {\ba{c}\mbox{ \footnotesize ~} \\ \mbox{extr}\\
  \mbox{\footnotesize{$\hat{f}(\hat{q},y),\hat{P}(\hat{q},y)$}}
  \ea}\!\!\!  f\left[ x(q),f(q,y),P(q,y), \hat{x}(\hat{q}),
  \hat{f}(\hat{q},y),\hat{P}(\hat{q},y) \right].
\end{equation}
The order of the extremum conditions is not binding, but, given the
physical meaning of the OPF $x(q)$, the maximum is to be taken
last. The free energy functional is \bml
\label{fe-functional-gen-prior} 
\begin{eqnarray}
  \label{fe-functional-all-gen-prior}
  f[\dots]
   & = &  \hat{f}_\rs [\dots] + \hat{f}_a^{(1)}[\dots] +
   \hat{f}_a^{(2)}[\dots]+
   f_i[\dots] + \alpha (f_\re [\dots] + f_a^{(1)}[\dots] + f_a^{(2)}[\dots]) , \\ 
  \label{fe-functional-s-gen-prior}
   \hat{f}_\rs [\dots] & = & \hat{f}(0,0), \\ 
  \label{fe-functional-a1-gen-prior}
  \hat{f}_a^{(1)}[\dots]& = & \int_0^{\hat{q}_{(1)}}\!\! \rd \hat{q}\ \int\rd y ~
    \hat{P} (\hat{q},y) 
    \left[ \ptl _{\hat{q}} \hat{f} (\hat{q},y) + \case{1}{2} 
   \ptl _y^2 \hat{f}(\hat{q},y) - \case{1}{2}\beta\hat{x}(\hat{q}) 
   \left( \ptl _y \hat{f}(\hat{q},y) \right)^2 \right], \\
  \label{fe-functional-a2-gen-prior}
  \hat{f}_a^{(2)}[\dots] & = & - \int\rd y~ \hat{P}(\hat{q}_{(1)},y) 
   \left[ \frac{1}{\beta} \ln \int \rd u~ w_0(u)~ e^{-\beta u y}
      + \hat{f} (\hat{q}_{(1)},y) \right], 
\end{eqnarray}
\eml where $f_\re [\dots]$, $f_a^{(1)}[\dots]$, $f_a^{(2)}[\dots]$, and
$f_i[\dots]$ are given by Eqs.\ (\ref{fe-functional-spher-e}),
(\ref{fe-functional-spher-a1}), (\ref{fe-functional-spher-a2}), and
(\ref{fei-entr}), respectively.  The remarkable symmetry of the above
expressions in the quantities with and without the $~\hat{}~$ mark is
the consequence of the fact that both the entropic and the energy
terms are essentially of the general Parisi form, the main difference
being in the starting point of the time variable and the initial
condition for the PDE\@.

The free energy functional $f[\dots]$ should be maximized in $x(q)$
and extremized over the other function arguments. Besides the
extremization of terms analogous to those appearing in Section
\ref{ssec-variation}, we have to calculate the functional derivatives
of $f_i[x(q),\hat{x}(\hat{q})]$. If $u^{-1}(u)$ is the inverse of some
function $u(t)$ then variation of the identity $u^{-1}(u(t))=t$ yields
the functional derivative of the inverse function as
\begin{equation}
  \label{func-deriv-inverse}
  \frac{\delta \,u^{-1}(u(t_1))}{\delta\, u(t_2)} = 
  - \frac{\delta(t_2-t_1)}{\dot{u}(t_1)}.
\end{equation}
This relation helps us to calculate the sought derivatives of the
interaction term \bml
\label{fi-func-deriv} 
\begin{eqnarray}
  \frac{\delta\, f_i[x(q),\hat{x}(\hat{q})]}{\delta\, x(q)} =
  \beta\,\hat{q}(x(q)) \label{fi-func-deriv1} \\ \frac{\delta\,
  f_i[x(q),\hat{x}(\hat{q})]}{\delta\, \hat{x}(\hat{q})} =
  \beta\,q(\hat{x}(\hat{q})).  \label{fi-func-deriv2}
\end{eqnarray}
\eml These, together with functional derivatives of the type
determined in Section \ref{ssec-variation}, lead to the stationarity
relations displayed in Section \ref{ssec-fe-stati-is} for intervals of
strictly increasing OPF-s, including the points
where the OPF-s exhibit a step.  Plateaus should be dealt with
in a manner similar to what was described in Section 
\ref{ssec-variation}.  Extremization in terms of the starting point in
time of the $\hat{~}$-ed PDE, $\hat{q}_{(1)}$, yields
\begin{equation}
  \label{func-deriv-q1}
  \beta\, \hat{x}(\hat{q}_{(1)})\, q_{(1)} 
  = \int\rd y~ \hat{P}(\hat{q}_{(1)},y)\, \hat{\chi}(\hat{q}_{(1)},y), 
\end{equation}
a condition which was not displayed in Section \ref{ssec-fe-stati-is}.

\subsection{On thermodynamical stability} \label{ssec-stability-indep}

In the case of independently distributed synapses the free energy
(\ref{fe-gen}) involves combined maximization and extremization.
Clearly, there are no stability requirements following from the 'extr'
condition.  On a simple example we now give the recipe for stability
calculations in the replicon sector for such a case.

Consider the two-variable function
\begin{equation}
  \label{demo-min1}
  F(x,\hat{x})= f(x) + \hat{f}(\hat{x})+x\hat{x},
\end{equation}
where $f$ and $\hat{f}$ are real functions. We are seeking
\begin{equation}
  \label{demo-min-extr}
  {\ba{c}\text{\footnotesize ~} \\ \text{min} \\ \text{\footnotesize
   $x$} \ea} 
  {\ba{c}\text{\footnotesize ~} \\ \text{extr} \\ \text{\footnotesize 
      $\hat{x}$}\ea} F(x,\hat{x}) .
\end{equation}
Extremum conditions for $x$ and $\hat{x}$ imposed simultaneously would
read as \bml
\label{demo-min2}
\begin{eqnarray}
  \label{demo-min2a}
  x & = & - \hat{f}^\prime(\hat{x}) , \\
  \label{demo-min2b}
  \hat{x} & =& - f^\prime(x) .
\end{eqnarray}
\eml Substitution of the stationary value (\ref{demo-min2b}) gives 
\begin{equation}
  \label{demo-min3}
  F(x)\equiv F(x,-f^\prime(x))=f(x)+\hat{f}(-f^\prime(x))-xf^\prime(x).
\end{equation}
The stationarity condition in terms of $x$ is
\begin{equation}
  \label{demo-min4}
  F^\prime(x)=-f^\pp(x)\left[x+\hat{f}^\prime(-f^\prime(x))\right]=0.
\end{equation}
The stationary point $x+\hat{f}^\prime(-f^\prime(x))=0$ is a minimum
of $F(x)$, if
\begin{equation}
  \label{demo-min5}
  \left. F^\pp(x)\right|_{x=-\hat{f}^\prime(-f^\prime(x))} = -
  f^\pp(x)\left[1 - \hat{f}^\pp(\hat{x}) f^\pp(x)\right]
   > 0 ,
\end{equation}
where (\ref{demo-min2a},\ref{demo-min2b}) are understood.  If we have
more-than-one-dimensional objects en lieu of $x$ and $\hat{x}$, and at
the saddle the Hessian matrices of $f$ and $\hat{f}$ can be
simultaneously diagonalized, then a similar formula holds, wherein the
appropriate eigenvalues of the Hessian at stationarity should be
substituted for $f^\pp(x)$ and $\hat{f}^\pp(\hat{x})$.

The above principle can be applied to the case of independently
distributed synapses.  There are two families of replicon eigenvalues,
$\lambda_\re (q_1,q_2,q_3)$ and
$\hat{\lambda}_\rs (\hat{q}_1,\hat{q}_2,\hat{q}_3)$ coming from the
energy and entropic term, respectively.  The contribution from the
energy term, $\lambda_\re (q_1,q_2,q_3)$, is the same as in the spherical
case, given by Eq.\ (\ref{repl-neur-spher-ener}).  Analogously, from
the $\hat{~}$-ed entropic term we have
\begin{equation}
    \hat{\lambda}_\rs (\hat{q}_1,\hat{q}_2,\hat{q}_3)= - \beta^2 \int\,
   \rd y_2\,\rd y_3\, \hat{\Gamma}_{\varphi \varphi \varphi} \left(
    \hat{q}_1;0,0;\hat{q}_2,y_2;\hat{q}_3,y_3 \right) \hat{\chi}
    \left(\hat{q}_2,y_2 \right)\, \hat{\chi} \left(\hat{q}_3,y_3 \right).
\label{repl-neur-indep-entr}
\end{equation}
Here we spell out the obvious, namely, the $\hat{~}$-ed PPDE gives
rise to a $\hat{~}$-ed Green function, see Section \ref{sssec-gf}, whence
the vertex function $\hat{\Gamma}_{\varphi \varphi \varphi}$ can
be defined.  Finally, based on (\ref{demo-min5}) the necessary
criterion of stability becomes
\begin{equation}
\label{repl-neur-indep-stabil}
        - \lambda_\re (q_1,q_2,q_3)\left[ 1 - \alpha
        \lambda_\re (q_1,q_2,q_3)\,
        \hat{\lambda}_\rs (\hat{q}_1,\hat{q}_2,\hat{q}_3) \right] \ge 0 .
\end{equation}
Here we have omitted a prefactor $\alpha$ and allowed equality for the
sake of possible Goldstone modes in a Parisi phase.  The stability
condition is of course understood at stationarity, which yields a
concrete $q=q(\hat{q})$ function.  That implies $q_i=q(\hat{q}_i)$, so
the overall replicon eigenvalue is parametrized by three independent
variables, as in the spherical case.


\section{Conclusions and Outlook}
\label{sec-conclusion}

The main messages of this paper were extensively discussed and
conclusions were advanced in Chs.\ \ref{intro} and \ref{history}, so we
only highlight a few moral issues here.

A sensitive question in approximating a CRSB state by finite $R$-RSB
is how good it will turn out to be in the end.  However, there is so
far no reliable {\em a priori} estimate of this error, as opposed to
say a series expansion, where the last power retained gives at least
asymptotically a bound for the error.  Sometimes there is a
qualitative indicator showing that a low order approximation is wrong.
Long known example is the ground state entropy of the SK model, which
was negative for finite $R$-RSB ans\"atze, a problem cured only by
Parisi's CRSB solution.  However, often macroscopic quantities are
quite well approximated with the RS or low $R$-RSB solution.  The main
advantage of the CRSB calculation \wrt the approximations is that the
latter may not be able to even qualitatively correctly predict
distributions of local, non-self-averaging quantities, like the
overlaps and local fields.  These are observables in numerical
simulations and can help to decide between candidate theories (see\eg
the numerical review \cite{mar98} on spin glasses).

On the technical side, the mathematical framework discussed in Chs.\ 
\ref{sec-PA-Gen}, \ref{sec-corr-stab}, and
\ref{sec-special-properties} relates to the general properties of CRSB
phases, irrespective of the storage problem of the neuron, upon which
its use was demonstrated subsequently.  It allows for a
non-perturbative description in a wide range of problems in disordered
systems, like long range interaction spin glasses other than the SK
model, and may be a starting point for the study of frustrated
phases\ie unsatisfiable situations, in optimization problems in
general.  

Among the notoriously difficult problems in artificial neural networks
is the problem of learning and generalization of unlearnable tasks
\cite{wat93,ev00}.  In the traditional scenario of equilibrium\ie
batch, learning from examples, an unlearnable problem is characterized
by the fact that there is a limited number of examples the network can
reproduce.  Beyond this limit of error-free learning, the
generalization ability might be further improved, but the minimal
training error is positive.  This is in close analogy with the region
beyond capacity in the storage problem, so it is a sensible assumption
that theoretical methods able to deal with imperfect storage may also
be of use for the description of learning the unlearnable.  Further
possible area of application is unsupervised learning \cite{ev00},
where no desired output is given, rather the properties of the
distribution of examples is to be extracted.  Again, if the network
can be saturated by the examples a complex phase appears, where
methods similar to those presented in this paper may be the key to the
solution.


\acknowledgments
 
The author acknowledges support by OTKA grant No.\ T017272.  Thanks
are due first and foremost to P. Reimann, the co-author of Refs.\ 
\cite{our-paper,our-paper-00}, with whom the work that forms the basis
of this paper was begun and in great part done.  It is regrettable
that, due to other obligations, he felt compelled to withdraw his name
from the Physics Reports project along the road.  But before that he
drafted an introductory chapter, from which much was retained in the
present Sections \ref{intro} and \ref{history}, and collected many
references.  He was responsible for the numerical evaluation of the
theoretical predictions, the text of and most of the work behind
Section \ref{ssec-numeric} is his contribution alone.  Those sections
are marked by an asterisk in the title.  His numerous comments on some
other parts greatly improved the presentation.  Enlightening
discussions with F.  P\'azm\'andi helped to raise and clarify fine
points of the theoretical framework, in particular the problem of a
discontinuity in the error measure potential.  The author remains
grateful to T.  B\'\i r\'o for introducing him to the subject of Lie
symmetries of partial differential equations.  As already stated in
Ref.\ \cite{our-paper-00}, the simulation was done on the PC cluster
of F.  Csikor and Z.  Fodor, built from grants OTKA-T22929 and
FKFP-0128/1997.

\eject
\appendix

\section{Abbreviations} 
\label{app-sec-abbrev}

This Appendix lists acronyms and abbreviations used throughout the
paper.  \vspace{\baselineskip}

\noindent
left-hand-side \dotfill \lhs  \\
right-hand-side \dotfill \rhs  \\
with respect to \dotfill \wrt  \\
de Almeida-Thouless [stability]\dotfill AT\\
Green function\dotfill GF \\
Order parameter function \dotfill OPF \\
Partial differential equation\dotfill PDE \\
Parisi's PDE, Eq.\ (\ref{ppde}) \dotfill PPDE \\
Replica symmetry breaking\dotfill RSB \\
$R$-step RSB\dotfill $R$-RSB \\
Continuous RSB\dotfill CRSB \\
Sherrington-Kirkpatrick [model] \dotfill SK \\
Sompolinsky's PDE, Eqs.\ (\ref{spde},\ref{spde-init}) \dotfill SPDE \\
Ward-Takahashi identity\dotfill WTI


\section{Derivation of the replica free energy}
\label{app-repl-fe}

The $n$-th moment of the partition function (\ref{partfunc}), with
(\ref{def-delta}) inserted as constraint, reads as
\begin{equation}
  \label{repl-partfunc}
   \left<Z^n\right> = \int \prod_a \rd^N\!  J_a\ w(\jv_a)
   \prod_{\mu}\rd y_a^{\mu}\ e^{-\beta V(y_a^{\mu})} 
   \left<\delta\left(y_a^{\mu}
       - N^{-1/2}\xi^\mu \sum_kJ_{ak}S_k^\mu \right)\right>. 
\end{equation}
The indices $k$, $\mu$, and $a$ run from $1$ to $N$, $M$, and $n$,
respectively. The Fourier transformation of the Dirac deltas
introduces the ancillary variables $x_a^{\mu}$ adjoint to $y_a^{\mu}$.
Average over the Gaussian distribution of patterns $S_k^\mu$, and over
the outputs $\xi^\mu$, which are $\pm 1$ equally likely, can be
performed straightforwardly.  In fact, since $S_k^\mu$ is scaled by
the vanishing factor $N^{-1/2}$, the same result would be obtained for
other distributions of $\xi^\mu S_k^\mu$ with zero mean and unit
variance independent of $\mu$ and $k$.  
\begin{eqnarray}
  \label{repl-partfunc2} \left<Z^n\right> & = & \int \left[ \prod_{a}
   \rd^N\!  J_a\ w(\jv_a) \,
   \prod_{\mu}\frac{\rd x_a^{\mu}\rd y_a^{\mu}}{2\pi} \right] \nn \\ &&
   \times \, \prod_{\mu} \exp \left(-\beta \sum_{a} V(y_a^{\mu}) +
   \sum_{a} i x_a^{\mu} y_a^{\mu} - \frac{1}{2N}
   \sum_{a,b}x_a^{\mu}x_b^{\mu} \sum_kJ_{ak} J_{bk} \right) .
\end{eqnarray}
If we substitute the overlaps defined in (\ref{syn-overlap}), the
product over $\mu$ gives the $M=\alpha N$-th power of $e^{-n\beta
  f_\re ({\sf Q})}$, where $f_\re ({\sf Q})$ is displayed in
(\ref{fe-spher1d}).  Our inserting the constraint (\ref{syn-overlap})
yields
\begin{equation}
  \label{repl-partfunc3} 
  \left<Z^n\right> =
  \int\left[ \prod_{a}\rd ^N\!  J_a\ w(\jv _a) \right] 
  \ e^{-N\alpha\beta f_\re ({\sf Q})} 
  \prod_{a<b}N\rd q_{ab}\ \delta\left(Nq_{ab} 
      - \sum_k J_{ak}J_{bk}\right). 
\end{equation}
For both the spherical (\ref{spher-prior}) and the independent (see
condition below (\ref{gen-prior})) prior distributions we have
$q_{aa}\equiv q_\rD=1$.  Fourier transformation of the Dirac deltas
introduces the variables $\tilde{q}_{ab}$ adjoint to $q_{ab}$, and we
have
\begin{eqnarray}
  \label{repl-partfunc4} \left<Z^n\right> & =& \int
  \left[\prod_{a<b}\frac {N}{2\pi}\rd q_{ab}\,\rd \tilde{q}_{ab}\right]
  \left[ \prod_{a}\rd ^N\!  J_a\ w(\jv _a) \right] e^{-N\alpha\beta
  f_\re ({\sf Q})} \nn \\ && \times \,
  \exp\left(iN\sum_{a<b}q_{ab}\tilde{q}_{ab} - i\sum_k \sum_{a<b}
  \tilde{q}_{ab} J_{ak}J_{bk}\right).
\end{eqnarray}

We shall see that for the prior densities of interest, after
integration by the synaptic coefficients, the exponential has the
overall coefficient $N$.  Then for $N\to\infty$ the saddle point
method can be applied, and it will turn out that the stationary value
of each $tilde{q}_{ab}$ is imaginary.  We presume that the prior
density is such that the integration path of $\tilde{q}_{ab}$ can be
distorted so as to go through the imaginary saddle point.  The path
can then be taken a straight line parallel or perpendicular to the
imaginary axis in a sufficiently large neighborhood of the saddle
point, depending on which orientation ensures a maximum at the saddle.
This procedure is typical if one integrates a fast oscillating
integrand --- then only an extremum, not specifically minimum,
condition should be satisfied.  If we succeed in determining the
saddle values of the $\tilde{q}_{ab}$-s as function of the $q_{ab}$-s,
we have to minimize in terms of $q_{ab}$.  We shall see that this can
be carried out for the spherical constraint (\ref{spher-prior}), but
we cannot explicitly determine the $\tilde{q}_{ab}$-s for general
independent synapses (\ref{gen-prior}).

In the case of the spherical constraint (\ref{spher-prior}) we shall
make use of the advance knowledge that the stationary values of
$\bar{q}_{ab}=i\tilde{q}_{ab}$ are real.  Let us insert the Fourier
transform of the Dirac deltas representing the spherical constraints,
thereby introducing the integration variables $\tilde{q}_a$, then
switch over to $\bar{q}_{a}=i\tilde{q}_{a}$ to obtain
\begin{eqnarray}
  \left<Z^n\right> & = & C_N \,
  N^\frac{n(n-1)}{2}  (2\pi)^{-\frac{n(n+1)}{2}} \int 
  \left[\prod_{a<b}\rd q_{ab}\,\rd \bar{q}_{ab}\right]
  \left[ \prod_{a}\rd ^N\!  J_a\, \rd \bar{q}_a \right] 
  e^{-N\alpha\beta f_\re ({\sf Q})} 
   \nn \\ 
\label{repl-partfunc-spher1}
 & & \times
  \exp\left(N\sum_{a<b}q_{ab}\bar{q}_{ab} +
   N\sum_a\bar{q}_a - \sum_{k,a} \bar{q}_a {J_{ak}}^2 
   - \sum_k \sum_{a<b} \bar{q}_{ab} J_{ak}J_{bk}\right).
\end{eqnarray}
We can introduce diagonal elements for the matrix ${\sf \bar{Q}}$ as
$\bar{q}_{aa}=2\bar{q}_a$.  Performing the Gaussian integrals over
$J_{ak}$ we obtain
\begin{eqnarray}
  \label{repl-partfunc-spher2} \left<Z^n\right> &=& C_N\,
  N^\frac{n(n-1)}{2} (2\pi)^\frac{n(N-n-1)}{2} \, \int
  \left[\prod_{a<b}\rd q_{ab}\right] \left[\prod_{a\leq b}\rd \bar{q}_{ab}
  \right] \nn \\ && \times \, \exp N \left( -\alpha\beta f_\re ({\sf Q})
  + \frac{1}{2} \text{Tr} {\sf Q}{\sf \bar{Q}} - \frac{1}{2}
  \text{lndet}{\sf \bar{Q}} \right) .  \end{eqnarray} Given the
  asymptotics of the prefactor
\begin{equation}
  \label{prefactor-asympt}
  \frac{\ln C_N}{N} \approx - \frac{1}{2} \ln 2\pi e , 
\end{equation}
in the large N limit we have by the saddle point method the
free energy
\bml
  \label{repl-free-energ-spher1}
\begin{eqnarray}
  \label{repl-free-energ-spher1a}
  f & = & \lim_{n\to 0} \frac{1}{nN\beta} (1 - \left<Z^n\right>) =  
  \lim_{n\to 0} \frac{1}{n}\
  {\ba{c}\text{\footnotesize ~} \\ \text{min}\\ 
    \text{\footnotesize $\sf Q$ }\ea} 
  {\ba{c}\text{\footnotesize ~} \\ \text{extr}\\ 
    \text{\footnotesize $\sf \bar{Q}$}\ea} 
  f({\sf Q},{\sf \bar{Q}}) ,  \\
  \label{repl-free-energ-spher1b}
  f({\sf Q},{\sf \bar{Q}}) & = & \frac{n}{2\beta}+ \alpha f_\re ({\sf Q}) 
  - \frac{1}{2\beta} \text{Tr} {\sf Q}{\sf \bar{Q}} + \frac{1}{2\beta} 
  \text{lndet}{\sf \bar{Q}} .
\end{eqnarray}
\eml 
By our using (\ref{der-lndet}) the extremum condition for the matrix
elements $\bar{q}_{ab}$ results in ${\sf \bar{Q}}^{-1}={\sf
Q}$. Substitution thereof into (\ref{repl-free-energ-spher1b}) gives
the spherical free energy (\ref{fe-spher1}).

Similar derivation yields the free energy (\ref{fe-gen}) for the prior
distribution (\ref{gen-prior}) of independent synapses. There we use
$\hat{q}_{a}=-i\beta^2\tilde{q}_{a}$ and obtain
\begin{eqnarray}
  \left<Z^n\right> & = & 
  \left(\frac{N}{2\pi}\right)^\frac{n(n-1)}{2} \int 
  \left[\prod_{a<b}\rd q_{ab}\,\rd \hat{q}_{ab}\right]
  e^{-N\alpha\beta f_\re ({\sf Q})}
  \exp\left(-N\beta^2\sum_{a<b}q_{ab}\hat{q}_{ab} \right)
   \nn \\ 
\label{repl-partfunc-gen}
 & & \times    \left\{\int \left[ \prod_{a}\rd J_a\, w_0(J_a) \right] 
  \exp\left(\beta^2 \sum_{a<b} \hat{q}_{ab} J_aJ_b\right)\right\}^N .
\end{eqnarray}
The ancillary matrix ${\sf \hat{Q}}$ has vanishing diagonal elements,
$\hat{q}_\rD =0$, with that in mind we recover the expression
(\ref{fe-gen}) for the free energy.  The ancillary matrix elements
cannot, in general, be eliminated as easily as in the spherical case.


\section{Derivation of the $R$-RSB free energy term} 
\label{app-sec-deriv-r-rsb}

This Appendix contains the few steps that lead to the $R$-RSB free
energy term, starting out from Eq.\ (\ref{temp1}).  The integrals
therein taken over the variables $x_a$-s yield Dirac-deltas, which fix
the values of the $y_a$-s.  The $j_r$ indices can be understood as
follows.  Assume as usual that each $m_r$ is a divisor of $n$.  The
ordered sequence of integers $1,\dots ,n$ are divided into $n/m_r$
``boxes'' each containing $m_r$ integers.  Then the index $j_r$
enumerates those boxes.  Given $1 \leq a\leq n $, for each $r$, the
$j_r(a)$ labels the box that contains $a$, that is,
\begin{equation}
  \label{j_r-def}
j_r(a) = \left[ \frac{a-1}{m_r} \right] + 1,
\end{equation}
where $[\dots ]$ denotes the integer part.  Then the coefficient of an
$x_a$ in the first term of the exponent in (\ref{temp1}) is
characterized by the $j_r(a)$-s.  That way we arrive at
\begin{equation}
  \label{temp2}
  e^{n\varphi[\varPhi(y),\qvs ,\mvs ]} =  \int
  \left[ \prod_{r=0}^{R+1}\prod_{j_r=1}^{n/m_r}\rD z_{j_r}^{(r)}\right]
  \prod_{a=1}^{n} \exp\varPhi\left(\sum_{r=0}^{R+1}
    z_{j_r(a)}^{(r)} \sqrt{q_r-q_{r-1}}\right)  .
\end{equation}
Note that $m_{R+1}=1$ and $a=j_{R+1}(a)$; we will substitute $j_{R+1}$
for $a$.  The integrals over $z_{j_{R+1}(a)}^{(R+1)}$ factorize as
\begin{eqnarray}
  \label{temp3} e^{n\varphi[\varPhi(y),\qvs ,\mvs ]} &= & \int \left[
  \prod_{r=0}^{R}\prod_{j_r=1}^{n/m_r}\rD z_{j_r}^{(r)} \right]
  \prod_{j_{R+1}=1}^{n/m_{R+1}} \int\rD z_{j_{R+1}}^{(R+1)}\nn \\ &&
  \times,\ \exp \varPhi\left(\sum_{r=0}^{R} z_{j_r(j_{R+1})}^{(r)}
  \sqrt{q_r-q_{r-1}} + z_{j_{R+1}}^{(R+1)}
  \sqrt{q_{R+1}-q_{R}}\right).
\end{eqnarray}
The functions $j_r(j_{R+1})$, $r\le R$, are step-like in that they are
constant for $m_R/m_{R+1}$ different $j_{R+1}$-s belonging to the same
box of length $m_R$.  Integrations over $z_{j_{R+1}}^{(R+1)}$-s
associated with the same box give identical results.  Different
integrals are characterized by different $j_{R}$-s, this can be given
as the new argument for the rest of the indices as $j_r(j_R)$, $r\le
R$.  We then have
\begin{eqnarray} 
   \label{temp4} e^{n\varphi[\varPhi(y),\qvs ,\mvs ]} & = &\int
   \left[ \prod_{r=0}^{R}\prod_{j_r=1}^{n/m_r}\rD z_{j_r}^{(r)}\right]
   \prod_{j_{R}=1}^{n/m_{R}} \nn \\ && \times \, \left[ \int\rD z_{R+1}
   \exp \varPhi\left(\sum_{r=0}^{R} z_{j_r(j_{R})}^{(r)}
   \sqrt{q_r-q_{r-1}} + z_{R+1} \sqrt{q_{R+1}-q_{R}} \right)\right]
   ^\frac{m_R}{m_{R+1}}.
\end{eqnarray}
Again, integration over a $z_{j_R}^{(R)}$ gives the same value for those
$j_R$-s that define the same $j_r(j_R)$, $r\le R-1$.  These can be
characterized by $j_{R-1}$, and one obtains
\begin{eqnarray}
 \label{temp5}
 e^{n\varphi[\varPhi(y),\qvs ,\mvs ]} & = &  
 \int\left[ \prod_{r=0}^{R-1}\prod_{j_r=1}^{n/m_r}\rD z_{j_r}^{(r)}\right]
  \prod_{j_{R-1}=1}^{n/m_{R-1}}\Bigg[ \int\rD z_{R} \Bigg[ 
   \int\rD z_{R+1}\nn \\
  & & \times \exp \varPhi\left(\sum_{r=0}^{R-1} z_{j_r(j_{R-1})}^{(r)}
   \sqrt{q_r-q_{r-1}} +\sum_{r=R}^{R+1} z_r \sqrt{q_r-q_{r-1}} 
  \right)\Bigg]^\frac{m_R}{m_{R+1}} \Bigg] ^\frac{m_{R-1}}{m_{R}} .
\end{eqnarray}
The expression can be rolled up by continuing the above reasoning and
we arrive at 
\begin{eqnarray} 
 \label{rrsb} e^{n\varphi[\varPhi(y),\qvs ,\mvs ]} & = &\int\rD z_0
 \Bigg [ \int\rD z_1 \Bigg [ \int\rD z_2 \dots \nn \\ && \times\, \Bigg [
 \int\rD z_{R+1}\exp \varPhi\left(\sum_{r=0}^{R+1} z_r \sqrt{q_r-q_{r-1}}
 \right)\Bigg ] ^\frac{m_R}{m_{R+1}} \dots \Bigg ]^\frac{m_{1}}{m_{2}}
 \Bigg ]^\frac{m_{0}}{m_{1}}.
\end{eqnarray}


\section{Derivation of the PPDE by continuation} 
\label{app-sec-duplantier}

To the author's knowledge Ref.\ \cite{dup80} is considered to be the
only publication on the derivation of the PPDE.  However, we were not
able to reproduce the derivation from that article, furthermore,
\cite{dup80} required $R\to\infty$ and $q_r-q_{r-1}\to 0$, conditions
which we did not find necessary to prescribe.

In essence, \cite{dup80} proposes an iteration in a direction that is
opposite to that of the recursion (\ref{rec-psi}).  We were unable to
reconstruct that, mostly because the starting term was not known.  In
other words, we evaluated the free energy term (\ref{temp1}) starting
from $r=R+1$, while \cite{dup80} did so from $r=0$ (in our notation).

When $q_r-q_{r-1}\to 0$ is assumed, our recursion yields the PPDE in
the spirit of Ref.\ \cite{dup80}.  We use the identity
\begin{equation}
  \label{diffop}
  \exp\left(\frac{c}{2}\frac{\rd^2}{\rd y^2}\right) F(y) 
  = \int \rD z~ F(y+z\sqrt{c})
\end{equation}
to rewrite (\ref{rec-psi1}) into 
\begin{equation}
  \label{rec-psiop}
    \psi_{r-1}(y) = e^{\frac{1}{2}(q_r-q_{r-1}) 
      \frac{\rd^2}{\rd y^2}}\,  \psi_r(y) ^ \frac{x_r}{x_{r+1}} .
\end{equation}
In order to produce a PDE from the recursion, the assumption of
ordering for $q_r$-s is necessary.  We can then relegate the
dependence on the index $r$ to dependence on the variable $q=q_r$.
Continuation is then performed by replacing $q_r$ by $q$, $\psi_r(y)$
by $\psi(q,y)$.  We allow for nontrivial limits $q_{(0)}$ and
$q_{(1)}$ as introduced in (\ref{q-extremities}).  The conditions
(\ref{ieq1}) and (\ref{ieq2b}) ensure monotonicity of $x(q)$.  If we
assume a smooth $x(q)$\ie that all $q_r-q_{r-1}\to 0$ and
$x_r-x_{r-1}\to 0$ for $1\leq r\leq R+1$, then an expansion of
(\ref{rec-psiop}) in the differences to lowest nontrivial order yields
for $\psi(q,y)$ the PDE (\ref{psi-pde}) in the interval
$(q_{(0)},q_{(1)})$.

As we found in Section \ref{sssec-ppde}, Eq.\ (\ref{psi-pde}) and,
equivalently, the PPDE (\ref{ppde}), stands even if $x(q)$ is not
smooth, with the right interpretation of (\ref{psi-pde}) at
discontinuities of $x(q)$.  On the other hand, the author gladly
acknowledges that the way he first obtained the PPDE for the general
free energy term (\ref{gen-parisi}) was in the spirit of the above
discussed derivation of Ref.\ \cite{dup80}.


\section{Multidimensional generalization of the PPDE} 
\label{app-high-dim}

We consider here the generalized free energy term
\begin{eqnarray}
\varphi[\varPhi(y),{\cal Q}] = \frac{1}{n} \ln \int & & \frac{\rd^{nK}\!
 x \, \rd^{nK}\! y}{(2\pi)^{nK}} \exp\sum\nolimits_{a=1}^n \varPhi
 (y_a^1,\dots,y_a^K) \nn \\ && \times \exp\left(
 i\sum\nolimits_{k=1}^K\sum\nolimits_{a=1}^n x_a^ky_a^k - \case{1}{2}
 \sum\nolimits_{k,l=1}^K\sum\nolimits_{a,b=1}^n x_a^k
 q_{ab}^{kl}x_b^l\right) ,
\label{gen-parisi-multi-d}
\end{eqnarray} 
where the order parameter matrix has now extra indices
\begin{equation} 
 [{\cal Q}]_{ab}^{kl} = q_{ab}^{kl}.
\label{Q-multi-d}
\end{equation} 
Such a situation occurs, for instance, in the treatment of
thermodynamical states in vector spin glasses, or, of the metastable
states in the SK model.  When counting the stationary states of the
Thouless-Anderson-Palmer equations, Bray and Moore \cite{bm81}
encountered Eq.\ (\ref{gen-parisi-multi-d}) with $K=2$ and a special
$\varPhi$.  They displayed the corresponding PPDE but did not pursue the
matter further.  Since Eq.\ (\ref{gen-parisi-multi-d}) is a
straightforward generalization of the Parisi term, we briefly give the
way how to evaluate it.  Also, we concisely formulate the calculation
of replica correlators.

The assumption of the Parisi structure for all individual submatrices
of $\cal Q$ with fixed $k,l$ can be cast into the form 
\begin{equation} 
  \label{pa-multi-d} {\cal Q} = \sum_{r=0}^{R+1} \left( {\sf Q}_r- {\sf
  Q}_{r-1}\right) {\sf U}_{m_r}
\otimes  {\sf I}_{n/m_r}.
\end{equation} 
Here 
\begin{equation} 
 [{\sf Q}_{r(a,b)}]^{kl} = q_{r(a,b)}^{kl} = q_{ab}^{kl}
\label{Q-r-multi-d} 
\end{equation} 
is the symmetric $K\times K$ matrix analog of (\ref{merger}).  The
quadratic form in the exponent in (\ref{gen-parisi-multi-d}) is now
\begin{equation} 
  \label{quad-multi-d} \sum_{r=0}^{R+1} \sum_{k,l=1}^K \left(
  q_r^{kl}-q_{r-1}^{kl}\right) \sum_{j_r=1}^{n/m_r}
  \sum_{a=m_r(j_r -1) +1}^{j_r m_r} x_a^k \sum_{b=m_r(j_r
  -1)+1}^{j_r m_r} x_b^l,
\end{equation}
with $q^{kl}_{-1}=0$. Let us diagonalize the difference between
subsequent ${\sf Q}_r$-s as
\begin{equation} 
 {\sf Q}_r-{\sf Q}_{r-1} = {\sf O}_r^T {\sf \Lambda}_r {\sf O}_r,
\label{L-multi-d}, ~~~~~~r=0,\dots ,R+1, 
\end{equation} 
where the orthogonal $K\times K$ matrix ${\sf O}_r$ is made up by
column eigenvectors of ${\sf Q}_r-{\sf Q}_{r-1}$ and ${\sf
\Lambda}_r$ is diagonal and has the real eigenvalues as diagonal
elements.  A derivation similar to that given in Section
\ref{ssec-finite-rsb} and Appendix \ref{app-sec-deriv-r-rsb} yields the
$R$-RSB term
\begin{eqnarray}
 \label{phirrsb-multi-d} \varphi\left.[\varPhi(\yv ),\{q^{kl}_r\},\xv]
 \right| _{n=0} &=& \frac{1}{x_1}\int\rD ^Kz_0 \ln \int\rD ^Kz_1 \Bigg
 [ \int\rD ^Kz_2 \dots \nn \\ && \times \,\Bigg [ \int\rD ^Kz_{R+1}\exp
 \varPhi\left(\sum_{r=0}^{R+1} \zv _r {\sf \Lambda}_r^{\hf} {\sf O}_r
 \right)\Bigg ] ^\frac{x_R}{x_{R+1}} \dots \Bigg
 ]^\frac{x_{1}}{x_{2}}.
\end{eqnarray}
Here ${\sf \Lambda}_r^{\hf}$ has the square root of the eigenvalues
(possibly also imaginary numbers, the sign being irrelevant) as
diagonal elements, $D^Kz$ denotes the $K$-dimensional Gaussian
integration measure, and $\zv _r$ is a $K$-dimensional vector. The
function $\varPhi(y^1,\dots,y^K)$ is naturally abbreviated by $\varPhi({\bf
y})$. The recursion
\bml
  \label{rec-psi-multi-d}
\begin{eqnarray}
  \label{rec-psi1-multi-d} \psi_{r-1}(\yv ) & = & \int\rD ^Kz~
  \psi_r\left(\yv +\zv {\sf \Lambda}_r^{\hf} {\sf O}_r\right) ^
  \frac{x_r}{x_{r+1}}, \\ \label{rec-psi2-multi-d} \psi_{R+1}(\yv ) &=&
  e^{\varPhi(\yvs )}
\end{eqnarray}
\eml
evaluates (\ref{phirrsb-multi-d}) as 
\begin{equation}
  \label{ferecg-multi-d} \left. \varphi [\varPhi(\yv ),\{q^{kl}_r\}
 ,\xv ]\right| _{n=0} = \frac{1}{x_1} \int\rD ^Kz~ \ln
 \psi_0\left(\zv {\sf \Lambda}_0^{\hf} {\sf O}_0\right).
\end{equation}

In order to produce a PDE we need to specify a time-like variable.
For practical purposes we consider the case when one diagonal element
is a known constant, say $q^{11}_{R+1}=1$.  Then we pick $q^{11}_r$ as
time variable, call its continuation $q$, and obtain the PDE for the
field $\psi(q,\yv )$ in $K$ spatial dimensions as
\bml
 \label{psi-pde-multi-d} \begin{eqnarray} \label{psi-pde1-multi-d}
 \ptl _q \psi & = & -\hf \nabla_y \dot{\sf Q} \nabla_y \psi +
 \frac{\dot x}{x} \psi \ln \psi \\ \label{psi-pde2-multi-d}
 \psi(1,\yv ) & = & e^{\varPhi(\yvs )}.
\end{eqnarray}
\eml
Here the dot means derivative in terms of $q$, of course $[\dot{\sf
Q}]^{11}=1$, and $q$ evolves from $1$ to $0$.  As in the case with one
spatial dimension, in the $q$-intervals $(q_{(1)},1)$ and
$(0,q_{(0)})$ we have $x(q)\equiv 1$ and $x(q)\equiv 0$, resp., where
$q_{(1)}=q_R$ and $q_{(0)}=q_0$.  Again, by introducing
\begin{equation}
  \label{phi-multi-d} \varphi (q,\yv ) = \frac{\ln\psi(q,\yv)}{x(q)}
\end{equation}
we obtain the $K$-dimensional PPDE as
\bml
 \label{ppde-multi-d} \begin{eqnarray} \label{ppde1-multi-d} \ptl _q
 \varphi & = & -\hf \nabla_y \dot{\sf Q} \nabla_y \varphi - \hf x
 (\nabla_y \varphi) \dot{\sf Q} \nabla_y \varphi\\
 \label{ppde2-multi-d} \varphi(1,\yv ) & = & \varPhi(\yv ).
\end{eqnarray}
\eml
Then the sought term is 
\begin{equation}
  \label{fe-multi-d} \left. \varphi [\varPhi(\yv ),\{q^{kl}_r\}
 ,\xv ]\right| _{n=0} = \varphi (0,{\bf 0}).
\end{equation}
The evolution in the interval  $(q_{(1)},1)$ can be solved explicitly
to give 
\begin{eqnarray}
  \label{phiR-multi-d} \varphi(q_{(1)},\yv ) &=& \int\rD ^Kz~
  \exp\varPhi\left(\yv +\zv {\sf \Lambda}_{R+1}^{\hf} {\sf
  O}_{R+1}\right) \nn \\ &=& \int \frac{\rd^Kv\, \rd^Kw}{(2\pi)^K}~
  \text{exp}\left[\varPhi(\vv ) + i\wv (\vv -\yv ) - \hf \wv ({\sf
  Q}_{R+1} - {\sf Q}_R) \wv \right],
\end{eqnarray}
that is the initial condition for further evolution in $(0,q_{(1)})$.
From the mathematical viewpoint, the problem of existence of the above
expression needs to be clarified for the specific $\varPhi$ in play.  It
typically occurs that a diagonal element of ${\sf Q}_{R+1}$ is known
to vanish, but for other $r$-s the same diagonal is positive.  In
general, ${\sf Q}_{R+1} - {\sf Q}_R$ is not necessarily a positive
definite matrix.  However, given the fact that Eq.\
(\ref{phiR-multi-d}) at $\yv ={\bf 0}$ is the RS free energy (where
$q_{(1)}$ is replaced by the RS value of $q$), on physical grounds we
surmise that the divergence of the integral is a rare threat.

In the present case there are $\hf K(K+1)$ OPF-s, namely, $x(q)$ and
$q^{kl}(q)$, $(k,l)\neq (1,1)$ and $k,l\leq K$.

Expectation values 
\begin{equation}
  \label{exp-multi-d}
          \lav A (\{x_a^k\}, \{y_a^k\}) \rav   
\end{equation}
we conveniently define by inserting the function $A$ in the integrand
of (\ref{gen-parisi-multi-d}) and omitting the $1/n\ln$ from in front
of the formula. The $n\to 0$ limit is understood.  As in one spatial
dimension, the GF ${\cal G}_\varphi(q_1,\yv _1;q_2,\yv _2)$ for
the multidimensional PPDE is a key help in calculating averages of
common occurrence. The GF is zero for $q_1>q_2$ and satisfies the PDE
\begin{equation}\label{gf-multi-d} \ptl _{q_1}
 {\cal G}_\varphi = -\hf \nabla_{y_1} \dot{\sf Q} \nabla_{y_1} {\cal
 G}_\varphi - \hf x(q_1)\, (\nabla_{y_1} \varphi(q_1,\yv _1)) \dot{\sf Q}
 \nabla_{y_1} {\cal G}_\varphi - \delta(q_1-q_2)\delta^K(\yv _1-\yv _2).
\end{equation}
Special significance is attached to
\begin{equation}\label{p-multi-d}
 P(q,\yv )={\cal G}_\varphi(0,{\bf 0};q,\yv ),
\end{equation}
a natural generalization of the $K=1$ field. Let us introduce  the
derivative fields
\bml
\label{deriv-fields-multi-d}
\begin{eqnarray}\label{mu-multi-d}
 \mu^k(q,\yv ) &=& \ptl _{y^k} \varphi(q,\yv ), \\
 \label{kappa-multi-d} \kappa^{kl}(q,\yv ) &=& \ptl _{y^k} \ptl
 _{y^l} \varphi(q,\yv ).
\end{eqnarray}
\eml
Then we can write the two-replica-correlator
\begin{equation}\label{two-corr-def-multi-d}
  \left.\frac{\ptl\varphi[\varPhi(y),{\cal Q}]}{\ptl
  q_{ab}^{kl}}\right|_{n=0} = - \lav x_a^k x_b^l \rav
   \equiv {C_x^{(2)}}^{kl}(q_{r(a,b)})
\end{equation}
as 
\begin{equation}\label{two-corr-multi-d}
  {C_x^{(2)}}^{kl}(q) = \int \rd^Ky P(q,\yv )\left[\mu^k(q,\yv )
  \mu^l(q,\yv ) + \theta(q-1^{-0}) \kappa^{kl}(q,\yv )\right].
\end{equation}
By use of this formula the stationarity conditions for a free energy
that contains a term like (\ref{gen-parisi-multi-d}) can be
immediately constructed.


\section{An identity between Green functions} 
\label{app-sec-GF-ident}

In this Appendix we prove the identity (\ref{gf-diff-vertex}).
The r.\ h.\ s.\ of
\begin{eqnarray}
  \ptl _q \Gamma_{\varphi\varphi\varphi} \left(
   q;\{q_i,y_i\}_{i=1}^3 \right) & = & \int\rd y\, \Big( \left[\ptl
   _q{\cal G}_{\varphi}(q_1,y_1;q,y)\right] {\cal
   G}_{\varphi}(q,y;q_2,y_2) \, {\cal G}_{\varphi}(q,y;q_3,y_3)
   \nn \\& & \; + {\cal G}_{\varphi}(q_1,y_1;q,y)
   \left[\ptl _q{\cal G}_{\varphi}(q,y;q_2,y_2) \right] \, {\cal
   G}_{\varphi}(q,y;q_3,y_3) \nn \\ & & \; + {\cal
   G}_{\varphi}(q_1,y_1;q,y)\, {\cal
   G}_{\varphi}(q,y;q_2,y_2) \, \left[ \ptl _q{\cal
   G}_{\varphi}(q,y;q_3,y_3)\right] \Bigr) 
\label{diff-Gamma}
\end{eqnarray}
can be expressed by our making use of the PDE-s for the participating
GF-s.  From  (\ref{gf-pde-backw}) we have
\begin{equation}
\ptl _q {\cal G}_{\varphi}(q_1,y_1;q,y) = \case{1}{2} \ptl _y^2 
{\cal G}_{\varphi}(q_1,y_1;q,y) - x(q) \ptl _y \left[ \mu(q,y)\, {\cal
  G}_{\varphi}(q_1,y_1;q,y) \right] + \delta(q-q_1) \delta(y-y_1),
\label{GF-pde1}  
\end{equation}
and for $i=2,3$ (\ref{gf-pde-forw}) holds as
\begin{equation}
\ptl _q {\cal G}_{\varphi}(q,y;q_i,y_i) = - \case{1}{2} \ptl _y^2 
{\cal G}_{\varphi}(q,y;q_i,y_i) - x(q)\,\mu(q,y)\,\ptl _y 
{\cal G}_{\varphi}(q,y;q_i,y_i) - \delta(q-q_i) \delta(y-y_i).
\label{GF-pde2}  
\end{equation}
Let us substitute the r.\ h.\ sides of the above PDE-s into
(\ref{diff-Gamma}). The sum of the terms linear in $x(q)$ turns out to
be a derivative by $y$, so -- under the plausible condition that the
GF-s decay for large $|y|$ -- integration by $y$ gives zero.  The
second derivatives in $y$ also cancel after partial integration but
for a remnant that yields
\begin{eqnarray}
  \ptl _q \Gamma_{\varphi\varphi\varphi} \left(
   q;\{q_i,y_i\}_{i=1}^3 \right) = && \int\rd y\, {\cal
   G}_{\varphi}(q_1,y_1;q,y) \left[\ptl _{y} {\cal
   G}_{\varphi}(q,y;q_2,y_2) \right] \,\left[\ptl _{y} {\cal
   G}_{\varphi}(q,y;q_3,y_3)\right] \nn \\ && + \delta(q-q_1) {\cal
   G}_{\varphi}(q_1,y_1;q_2,y_2) {\cal G}_{\varphi}(q_1,y_1;q_3,y_3)
   \nn \\ && - \delta(q-q_2) {\cal G}_{\varphi}(q_1,y_1;q_2,y_2) {\cal
   G}_{\varphi}(q_2,y_2;q_3,y_3) \nn \\ && - \delta(q-q_3) {\cal
   G}_{\varphi}(q_1,y_1;q_3,y_3) {\cal G}_{\varphi}(q_3,y_3;q_2,y_2)
   \label{diff-Gamma-before-final}
\end{eqnarray}
Eq.\ (\ref{gf-theta-eta}) relates derivatives of $ {\cal G}_{\varphi}$
and $ {\cal G}_{\mu}$, whence we obtain (\ref{gf-diff-vertex}) for
$q_1<q<q_2$ and $q_1<q<q_3$.


\section{PDE-s for high temperature} 
\label{app-sec-highT-pde}

Here we record the calculation leading to the lowest order nontrivial
correction for the distribution of local stabilities at high
temperatures.  Assuming $P(q,y) = P_0(q,y) + \beta\, P_1(q,y) +
O(\beta^2)$ and expanding the SPDE we obtain
\bml
\label{hT-spde}
\begin{eqnarray}
         \ptl _q P_0  & = & \hf \ptl _y ^2 P_0, \;\; 
        P_0(0,y) = \delta(y),
  \label{hT-sdpe0} \\
        \ptl _q P_1  & = & \hf \ptl _y ^2 P_1  + x \ptl _y (P_0 m_0), \;\;
        P_1(0,y)\equiv 0.   
\label{hT-spde1}
\end{eqnarray}
\eml 
Here $m_0(q,y)$ is the lowest order approximation for the field
$m(q,y)$ in (\ref{intro-m}), thus it satisfies (\ref{deriv-ppde-ener})
with $\beta=0$\ie evolves according to pure diffusion.  Using its
initial condition $m_0(1,y)=V^{\prime}(y)$ we get
\begin{equation}
        m_0(q,y) = \int\rd y_1\, G(y-y_1,1-q)\, V^{\prime}(y_1).
\label{hT-m}
\end{equation}
The zeroth order probability field is obviously
\begin{equation}
        P_0(q,y) = G(y,q),
\label{hT-P0}
\end{equation}
while the next correction can be obtained from (\ref{hT-spde1}) using
the Gaussian GF for pure diffusion.  This gives (in case of ambiguity
$q^{-0}$ should be understood in the upper limit of integration)
\begin{eqnarray}
  P_1(q,y)  & = & \int_0^q\rd q_1 \int\rd y_1\, G(y-y_1,q-q_1)\, x(q_1) 
  \, \ptl _{y_1} G(y_1,q_1)\, m_0(q_1,y_1) \nn \\
  & = & - \int_0^q\rd q_1 \int\rd y_1\, (\ptl _{y_1} G(y-y_1,q-q_1))\, x(q_1)
  \, G(y_1,q_1)\, m_0(q_1,y_1) \nn \\
 & = &   \ptl _{y} \int_0^q\rd q_1\, x(q_1)\int\rd y_1\rd y_2\, V^\prime (y_2)\,
 G(y_1,q_1)\, G(y-y_1,q-q_1)\, G(y_2-y_1,1-q_1).
\label{hT-P1}
\end{eqnarray}
In the last Eq.\ the formula (\ref{hT-m}) for $m_0$ was also
substituted.    We note the elementary identity
\begin{equation}
\label{3-Gauss-ident}
\int\rd y \prod_{i=1}^3 G(y-y_i,q_i) =  
\frac{1}{2\pi \sqrt{\sigma}} \exp\left( - \frac{A}{2\sigma} \right), 
\end{equation}
where
\bml 
\label{3-Gauss-ident-more} 
\begin{eqnarray}
A & = & y_1^2(q_2+q_3)+ y_2^2(q_1+q_3) + y_3^2(q_1+q_2) - 2y_1y_2q_3 -
2 y_2y_3q_1 - 2 y_3y_1q_2,
\label{3-Gauss-ident2} \\
\sigma & = & q_1q_2 +  q_2q_3 + q_3q_1.    
\label{3-Gauss-ident3}
\end{eqnarray}
\eml
Hence, at $q=1$, we obtain 
\begin{equation}
 P_1(1,y) = \ptl _{y} \int_0^1\rd q\, x(q) \int\rd y_1\, V^{\prime} (y_1) 
 \frac{1}{2\pi\sqrt{1-q^2}}\, \exp \left(- \frac{y^2+y_1^2 - 2yy_1q}
  {2(1-q^2)}\right). 
\label{hT-P1-2}
\end{equation}
This is identical to the function $\rho_1(y)$ given in Eq.\
(\ref{hT-rho1}).


\section{Longitudinal stability for high temperatures} 
\label{app-stab-high-T} 

Below we show that the linear operator displayed in (\ref{hT-long})
has all negative eigenvalues on the space of smooth functions
$\zeta(q)$ with $\zeta(0)=\zeta(1)=0$.  Consider the eigenvalue
problem,
\begin{equation}
  \label{hT-long-ev1}
  \int_0^1\rd q_2\, \zeta(q_2) \int_0^{\min(q_1,q_2)}
   \frac{\rd\bar{q}}{D(\bar{q})^3}  
   = \lambda\,  \zeta(q_1),
\end{equation}
where we omitted the factor $-1$ on the \rhs of (\ref{hT-long}), so
the positivity of the $\lambda$-s is to be proven.  The \lhs separates
as 
\begin{equation}
  \label{hT-long-ev2}
\int_0^{q_1}\rd q_2\, \zeta(q_2) \int_0^{q_2}
\frac{\rd\bar{q}}{D(\bar{q})^3} + \int_{q_1}^1\rd q_2\, \zeta(q_2)
\int_0^{q_1} \frac{\rd\bar{q}}{D(\bar{q})^3}.
\end{equation}
The first term is equivalently
\begin{equation}
  \label{hT-long-ev3}
\int_0^{q_1} \frac{\rd\bar{q}}{D(\bar{q})^3} \int_{\bar{q}}^{q_1}\rd
q_2\, \zeta(q_2), 
\end{equation}
which concatenates with the second term in (\ref{hT-long-ev2}) to
\begin{equation}
  \label{hT-long-ev4}
\int_0^{q_1} \frac{\rd\bar{q}}{D(\bar{q})^3} \int_{\bar{q}}^{1}\rd q_2\,
\zeta(q_2). 
\end{equation}
Introducing
\begin{equation}
  \label{hT-long-ev5}
\xi(q) = \int_{q}^{1}\rd q_2\, \zeta(q_2) ,
\end{equation}
we obtain after differentiation of the eigenvalue problem
(\ref{hT-long-ev1}) the equivalent form
\begin{equation}
  \label{hT-long-ev6}
\frac{\xi(q)}{D(q)^3} = - \lambda\, \xi^{\pp}(q).
\end{equation}
This equation may have a solution that vanish at the boundaries only
if $\lambda>0$.  Indeed, one can try to solve (\ref{hT-long-ev6}) by
the ``shooting method'' starting from $\xi(0)=0$ and attempting to
reach $\xi(1)=0$.  Then the sign of the curvature of $\xi(q)$ may not
be the same as the sign of $\xi(q)$ within the whole interval $(0,1)$,
or else $\xi(1)=0$ will never be reached.  Since $D(q)>0$ for $q<1$,
this implies $\lambda>0$. Thus we have demonstrated that the Hessian
(\ref{hT-long}) is negative definite.


\eject

\end{document}